\documentclass[12pt]{article}

\usepackage{epsfig,multicol,multirow}
\usepackage{amsmath,subfigure,latexsym,amssymb,fps}

\def\lsi{\raise0.3ex\hbox{$<$\kern-0.75em\raise-1.1ex\hbox{$\sim$}}}
\def\gsi{\raise0.3ex\hbox{$>$\kern-0.75em\raise-1.1ex\hbox{$\sim$}}}
\def\backder{\raise1.4ex\hbox{$\leftarrow$\kern-0.75em\raise-1.4ex\hbox{$\partial$}}}
\def\forder{\raise1.4ex\hbox{$\rightarrow$\kern-0.75em\raise-1.4ex\hbox{$\partial$}}}

\newcommand{\excleq}{\mathop{\stackrel{!}{=}}}

\newcommand{\backderi}{\mathop{\backder}}

\newcommand{\lsim}{\mathop{\lsi}}
\newcommand{\gsim}{\mathop{\gsi}}

\newcommand{\be}{\begin{equation}}
\newcommand{\ee}{\end{equation}}
\newcommand{\nn}{\nonumber}
\newcommand{\bea}{\begin{eqnarray}}
\newcommand{\eea}{\end{eqnarray}} 
\newcommand{\eps}{\epsilon}
\newcommand{\la}{\langle}
\newcommand{\ra}{\rangle}
\newcommand{\uno}{1 \!\! 1}
\newcommand{\Z}{Z \!\!\! Z}
\newcommand{\R}{{\kern+.25em\sf{R}\kern-.78em\sf{I} \kern+.78em\kern-.25em}}
\newcommand{\RR}{{\kern+.25em\sf{R}\kern-.6em\sf{I} \kern+.6em\kern-.25em}}
\newcommand{\N}{{\kern+.25em\sf{N}\kern-.78em\sf{I} \kern+.78em\kern-.25em}}
\newcommand{\C}{{\kern+.25em\sf{C}\kern-.45em\sf{I} \kern+.45em\kern-.25em}}

\newcommand{\tgh}{{\rm tgh}}

\makeatletter
\@addtoreset{equation}{section}
\makeatother

\begin{document}

\begin{flushright}
HU-EP-06/41 \\
DESY-06-208
\end{flushright}
\vspace*{4mm}

\begin{center}

{\Large\bf
Optimised Dirac Operators on the Lattice:} \\
\vspace*{6mm}
{\Large\bf 
Construction, Properties and Applications}

\vspace*{1.4cm}

Wolfgang Bietenholz \\

\vspace*{1cm}

Institut f\"{u}r Physik \\

\vspace*{1mm}

Humboldt-Universit\"{a}t zu Berlin \\

\vspace*{1mm}

Newtonstr.\ 15, D-12489 Berlin, Germany \\

\vspace*{5mm}

and \\

\vspace*{5mm}

John von Neumann-Institut f\"{u}r Computing (NIC) \\

\vspace*{1mm}

Deutsches Elektronen-Synchrotron (DESY) \\

\vspace*{1mm}

Platanenallee 6, D-15738 Zeuthen, Germany \\

\end{center}

\vspace*{9mm}


\noindent
We review a number of topics related to block
variable renormalisation group transformations of quantum fields on the
lattice, and to the emer\-ging perfect lattice actions. We first illustrate
this procedure by considering scalar fields. Then we proceed to lattice 
fermions, where we discuss perfect actions for free fields,
for the Gross-Neveu model 
and for a supersymmetric spin model. We also consider the extension to perfect
lattice perturbation theory, in particular regarding the axial anomaly
and the quark gluon vertex function. Next we deal with properties and
applications of truncated perfect fermions, and their chiral correction
by means of the overlap formula. This yields a formulation of lattice
fermions, which combines exact chiral symmetry with an optimisation of
further essential properties. We summarise simulation results for these
so-called overlap-hypercube fermions in the two-flavour Schwinger model
and in quenched QCD. In the latter framework we establish a link to Chiral
Perturbation Theory, both, in the $p$-regime and in the $\epsilon$-regime.
In particular we present an evaluation of the leading Low Energy Constants
of the chiral Lagrangian --- the chiral condensate and the pion decay
constant --- from QCD simulations with extremely light quarks.

\newpage

\begin{center}

{\Large\bf Motivation and Overview}

\end{center}

\vspace*{4mm}

Over the recent decades quantum field theory has been established as the
appropriate formalism for particle physics, as far as it is 
explored experimentally.
Its treatment by perturbation
theory led to successful results, for instance in Quantum Electrodynamics 
(QED), in the electroweak sector of the Standard Model
and in Quantum Chromodyna\-mics (QCD) at high energy.
However, there are still many open questions,
which require results at finite coupling strength --- beyond the range
of perturbation theory --- such as numerous aspects of QCD at low 
and moderate energy.

A method is known which has the potential to provide
fully non-perturbative results for a number of field theoretic
questions. This me\-thod applies Monte Carlo simulations to lattice
regularised quantum field theories. 
The generic uncertainty of perturbation theory
--- uncontrolled contributions beyond the calculated order
--- disappears in this approach. 
However, one has to deal with statistical errors, as well
as ambiguities in the extrapolation to the continuum
and to a large volume.

Simulation results are obtained at finite lattice spacing, 
which causes systematic artifacts in the numerically measured
observables. The stability of dimensionless ratios of observables
under the variation of the lattice spacing is denoted as the 
scaling behaviour. Its quality, which is vital for the reliability 
of the continuum extrapolation, depends on the way in which the 
lattice regularisation is implemented. This work deals with
renormalisation group techniques to improve the scaling behaviour
compared to the standard lattice formulations, 
which describes, for instance, derivatives simply by
differences between nearest neighbour lattice sites.
In contrast to Symanzik's program, this technique
does not attempt to correct a specific order in the lattice spacing,
but it directly addresses a finite cutoff.
We describe the renormalisation group approach in detail and
present a variety of results that it led to, in particular
for fermionic systems.

The symmetries of a model under consideration are a key aspect
for a controlled continuum extrapolation. A great virtue of the lattice
regu\-larisation is the conservation of exact gauge symmetries.
But global symmetries are often explicitly broken by the
lattice structure, for instance the continuous Poincar\'{e}
invariance. The question, how well --- and if --- they are restored
as we approach the continuum limit is a notorious issue, which is
related to the scaling behaviour. Again it depends on the
features of the lattice formulation, i.e.\ on the extent of
the explicit symmetry breaking due to a finite lattice spacing.
The renormalisation group technique provides a tool to improve
the symmetries on the regu\-larised level --- in principle they
can even be implemented exactly --- which renders the continuum
limit smoother and safer. This property is particularly relevant
for the (approximate) chiral symmetry of (almost) massless fermions.
The chiral symmetry is essential for instance in QCD at low energy,
and its discussion will take a central r\^{o}le in this work.
Here we also need a chiral extrapolation, in addition
to the limits that we mentioned already.\\

As an introduction, we summarise in Section 1 some basic
aspects of quantum field theory in the functional integral formulation.
In particular we sketch the road from classical mechanics to Euclidean
quantum field theory, with quantum mechanics and classical
field theory as intermediate steps.
We focus on the lattice regularisation, which we first introduce for
the case of scalar fields. 
This allows us to summarise the notions and notations used throughout
this work.

In Section 2 we describe --- still for scalar fields --- the concept
of block variable renormalisation group transformations. Under iteration
they lead to a perfect lattice action, which is free of any cutoff 
artifacts. We also encounter approximations
to a perfect action, which are needed for practical purposes, such as
the classical perfection and the truncation of the couplings.
Still the lattice artifacts can be kept small for such approximations,
as we illustrate for the dispersion relation, the topological
susceptibility of a quantum rotor and for thermodynamic quantities.

In Section 3 we proceed to fermionic quantum field theories, where
we start again with a few generalities.
We review the traditional formulations of
lattice fermions and describe the doubling problem.
It is related to the difficulty to keep track of the chiral symmetry
in a regularised system --- an obstacle, which obstructs other
regularisation schemes as well.
Therefore the existence of light quarks is an amazing feature of
Nature, which cannot be described easily in a natural way.
In that context, we discuss a brane world scenario as a possible 
solution to this hierarchy problem.

Section 4 applies the concept of perfect actions to lattice fermions,
which takes us to the main topic of this work. For free fermions,
we demonstrate that this approach provides both, a solution to the doubling
problem and at the same time an exact scaling behaviour. 
Depending on the choice of the
renormalisation group transformation, we can implement either locality 
or standard chirality in the perfect lattice action.
However, if we insist on locality, the resulting Dirac operator
still obeys the Ginsparg-Wilson
relation, which guarantees a lattice modified but exact chiral symmetry.

In principle, exact scaling and chirality can also be realised
at finite lattice spacing
in the interacting case, where, however, perfect actions can in general 
not be constructed explicitly. An exception is the Gross-Neveu model in
the limit of a large number of flavours. Here we present a perfect action
for staggered fermions, and we approve the perfect scaling for the ratio 
of the chiral condensate and the dynamically generated fermion mass.
The concept we are dealing with also reaches out to
perfect lattice currents.
With that ingredient, perfect actions can even capture 
exact supersymmetry on the lattice.

In Section 5 we consider perfect lattice perturbation theory. We give 
results for the anharmonic oscillator and the Yukawa term, which couples
fermions to a scalar field. In lattice gauge theory we show that
the perfect lattice action retrieves correctly the axial anomaly, and
we discuss the quark gluon vertex function in QCD. 

For practical applications, i.e.\ for the applicability in simulations,
the couplings have to be truncated.
In Section 6 we describe our truncation scheme for the perfect fermion to
a so-called hypercube fermion, which has been simulated successfully
in the Schwinger model. In QCD it has been used for the spectral
functions at finite temperature, and --- together with
a truncated perfect vertex function --- in the evaluation of the 
charmonium spectrum. For truncated perfect fermions,
the scaling behaviour and chirality
are not exact anymore, but the latter can be corrected again by
inserting the hypercube fermion into the overlap formula.

This procedure yields the ``overlap hypercube fermion'', which is
an exact solution to the Ginsparg-Wilson relation. 
Its construction and properties are presented in Section 7. 
Similarly we can arrange for a modi\-fied but exact parity symmetry for
lattice fermions in three dimensions. In two dimensions we review
simulations results for overlap hypercube 
fermions in the two-flavour Schwinger model, which reveal an excellent
scaling behaviour. Here and also in QCD we further observe a strongly
improved level of locality and approximate rotation symmetry
compared to the standard overlap fermion.

Section 8 finally presents simulation results with Ginsparg-Wilson 
fermions in QCD, using the overlap hypercube fermion as well as 
the standard formulation of overlap fermions, both in the quenched 
approximation. This enables simulations near the chiral limit. Here
our main goal is a connection to Chiral Perturbation Theory. 
This is an effective theory of strong interactions at low energy, which 
provides a variety of successful predictions.
However, its effective Lagrangian involves free parameters denoted as the Low 
Energy Constants, which play an important r\^{o}le in the physics of light
mesons. Their theoretical determination can only emerge from
QCD as the fundamental theory. 
This is a challenge for lattice simulations, and the
principal issue of Section 8.

We measured light meson masses in the $p$-regime (characterised by
a large volume), and we
reveal the difficulties to evaluate Low Energy Constants in that setting.
Then we focus our interest on the $\epsilon$-regime, which deals with a 
small volume. 
In the $\epsilon$-regime, the topological sectors play an extraordinary r\^{o}le.
Hence we first give results for the distribution of topological charges
and the resulting susceptibility, which is relevant for the mass
of the $\eta'$ meson.
Next we describe a 3-loop calculation which confirms the perturbative
renormalisability of the $\epsilon$-expanded effective theory.
We then apply various techniques to 
extract the leading Low Energy Constants: the chiral condensate 
--- which is the order parameter of chiral symmetry breaking ---
and the phenomenologically known
pion decay constant. In particular, the density of low lying eigenvalues
of the Dirac operator is fitted to predictions by chiral Random Matrix
Theory. The axial-vector current correlator, as well as the zero-mode
contributions to the pseudoscalar density correlation, are confronted with
formulae 
of quenched Chiral Perturbation Theory.
We will see that these methods do have the potential to evaluate the
Low Energy Constants with their phenomenological values --- which correspond
to the large volume limit --- even in the $\epsilon$-regime. However, the
final results have to await the feasibility of dynamical QCD
simulations with chiral quarks.

Section 9 is dedicated to concluding remarks, summarising
the status  of the fields of research that we addressed,
along with an outlook on future perspectives.

\newpage

\tableofcontents

\newpage

\section{Introduction}

\subsection{From classical mechanics to quantum mechanics}

In {\em classical mechanics}, the trajectory $\vec x (t)$ of a point
particle between fixed endpoints $\vec x (0)$ and $\vec x (T)$ is ---
in simple situations --- determined by the principle of least action,
which imposes the condition $\delta S =0$. The action $S[x]$ is a
functional of the conceivable particle paths $\vec x (t)$,
\be
S [x] = \int _{0}^{T} dt \, L ( \vec x , \dot {\vec x}) \ ,
\ee
where $L$ is the Lagrange function. A simple form of it reads
\be
L ( \vec x , \dot {\vec x}) = \frac{m}{2} \dot {\vec x}(t)^{\, 2} 
- V(\vec x (t)) \ ,
\ee
with the particle mass $m$ and a potential $V$ (which we assume to be
velocity independent). The variational condition $\delta S =0$
corresponds to Newton's equation of motion, 
$m \ddot {\vec x} = -\nabla V$, at each instant $t \in [0,T]$. \\

Let us consider this transition in {\em quantum mechanics.}
In contrast to classical mechanics, we now deal with a transition amplitude,
which picks up contributions from all possible paths connecting the fixed
endpoints. Hence the path in between is not determined.
These contributions are summed up coherently,
\be  \label{transamp}
\langle \vec x (T) \vert \vec x (0) \rangle = 
\int {\cal D}x \, \exp \Big( \frac{i}{\hbar} S [x] \Big) \ .
\ee
This expression represents a path integral (or functional integral),
where the functional measure ${\cal D}x$ symbolises the summation
over all possible paths (which formally requires an infinite dimensional
integral) \cite{PIQM}. 

In the (hypothetical) limit $\hbar \to 0$ solely the classical path 
(which we assume to be unique) contributes,
whereas the additional contributions for $\hbar > 0$ correspond to the
quantum effects. However, if a path far from the classical one is varied,
the phase in eq.\ (\ref{transamp}) tends to rotate rapidly, so that
such contributions almost cancel. As long as $\hbar$ is small 
compared to the action shift caused by path variations on the scale of interest,
it is the vicinity of the classical path
that dominates the transition amplitude (\ref{transamp}).

This situation has a historically older counterpart in optics, where the 
classical and the quantum mechanical description correspond to the principles
by Fermat and by Huygens, respectively.

In order to attribute an explicit meaning to the functional measure
${\cal D}x$, we divide the period $T$ into $N$ equidistant intervals
of length $a = T/N$. In this discretised system, the path integral
is given by $N-1$ integrals over the possible positions at the times
$t_{j}= j \cdot a$, $j = 1 \dots N-1$. The expression (\ref{transamp})
is then understood as the continuum extrapolation $a \to 0$
(which, at fixed $T$, corresponds to $N\to \infty$),
\be
\prod_{j=1}^{N-1} \int_{{\RR}^{3}} d^{3}x_{j} \ \dots \quad
 \ ^{ a \to 0}_{\longrightarrow} \qquad \int {\cal D} x \ \dots
\ee

\subsection{Classical field theory}

In {\em field theory} we do not consider particle paths $\vec x (t)$,
but instead fields $\phi( x)$, where $x = (t , \vec x )$ 
is a point in space-time. The (classical) field takes its value in some
abstract space, like $ \R^{n}$ or $\C^{n}$, for example.
Now space and time are treated on an equal footing
(up to the signature in the metrics), which is a prerequisite for
covariance. Moreover, the number of degrees of freedom is extended
drastically: before there were just three of them (in each time point $t$),
but now there is a degree of freedom for each field component
in each single space-time point $x$.

We assume in each point $x$ a Lagrange density 
${\cal L} (\phi , \partial_{\mu} \phi )$ to be defined
($\mu = 0, \dots ,3$), which we denote as the {\em Lagrangian.}
The field theoretic action is given by
\be
S [ \phi ] = \int d^{4}x \, {\cal L} (\phi , \partial_{\mu} \phi ) \ .
\ee
Now an action value is obtained for each field configuration
$\phi (x)$. This means that $S$ is a functional of the fields involved, 
which take the r\^{o}le of the paths in the mechanical system.

The simplest case is a neutral scalar field $\phi(x) \in \R$.
If this field describes free scalar particles of mass $m$,
its Lagrangian reads\footnote{For convenience, 
we set the speed of light to $c=1$.} 
\be   \label{freescallag}
{\cal L} (\phi , \partial_{\mu} \phi ) = \frac{1}{2} \partial_{\mu} \phi (x)
\partial^{\mu} \phi (x) - \frac{m^{2}}{2} \phi(x)^{2} \ .
\ee
Assembling the Lagrangian ${\cal L}$ only by
covariant terms --- as it is the case in eq.\ (\ref{freescallag})
--- ensures that we are dealing with relativistic field theories.

In {\em classical field theory} the configuration is determined by again
enforcing the variational condition $\delta S = 0$. For a neutral
scalar field, this implies
\be
\frac{\partial {\cal L}}{\partial \phi} - 
\partial_{\mu} \frac{\partial {\cal L}}{\partial (\partial_{\mu}\phi)} = 0 \ ,
\ee
which translates for the Lagrangian (\ref{freescallag})
into the Klein-Gordon equation of motion for the scalar field,
\be  \label{KGeq}
[ \, \partial_{\mu} \partial^{\mu} - m^{2} \, ] \, \phi (x) = 0 \ .
\ee
In simple situations, the variational principle and the boundary
conditions fix the classical field configuration $\phi (x)$ everywhere
in space-time.\\

As a well-known example, {\em electrodynamics} deals
with vector fields ${\cal A}_{\mu}(x)$, which represent the electromagnetic 
potentials. The Lagrangian
\be  \label{emlag}
{\cal L} = - \frac{1}{4} F_{\mu \nu} F^{\mu \nu} + j_{\mu} {\cal A}^{\mu} \ ,
\quad F_{\mu \nu} = \partial_{\mu} {\cal A}_{\nu} - \partial_{\nu} 
{\cal A}_{\mu} \ ,
\ee
is constructed from the (gauge invariant) field strength tensor
$F_{\mu \nu}$, and we 
added an external,
electrically charged current $j_{\mu}(x)$. Now the condition $\delta S = 0$
leads to the inhomogeneous Maxwell equations 
$$
\partial_{\mu} F^{\mu \nu} = j^{\nu} \ , 
$$
from which we infer that the current classically
obeys the continuity equation $\partial_{\nu} j^{\nu} = 0$.
(On the other hand, the homogeneous Maxwell equations are already
encoded in the use of potentials.)

\subsection{Quantum field theory}

The transition from classical field theory to {\em quantum field theory}
can be performed in analogy to the quantisation of the mechanical system
in Subsection 1.1. Since the r\^{o}le of paths in that case is now taken by
configurations, we quantise the field theoretic system by including
contributions of all possible field configurations. To render such a
huge summation well-defined, we introduce again a discretisation.
Since the fields take their values in each space-point $x$, we now need
a space-time lattice, which we choose to be hypercubic, and we denote the
lattice spacing again by $a$. Thus the lattice consists of the sites
\be
\left\{ \ x  \ \Big\vert \
\frac{x_{\mu}}{a} \in \Z \ , \ \forall \mu \right\} \ .
\ee
If we stay with the example of a neutral scalar field, then all the
configurations are summed over as follows,
\be
\prod_{x \, \in \, {\rm lattice}} \int_{- \infty}^{\infty} d \phi_{x} \ \dots 
\qquad ^{ a \to 0}_{\longrightarrow} \qquad \int {\cal D} \phi \ \dots
\ee
In this summation, we are going to attach a phase factor
$\exp ( \frac{i}{\hbar} S [ \phi ] )$ to each configuration,
similar to eq.\ (\ref{transamp}).
On the right-hand-side we indicate again the continuum limit,
the details of which will be of prominent interest in this work.

A configuration which corresponds to the lowest possible energy
is denoted as a vacuum $\Omega$. Similar to eq.\ (\ref{emlag}) we add
an external source field $J(x)$, which now couples to the field
$\phi (x)$. Then the vacuum-to-vacuum transition amplitude is defined as
\be  \label{ZJ}
Z[J] = \langle \Omega \vert \Omega \rangle_{J} =
\int {\cal D} \phi \, \exp 
\Big( \frac{i}{\hbar} ( S [ \phi ] + J \phi ) \Big) \ ,
\ee
where we use continuum notation, and $J \phi = \int d^{4}x \, J(x) \phi (x)$.


Let us assume for simplicity the solution of the equation $\delta S =0$
to be unique. Then it is again the vicinity of this classical configuration
which contributes in a dominant way (on an action scale where $\hbar $ is small);
also here the contributions at large $| \delta S |$ are mostly washed out by the
rapidly rotating phase.

The convergence of the sum over the configurations can be accelerated
drastically if we perform a Wick rotation $t \to -i t$ to arrive at
{\em Euclidean space.} There we denote a point as 
$x = (\vec x, x_{4})$, $x_{4}$
being the Euclidean time, and the above quantities turn into
\bea
{\cal L}_{\rm E} (\phi, \partial_{\mu}\phi) &=& 
\frac{1}{2} \partial_{\mu} \phi \partial_{\mu} \phi + V (\phi ) \ , \quad
S_{\rm E} [ \phi ] = \int d^{4}x \, {\cal L}_{E} (\phi, \partial_{\mu}\phi)
\ , \nn \\  \label{partfunc}
Z_{\rm E} & = & Z_{\rm E}[J=0] = \int {\cal D} \phi \, \exp 
\Big( - \frac{1}{\hbar} S_{\rm E} [ \phi ] \Big) \ . 
\eea
${\cal L}_{E}$ and $S_{\rm E}$ are the Euclidean Lagrangian and action.
$V( \phi )$ is some potential, which is --- for instance ---
quadratic in the free case, as we saw in eq.\ (\ref{freescallag}).
In Euclidean space we only write lower indices, and doubled indices
are summed over from $1$ to $4$ with the metric tensor $\delta_{\mu \nu}$.

Now the contributions by configurations deviating from the action minimum
are suppressed exponentially, which speeds up the convergence of the functional
integral tremendously.\footnote{Of course, 
the Wick rotation could have been performed
earlier in quantum mechanics, where it accelerates the convergence
of the path integral as well.}
This property is highly welcome if we try to evaluate
functional integrals approximately by summing over a small but (as far as
possible) representative set of random configurations. This is the method 
used in numerical simulations, which we will be concerned with
later. If conclusive simulations are feasible, they provide
in most cases the only access to functional integral results beyond 
perturbative, semi-classical or effective approximations, i.e.\ to 
actual functional integrals at finite interaction parameters.

In the terminology of statistical mechanics, $Z_{\rm E}$ is a partition
function. Then $\hbar$ takes a r\^{o}le analogous to the
temperature, which controls
the extent of field fluctuations around an action minimum.\footnote{In 
this sense, the variation of $\hbar$ does have a 
realistic interpretation, although it is fixed in Nature.}
In the limit $\hbar \to 0$ only the latter contributes  (the system
``freezes'' to the classical configuration), so this limit leads
back to the classical field theory of Subsection 1.2.
Once more there is an analogy to the point mechanics in Subsection 1.1.

In quantum field theory, the fluctuations around the vacuum are
essential; they record the occurrence of particles, deviating from
a vacuum state $\Omega$. 

In view of the statistical interpretation, we can build expectation
values, and these are the quantities that contain the physical information.
The vacuum expectation value of some observable ${\cal O}(\phi )$ is
given by
\be
\langle {\cal O}(\phi ) \rangle :=
\langle \Omega | {\cal O}(\phi ) | \Omega \rangle =
\frac{1}{Z[0]} \int {\cal D} \phi \, {\cal O}(\phi ) \, \exp 
\Big( \frac{i}{\hbar} ( S [ \phi ] ) \Big) \ ,
\ee
so that eq.\ (\ref{ZJ}) fixes the normalisation $\langle 1 \rangle = 1$.
In particular, the Euclidean 2-point function takes the form
\bea
G_{2}(x - y) &=& \langle \phi (x) \phi (y) \rangle = 
\frac{\hbar^{2}}{Z_{\rm E}}
\frac{\partial}{\partial J(x)} \frac{\partial}{\partial J(y)}
Z_{\rm E} [J] \Big\vert_{J=0} \nn \\
&=& \frac{1}{Z_{\rm E}} \int {\cal D} \phi \, \phi(x) \phi(y)
\exp \Big( - \frac{1}{\hbar} S_{\rm E}[ \phi] \Big) \ .
\eea
Here we assumed the condensate (or 1-point function) $\langle \phi \rangle$
to vanish, hence $G_{2}(x -y)$ 
coincides with the connected correlation function (with the general
form $\langle \phi (x) \phi (y) \rangle - 
\langle \phi (x) \rangle \langle \phi (y) \rangle$ ).
It characterises the correlation over a temporal separation 
$\Delta t = x_{4} - y_{4}$ and a spatial distance $\vec x - \vec y$.
If one Fourier transforms the distance, one usually obtains an
exponential decay in $\Delta t$,
\be
G_{2}(\vec p , \Delta t ) \propto e^{- E(\vec p\, ) \, \Delta t} \ ,
\ee
where $E(\vec p \, )$ is the energy of the particle involved; in particular
$E( \vec 0 )$ is its mass. 

Similarly we may extract further information of physical interest by
evaluating higher $n$-point functions
\be
G_{n}(x^{(1)}, \dots ,x^{(n)}) = \langle \phi (x^{(1)}) 
\cdots \phi (x^{(n)})  \rangle \ ,
\ee
or their connected part, which is often of primary interest,
\be
G_{n}^{\rm (c)}(x^{(1)}, \dots ,x^{(n)}) = 
(- \hbar)^{n} \frac{\partial^{n}}{\partial J(x^{(1)}) \dots
\partial J(x^{(n)}) } \ln Z_{\rm E} [J] \Big\vert_{J=0} \ .
\ee
Here all the $x^{(i)}$ are Euclidean space-time points.

In the further Sections we will stay in Euclidean space
(unless it is specified otherwise), 
and we will from now on suppress the subscript ``E''. 
The use of the Euclidean signature is justified because the
expectation values --- which provide the physical observables ---
can be carried over to Minkowski space, if four conditions
are fulfilled. These conditions 
are known as the Osterwalder-Schrader axioms \cite{OSax}.
Two of them (``analyticity'' and ''regularity'')
are rather technical, while ``$O(4)$ invariance'' and ``reflection
positivity'' have a physical interpretation. Note also that
$n$-point functions in Minkowski space require a time ordering.
If we deal with Euclidean lattices, we assume first a
continuum limit to be taken, and then the transition to
Minkowski space to be justified.

From now we use on natural units, $\hbar = c =1$, and --- 
when it is specified --- also lattice units, which set in addition
the lattice spacing $a=1$.
In Sections 7 and 8 we identify the spacing in lattice QCD with a physical
scale, which then attaches physical units to all dimensional quantities 
involved.

Derivations and details of the basic features that we have sketched
in this Introduction can be found at
numerous places in the literature. Due to their established status, we
hardly indicated references so far. At this point we would
like to attract attention to Ref.\ \cite{Roep}, 
which covers the subjects hinted at in Section 1 with great
precision. This also includes an explanation of the
Osterwalder-Schrader axioms and a comprehensive list of
references on the functional integral formulation of quantum physics.

\section{Renormalisation Group Transformations and Perfect Lattice Actions}

\subsection{Block variable transformations}

In Section 1 we have introduced the partition function $Z$ and 
its functional derivatives as the quantities of interest.
They are well-defined on a Euclidean lattice, which restricts the momenta
to the Brillouin zone
\be  \label{BZ}
B = \ \left( - \frac{\pi}{a} , \frac{\pi}{a} \, \right]^{d} \ ,
\ee
i.e.\ it naturally introduces a momentum cutoff $\Lambda = \pi /a$.

For a neutral scalar field, $\phi \in \R \, $, the source-free partition 
functions takes the form
\be
Z = \int \prod_{x} d\phi_{x} \, e^{-S[\phi ]} \ .
\ee
Also the action $S[ \phi ]$ is affected by the discretisation.
The standard form for a free lattice scalar field reads
\bea
S [ \phi ] &=& a^{d} \sum_{x} \Big[ \, \frac{1}{2 a^{2}} \sum_{\mu} 
\Big( \phi_{x + a \hat \mu} - \phi_{x} \Big)^{2} + \frac{m^{2}}{2}
\phi_{x}^{2} \, \Big] \nn \\  \label{standardscal}
&=& \Big( \frac{a}{2 \pi} \Big)^{d} \int_{B} d^{d}p \, 
\frac{1}{2} \phi(-p) \, \left[ \hat p^{2} + m^{2} \right] \, \phi (p) \ ,
\eea
where $\hat \mu$ is the unit vector in $\mu$-direction, and
\be
\hat p_{\mu} := \frac{2}{a} \sin \frac{a p_{\mu}}{2} \ , \qquad
\hat p^{2} = \sum_{\mu =1}^{d} \, \hat p_{\mu}^{2} \ .
\ee
As this modified momentum shows, 
the lattice structure introduces artifacts on a scale fixed by 
the cutoff $\Lambda$.
\footnote{In this case, the artifacts occur in $O( \Lambda^{2})$,
which is generic for bosonic systems.}

Now we would like to alter the lattice action in a way that
moves the cutoff effects to higher energies. This can be achieved
by a {\em renormalisation group transformation} (RGT) \cite{WilKog} to a new 
lattice field $\phi '$ living on a coarser lattice $\{ x' \}$, for instance
with spacing $2a$. We can choose the sites $x'$ as the centres of 
disjoint unit hypercubes on the fine lattice $\{ x \}$. Then the action
$S'$ for the lattice field $\phi '$ can be formulated as
\be  \label{scalRGT}
e^{- S' [ \phi ']} = \int \prod_{x} d \phi_{x} 
\exp \left[ - S [ \phi ] - \alpha \sum_{x'} \Big( \phi '_{x'} -
\frac{b}{2^{d}} \sum_{x \in x'} \phi_{x} \Big)^{2} \right] \ ,
\ee
where the sum $x \in x'$ runs over the $2^{d}$ sites on the fine lattice
in the unit hypercube with centre $x'$. $\alpha > 0$ and $b$ are
RGT parameters, which will be commented on below.

The RGT (\ref{scalRGT}) leaves the partition function invariant
(up to a constant factor),\footnote{This constant factor does not 
have any impact on physical 
properties, since it drops out in all expectation values.}
\bea
Z' &=& \int \prod_{x'} d\phi '_{x'}  \, e^{-S' [ \phi ']} \nn \\
   &=& \prod_{x',x} \int d \phi_{x} \, e^{-S [ \phi ]}
\int d \phi'_{x '} \, \exp \Big[ - \alpha \sum_{x'} ( \phi '_{x'} -
\frac{b}{2^{d}} \sum_{x \in x'} \phi_{x})^{2} \Big] \nn \\
   &=&  Z \ \cdot \ {\rm const.}
\eea
Also the $n$-point functions are transferred to the coarse lattice without 
any damage, for instance
\be
\langle \ \phi'_{x'} \, \phi'_{y'} \ \rangle =
\Big\langle \Big( \sum_{x \in x'} \phi_{x} \Big) 
\Big( \sum_{y \in y'} \phi_{y} \Big) \Big\rangle \ .
\ee

If we now consider the situation in terms of lattice units, we set on the
fine lattice $a=1$ and on the coarse lattice $a' = 2a =1$. In the transition
from the former to the latter lattice units,
the fields and the parameters are re-scaled according to their dimensions.
But the use of the blocked actions $S' [ \phi ']$ guarantees that the lattice
artifacts --- in particular the discretisation errors in the kinetic term ---
are still those of the fine lattice. Hence their scale is $\Lambda =
2 \Lambda '$, in contrast to the standard action on the coarse lattice.

This blocking variable RGT can be iterated, and --- for the RGT parameter
$b = 2 ^{d/2 -1}$ --- the lattice 
action converges to a finite fixed point \cite{BeWi}\footnote{For 
some field of dimension $[{\rm Mass}]^{\gamma}$ and
a blocking factor $n$, the corresponding parameter 
multiplying $\sum_{x\in x'} \dots $
has to be set to $n^{\gamma -d} $ for the sake of 
convergence under RGT iterations. This factor compensates the re-scaling
at the end of each step.}
$S^{*}$,
\be
S \longrightarrow S' \longrightarrow S'' 
\longrightarrow \quad \dots \quad \longrightarrow S^{*} \ .
\ee
The scale for the lattice artifacts is unchanged, hence it diverges in
lattice units,
\be
\Lambda \longrightarrow 2 \Lambda' \longrightarrow 4 \Lambda '' 
\longrightarrow \quad \dots \quad \longrightarrow \infty \ .
\ee
Therefore, $S^{*}$ is free of any cutoff artifacts; it is a 
{\em perfect lattice action}.

\subsection{Blocking from the continuum}

Let us now generalise the blocking factor to $n \in \{ 2,3, \ \dots \ \}$,
so that $a' = n \cdot a$. The limit $n \to \infty$ means that we perform
a {\em blocking from the continuum}; on the scale of the blocked lattice units,
the initial lattice appears continuous. Thus we arrive at the perfect action
in one single step, $S \to S^{*}$. The corresponding transformation reads
\be  \label{trafocont}
e^{-S^{*}[\phi ]} = \int {\cal D} \varphi \,
\exp \left( - S[\varphi ] 
- \alpha \sum_{x} \Big( \phi_{x} - \int_{C_{x}} d^{d}u \, \varphi (u)
\Big)^{2} \right) \ ,
\ee
where now $\phi$ is the final lattice field on the sites $x$, while
$\varphi$ is a continuum field, $S$ is the continuum action,
and $C_{x}$ is the unit hypercube (in final
lattice units) with centre $x$.~\footnote{Also the continuum 
field $\varphi$ is expressed in the (upcoming) lattice units.}
Since the RGT does not alter physical
properties, we see here directly that $S^{*}$ captures continuum physics
on the lattice, without any possibility for lattice artifacts to sneak in.

Let us now look at the explicit form of the free scalar propagator
for the standard lattice action (\ref{standardscal})
and for the perfect action $S^{*}[\phi ]$ in momentum
space \cite{BeWi,WBscal} (in lattice units),
\begin{eqnarray}
\frac{1}{2} G_{2}(p)_{\rm standard} &=& \frac{1}{2}
\langle \phi( -p) \phi (p) \rangle_{\rm standard} =
\frac{1}{ \hat p^{2} + m^{2}} \ , \nn \\
\frac{1}{2} G_{2}(p)_{\rm perfect} & = & \frac{1}{2}
\langle \phi( -p) \phi (p) \rangle_{\rm perfect} =
\sum_{l \in \Z^{d}} \frac{\Pi^{2}(p + 2\pi l )}{(p + 2\pi l)^{2}
+ m^{2} } + \frac{1}{\alpha} \ , \nn \\ 
{\rm where} && \Pi (p) := \prod_{\mu =1}^{d} \frac{\hat p_{\mu}}{p_{\mu}} \ .
\label{perfscalprop}
\end{eqnarray}
We see that the perfect propagator consists of the continuum
propagator with an analytic factor $\Pi^{2}$ and all $2 \pi$ periodic
copies, plus a constant term. The latter vanishes in the limit
to a $\delta $-function RGT, $\alpha \to \infty$.

The function $\Pi (p)$ ensures that the sum over the
integers $l_{\mu}$ converges. In the exponential transformation term
in eq.\ (\ref{trafocont}) we have used a step function shape for the
integration of $\varphi$ (1 inside $C_{x}$, 0 otherwise).
This shape could be varied, which implies different forms
of the function $\Pi$ (without danger for the converges of the
sum over the copies of the Brillouin zone). For instance, the
generalisation to B-spline blocking functions is discussed in 
Ref.\ \cite{WBscal}.

The function $\Pi (p)$ in eq.\ (\ref{perfscalprop})
is also affected by the hypercubic structure
of the lattice; as an example, we could stay with the step function shape
and consider a 2d triangular lattice instead, which leads to a
different function in the enumerator \cite{WBscal}
(in this case, the lattice cells to be integrated over are the 
hexagons of the dual lattice). 
However, it always has to be analytic,\footnote{We assume the 
analytic continuation to be inserted
at the removable singularities.}
hence it does not affect the {\em dispersion relation} to be extracted from the
perfect propagator.\footnote{Of course, one always considers the 
branch with the lowest energy.}
The latter coincides indeed with the continuum
dispersion, which also means that it displays exact and continuous
rotation invariance (which turns into exact Lorentz invariance in
Minkowski space). We emphasise that this symmetry can be implemented
{\em in the phy\-sical observables} (not in the form of the propagator),
irrespective of the lattice structure.

In coordinate space we write the perfect action in the form of a discrete
convolution 
\be
S^{*}[\phi ] = \sum_{x,y} \phi_{x} \rho (x-y) \phi_{y} \ ,
\ee
where $\rho (x -y) $ is the inverse Fourier transform of
$G_{2}(p)_{\rm perfect}^{-1}$.
The decay of $| \rho (x-y) |$ is exponential in $|x -y|$
for any mass $m$ (for increasing $m^{2}$ the decay is accelerated).
Explicit examples are shown in Ref.\ \cite{WBscal}. 
This means that the perfect lattice action is {\em local}.
Generally, locality is reputed as a vital requirement to ensure that 
a lattice action has a sensible continuum limit.

However, in $d \geq 2$ the couplings in $\rho$ extend to infinite 
distances $| x - y|$, in contrast to the ultralocal standard 
action.\footnote{Ultralocality means that the couplings drop to zero beyond 
a finite number of lattice spacings.\label{ultralocfn}}
For practical purposes this set of couplings has to be truncated.
To this end, we first identified the value of $\alpha$
which optimises the level of locality, 
i.e.\ the rapidity of the exponential decay. This optimal value depends
on $m$; a good approximation is \cite{WBscal}
\be  \label{alphamscal}
\alpha_{\rm optimal} (m) \simeq \frac{\sinh m -m}{m^{3}} \ ,
\ee
which is derived from the property that it only couples
nearest neighbours in $d=1$.
Then we may truncate $\rho$ to a unit hypercube --- i.e.\ we enforce
that couplings $\rho (x-y) \neq 0$ only occur if 
$| x_{\mu} - y_{\mu}| \leq a$, $\forall \mu$ --- by imposing periodic
boundary conditions over 3 lattice spacings.
This method yields a perfect action in a lattice
volume $3^{4}$, which we then use as an truncated approximation
to the perfect action on larger lattices. Unlike other truncation schemes,
this one guarantees for instance the correct normalisation of the couplings.

\begin{figure}[h!]
\vspace*{1cm}
  \centering
  \includegraphics[angle=270,width=0.47\linewidth]{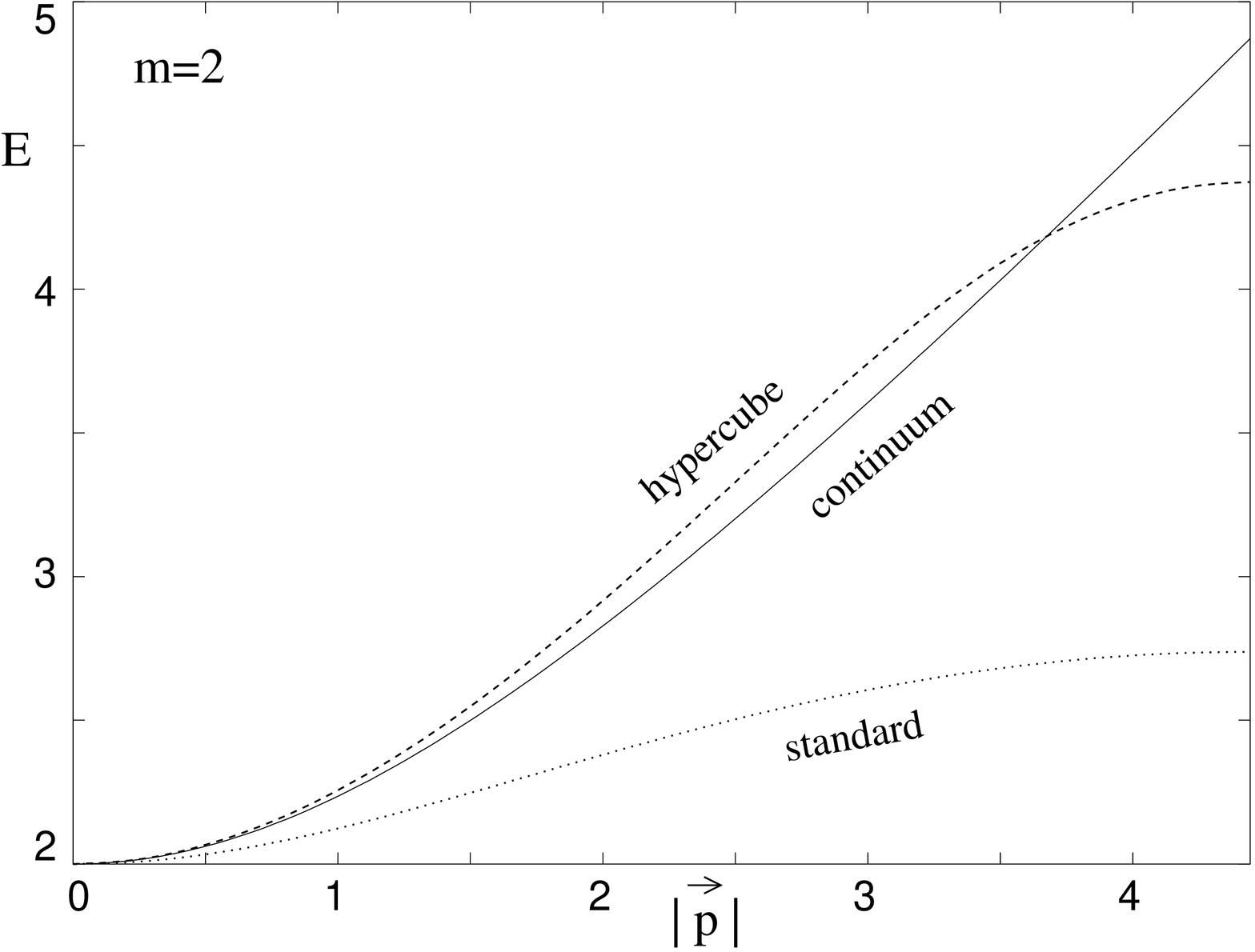}
  \includegraphics[angle=270,width=0.5\linewidth]{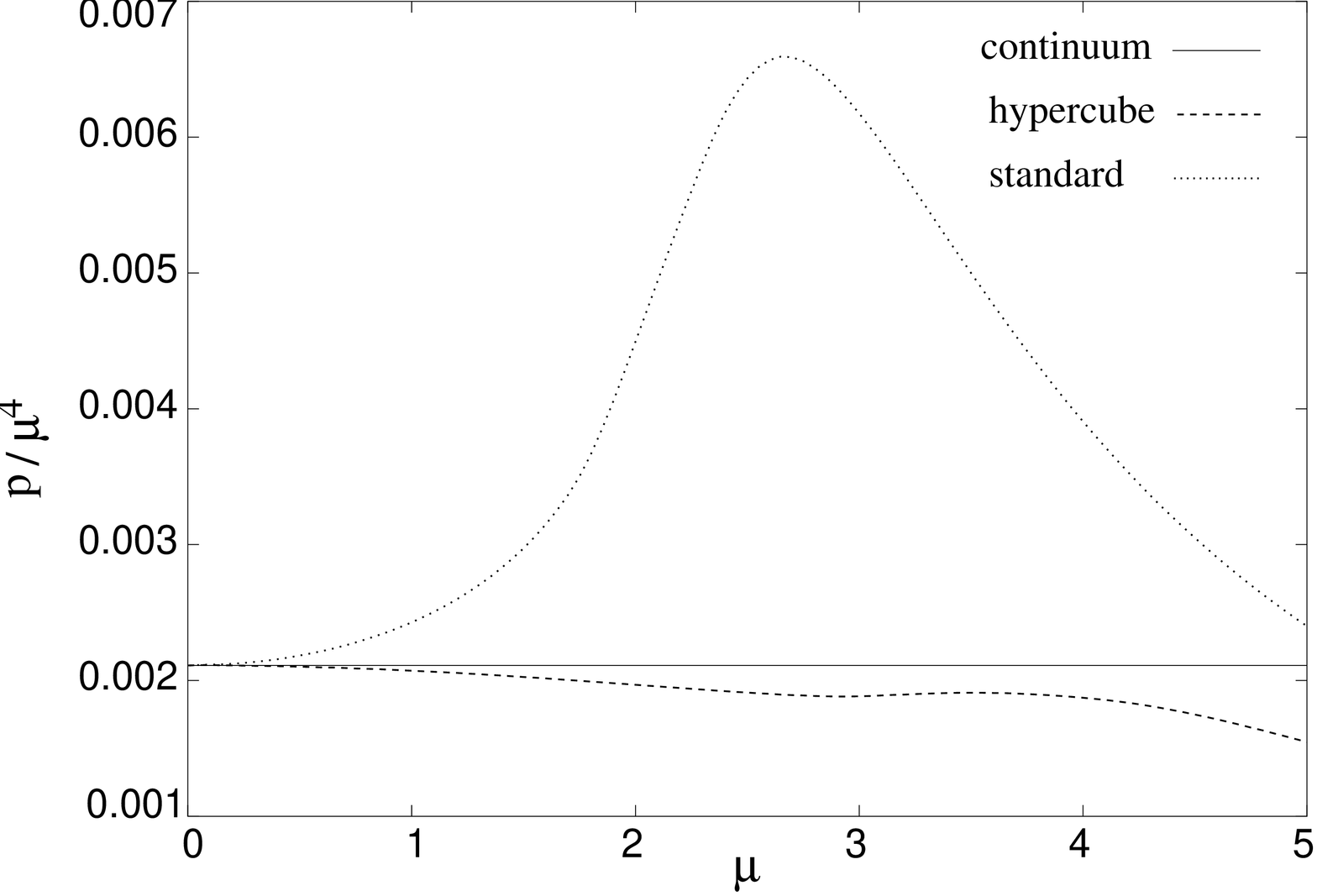}
\caption{{\it On the left: The dispersion relation for a free scalar particle
of mass $m=2$ for the perfect lattice action (which coincides with
the continuum dispersion), the hypercubic, truncated perfect action 
and the standard action. For increasing momenta, the magnitude of the
lattice artifact rises rapidly for the standard action, whereas
it remains modest for the truncated perfect action.
\newline
On the right: the corresponding scaling test for massless lattice scalars
with respect to the ratio between the pressure $P$ and the fourth power
of the chemical potential 
$\mu$. As $\mu$ increases, lattice artifacts cause a deviation from the
continuum value $P / \mu^{4} = 1 / ( 48 \pi^{2})$. This deviation
is large for the standard action, but harmless for the truncated 
perfect hypercube action.
\newline
In both plots all quantities are given in lattice units.}}
\label{scalscale}
\end{figure}

Figure \ref{scalscale} illustrates as examples
the dispersion relation $E(\vec p \, )$ 
(for momenta $\vec p \propto (1,1,0)$)
for the free scalar particle of mass $m=2$,
as well as the thermodynamic ratio $P/\mu^{4}$ at $m=0$ (where $P$ 
is the pressure and $\mu$ the chemical potential).\footnote{The inclusion
of a chemical potential in a perfect lattice action will be commented
on later in the fermionic context (Subsection 4.1).}

\subsection{Classically perfect actions}

For interacting theories, the perfect action can in general not be
computed explicitly, since this requires carrying out a functional
integral. P.\ Hasenfratz and F.\ Niedermayer \cite{HasNie} suggested a 
feasible simplification, which evaluates the RGT steps in the classical
approximation. This idea has revived and boosted the RGT method
in lattice field theory. In our case, this classical RGT step takes 
the form
\be
S'[\phi '] = \ ^{\rm min}_{~ \phi} \ \Big\{ S[ \phi ] + \alpha
\sum_{x'} \Big( \phi'_{x'} - \frac{1}{n^{d/2+1}} \sum_{x \in x'}
\phi_{x} \Big)^{2} \Big\} \ .
\ee
Iteration leads also here to a fixed point --- the {\em classically perfect
action.} With a suitable parameterisation ansatz, it can be determined
numerically to some approximation by inserting a set of configurations
$\phi '$ on the coarse lattice and performing the minimisation. 
The parameters which are used in the ansatz for the action are then
tuned until one obtains optimal approximate invariance under this 
transformation, i.e.\ an approximate classical fixed point action.
This procedure is particularly promising for asymptotically free
theories: there the weak field expansion of the classical
fixed point equation corresponds in leading order the 
(soluble) case of free fields. 
The pioneering work \cite{HasNie} for this method evaluated
and simulated a classically perfect action with a large number of
parameters in the 2d $O(3)$ model (a non-linear $\sigma$-model). 
A subtle scaling test (suggested in Ref.\ \cite{LWW})
revealed practically no lattice
artifacts at all down to $\xi / a \simeq 5$ (where $\xi$ is the correlation 
length). This is in contrast to the standard action, where
scaling artifacts are visible even at $\xi / a \simeq 15$ \cite{HasNie}. 

Later applications of classically perfect actions include 
topological aspects of the 2d $O(3)$ model \cite{O3topo}, the 2d
$CP(3)$ model \cite{Ruedi}, pure $SU(2)$ \cite{SU2} and $SU(3)$ 
\cite{SU3} gauge theory in $d=4$, the two-flavour Schwinger model \cite{Pany}
and finally QCD \cite{Bern}.

\begin{figure}[h!]
\begin{center}
 \includegraphics[angle=0,width=.55\linewidth]{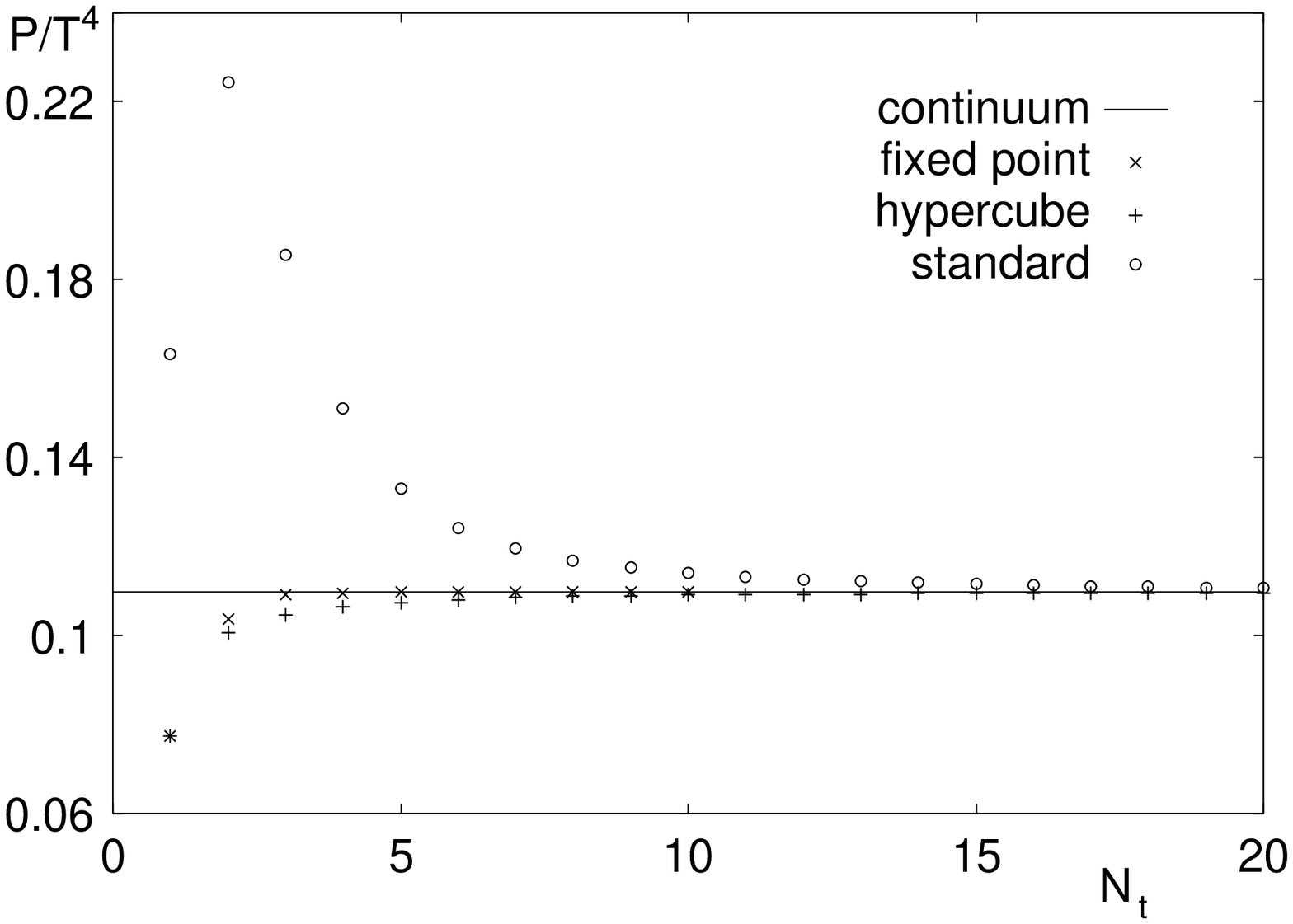}
 \includegraphics[angle=270,width=.55\linewidth]{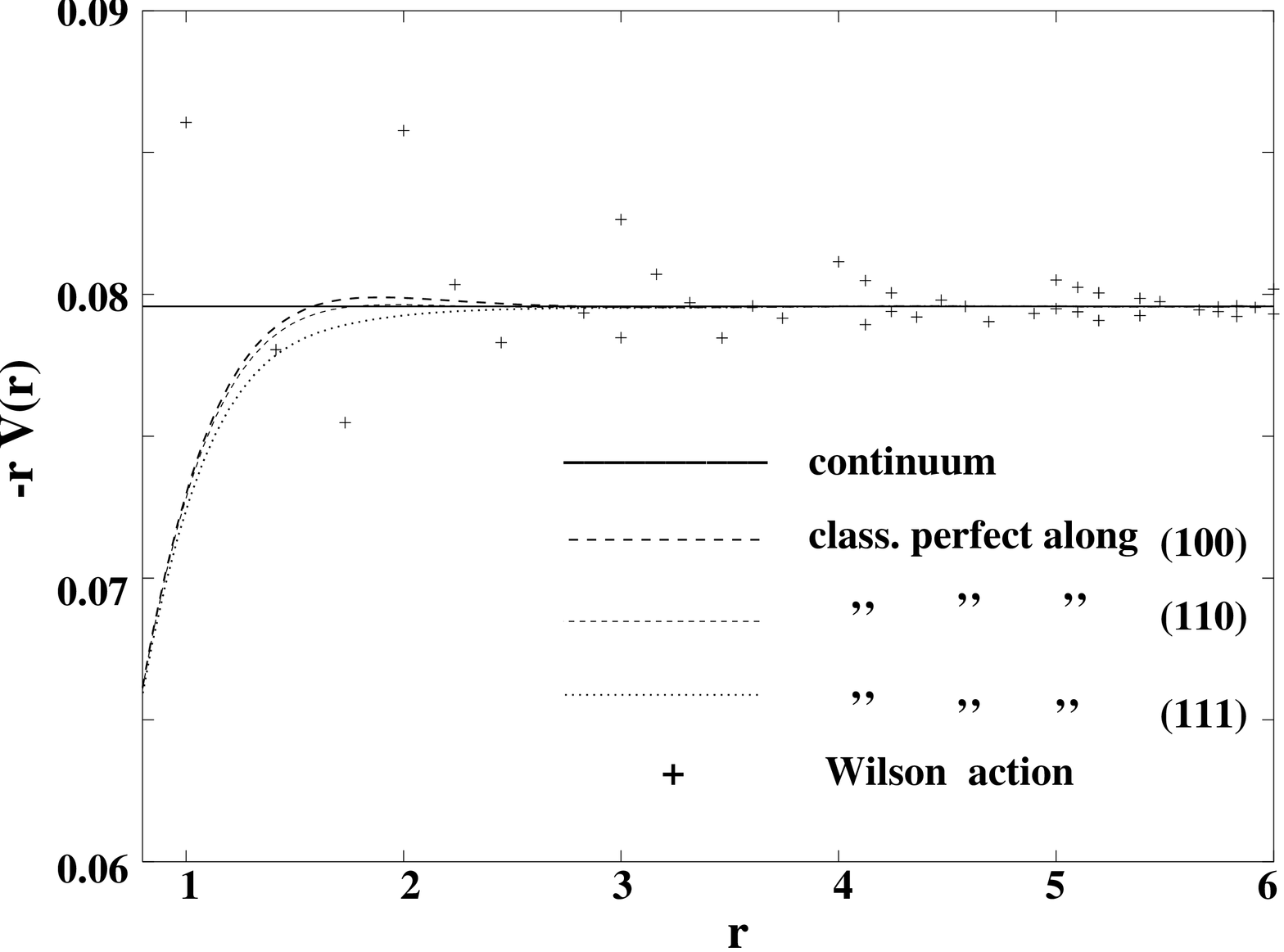}
\end{center}
\caption{{\it On top: the scaling ratio 
${\rm (pressure)} / {\rm (temperature)}^{4}$ for finite numbers $N_{t}$ 
of lattice sites in the temporal direction. A decreasing number $N_{t}$ 
corresponds to a coarser lattice, which amplifies the artifacts, in particular
for the standard action.
(The continuum value is given by the Stefan-Boltzmann law, $P /T^{4}
= \pi^{2} /90$.)
\newline
Below: The (re-scaled) static quark-antiquark potential $V(r)$
at different distances. Wilson's standard formulation is only defined
at discrete distances and exhibits significant artifacts at short $\, r$. 
The classically perfect potential captures all distances and suffers
much less from lattice artifacts.}}
\label{Coulomb}
\end{figure}

In Figure \ref{Coulomb} we show a comparison --- involving classically 
perfect actions --- for the scaling 
of the thermodynamic ratio $P/T^{4}$ (where $T$ is the temperature) 
for free scalars \cite{WBscal} (on top), and for the static quark-antiquark
potential \cite{QuaGlu} (below).

\begin{figure}[h!]
  \centering \hspace*{-4mm}
  \includegraphics[angle=0,width=0.54\linewidth]{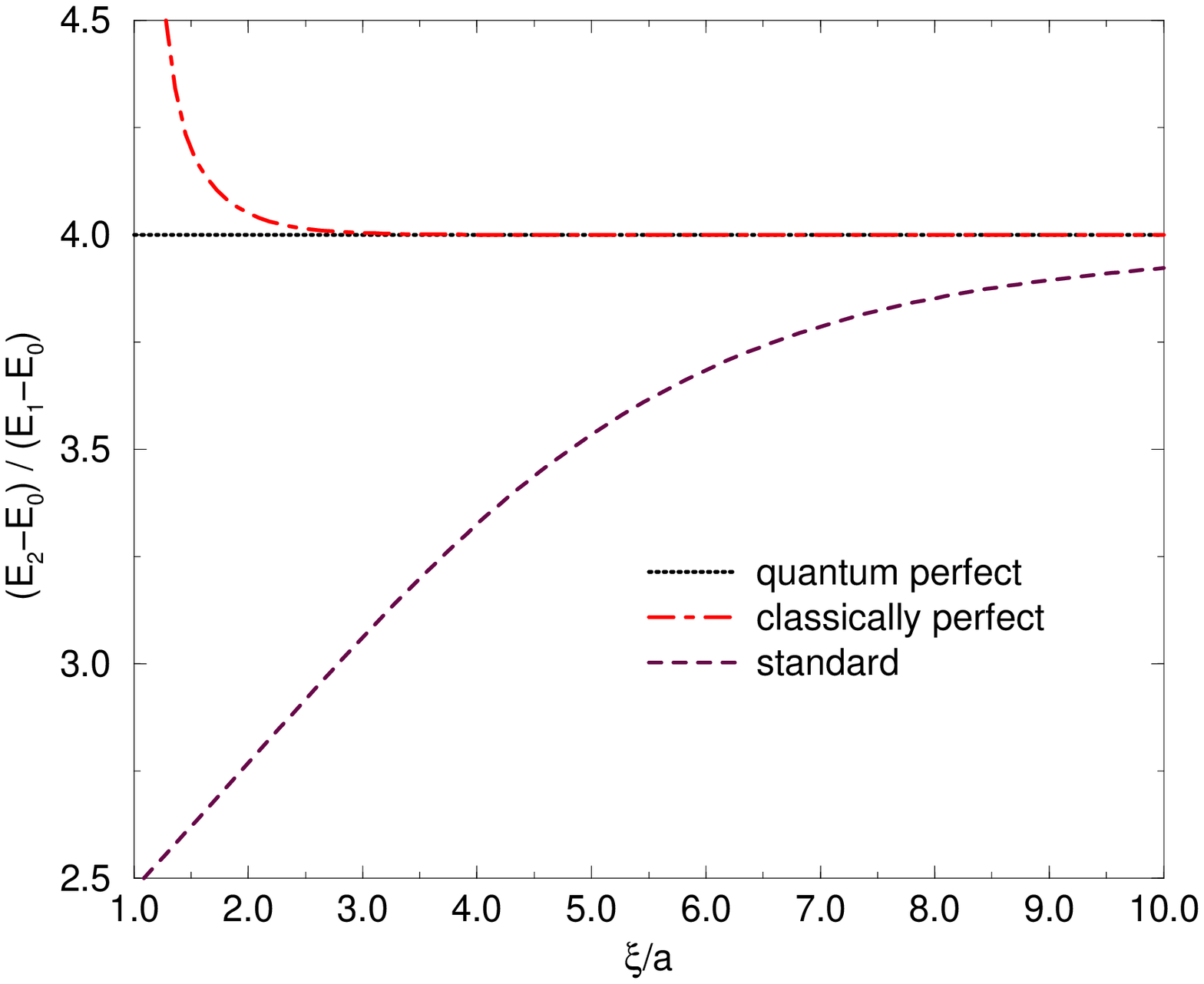} \hspace*{-11mm}
  \includegraphics[angle=0,width=0.54\linewidth]{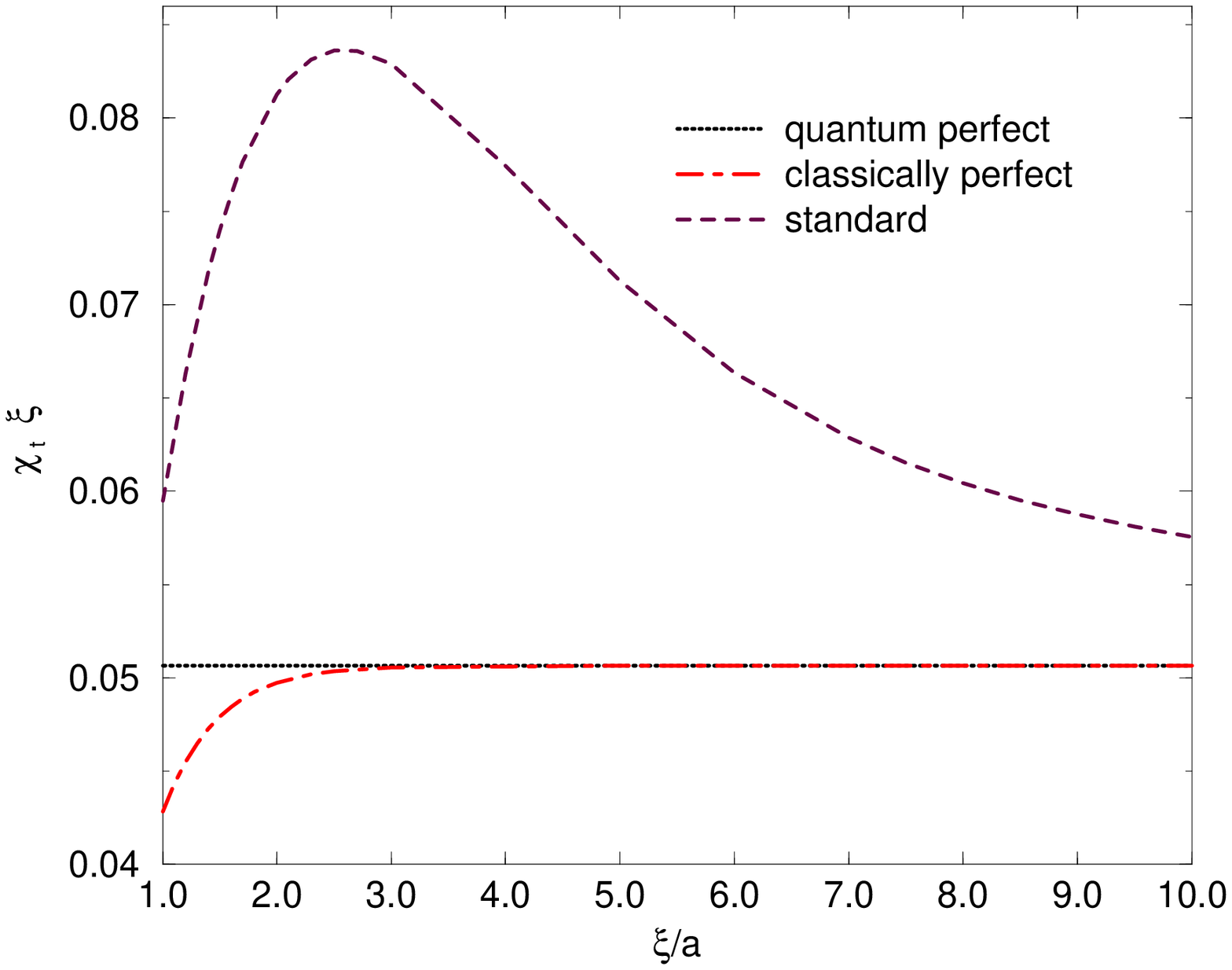}
\caption{{\it The scaling behaviour of a free scalar particle on a circle.
We show two scaling quantities as functions of the correlation length
in lattice units: the ratio between the first two
energy gaps (on the left) and the topological 
susceptibility (on the right).}}
\label{toposcal}
\end{figure}

In Ref.\ \cite{qrot} we studied a free scalar particle on a circle
(a quantum rotor) with a discrete Euclidean time and periodic 
boundary conditions over a period $T$. 
We considered the scaling of the ratio between the first
two energy gaps 
and of the topological susceptibility 
(scaled by the correlation length $\xi$),
\be  \label{topsus}
\frac{E_{2} - E_{0} } {E_{1} - E_{0} } \qquad {\rm and} \qquad
\chi_{t} = \frac{1}{T} \langle \nu^{2} \rangle \ .
\ee
The latter is based on the expectation value of the squared 
winding number $\nu$, which is the simplest case of a topological 
charge. These results are plotted in Figure~\ref{toposcal}.

It is remarkable that also the continuum topology can be
represented exactly on the lattice, thanks to the formulation with
perfect actions and operators. Generally, we build (classically) perfect
operators from the lattice fields obtained by (classical) blocking
\cite{QuaGlu}.

In contrast, for the standard action 
it is not even obvious how to define topological sectors (since
all lattice configurations can be continuously deformed into one
another). We use it here with the geometrical definition of the 
topological charge \cite{geocharge}, which is the best option, 
but we observe strong scaling artifacts.\footnote{In Figure \ref{toposcal} 
we use the continuum correlation length as the scale. If one inserts instead
for the standard action the correlation length as obtained from
standard action simu\-lations, the artifacts are reduced,
but the hierarchy in the scaling quality persists \cite{Boyer}.}
On the other hand, the perfect formulation keeps track of each detail in the
intervals between the discrete time points, since it emerges from blocking 
transformations. This means that
any winding number between nearest neighbour time sites is included
as a possibility in the expectation value $\langle \nu^{2} \rangle$.
(Of course, a large number of windings is strongly suppressed by
the kinetic term in the exponent of the Boltzmann factor,
cf.\ eq.\ (\ref{partfunc})). The classically perfect action still 
approximates the continuum value of $\, \chi_{t} \, \xi \,$ 
to a very good approximation.

\section{Fermions}

\subsection{The Dirac equation}

For convenience we temporarily return to Minkowski space
for the Subsections 3.1 to 3.3, which contain
introductory remarks on fermions.

Let us go back to quantum mechanics
as a renewed starting point.
Taking the relativistic energy-momentum relation
$E^{2} = \vec p^{\, 2} + m^{2}$ as a guide-line, one arrives
at an obvious ansatz for a relativistic Schr\"{o}dinger equation,
\be
[ \partial_{\mu} \partial^{\mu} - m^{2} ] \, \Psi = 0 \ .
\ee
This is the Klein-Gordon equation, which
we already encountered in eq.\ (\ref{KGeq}) in the context
of classical field theory. An apparent
problem with it, which worried the pioneers of quantum mechanics, 
is the occurrence of negative energies. P.A.M.\ Dirac
wanted to avoid them by linearising this equation with the ansatz
\be  \label{Diraceq}
[ \, i \gamma_{\mu} \partial^{\mu} - m \, ] \, \Psi = 0 \ .
\ee
In order to reproduce the relativistic energy, the coefficients 
$\gamma_{\mu}$ have to obey the anti-commutation relation
\be  \label{antigam}
\{ \gamma_{\mu} , \gamma_{\nu} \} = 2 g_{\mu \nu} \ , 
\quad g \equiv {\rm diag}(1,-1,-1,-1) \ .
\ee
Therefore, these coefficients $\gamma_{\mu}$ in the Dirac equation 
(\ref{Diraceq}) have to be (at least) $4 \times 4$
complex matrices in $d=4$. Thus the {\em spinor} $\Psi$ has four 
components,\footnote{In two dimensions, we can live with
$2\times 2$ matrices $\gamma_{\mu}$ and 2-component spinors.}
\be
\Psi = \left( \begin{array}{c}
\psi_{1} \\ \psi_{2} \\ \psi_{3} \\ \psi_{4} \end{array} \right) \ .
\ee
Actually this linearisation does {\em not} overcome the negative 
energy eigenvalues. Nevertheless this ansatz was extremely successful;
for instance, it led to the prediction of the positron just before
its experimental discovery in 1931. In fact, the spinor $\Psi$ captures a
spin-1/2 particle plus its antiparticle.

Later on, relativistic quantum mechanics considered the Dirac equation
appropriate for fermions, and the Klein-Gordon equation for bosons.

\subsection{Fermionic field theory}

In the functional integral formulation of fermionic field theory, 
the Dirac operator is still present as the central ingredient 
in the Lagrangian. For free fermions of mass $m$, the partition function
and the action are written as
\bea
Z &=& \int {\cal D} \bar \Psi {\cal D} \Psi \, e^{i S [ \bar \Psi , \Psi ]} \ ,
\nn \\
S [ \bar \Psi , \Psi ] &=& \int d^{4}x \, \bar \Psi (x) 
\, [ i \gamma_{\mu} \partial^{\mu} - m ] \, \Psi (x) \ ,  \label{fermiZ}
\eea
where $\bar \Psi (x) = ( \bar \psi_{1}(x), \bar \psi_{2}(x),\bar \psi_{3}(x),
\bar \psi_{4}(x))$ and $\Psi (x)$ are spinor fields.
Application of the variational principle $\delta S =0$ leads to the
Dirac equation (\ref{Diraceq}) for $\Psi$, and to the adjoint Dirac equation
\be
\bar \Psi \, [ \, i \gamma_{\mu} \!\! \backderi \ ^{\!\! \mu} + m \, ] = 0 \ .
\ee

In the light of the Spin-Statistics Theorem, fermion field components
anti-commute, hence one describes them by {\em Grassmann variables.}
A set of Grassmann variables $\{ \eta_{i} \}$, $i=1,2,\dots $
(as it is used here for the components of $\bar \Psi$ and of $\Psi$
in a specific point $x$) obeys the relations
\be  \label{grass}
\{ \eta_{i} , \eta_{j} \} = 0 \ , \quad
\int d \eta_{i} \, \eta_{j} = \delta_{ij} \ .
\ee
A striking difference from the Dirac algebra
(\ref{antigam}) is the pro\-perty $\eta_{i}^{2} = 0$.
The integration rule is motivated by the analogy to the
translation invariance of the real, unbounded integral.
The Grassmann integral has no bounds, and its effect is equivalent
to differentiation.
It provides the basis for the functional integral in eq.\ (\ref{fermiZ})
\cite{Berezin}, which we will make explicit in Subsection 3.4.

Interactions can be included for instance by adding a 4-Fermi term
$(\bar \Psi (x) \, \Psi(x) )^{2}$ to the Lagrangian, 
which we will consider in Subsections 3.5, 4.3 and 5.2.
Another type of interaction is generated by coupling the
fermions to a {\em gauge field} ${\cal A}_{\mu}$ through a covariant 
derivative, which turns the Dirac operator and the partition function into
\bea  \label{DAeq}
D({\cal A}) &=& i \gamma_{\mu} \left[ \partial^{\mu} - 
g {\cal A}^{\mu}(x) \right] - m \ , \\
Z &=& \int {\cal D} \bar \Psi {\cal D} \Psi {\cal D} {\cal A} \,
\exp \Big( i \int d^{4} x \, \bar \Psi (x) D({\cal A}) \Psi (x) + 
i S[{\cal A}] \Big) \ . \nn
\eea
$S[{\cal A}]$ represents the pure gauge action; it could be for instance
the Abelian gauge action obtained by integrating the Lagrangian
(\ref{emlag}). In that framework,
the term \, $g \bar \Psi \gamma_{\mu} \Psi$ \, takes the r\^{o}le
of the external, charged current $j_{\mu}$ 
(and $g$ is the gauge coupling). 
Fermionic $n$-point functions are defined in analogy to the bosonic case
(see Subsection 1.3), but the order matters, of course.

\subsection{Chiral symmetry}

Due to the anti-commutation rule (\ref{antigam}), the matrix
\be  \label{gamma5}
\gamma_{5} := i \gamma_{0} \gamma_{1} \gamma_{2} \gamma_{3} \quad
{\rm obeys} \quad \{ \gamma_{5} , \gamma_{\mu} \} = 0 \quad {\rm and} 
\quad \gamma_{5}^{2} = \uno \ .
\ee
Therefore, the operators
\be
P_{\pm} := \frac{1}{2} ( \uno \pm \gamma_{5} )
\ee
are complementary projectors ($P_{\pm}^{2} = P_{\pm}$ ,
$P_{+} + P_{-} = \uno$). They can be used to decompose the spinor fields 
into their so-called left-handed and right-handed parts,
\be  \label{hands}
\Psi_{L,R}(x) = P_{\pm} \Psi(x) \ , \quad 
\bar \Psi_{L,R}(x) = \bar \Psi (x) P_{\mp} \ .
\ee
In these terms, the fermionic part of the Lagrangian in eq.\
(\ref{DAeq}) reads
\be  \label{Lagm}
{\cal L} = \bar \Psi_{L} D({\cal A})_{m=0} \Psi_{L} 
+ \bar \Psi_{R} D({\cal A})_{m=0} \Psi_{R}
- m ( \bar \Psi_{L} \Psi_{R} + \bar \Psi_{R} \Psi_{L}) \ .
\ee
In the chiral limit $m \to 0$ the left-handed and the right-handed
parts decouple completely. This property corresponds to the relation
\be
\{ D_{m=0} \, , \gamma_{5} \} = 0 \ ,
\ee
which manifests itself in a global symmetry, namely the invariance of
${\cal L}$ under the ``chiral rotation''
\be  \label{chiralrot}
\bar \Psi \to \bar \Psi \, e^{i \alpha \gamma_{5}} \ , \quad
\Psi \to e^{i \alpha \gamma_{5}} \, \Psi \
\ee
for an arbitrary parameter $\alpha$.

Obviously, the term that enters the Lagrangian (\ref{Lagm}) for
a fermion mass $m \neq 0$ breaks the chiral symmetry explicitly;
the chiral rotation (\ref{chiralrot}) transforms the mass term as
\be
m \bar \Psi \Psi \ \to \ m \bar \Psi e^{2i \alpha \gamma_{5}} \Psi \ .
\ee

In general, global symmetries --- such as the chiral invariance ---
are only realised approximately in Nature,\footnote{An exception
is CPT invariance, which is assumed to be exact \cite{Jost}
(this was first postulated by W.\ Pauli in 1955).
Its breaking would imply the violation of Lorentz invariance \cite{Green},
which appears as a global/local symmetry in special/general relativity.} 
hence a breaking by a mass term is not necessarily a problem.
However, by the introduction of gauge fields one arrives at 
local symmetries, and they have got to be exact. 
In a vector theory, the gauge fields couple in the
same way to the left-handed and to the right-handed fermions.
This is the case for the gluon fields in QCD.
Then the fermion mass is allowed --- quark masses can be
inserted into the QCD Lagrangian.

On the other hand, the electroweak sector of the Standard Model
is an example for a
chiral gauge theory, where the gauge fields couple in different
ways to the left-handed and to the right-handed fermions.
Then we have to require their invariance under independent
transformations, which forbids explicit mass terms in ${\cal L}$.
In that framework, the masses of fermions (and also
those of gauge fields) can only be generated dynamically.
It takes Yukawa couplings to a Higgs field and 
spontaneous symmetry breaking to arrive at massive quarks 
and leptons (and massive gauge bosons $W^{\pm}$ and $Z^{0}$).\footnote{We
do not consider renormalisation effects at this point.}

\subsection{Fermions on a Euclidean lattice}

We return to Euclidean space, where the $\gamma$-matrices obey
\be  \label{gamma5E}
\{ \gamma_{\mu} , \gamma_{\nu} \} =2 \delta_{\mu \nu } \ , \quad
\gamma_{5} := \gamma_{1} \gamma_{2} \gamma_{3} \gamma_{4} \ , \quad
\{ \gamma_{\mu} , \gamma_{5}\} = 0 \ , \quad \gamma_{5}^{2} = \uno \ .
\ee
We choose them to be Hermitian.
We write a bilinear fermionic lattice action --- such as the action
for free fermions, or for fermions interacting through gauge fields
--- in the form
\be
S = \sum_{i,j =1}^{N} {\bf \bar \Psi}_{i} M_{ij} {\bf \Psi}_{j}
\equiv {\bf \bar \Psi} M {\bf \Psi} \ .
\ee
Here the components ${\bf \bar \Psi}_{i}, \, {\bf \Psi}_{i} $ run over all
the lattice sites, and on each site over all internal degrees of
freedom (spinor and flavour indices, and for instance 
in QCD also colour indices). It is easy to see that
the partition function is given by the celebrated fermion determinant,
\be  \label{fermidet}
Z = \int \prod_{i=1}^{N} d{\bf \bar \Psi}_{i} d{\bf \Psi}_{i} \
e^{-{\bf \bar \Psi} M {\bf \Psi}} = {\rm det} M \ .
\ee
This expression also attaches an explicit meaning to the Grassmann
functional integrals.\footnote{It is entertaining to compare this result 
to the expression for a complex scalar field, 
$ \int \prod_{i} d {\rm Re} \Phi_{i} d {\rm Im} \Phi_{i} \,
\exp (- \Phi^{\dagger} M \Phi ) \propto ( {\rm det} M)^{-1}$, or
$1 / \sqrt{{\rm det } M}$ for a neutral scalar field.}  
(The order of the single Grassmann integrals
matters, since the rule in eq.\ (\ref{grass}) refers particularly
to the innermost integral.)

As an example, we consider a free, massless fermion with the 
Euclidean continuum action
\be  \label{contfermact}
S [\bar \psi , \psi ] = \int d^{4}x \, \bar \psi (x) \gamma_{\mu} 
\partial_{\mu} \psi(x) \ .
\ee
On a lattice with unit spacing ($a=1$) the simplest discretisation ansatz 
reads
\bea
S [\bar \Psi , \Psi ] &=& \sum_{x} \bar \Psi_{x} \gamma_{\mu} \frac{1}{2}
( \Psi_{x + \hat \mu} - \Psi_{x - \hat \mu} ) \nn \\
&=& \frac{1}{(2 \pi )^{4}} \int_{B} d^{4}p \, \bar \Psi (-p)
i \gamma_{\mu} \sin p_{\mu} \Psi (p) \ .
\eea
This formulation is known as the {\em naive lattice fermion.}
As the name suggests, there is a serious problem with it:
the propagator
\be
G_{2}(p)_{\rm naive} = \frac{1}{i \gamma_{\mu} \sin p_{\mu}}
\ee
has inside the (first) Brillouin zone (\ref{BZ}) not only the physical
pole at $p=0$, but it has poles whenever
$p_{\mu} \in \{ 0 , \pi \}$, $\mu =1 \dots 4$. Hence there are 16 poles
(in general, $2^{d}$ poles) instead of the one that we have ordered.
This effect is known as the {\em fermion doubling problem} --- it is due to
the occurrence of a linear derivative. In fact, these
doublers distort physical properties regardless how fine the lattice 
might be, hence this formulation cannot be applied.
For instance, among the 16 species the chiralities are
equally distributed, which makes it impossible to construct a chiral
gauge theory \cite{NoGo}. Moreover, the trouble also affects vector theories,
since these species contribute to the axial anomaly
with alternating signs, hence doubled lattice fermions cannot reproduce
a non-vanishing axial anomaly either \cite{KaSmit81}.

So we have to consider further options for the lattice action
\be
S = \sum_{xy} \bar \Psi_{x} D_{xy} \Psi_{y} \ .
\ee
We recall that locality is in general a requirement for a controlled continuum 
limit (in view of the extension to the interacting case).
In coordinate space, a local lattice Dirac operator $D$ has to be bounded 
as\footnote{Different definitions of locality appear in the lattice literature,
but the condition of an exponential decay 
--- which we referred to already for scalar fields ---
is the relevant one, because it guarantees a safe continuum limit.}
\be
| D_{xy} | \leq c_{1} e^{- c_{2}| x -y|} \quad , \qquad c_{1}, \ c_{2} > 0 \ .
\ee
In momentum space this means that $D(p) = G_{2}(p)^{-1}$ has to be analytic.

It is not easy to find a satisfactory solution to the doubling problem.
This statement was made precise by the {\em Nielsen-Ninomiya No-Go Theorem}
\cite{NoGo}. Putting aside technical details,\footnote{The proof 
requires some additional assumptions --- like
lattice translation invariance --- but they are not especially tricky.}
it essentially states that an undoubled lattice fermion cannot 
be chiral and local at the same time. 

For an intuitive and simplified illustration, we write a rather general
ansatz for a chiral lattice Dirac operator for free fermions,
\be
D(p) = i \rho_{\mu}(p) \gamma_{\mu} \ , \quad
\rho_{\mu} (p) = p_{\mu } + O(p^{2}) \ .
\ee
The leading momentum order of $\rho_{\mu}$ is required by
the correct continuum limit 
(which is determined by small momenta in lattice units).
We may consider the specific momenta $p = ( p_{1}, 0,0,0)$,
so that 
$$
D(p_{1}, 0,0,0) = i \rho_{1}(p_{1}) \gamma_{1} \ ,
$$
with the physical zero at $p_{1}=0$.
Since $2\pi$ periodicity is mandatory, and since locality requires an analytic
function $\rho_{1}(p_{1}) = p_{1} + O(p_{1}^{2})$, 
at least one additional zero (generally: an odd
number of them) is inevitable inside the Brillouin range 
$p_{1} \in \ ( -\pi , \pi ]$, even if one deviates from the naive form
$\rho_{1}(p_{1}) = \sin p_{1}$.

Many suggestions have been made to circumvent this problem by
breaking one of the desired properties on the lattice, hoping this would 
not affect the continuum limit. We do not review all these efforts,
but we mention as an example the
SLAC fermion \cite{SLAC}. In the above conside\-ration, it sets
$\rho_{1}(p_{1}) = p_{1}$ in $p_{1} \in \ ( -\pi , \pi ]$, which is
then periodically continued (and the same for the other momentum components).
Due to the jumps at the edges 
of the Brillouin zone this formulation is non-local. The hope to get
away with this was crushed by Karsten and Smit, who showed that this
formulation is inconsistent at the one-loop level of gauge theory,
where it fails to reproduce Lorentz symmetry in the continuum 
limit \cite{KarSmi}.\footnote{However, this conceptual problem
at the one-loop level is specific to gauge interactions. The
SLAC fermion may still be in business for instance in supersymmetric
spin models without gauge fields \cite{Jena}.} \\

The standard lattice fermion formulation, which has been used most 
in simulations --- in QCD in particular --- was put 
forward by K.G.\ Wilson in 1979 \cite{Wilfer}.
The free {\em Wilson operator} reads
\be  \label{Wilsonop}
D_{{\rm W}, xy} 
= \frac{1}{2} \sum_{\mu} \Big[ \gamma_{\mu} (\delta_{x,y - \hat \mu}
- \delta_{x,y + \hat \mu}) - (\delta_{x,y - \hat \mu} + \delta_{x,y + \hat \mu}
- 2 \delta_{x,y}) \Big] + m \delta_{x,y} \ .
\ee
Wilson added the term in the second round bracket to the naive form
that we considered before. This term represents a Laplacian operator,
which is discretised in the simplest way.\footnote{The Wilson term 
can also be multiplied by some independent coefficient
(the Wilson parameter), but this generalisation is not particularly fruitful.}
In fact it avoids the fermion doubling 
by sending the doublers to the cutoff energy.
There is no doubt that this operator is local, and the Wilson term
is $O(a)$ suppressed, so one could hope that it does not distort the
continuum limit.

However, due to this extra term the chiral symmetry is broken expli\-citly,
$\{ D_{{\rm W}, m=0} \, , \gamma_{5} \} \neq 0$.
As interactions are switched on, this causes numerous problems.
In particular, a gauge field can be added as a set of link variables in 
the gauge group, which provides invariance under gauge transformation
of the matter fields on the sites. One often writes this compact
link variable as
\be
U_{\mu ,x} = \exp \Big( i \int_{x}^{x+\hat \mu} dy_{\mu} \, 
{\cal A}_{\mu}(y) \Big) \, \in \, {\rm \{ \, gauge ~ group \, \} } \ ,
\ee
which indicates a connection to the (non-compact) continuum gauge field
${\cal A}_{\mu}$. For non-Abelian gauge groups
this exponential is formulated as a path ordered product \cite{latbooks}.
Such a gauge field suppresses the terms $\delta_{x,y \pm \mu}$ in the
Wilson term, but not its last entry $2 \delta_{x,y}$. This different
treatment gives rise
to additive mass renormalisation. If one tries to approach the chiral
limit, where the renormalised fermion mass vanishes, 
one has to fine tune the bare mass to some value $m < 0$,
which compensates for the additive renormalisation.

A further (related) inconvenience for interacting Wilson fermions
is that the lattice artifacts can appear in $O(a)$ already\footnote{For 
the free Wilson fermions, the scaling artifacts are of $O(a^{2})$.}
(unless one adds another term to cancel the $O(a)$ artifacts ---
following Symanzik's program ---
which requires fine tuning again \cite{clover}).\\

Another formulation, which has been considered standard over the past
decades, and which is regularly applied in simulations, is known as the
{\em staggered fermions} (or Kogut-Susskind fermions)
\cite{KoSus}. An elegant way to construct
them starts from the naive action on a unit lattice,
\be
S[\bar \Psi, \Psi ] = \sum_{x} \Big[ \frac{1}{2} \sum_{\mu =1}^{d} 
\left( \bar \Psi_{x} \gamma_{\mu} U_{\mu ,x} \Psi_{x + \hat \mu} 
 - \bar \Psi_{x + \hat \mu} \gamma_{\mu} U_{\mu ,x}^{\dagger} \Psi_{x} \right)
+ m \bar \Psi_{x} \Psi_{x} \Big] \ ,
\ee
and performs the substitutions \cite{KawaSmi}
\be
\bar \Psi_{x} ' = \bar \Psi_{x} \gamma_{1}^{x_{1}} \dots 
\gamma_{d}^{x_{d}} \ , \quad
\Psi_{x} ' = \gamma_{d}^{x_{d}} \dots \gamma_{1}^{x_{1}} \Psi_{x} \ .
\ee
This leaves the mass term invariant and renders also the kinetic term
diagonal in the spinor space. Hence one may reduce the transformed spinors 
to a single component $\bar \chi$, $\chi$, and one obtains
\bea
S [\bar \chi , \chi ] &=& \sum_{x} \Big[ \frac{1}{2} \sum_{\mu =1}^{d} 
\Gamma_{\mu ,x} \left( \bar \chi_{x} U_{\mu ,x} \chi_{x + \hat \mu} 
 - \bar \chi_{x + \hat \mu} U_{\mu ,x}^{\dagger} \chi_{x} \right)
+ m \bar \chi_{x} \chi_{x} \Big] \ , \nn \\
\Gamma_{\mu ,x} &:=& (-1)^{x_{1} + x_{2} + \dots + x_{\mu -1}} \ .
\label{stagact}
\eea
This structure distinguishes two sublattices by
the criterion if $\sum_{\mu =1}^{d} x_{\mu}$ is even or odd, i.e.\
by the sign term
\be  \label{epsistag}
\epsilon (x) = (-1)^{x_{1} + \dots + x_{d}} \ .
\ee
The link variables $U_{\mu ,x}$ always connect sites belonging to distinct
sublattices. For $m=0$, 
the action (\ref{stagact}) is invariant under the transformations
\be
\bar \chi_{x} \to e^{\alpha \epsilon (x)} \bar \chi_{x} \ , \quad
\chi_{x} \to e^{\alpha \epsilon (x)} \chi_{x} \ ,
\ee
which amounts to a remnant chiral symmetry $U(1)_{\rm e} \otimes U(1)_{\rm o}$,
where $\epsilon (x)$ adopts the r\^{o}le of $\gamma_{5}$
(the subscripts refer to the even/odd sublattice).
The $2^{d}$ components on the corners of the disjoint unit hypercubes
are nowadays denotes as ``tastes'' (the earlier literature also called
them ``pseudoflavours'').
As long as one finally assembles exactly 4 flavours from them, this 
remnant symmetry is sufficient to avoid additive mass renormalisation 
and $O(a)$ scaling artifacts. 

Recently it became fashion to try to build single flavours with staggered
fermions by taking the fourth root of the fermion determinant (\ref{fermidet}). 
However, it is likely that this formulation is non-local, and --- even if someone
is willing to accept that --- additive mass renormalisation sets in again
(see e.g.\ Refs.\ \cite{AHH}). The question if this formulation might
provide correct results even if it is non-local is still debated \cite{SRS}.

\subsection{Are light fermions natural ?}

We would like to stress that keeping track of
the chiral symmetry  is {\em not} a specific problem of the lattice 
regu\-larisation. It should rather be viewed as a generic and deep
problem, which plagues other regularisations as well.
For instance, in dimensional regularisation \cite{BolGiam} one
performs computations
in $4 + \varepsilon$ dimensions (in the sense of distribution theory) 
and sends $\varepsilon \to 0$ at the end. This is the most popular
regularisation scheme for perturbative calculations, but it is a longstanding
problem to find a generally reliable rule for
handling $\gamma_{5} = \prod_{\mu =1}^{d} \gamma_{\mu}$ on the
regularised level (or $i \prod_{\mu =0}^{d-1} \gamma_{\mu} $ in the
Minkowski signature), generalising eq.\ (\ref{gamma5E}) (resp.\ 
(\ref{gamma5})).
A careful analysis of this issue can be found 
in Ref.\ \cite{Fred}. \\

On the conceptual level, this observation means that
the existence of light fermions in our world
actually appears to be unnatural. Nature must be non-perturbative, 
so we can only refer to a non-perturbative regularisation
scheme when addressing this question, which means essentially the
lattice.\footnote{A conceivable alternative might be the formulation
on a ``fuzzy sphere'' \cite{Bal}. However, even simulations
of models without fermions \cite{fuzzysimu} 
show that it is not obvious to recover the 
desired conti\-nuum limit in these formulations.}
It is possible to formulate light lattice 
fermions --- for instance light quarks in lattice QCD ---
but only with tedious and sophisticated constructions (see Section 7),
which do not appear to mimic a conceivable mechanism in Nature.
What could be an acceptable mimic is something very simple like the 
Wilson fermion (\ref{Wilsonop}), which, however, pushes the fermion mass
to the cutoff scale due to a strong additive mass renormalisation ---
unless a negative bare mass is fine tuned, which appears unnatural 
again.\footnote{In the light of the properties mentioned in the last paragraph 
of Subsection 3.4, the staggered fermion formulation cannot really
be considered a solution to this problem either.}

Nevertheless there do exist in particular two light quark flavours,
\be
m_{u}, \, m_{d} \ll \Lambda_{\rm QCD} \ .
\ee
The question how this is realised in Nature
is a hierarchy problem that is not properly understood. 
The situation is different in pure Yang-Mills theory, for example, 
where (regularised) glueball masses can be made arbitrarily
light thanks to asymptotic freedom and the absence of additive mass
renormalisation. But for quarks it does not work
in this simple way, due to the problems to keep track of an approximate
chiral symmetry in a regularised system, such that the fermion mass is far
below the cutoff. In a broader framework, this hierarchy problem
raises the question why hadron masses are far below the Planck scale,
and therefore why we do not just consist of gluons.

In Ref.\ \cite{branes} we studied the question if this problem 
could be solved (qualitatively) in a {\em brane world.}
As a toy model, our target theory was the 2d Gross-Neveu model 
\cite{GNorig} with the (continuum) action 
\be  \label{act2d}
S [\bar \Psi , \Psi ] = \int d^{2}x \Big[ \bar \Psi \gamma_{\mu}
\partial_{\mu} \Psi - \frac{g}{2N} (\bar \Psi \Psi)^{2} \Big] \ ,
\ee
where we suppress the flavour index $1 \dots N$. It enjoys a discrete
chiral $Z(2)$ symmetry
\be  \label{chi-sym}
(\bar \Psi_{L}, \Psi_{L}) \to \pm (\bar \Psi_{L}, \Psi_{L}) \ , \quad
(\bar \Psi_{R}, \Psi_{R}) \to \mp (\bar \Psi_{R}, \Psi_{R}) 
\ee
(the chiral components are defined in eq.\ (\ref{hands})).
With an auxiliary scalar field $\Phi $ 
the action (\ref{act2d}) is equivalent to
\be
S [\bar \Psi , \Psi , \Phi ] = \int d^{2}x \Big[ \bar \Psi \gamma_{\mu}
\partial_{\mu} \Psi - \Phi \bar \Psi \Psi + \frac{N}{2g} \Phi^{2} \Big] \ ,
\ee
as we see by integrating out the field $\Phi$.
The sign of $\Phi$ flips under a $Z(2)$ chiral transformation (\ref{chi-sym}).

In the limit $N \to \infty$, $\Phi$ freezes to a constant \cite{GNorig}, 
and $\bar \Psi , \Psi$ can be integrated out. The resulting fermion
determinant gives rise to an effective potential,
\be
\int D \bar \Psi D \Psi \, e^{-S[\bar \Psi , \Psi , \Phi ]} =
e^{-N\cdot V \cdot V_{\rm eff}(\Phi )} \ .
\ee
In a large volume $V$, the minima $\pm \Phi_{0}$ of $V_{\rm eff}$ obey 
the gap equation
\be  \label{gapeq}
\frac{1}{g} = \frac{1}{\pi} \int_{0}^{\Lambda_{2}} dk \,
\frac{k}{k^{2} + \Phi_{0}^{2}} \quad
\qquad ( k = \sqrt {k_{1}^{2} + k_{2}^{2}} )
\ .
\ee
At weak coupling $g \ll 1$ we are dealing with
a cutoff $\Lambda_{2} \gg \Phi_{0}$ and
\be  \label{asymfree}
m = \Phi_{0} = \Lambda_{2} \, e^{-\pi /g}
\ee
represents the  fermion mass, which is generated by the spontaneous
breaking of the $Z(2)$ symmetry (\ref{chi-sym}).
The exponent in eq.\ (\ref{asymfree}) expresses asymptotic freedom. \\

Let us proceed to three dimensions, where the action
\be  \label{act3d}
S [\bar \Psi , \Psi ] \! = \!\! \int \!\!
d^{3}x \Big[ \bar \Psi \gamma_{\mu}
\partial_{\mu} \Psi + \bar \Psi \gamma_{3} \partial_{3} \Psi
- \frac{G}{2N} (\bar \Psi \Psi)^{2} \Big] \ , \quad (\mu =1,2)
\ee
still has a $Z(2)$ symmetry,
\bea
(\bar \Psi_{L}, \Psi_{L})\vert_{(\vec x, x_{3})} & \to & 
\pm (\bar \Psi_{L}, \Psi_{L})\vert_{(\vec x, -x_{3})} \ ,
\nn \\  \label{Z2sym}
(\bar \Psi_{R}, \Psi_{R})\vert_{(\vec x, x_{3})} & \to &
\mp (\bar \Psi_{R}, \Psi_{R})\vert_{(\vec x, -x_{3})} \ ,
\eea
which turns into the discrete chiral symmetry (\ref{chi-sym}) 
after dimensional reduction. The 3d gap equation reads
\be  \label{3dgap}
\frac{1}{G} = \frac{1}{(2 \pi )^{3}} \int d^{3}k \, \frac{2}{k^{2}+
\Phi_{0}^{2}} \ ,
\ee
and for a cutoff $\Lambda_{3} \gg \Phi_{0}$ one identifies a critical
coupling $G_{c} = \pi^{2}/\Lambda_{3}$. At $G > G_{c}$ we are in a
phase of broken $Z(2)$ symmetry with 
\be
\Phi_{0} = 2\pi \Big( \frac{1}{G_{c}} - \frac{1}{G} \Big) \ ,
\ee
whereas weak coupling ($G \le G_{c}$) corresponds to a symmetric 
phase ($\Phi_{0}=0$). \\

%

Canonical dimensional reduction from 3 to 2 dimensions works in the usual way if 
we start from the 3d symmetric phase, and it leads to light 2d fermions 
\cite{branes}. However, this is not satisfactory
in view of our motivation: for instance a non-perturbative treatment
at finite $N$ (on the lattice) should not start from the symmetric phase,
because this just shifts the problem of fine tuning to $d=3$.
Therefore we focus on {\em dimensional reduction from the broken phase}. \\

We denote the (periodicity) extent of the third direction by $\beta$,
and $\xi = 1/m$ is the correlation length. 
Starting from the 3d broken phase, the limit
\ $^{\, \lim}_{\beta \to 0} \ \beta /\xi = 2 \ln (1 + \sqrt{2})$
does not provide light fermions.  Hence we proceed differently
and generate a light 2d fermion as the $k_{3}=0$ mode on a brane.
For the latter we make the ansatz $\Phi (x_{3}) = 
\Phi_{0} \, \tgh (\Phi_{0} x_{3})$, which is inspired by Refs.\ \cite{DHN}.
We choose $x_{2}$ as the time direction, hence the Hamiltonian reads
\be
\hat H = \gamma_{2} [ \gamma_{1} \partial_{1} + \gamma_{3} \partial_{3} 
- \Phi (x_{3}) ] \ .
\ee
The ansatz $\Psi(x_{3}) e^{ik_{1}x_{1}} e^{-iEt}$
(and the chiral representation for $\gamma_{i}$) 
reveals one localised eigenstate of $\hat H$,
\be
\Psi_{0}(x_{3}) = \sqrt{\frac{\Phi_{0}}{2}} \left(
\begin{array}{c}
0 \\ \cosh^{-1}(\Phi_{0}x_{3}) \end{array} \right)
\ee
with energy $E_{0} = -k_{1} >0$, i.e.\ a left-mover.
(On an anti-brane $-\Phi (x_{3})$
one obtains a right-mover with $E_{0}=k_{1}>0$
and exchanged components in $\Psi_{0}(x_{3})$).

In addition there are bulk states (not localised in $x_{3}$),
\be
\Psi_{k_{3}}(x_{3}) =
\frac{e^{ik_{3}x_{3}}}{\sqrt{2E(E+k_{1})}} \left( \!\!
\begin{array}{c}
i (E + k_{1}) \\ 
\Phi_{0} \tgh (\Phi_{0} x_{3}) -i k_{3}
\end{array} \!\! \right)
\ee
with $E = \pm \sqrt{\vec k^{2} + \Phi_{0}^{2}}$,
which form together with $\Psi_{0}$ an orthonormal 
basis for the 1-particle Hilbert space.

To verify the consistency of the brane profile we have to consider
the chiral condensate $- \bar \Psi \Psi$. $\Psi_{0}$ does not contribute to it,
and if we sum up the bulk contributions of $E<0$ we reproduce exactly
$\Phi (x_{3})$, which confirms the self-consistency of this single brane
world.

In addition we are free to fill some of the $\Psi_{0}$ states.
Those with $E_{0} < \Phi_{0}$ are confined
to the $(1+1)$-d world, whereas states with $E_{0} \geq \Phi_{0}$ can escape
in the 3-direction. For the low energy observer on the brane this event
appears as a fermion number violation.\\

We now want to include both, $\Psi_{L}$ and $\Psi_{R}$, to be localised
on a brane and an anti-brane, and $\beta$ now denotes their separation. 
For the corresponding profile we make the ansatz
\bea
\Phi (x_{3}) &=& \Phi_{0} ( a [ \tgh_{+} - \tgh_{-}] -1 ) \ , \nn \\
\tgh_{\pm} & := & \tgh (a \Phi_{0} [x_{3} \pm \beta /2 ])\ , \
a \in [0,1] \ .  \label{profileq}
\eea
The ansatz for a bound state with the same form as on single branes,
\be
\Psi_{0}(x_{3}) = c \left(
\begin{array}{c} \alpha_{1} \cosh^{-1}(a \Phi_{0}[x_{3} - \beta /2]) 
\\ \alpha_{2} \cosh^{-1} (a \Phi_{0}[x_{3} + \beta /2])
\end{array} \right) , 
\label{ansatzwaw}
\ee
implies the condition $\tgh (a \Phi_{0} \beta ) = a$.
Hence the parameter $a$ controls the brane separation, such that $a\to 0$
and $a \to 1$ correspond to $\beta \to 0$ and $\beta \to \infty$,
respectively.

\begin{figure}[h!]
  \centering
  \includegraphics[angle=270,width=0.6\linewidth]{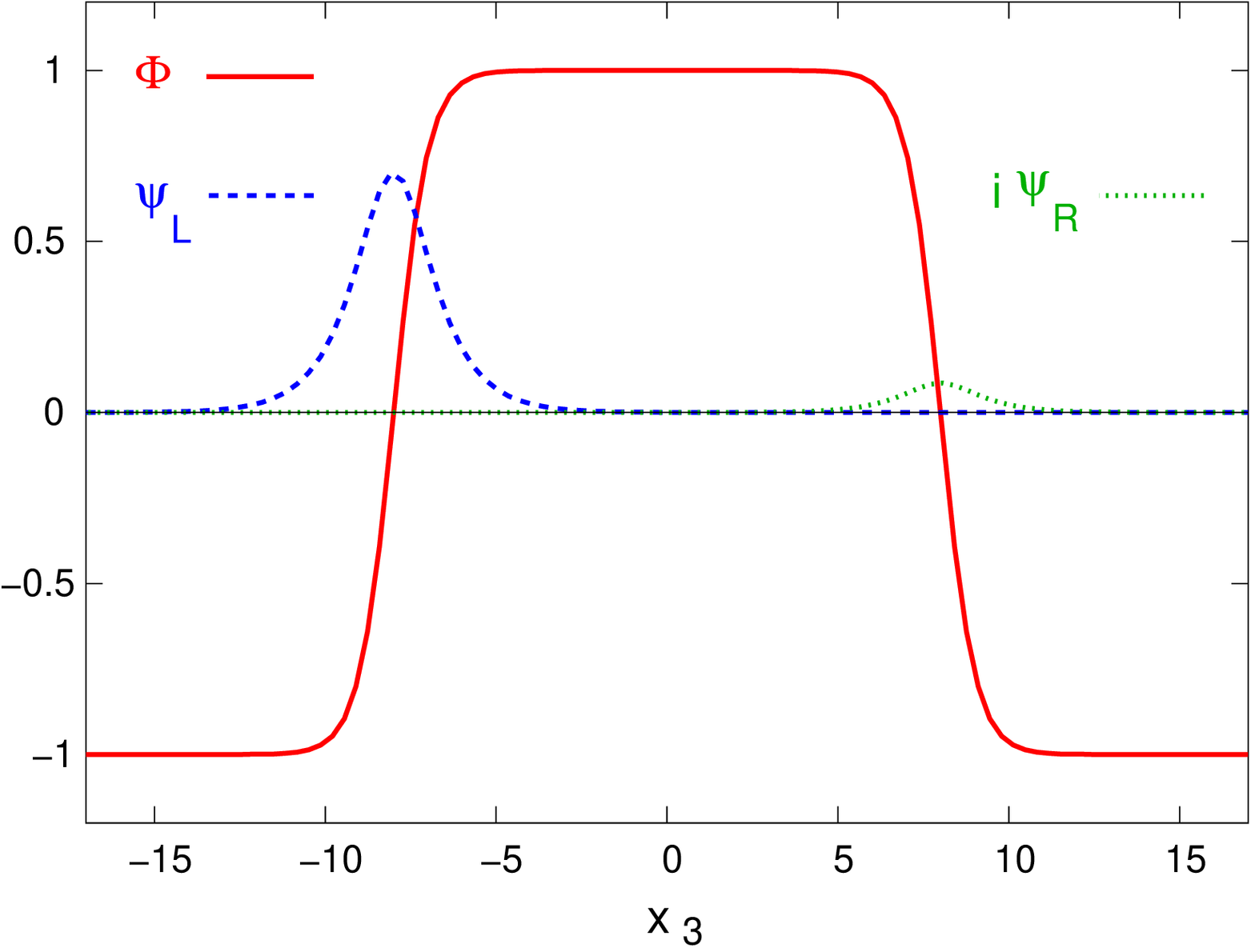}
  \includegraphics[angle=270,width=0.6\linewidth]{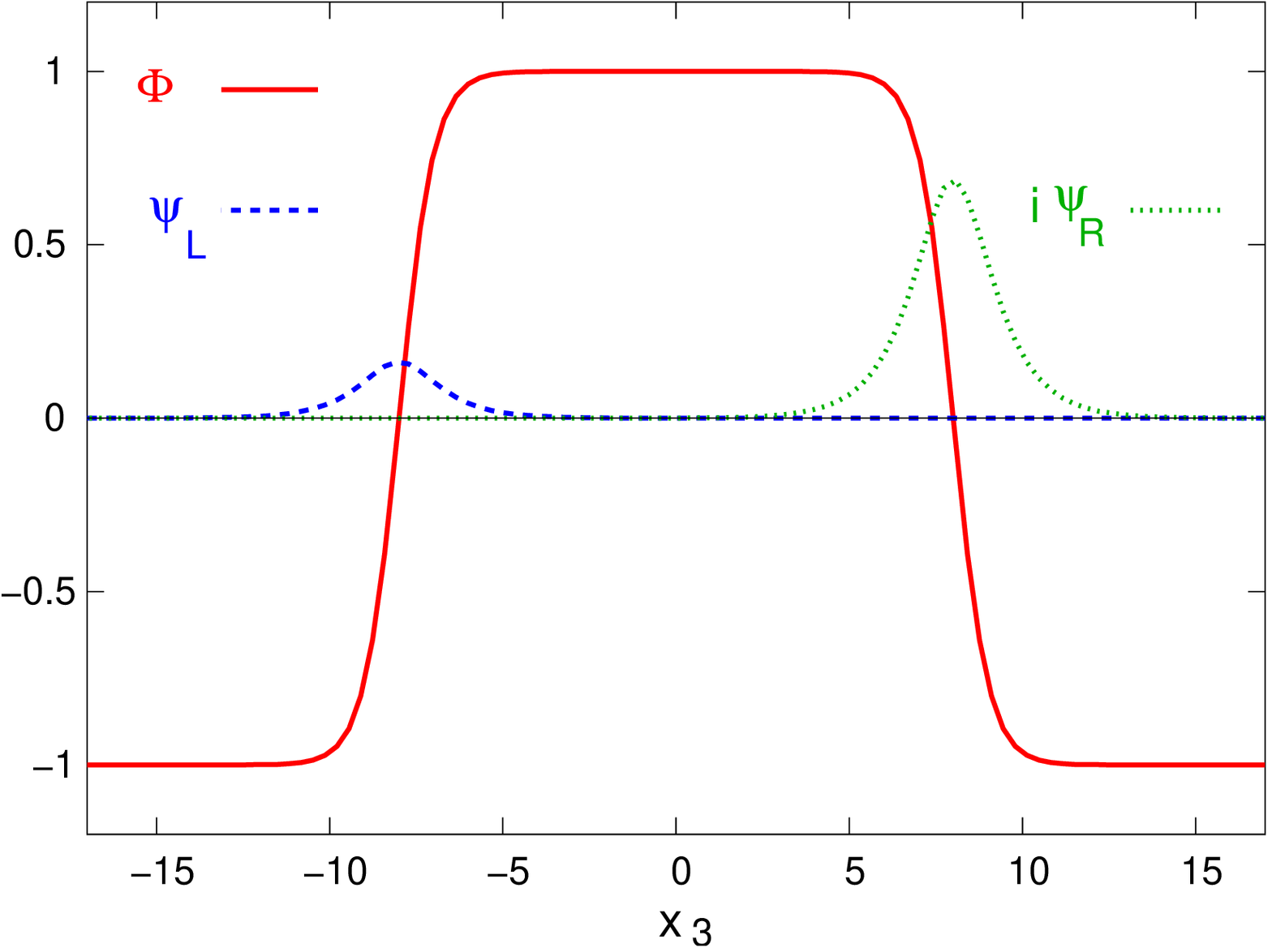}
\caption{{\it We show the brane anti-brane profile (\ref{profileq})
in the 3-direction of our toy brane world \cite{branes}. The 2-direction is
the (Euclidean) time, and the left-handed (right-handed) fermion moves
to the left (right) in the one spatial direction inside the brane world.
In the plot above it does so with a momentum $p_{1} = -4m$ and below with
$p_{1} = 2 m$, where $m$ is the fermion mass. The latter arises from the
communication between the left- and right-handed components, which 
are localised on the brane and anti-brane. As they drift apart, $m$
decreases exponentially in the distance, hence a low energy observer
on the brane perceives them on top of each other as a point-like
Dirac spinor. }}
\label{branefig} 
\end{figure}

The Dirac equation in this background still has an analytic
solution, which is given by the ansatz (\ref{ansatzwaw}) with
\bea
&& \hspace*{-5mm} c = \frac{1}{2} \sqrt{ \frac{a \Phi_{0}}
{E_{0} ( E_{0} + k_{1})} } \ , \ \
E_{0} = \pm \sqrt{k_{1}^{2} + m^{2}} \ , \nn \\
&& \hspace*{-5mm} \alpha_{1} = -i (E_{0} + k_{1}) \ , \ \
\alpha_{2} = m = \sqrt{1 - a^{2}} \, \Phi_{0} \ .
\eea
The resulting $\Psi_{0}(x_{3})$ represents a 
Dirac fermion with components $\Psi_{L}$,
$\Psi_{R}$ localised on the brane resp.\ the anti-brane.
For a fast motion to the left (right) we have $0 < E_{0} \simeq
-k_{1} \ (+k_{1})$, so that the lower (upper) component dominates.
This situation is sketched in Figure \ref{branefig}.
The mass $m$ measures the extent of the $L,R$ mixing. The limit $a \to
0$ does not provide a light fermion ($m= \Phi_{0}$), but
the opposite limit $a \to 1$ achieves this, since the $L,R$
mixing is suppressed as
\be
m \simeq 2 \, \Phi_{0} \, e^{- \beta \Phi_{0}} \ .
\ee
Counter-intuitively, {\em large} $\beta$ implies $\xi \gg \beta$ and 
therefore dimensional reduction. A low energy observer in $d=1+1$ now
perceives a point-like Dirac fermion composed of $L-$ and $R-$modes.
On the other hand, a high energy observer in $d=2+1$ refers to the scale
$\Phi_{0}$ (the 3d fermion mass) and observes a Dirac fermion with 
strongly separated $L-$ and $R-$constituents.

Also the bulk states can be determined analytically,
\bea
\Psi_{k_{3}}(x_{3}) &=& \frac{e^{ik_{3}x_{3}}}{\sqrt{U}} \left(
\begin{array}{c}
~~~ i (E+k_{1}) [a \Phi_{0} \tgh_{-} - ik_{3}] \\
- (\Phi_{0} + ik_{3}) [a \Phi_{0} \tgh_{+} -ik_{3}]
\end{array} \right) \ , \\
U &:=& 2E (E+k_{1})(k_{3} + a^{2} \Phi_{0}^{2}) \ , \
E = \pm \sqrt{ k_{1}^{2} + k_{3}^{2} + \Phi_{0}^{2} } \ . \nn
\eea
Summing up again their $E<0$ contributions to $- \bar \Psi \Psi$ yields
\be
- \frac{G}{N} \int dk_{3} \Psi_{k_{3}} \bar \Psi_{k_{3}} \Big\vert_{E<0}
= \Phi(x_{3}) + {\cal C} \ ,
\ee
i.e.\ the desired result up to a term ${\cal C}$, which is given
explicitly in Ref.\ \cite{branes}.
It has to be cancelled by occupying bound states
in $\Psi_{0}$, which do contribute this time. 
This requires all the bound states
with energies $E_{0} \le E_{F}$ to be filled. The Fermi energy turns
out to be $E_{F} = \Phi_{0}$, i.e.\ 
exactly the threshold energy for the escape into the third dimension.
Hence this brane anti-brane world does contain naturally
light fermions, but it is completely packed with them, so its
physics is blocked by Pauli's principle. \\

Since this brane anti-brane world is not topologically stable,
we also checked if the brane and anti-brane repel or attract
each other, which could lead to disastrous scenarios.
However, it turns out that the brane tension energy per fermion
does not depend on the brane
separation, so this toy world is indeed stable \cite{branes}.

Finally we studied the possibility of adding a fermion mass term
$M \bar \Psi \Psi$ to the Lagrangian, so that the $Z(2)$ symmetry is
also {\em explicitly} broken in $d=3$ (which is actually realistic for a 
lattice formulation at finite $N$). This lifts the degeneracy of the
minima of $V_{\rm eff}(\Phi )$. If we still insert the profile
(\ref{profileq}), the condition 
for $- \bar \Psi \Psi$ requires the bound fermion states to be
filled even beyond $\Phi_{0}$, hence in this case there is no
stable configuration at all.

One might also start from the symmetric phase and add a mass term
to construct a somehow natural starting point. However, such a mass is
simply inherited by the dimensionally reduced model (while the cutoff
keeps the same magnitude), hence this does not
solve the hierarchy problem under consideration.\\

%
%
%

We drop this mass term again and summarise Subsection 3.5 by repeating
that the construction of naturally light fermions is basically successful, 
but unfortunately this world does not enjoy any flexibility for physical 
processes. However, we assumed translation invariance in the 2d world so far.
That symmetry may be broken at sufficiently large chemical potential,
so that the chiral condensate prefers a kink anti-kink pattern, rather
than a constant \cite{Theis}. This could possibly provide the
missing flexibility for a lively brane world of that kind.

\section{Perfect Actions for Lattice Fermions}

\subsection{Free fermions}

In the previous Section we have described the severe
conceptual difficulties with the formulation of fermions
on the lattice. We now proceed to the application of the RGT
technique --- described in Section 2 --- to lattice
fermions. This is going to reveal how the perfect action
handles --- and solves ---
the problems of species doubling and chiral symmetry.

We start with the free fermion and apply immediately the blocking from 
the continuum (introduced in Subsection 2.2), which is most efficient
for analytic calculations.
In analogy to eq.\ (\ref{trafocont}) we now relate
lattice spinor fields $\bar \Psi_{x}$, $\Psi_{x}$ to their counterparts
in the continuum,
\be  \label{relatedto}
\bar \Psi_{x} \sim \int_{C_{x}} d^{d}u \, \bar \psi (u) \ , \quad
\Psi_{x} \sim \int_{C_{x}} d^{d}v \, \psi (v) \ .
\ee
This relation is imposed by the RGT, which leads to the perfect lattice
action $S[ \bar \Psi , \Psi ]$ for free lattice fermions,
\bea
&& \hspace*{-1cm} e^{-S [ \bar \Psi , \Psi ]} = 
\int {\cal D} \bar \psi {\cal D} \psi \, e^{-s [ \bar \psi , \psi ]} \times
\nn \\
&& \hspace*{-1cm} \exp \Big\{ - \sum_{xy} \Big[ \bar \Psi_{x}^{i} -
\int_{C_{x}} d^{d}u \, \bar \psi^{i}(u) \Big] (R^{-1})_{xy}^{ij}
\Big[ \Psi_{y}^{j} - \int_{C_{y}} d^{d}v \, \bar \psi^{j}(v) \Big] \Big\} \ , 
\quad \ \label{fermRGT}
\eea
where $s [ \bar \psi , \psi ] = \int d^{d}u \, \bar \psi (u) 
[ \gamma_{\mu} \partial_{\mu} + m] \psi (u)$ is the continuum action 
(cf.\ eq.\ (\ref{contfermact})). On the lattice (of spacing $a=1$)
we arrive at the following perfect action and propagator
\bea
S [ \bar \Psi , \Psi ] &=& \frac{1}{(2 \pi )^{d}} \int_{B}
d^{d}p \, \bar \Psi (- p) G(p)^{-1} \Psi (p) \ , \nn \\
G(p) &=& \sum_{l \in \Z^{d}} \frac{\Pi( p + 2 \pi l )^{2}}
{i \gamma_{\mu} ( p_{\mu} + 2\pi l_{\mu}) + m} + R(p) \ , \label{perfermiprop}
\eea
where the function $\Pi (p)$ is defined in eq.\ (\ref{perfscalprop})
(also here it could be generalised).
This formula has been computed in various ways 
\cite{Wie92,Lat93,QuaGlu}.\footnote{Later on it turned out that
the perfect propagator $G(p)$
was already discussed in the Ref.\ \cite{GiWi}. However,
that work was forgotten until it was accidentally
re-discovered by P.\ Hasenfratz  in 1997 \cite{Has97}.}
It incorporates  the continuum propagator, its periodic copies
and the blocking term, in full analogy to the perfect action for
free scalars in eq.\ (\ref{perfscalprop}).

Let us now discuss the r\^{o}le of the blocking term, which we
have generalised from the constant $1 / \alpha$ in the scalar case
(eq.\ (\ref{scalRGT})) to the form $R_{xy}^{ij}$. 
For sure we have to require $R$ to be
{\em local}. Thus it cannot disturb the pole structure of $G(p)$.
Hence the formulation is free of doublers, and the dispersion
relation\footnote{We repeat that one always considers the 
branch with the lowest energy, cf.\ Subsection 2.2.} 
coincides with the continuum.

In the limit $R \to 0$ we perform a {\em $\delta$-function blocking,}
as we mentioned for the scalar fields before. Then the 
relations (\ref{relatedto}) turn into equations. In this case
(or more generally, whenever $\{ R , \gamma_{5} \}$ vanishes), 
$G(p)_{m=0}$ --- and therefore also the Dirac 
operator $D(p)_{m=0}$ --- anti-commutes exactly with $\gamma_{5}$.
Then we have chirality, i.e.\ invariance under the
global transformation (\ref{chiralrot}),
just as in the continuum. Hence the question
arises in which way a contradiction to the (mathematically rigorous)
Nielsen-Ninomiya Theorem
\cite{NoGo} is avoided. The answer is that in this case the Dirac
operator is {\em non-local} \cite{Wie92,Lat93}: 
it does not decay exponentially, but only as
\cite{QuaGlu}
\be
D(r)_{m=0} \propto \frac{1}{r^{d-1}} \ .
\ee
As soon as we proceed to some non-vanishing, local term $R$, which obeys
\be  \label{Rnonzero}
\{ R , \gamma_{5} \} \neq 0 \ ,
\ee
locality is restored. However, this obviously leads to 
\be  \label{nonanticom}
\{ D(p)_{m=0}, \gamma_{5} \} \neq 0 \ ,
\ee
hence we do not have chirality in the standard form (\ref{chiralrot})
anymore.

Still, this breaking of the chiral symmetry can only be superficial:
we know that the RGT does not distort any physical properties, hence
the chirality of the continuum must be preserved in the physical
obser\-vables, despite the relation (\ref{nonanticom}), as
we emphasised at numerous occasions \cite{perfaxial,Lat95,QuaGlu,Buckow}.
Therefore this must be a specifically harmless anti-commutator. Indeed it gave 
the crucial clue for a general criterion for the form of such a non-vanishing
anti-commutator \cite{Has97}, which is still compatible with chiral symmetry 
in a lattice modified form \cite{ML98}. This criterion is now denoted as the 
Ginsparg-Wilson relation (since it was already mentioned
in Ref.\ \cite{GiWi}), which we will discuss in Section 7.

We assume $R$ to have the structure of a Dirac scalar,
and we move to coordinate space. The perfect action 
for the free fermion is given in the form
\bea
S [ \bar \Psi , \Psi ] &=& \sum_{x,y} \bar \Psi_{x} D_{xy} \Psi_{y} \ , 
\nonumber \\
D_{xy} &=& (G^{-1})_{xy} = \gamma_{\mu} \rho_{\mu} (x-y)
+ \lambda (x-y) \ , 
\label{perfactcoor}
\eea
i.e.\ it consists of a vector term plus a scalar term.
We consider the local case (\ref{Rnonzero}), where $\rho_{\mu} (x-y)$
and $\lambda (x-y)$ decay exponentially in the distance $|x-y|$.
For practical purposes we need a truncation in these couplings,
and we follow again the scheme of Section 2: we first optimise
$R$, for the case $R_{xy}^{ij} = \rho \, \delta_{xy} \delta^{ij}$. 
An analytic calculation in $d=1$ suggests the choice
\be  \label{rhom}
\rho (m) = \frac{e^{m} - m -1}{m^{2}} \ .
\ee
Only for this form of $\rho (m)$ the 1d couplings are limited to
nearest neighbour sites, i.e.\ they take the structure of $D_{\rm W}$.
In $d\geq 2$ couplings over all distances are inevitable, but the choice
(\ref{rhom}) still provides practically optimal locality, i.e.\
optimally fast exponential
decays of the functions $\rho_{\mu}(x-y)$ and $\lambda (x-y)$;
this is illustrated in Ref.\ \cite{QuaGlu}.

As a truncation scheme, we computed for this function
$\rho (m)$ the couplings of perfect actions in a 
periodic $3^{4}$ lattice, and applied these couplings in larger volumes 
\cite{Lat96}. This yields the free {\em hypercube fermion} (HF), 
which still has the structure of eq.\ (\ref{perfactcoor}), but now
with strictly limited supports for the ingredients to the Dirac
operator,
\bea
&& \hspace*{-1cm} D_{{\rm HF},xy} 
= \gamma_{\mu} \rho_{\mu} (x-y) + \lambda (x-y) \ , \nn \\
&& \hspace*{-1cm} {\rm supp} [\rho_{\mu} (x-y)], \ {\rm supp}[ \lambda (x-y)]
\subset \Big\{ x,y \ \Big\vert \ | x_{\mu} - y_{\mu}| \leq 1 \ , \ \forall \mu 
\Big\} \ . \quad \label{HFparam}
\eea
Tables for the explicit couplings for such HFs at various masses are
given in Ref.\ \cite{Lat96}; in Table \ref{HFtab} we display here the HF 
couplings at $m=0$ to an extended precision of 16 digits.

\begin{table}
\centering
\begin{tabular}{|c||c|c|}
\hline
$r$ & $\rho_{1}(r)$ & $\lambda (r)$ \\
\hline
\hline
$(0,0,0,0)$ & 0 & $1.852720547165511$ \\
\hline
$(1,0,0,0)$ & $0.1368467943177540$ & $-0.060757866428667176$ \\
\hline
$(1,1,0,0)$ & $0.032077284302446526$ & $-0.030036032105554878$ \\
\hline
$(1,1,1,0)$ & $0.011058131255574036$ & $-0.015967620416694967$ \\
\hline
$(1,1,1,1)$ & $0.0047489906005042248$ & $-0.0084268119917885868$ \\
\hline
\end{tabular}
\caption{{\it The couplings of the free, massless HF with the parameterisation
of eq. (\ref{HFparam}). $\rho_{\mu}(r)$ is anti-symmetric in 
$r_{\mu}$ and symmetric in all other components $r_{\nu}$, while
$\lambda (r)$ is symmetric in all directions.}}
\label{HFtab}
\end{table}

\begin{figure}[h!]
\vspace*{-3mm}
  \centering
  \includegraphics[angle=0,width=0.6\linewidth]{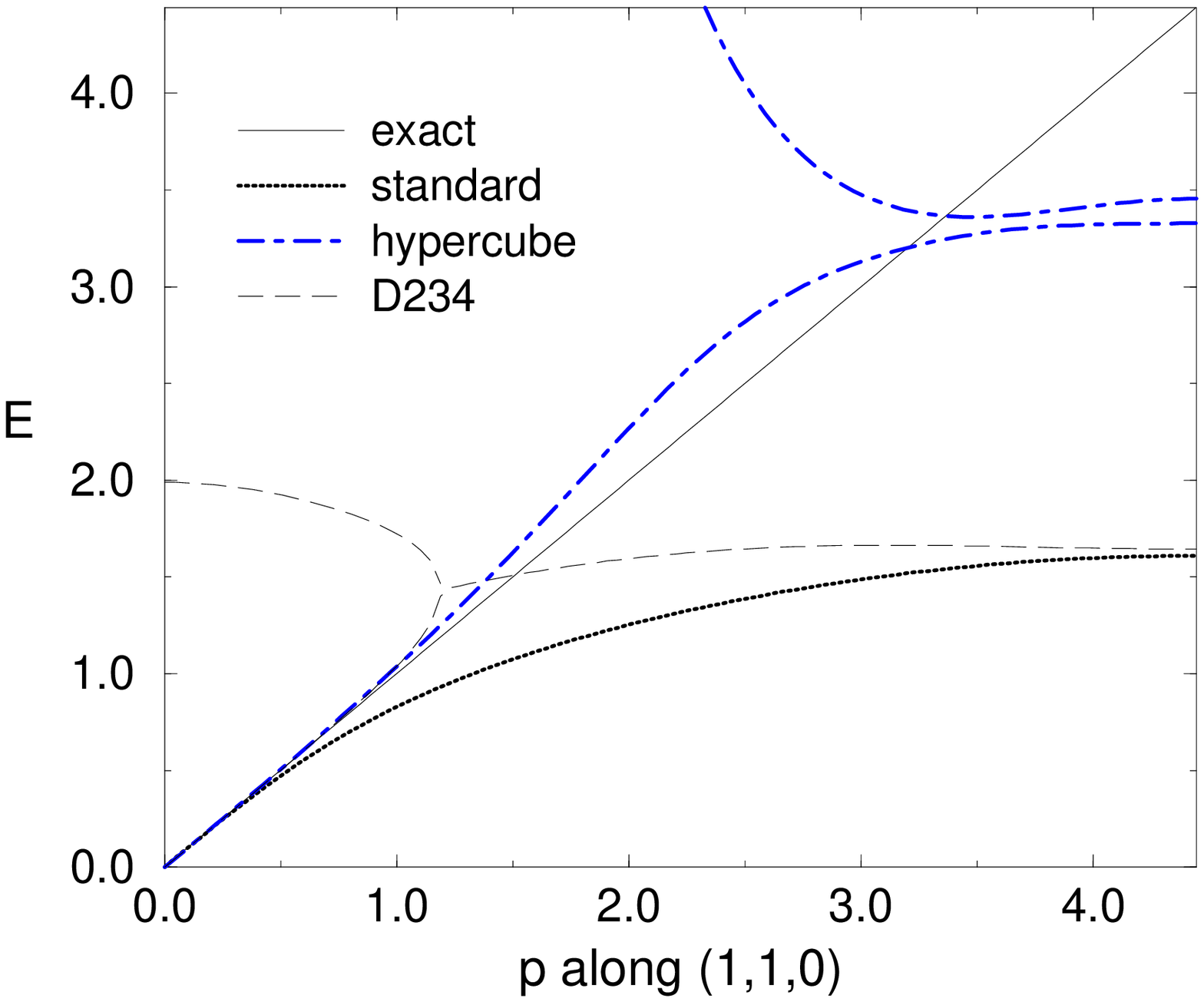}
  \includegraphics[angle=0,width=0.6\linewidth]{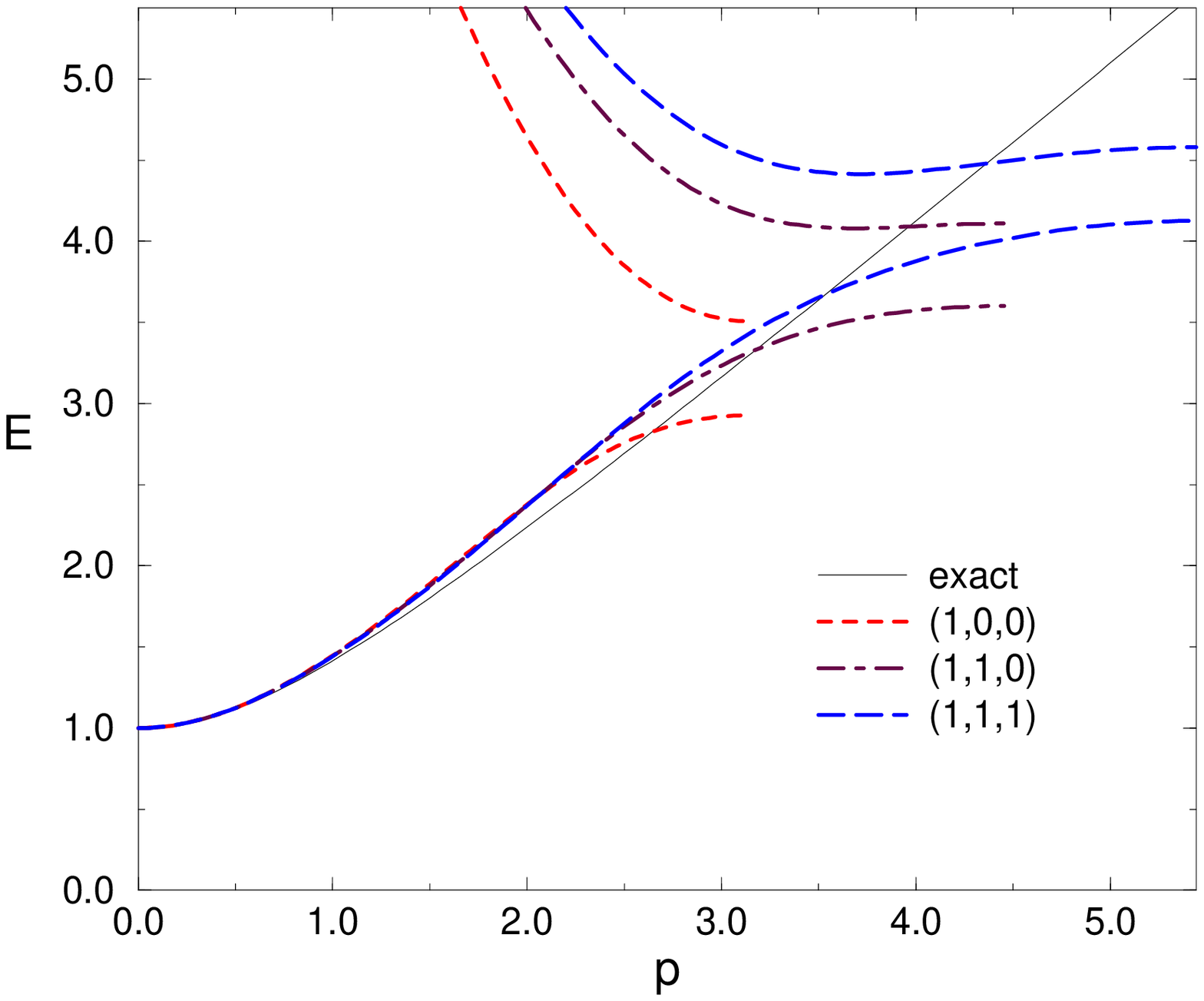}
\vspace*{-4mm}
\caption{{\it On top: 
The dispersion relation for free, massless lattice fermions
in $d=4$ for spatial momenta $\vec p \propto (1,1,0)$ (as an example).
For the perfect fermion the dispersion coincides with the exact
dispersion in the continuum, and the HF dispersion follows it closely.
The standard Wilson fermion deviates strongly at increasing momenta,
while the Symanzik improved D234 fermion behaves well up to
$| \vec p \, | \approx 1$, before it hits a doubler coming down from
higher energy. \newline
Below: Dispersion relation for the free HF at mass $m=1$. Here we show
the energy $E$ for various directions of the momentum $\vec p$ 
($p = | \vec p \, |$) to illustrate that they all follow closely the
continuum dispersion over a sizable part of the Brillouin zone.}}
\vspace*{-1mm}
\label{fermscale}
\end{figure}

After truncation, the scaling 
behaviour is still by far superior to the Wilson fermion, and also
to the so-called D234 fermion \cite{D234}, which is improved to the leading
order in the lattice spacing, following Symanzik's program.
A comparison of the dispersion relations at mass $m=0$ and $1$ is
shown in Figure \ref{fermscale}.
We see a striking improvement for the truncated perfect action.

\begin{figure}[h!]
  \centering
  \includegraphics[angle=0,width=0.6\linewidth]{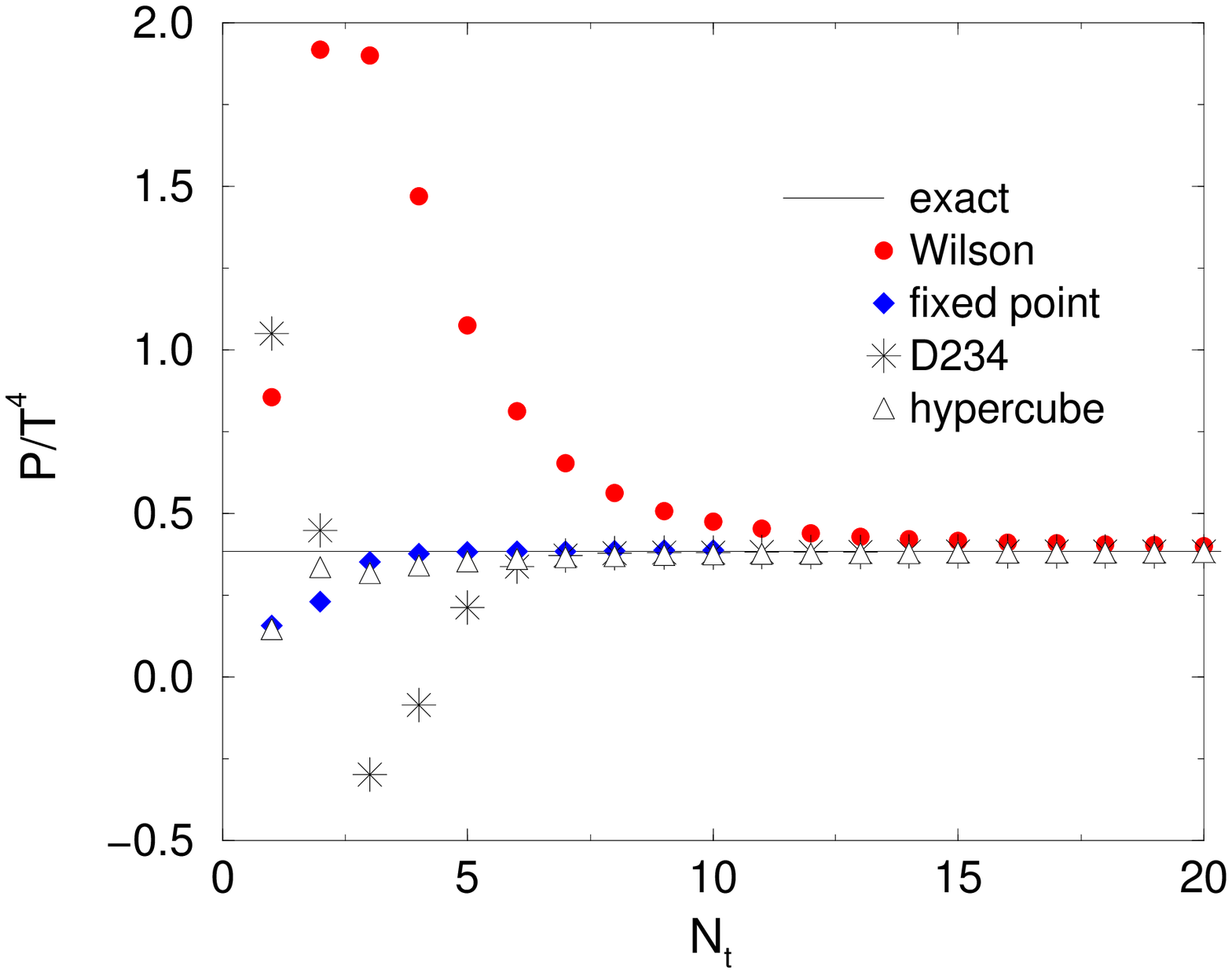}
\caption{{\it The ratio between ${\rm pressure}$ and ${\rm (temperature)}^{4}$
for various types of free lattice fermions, compared to the Stefan-Boltzmann
law in the continuum. The RGT improved actions converge much faster
to this value for decreasing temperature (increasing $N_{t}$)
than the Wilson action or the D234 action.}}
\label{presferm}
\end{figure}

\begin{figure}[h!]
  \centering
  \includegraphics[angle=270,width=0.48\linewidth]{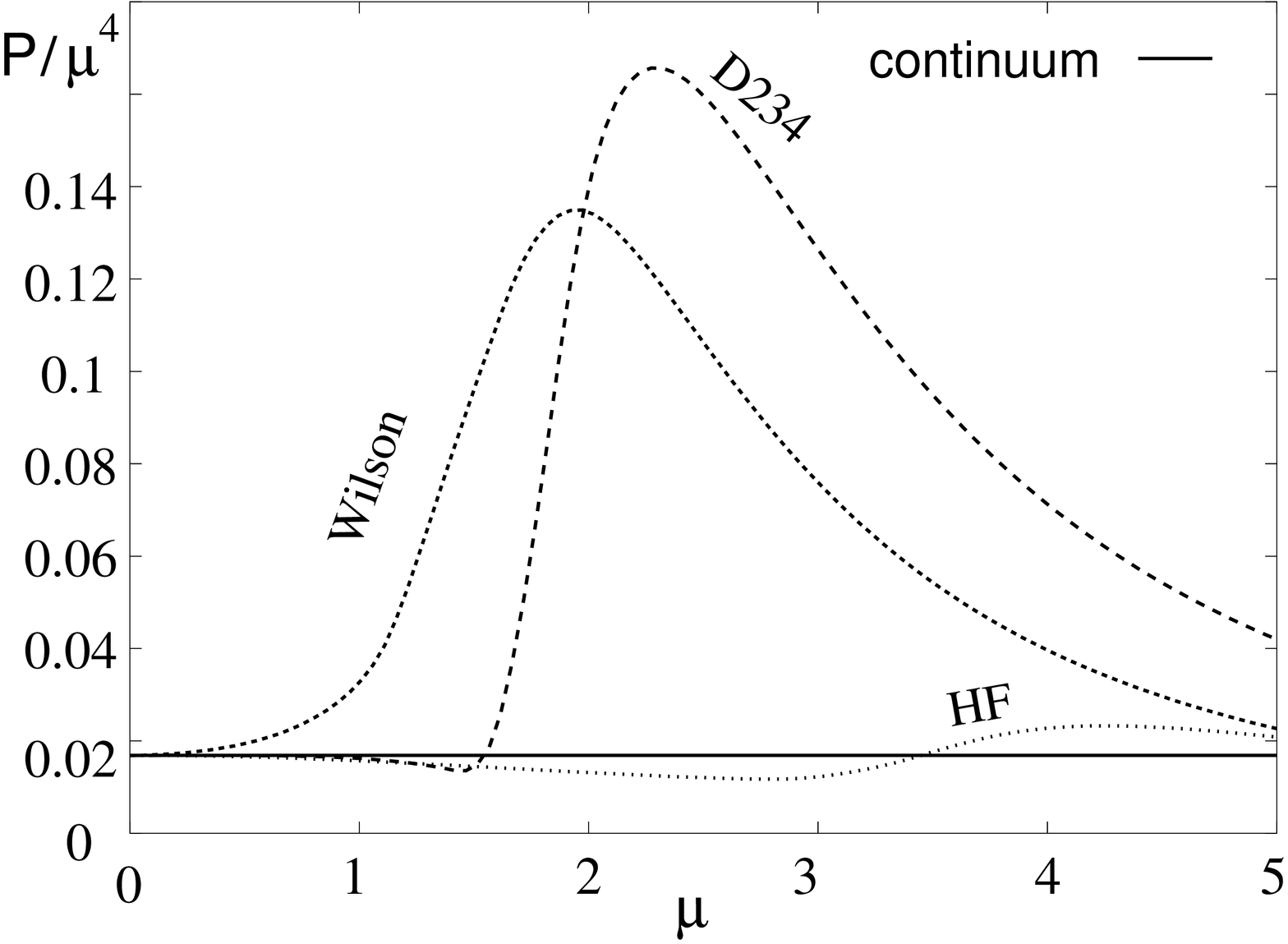}
  \includegraphics[angle=270,width=0.48\linewidth]{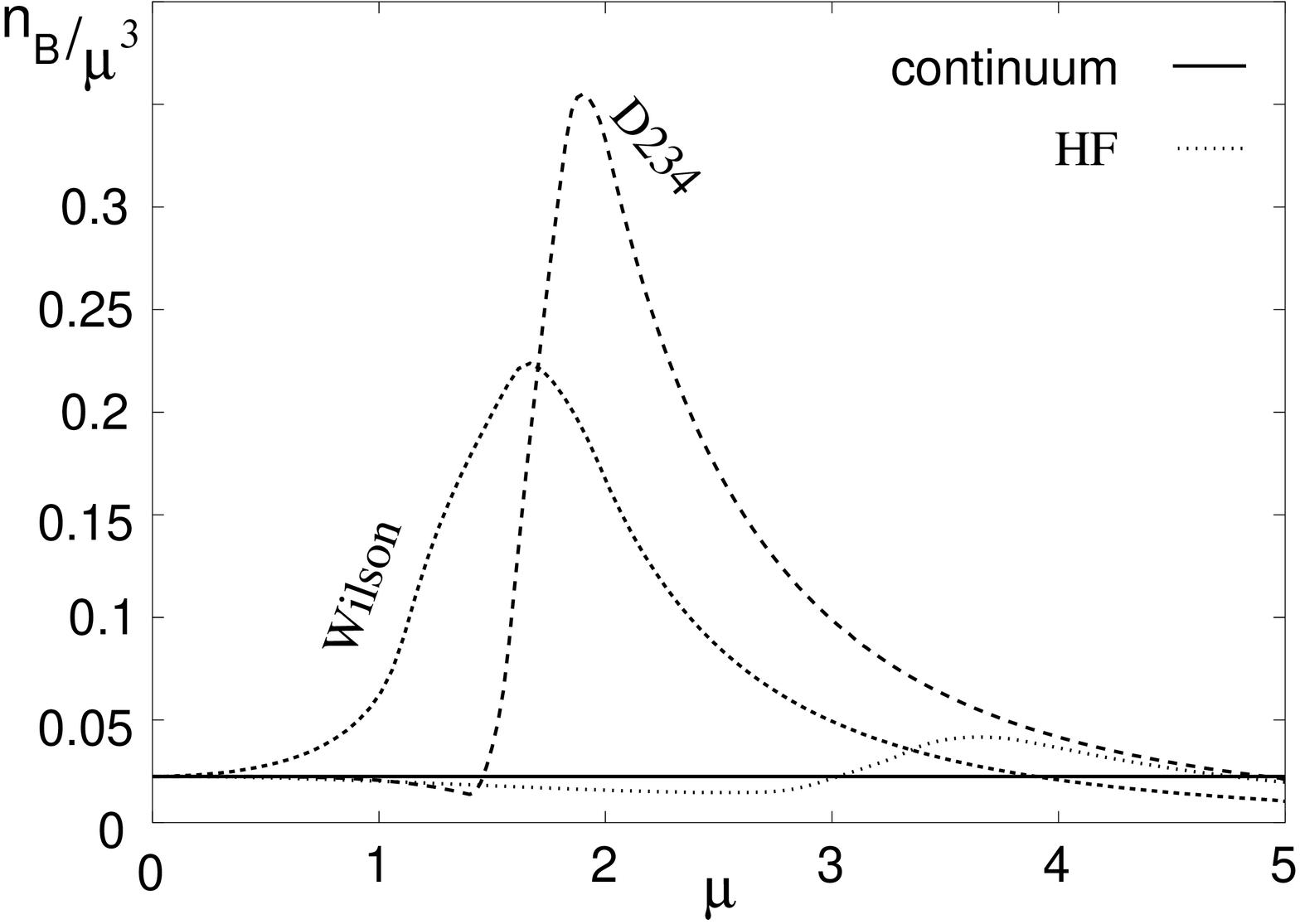}
\caption{{\it The ratios $P/\mu^{4}$ and $n_{B}/\mu^{3}$, for the
pressure $P$, the baryon density $n_{B}$ and the chemical potential
$\mu$, at zero temperature, for various types of free massless lattice 
fermions. For the truncated perfect HF these ratios converge 
very fast to the continuum values as $\mu$ decreases, in contrast to
the Wilson fermion and the D234 fermion.}}
\label{chempotferm}
\end{figure}

This trend is also confirmed for the thermodynamic quantities plotted
in Figures \ref{presferm} and \ref{chempotferm}. 
The pressure $P$ at finite temperature $T$ is obtained 
by imposing periodic boundary
conditions in the time direction over $N_{t}$ lattice points. The 
corresponding data in Figure \ref{presferm} are evaluated with the formula
\bea
\frac{P}{T^{4}} & = & \frac{N_{t}^{4}}{(2 \pi )^{3}} 
\int_{-\pi}^{\pi} d^{3}p \,
\Big[ \frac{1}{N_{t}} \sum_{n=1}^{N_{t}} \ln 
{\rm det} D ( \vec p , p_{4,n}) 
\Big\vert_{p_{4,n}= 2\pi (n-1/2) /N_{t}} \nn \\
&& \hspace*{2.8cm} - \frac{1}{2\pi} \int_{-\pi}^{\pi} dp_{4} \, 
\ln {\rm det} D(\vec p , p_{4}) \Big] \ .
\eea
In this case we find a deviation from the continuum
value $P /T^{4} = 7 \pi^{2} /180$ 
even for the perfect action, because its perfection is
designed specifically for zero temperature.

Figure \ref{chempotferm} deals with the inclusion of the chemical
potential $\mu$. This is achieved by the prescription worked out
in Refs.\ \cite{chempot}. The key observation is that starting from
any prefect lattice action at $\mu =0$ and performing consistently the 
substitutions
\be  \label{chempotsub}
\bar \Psi ( \vec x , x_{4} ) \to e^{ - \mu x_{4}} 
\bar \Psi ( \vec x , x_{4} ) \ , \quad
\Psi ( \vec x , x_{4} ) \to e^{ \mu x_{4}} 
\Psi ( \vec x , x_{4} ) \ ,
\ee
one obtains in fact a perfect action at finite $\mu$.
(Also classical perfection is preserved under these substitutions.)
In our perfect propagator in momentum space (\ref{perfermiprop}), 
this substitution can be implemented by shifting $p_{4} \to p_{4} + i \mu$.
Then one obtains the pressure and the baryon density (one third of the 
fermion density) at $T=0$ as
\bea
P(\mu ) &=& \frac{1}{(2 \pi )^{4}} \int_{B} d^{4}p \, 
\Big[ \ln {\rm det} D (\vec p , p_{4}) -
\ln {\rm det} D (\vec p , p_{4} + i \mu) \Big] \ , \nn \\
n_{B} &=& \frac{1}{3} \frac{\partial}{\partial \mu} P (\mu ) \ .
\eea
The scaling is then measured by the deviations from the continuum values
$P/ \mu^{4} = 1 / ( 6 \pi^{2})$ and $n_{B} / \mu^{3} = 2 / (9 \pi^{2})$.
Lattice artifacts are amplified for increasing chemical potential $\mu$.
We see that they remain modest over a broad range (i.e.\ up
to coarse lattices) for the truncated perfect action, in contrast to
the Wilson fermion and the D234 fermion. 
For large $\mu$ the fermion density turns into a constant plateau
for the usual lattice fermion formulations, and one might believe
that this is inevitable due to Pauli's principle. However, the height
of this plateau depends on the coupling range of the lattice Dirac 
operator, and it rises to infinity for the (untruncated) perfect action 
--- hence the RGT is able to solve this problem as well \cite{chempot}.

\subsection{Perfect staggered fermions}

As we pointed out in the previous Subsection, 
the full standard chiral symmetry cannot
be preserved in a perfect and local lattice action.
On the other hand, staggered fermions only have a remnant chiral 
symmetry --- see Subsection 3.4 ---
which raises the question if that symmetry can
persist in a perfect and local staggered fermion formulation.
In fact, this $U(1)_{\rm o} \otimes U(1)_{\rm e}$ symmetry
can be preserved under the RGT, if the block variables are constructed
such that they do not mix any of the $2^{d}$ tastes. A corresponding
blocking scheme with overlapping blocks was first proposed in 
Ref.\ \cite{KMS}. By its iteration we constructed a perfect action
for free staggered fermions \cite{Lat93}, which does fulfil the 
$U(1)_{\rm o} \otimes U(1)_{\rm e}$ symmetry
exactly, and which is manifestly local --- the Nielsen Ninomiya Theorem
does not exclude this remnant chiral symmetry.\footnote{Another
method, where different tastes contribute to a block variable,
has been applied recently \cite{Shamirstag}
to study the fourth root approach (cf.\ Subsection 3.4).
That RGT drives the rooted staggered fermion to a sensible perfect
action. However, the same is true for instance for the SLAC fermion
\cite{Brasil}, although the latter is incorrect under
gauge interaction \cite{KarSmi}, as we mentioned before
(in Subsection 3.4).}
In $d=2$ the corresponding free perfect action for the four massless 
tastes reads
\bea  \label{stagfix}
S [ \bar \chi , \chi ] &=& \sum_{x,y} \sum_{i,j=1}^{4} \bar \chi^{i}_{x}
[\alpha^{-1}]^{ij}(x-y) \chi^{j}_{y} \ , \\
\tilde \alpha (p) &=& d(-p) \alpha (p) d(p) = \left( \begin{array}{cccc}
0 & \tilde \alpha_{1}(p) & \tilde \alpha_{2}(p) & 0 \\
\tilde \alpha_{1}(p) & 0 & 0 & - \tilde \alpha_{2}(p) \\
\tilde \alpha_{2}(p) & 0 & 0 & ~~~ \tilde \alpha_{1}(p) \\
0 & \tilde \alpha_{1}(p) & \tilde \alpha_{2}(p) & 0
\end{array} \right) \nn \\
\tilde \alpha_{\mu}(p) &=& 
2 \sum_{l \in \Z^{2}} \frac{p_{\mu} + 2 \pi l_{\mu}}
{( p + 2\pi l)^{2}} (-1)^{l_{\mu}} \, \prod_{\nu =1}^{2}
\Big( \frac{ \sin (p_{\nu}/2)}{ p_{\nu}/2 + \pi l_{\nu}} \Big)^{2} +
c \sin ( k_{\mu}/2) \ , \nn
\eea
where $d(p)$ is a matrix of phase factors, which arrange for the
shifts to the appropriate lattice sites (it is given explicitly
in Ref.\ \cite{GN}, which denotes it as $D(p)^{1/2}$).
$c$ is an arbitrary (real) RGT parameter, which we tuned again for
optimal locality. In this case, the analytic optimisation in $d=1$ 
yields $c=1/2$. Ref.\ \cite{BBCWstag} discusses the extension of this
action to $d=4$, as well as the generalisation to a finite fermion 
mass $m$, which fills in diagonal elements in the above matrix
and changes the locality optimal RGT term.

Also this result can be derived efficiently by blocking from the
continuum, if the overlapping integration cells are treated carefully.
That method also allows for a blocking of non-compact gauge fields,
which is consistent in the sense that the link variables never connect
fermionic variables on the same sublattice \cite{BBCWstag}.

\subsection{Application to the Gross-Neveu model}

We return to the Gross-Neveu model that we described previously in
Subsection 3.5. More precisely we now consider its lattice formulation
in terms of staggered fermions. Again we replace the 4-Fermi term 
by a Yukawa coupling\footnote{By a Yukawa term we mean a product 
of a bosonic field and fermionic fields $\bar \psi$, $\psi$
that contributes to the Lagrangian, as it also appears in the
Standard Model.}
to an auxiliary scalar field $\phi$. Since $\phi$
is taste-free, it is adequate to put its lattice variables on 
the cell centres $z$ of the fermionic lattice \cite{JMP}. The standard
formulation then couples $\phi_{z}$ in the same manner
to the $2^{d}$ taste variables
located on the corners of the cell with centre $z$.

As in Subsection 3.5 we considered
the large $N$ limit, where the field $\phi_{z}$ freezes 
to a constant $\phi^{(0)}$. Then the fermions can be integrated out,
so that the RGT can be computed explicitly. In Ref.\ \cite{GN} we derived the
perfect staggered fermion action for this case. 
To analyse the scaling behaviour, we evaluated two quantities of dimension
mass for the staggered standard action and for the perfect action:

\begin{itemize}

\item First we computed the chiral condensate 
$\langle \chi \bar \chi \rangle$.
For the perfect action $S [ \bar \chi , \chi , \phi]$
this was achieved by a perturbation
\be
S_{\epsilon} [ \bar \chi , \chi , \phi] = 
S [ \bar \chi , \chi , \phi] + \epsilon X [ \bar \chi , \chi , \phi] \ .
\ee
The operator $X$ has the standard lattice form 
$\sum_{x} \chi_{x} \bar \chi_{x}$,  and its perfect lattice form
was computed again by the RGT technique, i.e.\ this perturbation
was included to $O(\epsilon )$ in the transformation.

\item From the gap equation (analogous to the continuum eq.\ (\ref{gapeq}))
\be
2 \phi^{(0)} = \frac{g}{(2\pi )^{2}} \int_{B} d^{2}p \,
\ln \, {\rm det} M(p, \phi^{(0)})
\ee
we extracted the fermion mass $m_{f}$, which is dynamically generated 
by the breaking of the discrete, remnant
chiral symmetry. $M$ is the fermion determinant (see eq.\
(\ref{fermidet})), either for the standard formulation or for the 
perfect formulation.

In this context, we also considered the {\em asymptotic scaling} by
investigating how closely $\phi^{(0)}(1/g)$ follows an exponential
behaviour. (This behaviour is known in the continuum version of this
model, see eq.\ (\ref{asymfree}),
and it characterises asymptotic freedom.)
Theoretically, asymptotic scaling does not need to be
improved by the perfect action, since it is in principle independent
from the scaling itself. Nevertheless we observed that it is significantly
improved as well \cite{GN}, in agreement with similar observations 
for truncated classically perfect actions for $SU(3)$ gauge theory
\cite{asyscalQCD}.

\end{itemize}

While these calculations involve lengthy expressions, the outcome for the
(dimensionless) ratio of these two terms, which represents our scaling
quantity, takes a simple form,
\be
\frac{\langle \chi \bar \chi \rangle}{m_{f}} \Big\vert_{\rm standard}
= \frac{2 \sinh (a m_{f}/2)}{a m_{f}} \quad , \qquad
\frac{\langle \chi \bar \chi \rangle}{m_{f}} \Big\vert_{\rm perfect}
= 1 \ .
\ee
Hence the {\em perfect scaling} is indeed confirmed, i.e.\ for the perfect 
action the considered scaling ratio takes the exact continuum value at any
lattice spacing $a$. In contrast, for the standard action this ratio
is only obtained in the limit $a \to 0$. We add that in this case also the
classically perfect action scales perfectly; artifacts are switched off
by the large $N$ limit \cite{GN}.

\subsection{Exact supersymmetry on the lattice}

Since the RGT technique enables us to transfer continuum properties
to the lattice without any damage in the physical observables, this
procedure can in principle also preserve exact supersymmetry (SUSY) on
the lattice \cite{SUSY}. This may appear surprising, because continuous 
SUSY seems to contradict the lattice
structure. For a review which presents a variety of approaches
to handle SUSY on the lattice we refer to Ref.\ \cite{Alessandra}, and
examples for further efforts to construct exact lattice SUSY 
are collected in Refs.\ \cite{SUSYlatexact}.

For an illustration of the perfect action treatment of SUSY,
we considered the simplest supersymmetric model \cite{Nicolai}: 
it is given in $d=2$ by the Lagrangian
\be
{\cal L}[ \psi , \varphi ] = \bar \psi \gamma_{\mu} \partial_{\mu}
\psi + \partial_{\mu} \varphi \partial_{\mu} \varphi \ ,
\ee
with a Majorana spinor $\psi$ and a neutral scalar field $\varphi$.
The action is invariant under simultaneous transformations with
\be  \label{SUSYtrafo}
\delta \psi = - \gamma_{\mu} \partial_{\mu} \varphi \, \varepsilon \ ,
\quad \delta \varphi = \bar \varepsilon \, \psi \ ,
\ee
where $\varepsilon$ is a two-component Grassmann vector.
The SUSY transformation generators form a closed algebra with the
translation operators,
\be  \label{SUSYalgebra}
[ \delta_{1}, \delta_{2} ] \varphi =
( \bar \varepsilon_{1} \gamma_{\mu} \varepsilon_{2} -
  \bar \varepsilon_{2} \gamma_{\mu} \varepsilon_{1} ) \partial_{\mu} 
\varphi \ .
\ee
By blocking from the continuum we transfer this model to the unit 
lattice and arrive at
\bea
S [ \Psi , \Phi ] &=& \frac{1}{(2\pi )^{2}} \int_{B} d^{2}p \, \Big[
\bar \Psi (-p) G (p)^{-1} \Psi (p) +
\Phi (-p) G_{s}(p)^{-1} \Phi (p) \Big] \ , \nn \\
 G_{s}(p) &=& \sum_{l \in \Z^{2}} \frac{\Pi (p + 2 \pi l)^{2}}
{(p_{\mu} + 2\pi l_{\mu})^{2} } + R^{s}(p) \ ,
\eea
where $G$ is the perfect fermion propagator (\ref{perfermiprop}),
and $G_{s}$ is a perfect scalar propagator (an obvious generalisation
of the form given in eq.\ (\ref{perfscalprop})).
If we perform the SUSY transformations (\ref{SUSYtrafo}) in the continuum, 
they are carried over to the blocked lattice fields,
\be  \label{SUSYtrafolat}
\delta \Psi_{x} = - \gamma_{\mu} \int_{C_{x}} \partial_{\mu} \varphi (u)
du \ \varepsilon  \ , \quad 
\delta \Phi_{x} = \bar \varepsilon \int_{C_{x}} \psi (u) du \ ,
\ee
which --- under these transformations --- obey exact SUSY too.

\begin{figure}[h!]
\begin{center}
 \includegraphics[angle=270,width=.8\linewidth]{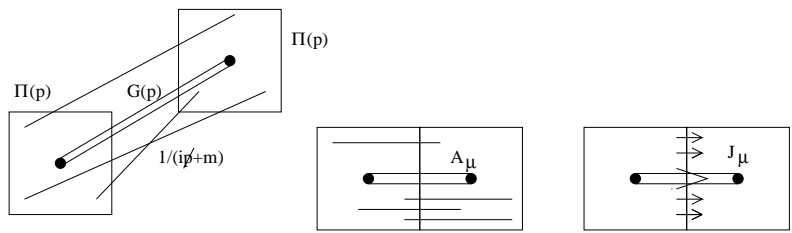}
\end{center}
\caption{{\it A cartoon of the schemes that we used to block various
quantities from the continuum: matter fields are blocked to the lattice 
by integrating the continuum field in a lattice cell, with the convolution 
function $\Pi$. The perfect propagator $G$ is obtained by integrating
all continuum propagators between points in the corresponding lattice cells,
as shown on the left (the formulae are given in Subsections 2.2, 4.1 and 4.2). 
In the centre we illustrate the blocking for
gauge fields, to be discussed in detail in Subsection 5.3.
Here we integrate all straight parallel transporters between continuum
points, which have the same relative position in adjacent lattice cells.
At last, a perfect current is obtained by integrating the continuum 
flux through the face between adjacent lattice cells (it will be
used again Subsection 5.3).}}
\label{blockingfig}
\end{figure}

In particular, we may treat the term $j_{\mu} = \gamma_{\mu} \varphi$
as a continuum current. As a general prescription
we block a continuum current to the lattice by integrating its
flux over the face $f_{\mu ,x}$ between adjacent lattice cells 
\cite{perfaxial},
\be  \label{perfcurrent}
J_{\mu ,x} = \int_{f_{\mu ,x}} d^{d-1}y \, j_{\mu}(y) \ .
\ee
This blocking scheme is illustrated in Figure \ref{blockingfig}
on the right-hand side.
The lattice divergence of the blocked current
is then equal to the conti\-nuum
divergence integrated over the corresponding lattice cells,
\be  \label{perfdiv}
\delta J_{x} = \sum_{\mu} ( J_{\mu, x + \hat \mu /2} -
J_{\mu, x - \hat \mu /2} ) 
= \int_{C_{x}} d^{d}y \, \partial_{\mu} j_{\mu}(y) \ .
\ee
In this way, the transformations
(\ref{SUSYtrafolat}) can be expressed solely in terms of lattice quantities,
i.e.\ the lattice current and the blocked lattice field \cite{SUSY}.
Accordingly, the algebraic relation (\ref{SUSYalgebra}) is now precisely
reflected in terms of lattice quantities as
\be
[ \delta_{1}, \delta_{2} ] \Phi =
( \bar \varepsilon_{1} \nabla_{\mu} J_{\mu} \varepsilon_{2} -
  \bar \varepsilon_{2} \nabla_{\mu} J_{\mu}\varepsilon_{1} ) \ ,
\ee
where $\nabla_{\mu}$ is the standard (symmetric) lattice derivative.

These properties also extend to the free 2d Wess-Zumino model,
which involves an additional scalar field that balances the fermionic
and bosonic degrees of freedom, and to the free 4d Wess-Zumino model,
where further field components are added.

In terms of classically perfect fields, the continuum SUSY transformations
can be carried over to the lattice also in the interacting case.
However, the explicit construction of the corresponding lattice 
terms is a challenging numerical project, which has not been 
carried out so far.

Still the lattice is hostile by its nature towards SUSY. In order
to simulate SUSY models nevertheless, also the discrete formulation
on a fuzzy sphere --- that we already mentioned in Subsection 3.5 ---
should be considered \cite{FuzzySUSY}. For corresponding
simulations we refer to Refs.\ \cite{FuzzySUSYnum,prep}.

\section{Perfect Lattice Perturbation Theory}

On the level of analytical calculations, the construction of
perfect lattice actions can be extended from the free
fields to perturbative interactions. As a first example,
we already sketched the computation of a perfect chiral
condensate in Subsection 4.3.

The method of blocking from the continuum is still applicable and
highly efficient for this purpose. One now blocks various fields
in such a way that all the continuum propagators between the
continuum points in the lattice cells are integrated over.
In the case of gauge interactions, also the gauge fields undergo
a blocking procedure, which can be made explicit most conveniently
for non-compact gauge fields, see Subsections 5.3 and 5.4, and for
illustrations Figures \ref{blockingfig} and \ref{qgblocking}.

\subsection{The anharmonic oscillator}

\begin{figure}[h!]
\begin{center}
 \includegraphics[angle=0,width=.6\linewidth]{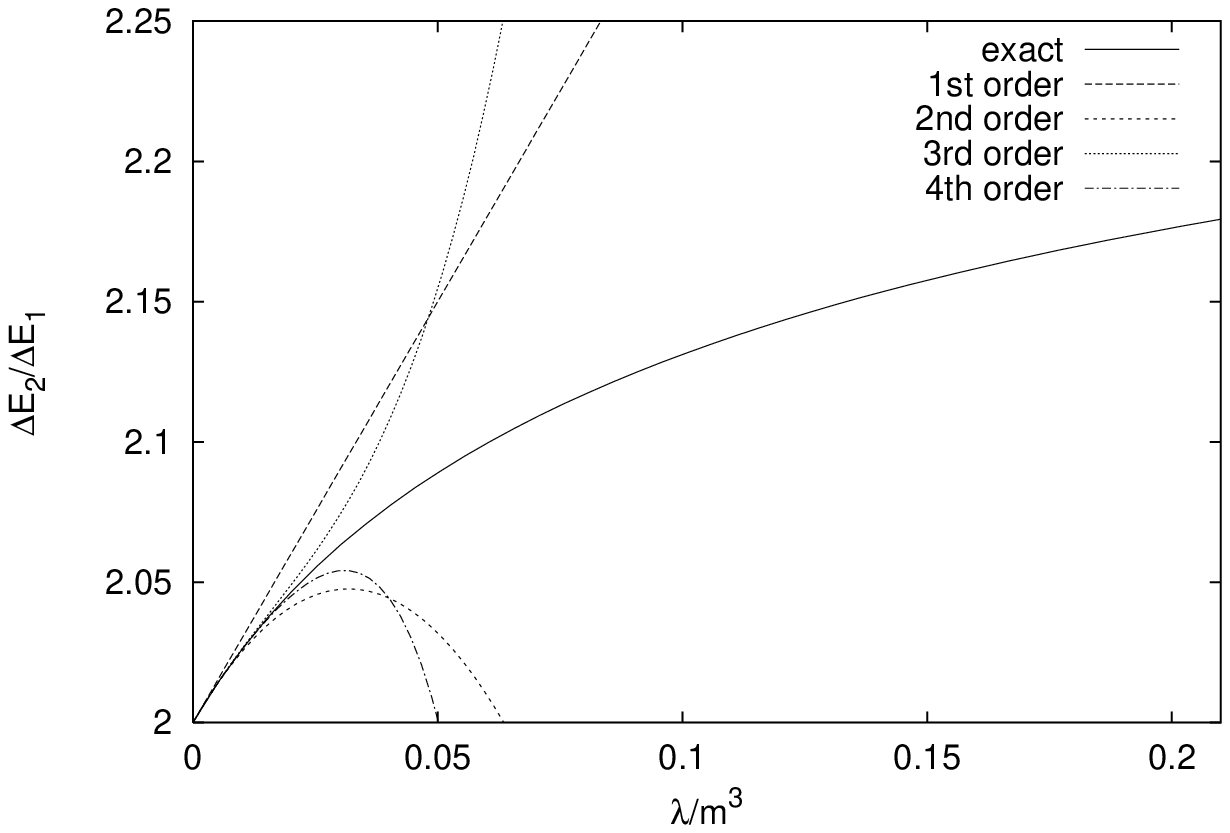}
 \includegraphics[angle=0,width=.6\linewidth]{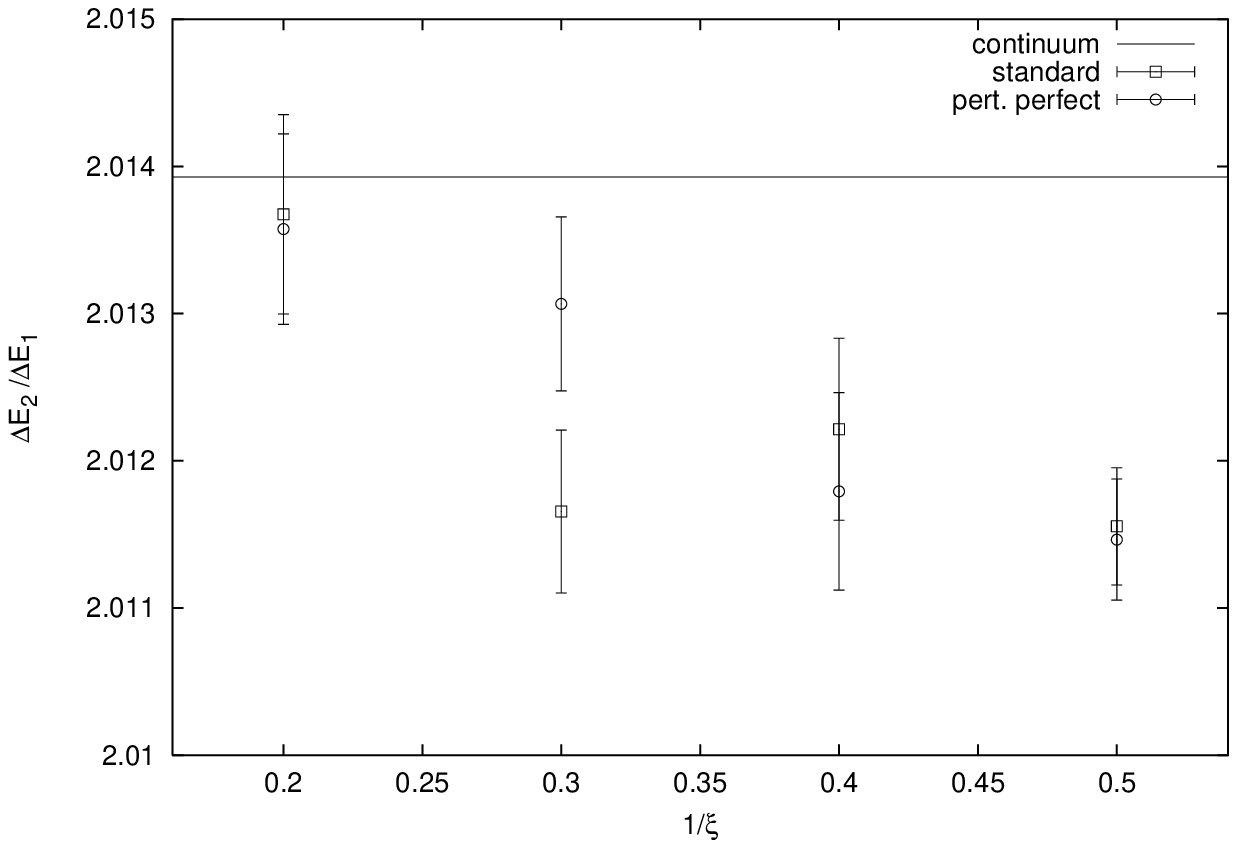}
\label{dE2dE1}
\end{center}
\caption{{\it Above: The ratio between the leading energy gaps,
$\Delta E_{2} / \Delta E_{1}$, against the perturbative predictions.
Higher orders do extend the range of the validity of perturbation
theory gradually, but even at fourth order it is still limited to
small interaction parameters ($\bar \lambda \lsim 0.03$). \newline
In the plot below we compare simulation results at $\bar \lambda = 0.005$
with the standard action and the $O (\bar \lambda )$
perfect action, for different correlation lengths $\xi$ in lattice 
units \cite{anharm}.
}}
\end{figure}

\begin{figure}[h!]
\begin{center}
 \includegraphics[angle=0,width=.6\linewidth]{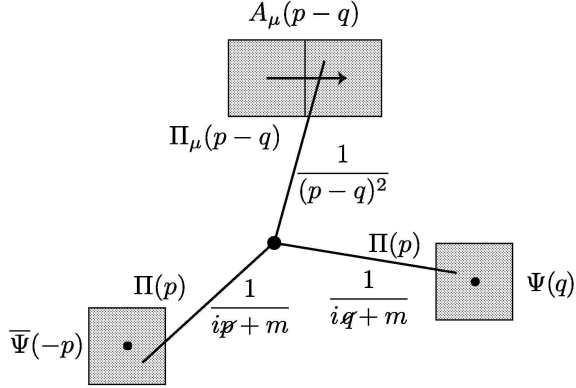}
\end{center}
\caption{{\it Overview over the construction scheme of the perfect
quark gluon vertex function: the lattice fermion fields $\bar \Psi$ and $\Psi$
and the gluon field $A_{\mu}$ are obtained by blocking from the continuum, 
cf.\ Figure \ref{blockingfig}. For a given point (as the one marked by
the central dot in this Figure) we then integrate the continuum propagators 
to all continuum fields involved in this blocking process.}}
\label{qgblocking}
\end{figure}

As a toy model from quantum mechanics, we considered the anharmonic
oscillator \cite{anharm}. We write its action in field theoretic notation as
\be
s [ \varphi ] = \int dt \, \Big[ \, \frac{1}{2} \dot \varphi (t)^{2}
+ \frac{m^{2}}{2} \varphi (t)^{2} + \lambda \varphi (t)^{4} \, \Big] \ ,
\quad (\varphi (t) \in \R ) \ .
\ee
As in the case of the quantum rotor (discussed in Refs.\ \cite{qrot,Boyer}
and reviewed in Subsection 2.3), we use the ratio between the first
two energy gaps, $\Delta E_{1} = E_{1} - E_{0}$ and
$\Delta E_{2} = E_{2} - E_{0}$, as a scaling quantity. 
In continuum perturbation theory,
the corresponding expansion can be found at many places in the literature,
e.g.\ in Ref.\ \cite{Caswell}. In terms of the dimensionless interaction
parameter $\bar \lambda := \lambda / m^{3}$ one obtains
\be
\frac{\Delta E_{2}}{\Delta E_{1}} ( \bar \lambda ) =
2 + 3 \bar \lambda - \frac{189}{4} \bar \lambda^{2} 
+ \frac{7857}{8} \bar \lambda^{3} - \frac{1569069}{64} 
\bar \lambda^{4} + O( \bar \lambda^{5} ) \ .
\ee
First we evaluated this ratio to a high precision by Metropolis 
Monte Carlo simulations and we compare it to the perturbative results
in various orders in Figure \ref{dE2dE1} (above). The latter approach
the correct result only laboriously in a small range for $\bar \lambda$,
even if we include the fourth order. This may serve as a caution to be
careful in general with extrapolations to finite interaction 
strength based on perturbation theory.

Next we calculated the perfect lattice action to $O(\bar \lambda )$.
We chose the RGT parameter so that the action at $\bar \lambda =0$
consists of nearest neighbour couplings only. This is possible
in 1d field theory (i.e.\ quantum mechanics) with the parameter
given in eq.\ (\ref{alphamscal}). We then extended the
blocking from the continuum to $O(\bar \lambda )$. This generates
additional 2-spin and 4-spin terms, which were written down explicitly
in momentum space \cite{anharm}. There inverse Fourier transform
yields a set of couplings that we computed for various parameters
$m^{2}$ up to a coupling distance of two lattice spacings.
This truncation is justified because the couplings undergo a fast decay, 
which speeds up for increasing $m^{2}$.

Finally we simulated the resulting perturbatively perfect action.
For the scaling test, we fixed $\bar \lambda = 0.005$, i.e.\
a value where Figure \ref{dE2dE1} (above) suggests the validity
of first order perturbation theory. The results 
at various correlation lengths are compared
to the outcome with the standard action
in Figure \ref{dE2dE1} (below). At a correlation length of $\xi =5$ 
(in lattice units), both actions perform very well, and below 2.5 
both suffer from similar scaling artifacts. In between, there appears
a window where the perturbatively perfect action seems
superior, as we observed at $\xi = 10/3$.

\subsection{The Yukawa term}

We computed a perturbatively perfect action in the framework
of the Gross-Neveu model with staggered fermions, cf.\ Subsections 
4.2 and 4.3,
but now for four tastes. In this case (without a large $N$ limit),
the auxiliary scalar field $\Phi$ is not
constant anymore, but we assumed it to be small. More precisely,
we absorbed the Yukawa coupling in $\Phi$ and considered its first order. 
In this approximation --- which describes the asymptotically free
system at high energy --- the perfect staggered action takes the form
\bea
S [ \bar \chi , \chi , \Phi ] &=& \sum_{xy,ij} \bar \chi_{x}^{i}
[\alpha^{-1}]_{xy}^{ij} \chi_{y}^{j} + \frac{1}{2} \sum_{z} \Phi_{z}^{2} \nn \\
& + & \sum_{xyz,ij}  \bar \chi_{x}^{i} \sigma^{ij}(x-z, y-z) \chi^{j}_{y}
\Phi_{z} \ ,
\eea
where $x,y$ run over the lattice which hosts the fermionic degrees
of freedom, whereas $z$ runs over the plaquette centres, 
and $i,j = 1 \dots 4$.
If we take the spacing between fermion components of the same taste as
the unit, $z$ is spaced by $1/2$. In momentum space
we write the interaction term as $\bar \chi (-p) \sigma(p,q)
\chi (-q) \Phi (p+q)$.
In the taste space, the shifted kernel $\tilde \sigma$ 
is a $4\times 4$ matrix, which 
only couples tastes of the same sublattice.
To be explicit, its
first order perturbatively perfect form reads \cite{GN}
(we use the notation of eq.\ (\ref{stagfix}))
\be
\tilde \sigma (p,q) = d(p) \alpha (p) \sigma(p,q) \alpha (-q) d(-q) =
\left( \begin{array}{cccc}
\tilde \sigma_{0} & 0 & 0 & - \tilde \sigma_{3} \\
0 & \tilde \sigma_{0} & \tilde \sigma_{3} & 0 \\
0 & - \tilde \sigma_{3} & \tilde \sigma_{0} & 0 \\
\tilde \sigma_{3} & 0 & 0 & - \tilde \sigma_{0}
\end{array} \right) \ ,
\ee
with the matrix elements
\bea
\tilde \sigma_{0}(p,q) &=& \sum_{l,m \in \Z^{2}} \sum_{n}
\frac{p_{\mu}^{(l,n)} q_{\mu}^{(m,n)}} { p^{(l,n)\, 2} q^{(m,n) \, 2} } \nn \\
& \times & 
\prod_{\nu =1}^{2} (-1)^{l_{\nu} + m_{\nu} + n_{\nu}}
\frac{\hat p_{\nu} \hat q_{\nu} ( \widehat{p+q} )_{\nu} }
{ p_{\nu}^{(l,n)} q_{\nu}^{(m,n)} [p_{\nu}^{(l,n)} + q_{\nu}^{(m,n)} ] } \ , 
\nn \\
\tilde \sigma_{3}(p,q) &=& \sum_{l,m \in \Z^{2}} \sum_{n}
\frac{\epsilon_{\mu \rho} 
p_{\mu}^{(l,n)} q_{\rho}^{(m,n)}} { p^{(l,n)\, 2} q^{(m,n) \, 2} } \nn \\
& \times & 
\prod_{\nu =1}^{2} (-1)^{l_{\nu} }
\frac{\hat p_{\nu} \hat q_{\nu} ( \widehat{p+q} )_{\nu} }
{ p_{\nu}^{(l,n)} q_{\nu}^{(m,n)} [p_{\nu}^{(l,n)} + q_{\nu}^{(m,n)} ] } \ ,
\eea
where $p_{\mu}^{l,n} = p_{\mu} + 4 \pi l_{\mu} + 2\pi n_{\mu}$, and
$n_{\mu} \in \{ 0,1 \}$. Note that these matrix elements are $4 \pi$
periodic, in accordance with the central positions of the auxiliary scalar
variables.

This Yukawa term identifies the direction of a ``renormalised trajectory''
(a line of perfect actions in parameter space) emanating from
the critical surface.\footnote{The endpoint of this trajectory is the
free perfect action that we identified before, due to the asymptotic freedom
of the Gross-Neveu model.}
The corresponding couplings in coordinate space can be evaluated numerically,
and they have been applied --- in a truncated form --- in lattice simulations
\cite{Erich}.

\subsection{Perfect gauge actions and the axial anomaly}

The attempts to formulate non-local lattice fermions with a finite gap
at the edge of the Brillouin zone were unsuccessful; we mentioned
the SLAC fermion in Subsection 3.4.
A refined approach was presented by C.\ Rebbi, who formulated a
non-local fermion with divergences at these edges instead \cite{Rebbi}.
However, the Rebbi fermion does not reproduce a non-zero axial anomaly,
as A.\ Pelissetto pointed out \cite{Peli}.

If we construct the perfect fermion for a $\delta$-function blocking RGT,
we obtain a non-locality of the same type as the Rebbi fermion \cite{Wie92,Lat93}, 
hence we wondered what happens to the axial anomaly in that case.

Once we deal with the $\delta$-function blocking, the relations 
(\ref{relatedto}) turn into equations. In momentum space they read
\be  \label{deltafermi}
\bar \Psi (p) = \sum_{l \in \Z^{d}} \Pi (p + 2 \pi l) \bar \psi (p) \ , \quad
\Psi (p) = \sum_{l \in \Z^{d}} \Pi (p + 2 \pi l) \psi (p) \ .
\ee
Analogously, we now block an Abelian gauge field ${\cal A}_{\mu}(x)$ from the
continuum to construct the non-compact link variable \cite{perfaxial,QuaGlu}
\be
A_{\mu, x} = \int_{C_{x - \hat \mu /2}} d^{d}u \, 
(1 + u_{\mu} - x_{\mu}) {\cal A}_{\mu}(u) +
\int_{C_{x + \hat \mu /2}} d^{d}v \, 
(1 - v_{\mu} + x_{\mu}) {\cal A}_{\mu}(v) \ . \ 
\ee
Here $x$ is a link centre on a unit lattice, so that we
integrate over adjacent lattice cells. 
This blocking scheme
is illustrated in the centre of Figure \ref{blockingfig}.
A gauge transformation
in the continuum, ${\cal A}_{\mu} \to {\cal A}_{\mu}
+ \partial_{\mu} \lambda$,
induces exactly a lattice gauge transformation
\be
A_{\mu ,x} \to A_{\mu ,x} + \Lambda_{x + \hat \mu /2} 
- \Lambda_{x - \hat \mu /2} \ , \quad 
\Lambda_{x} = \int_{C_{x}} d^{d}y \, \lambda (y) \ ,
\ee
which shows that this lattice gauge field is covariant.
In momentum space it takes the form
\bea
A_{\mu}(p) &=& \sum_{l \in \Z^{d}} \Pi_{\mu}(p+2 \pi l)
(-1)^{l_{\mu}} \, {\cal A}_{\mu}(p + 2\pi l) \ , \nn \\
{\rm where} && \Pi_{\mu}(p) := \frac{\hat p_{\mu}}{p_{\mu}} \, \Pi (p)
\eea
is anti-periodic over the Brillouin zone in the $\mu$-direction
(and periodic in the other directions).
It is convenient to start from a continuum action
in the Landau gauge, which leads to the perfect lattice
gauge action \cite{QuaGlu}
\bea
S [A] &=& \frac{1}{(2 \pi)^{d}} \int_{B} d^{d}p \, \frac{1}{2}
A_{\mu}(-p) G^{(g)}_{\mu}(p)^{-1} A_{\mu}(p) \ , \nn \\
G^{(g)}_{\mu}(p) &=& \sum_{l \in \Z^{d}} 
\frac{\Pi_{\mu}(p + 2 \pi l)^{2}}{(p + 2\pi l)^{2}} + R^{(g)}(p) \ ,
\eea
where the term $R^{(g)}$ smears the RGT in analogy to the treatment
of the matter fields in eqs.\ (\ref{scalRGT}) and (\ref{fermRGT}).
As a remarkable property, the specific choice
\be
 R^{(g)}(p) = \frac{1}{6} - \frac{1}{72} \hat p_{\mu}^{2}
\ee
yields for an Abelian gauge field in $d=2$ the standard plaquette
action, which is therefore perfect already \cite{QuaGlu}. 
This property is similar to the standard lattice scalar and 
the Wilson fermion in $d=1$, which are also perfect in the
non-interacting case, as the choice of suitable RGT para\-meters confirms
(this was pointed out previously in Subsections 2.2 and 4.1). 
Again we use this property as a tool to optimise the
RGT in view of locality also in higher dimensions.

In Ref.\ \cite{perfaxial} we considered a more general perfect lattice
gauge action, without previous gauge fixing in the continuum.
There we also proceeded to perturbation theory to the first order
in the gauge coupling $g$. In this case, the blocking RGT of the
fermion field is extended from eq.\ (\ref{deltafermi}) to the form
\bea  \label{fermiblockgauge}
\Psi_{i} (p) &=& \sum_{l \in \Z^{d}} \Pi (p + 2\pi l) \psi_{i} (p + 2\pi l)
+ \frac{g}{(2 \pi )^{d}} \sum_{l \in \Z^{d}}  \int d^{d} q \, \nn \\
&\times & K_{\mu}(p + 2\pi l , q + 2\pi l) 
{\cal A}_{\mu}^{c} (p-q) \lambda^{c}_{ij} \psi_{j} (q + 2\pi l) \ ,
\eea
and correspondingly for $\bar \Psi_{i}$. 
Eq.\ (\ref{fermiblockgauge}) refers to a $SU(N)$ (or $U(N)$) gauge field,
where $\lambda^{c}$ are Hermitian generators. However, in the
current Subsection we will deal with the Abelian gauge field, where the 
last factor simplifies to ${\cal A}_{\mu} (p-q) \psi (q + 2\pi l)$.

The kernel $K_{\mu}$ has to be regular, and gauge covariance requires
\be  \label{kernelcond}
(p_{\mu} - q_{\mu}) K_{\mu}(p,q) = \Pi (p-q) \Pi(q) - \Pi (p) \ .
\ee
To this order, the perfect action includes --- in addition to the
pure fermion and pure gauge action --- an interaction term of the
structure
\be  \label{vertex}
V [ \bar \Psi , \Psi , A ] = \frac{g}{(2 \pi )^{2d}} \int_{B^{2}} d^{d}p \,
d^{d}q \, \bar \Psi_{i} (-p) V_{\mu}(p,q) A_{\mu}^{c}(p-q) \lambda_{ij}^{c} 
\Psi_{j} (q) \ .
\ee
The explicit form of $V_{\mu}(p,q)$ is rather lengthy; it is written down
for the gauge group $U(1)$ in Ref.\ \cite{perfaxial}, and for
QCD in Ref.\ \cite{QuaGlu}.
Gauge invariance requires that the vertex function $V_{\mu}$ obeys
the lattice Ward identity
\be  \label{Ward}
( \widehat{p-q} )_{\mu} V_{\mu}(p,q) = G(q)^{-1} - G(p)^{-1} \ ,
\ee
where $G$ is the perfect fermion propagator (\ref{perfermiprop}).
 
To investigate the axial anomaly, we now block the axial continuum current
\be
j_{\mu}^{5} (p) = \frac{1}{(2 \pi )^{d}} \int d^{d}q \, \bar \psi (p-q)
\gamma_{\mu} \gamma_{5} \psi (q)
\ee
to the lattice. Following the prescription (\ref{perfcurrent}) we obtain
a lattice current $J_{\mu ,x}^{5}$. In momentum space it takes the form
\be
J_{\mu ,x}^{5}(p) = \sum_{l \in \Z^{d}} j_{\mu}^{5}(p + 2\pi l)
\Pi_{\neg \mu} (p+2\pi l) (-1)^{l_{\mu}} \ , \ \,
\Pi_{\neg \mu} (p) := \prod_{\nu \neq \mu} \frac{\hat p_{\nu}}{ p_{\nu}} \ .
\ee
In the Schwinger model (two dimensional QED \cite{Schwinger}), 
the continuum current in a gauge background
is known to obey the relation
\be
\langle j_{\mu}^{5} (p) \rangle_{\cal A} = \frac{g}{\pi} \, \frac{p_{\mu}}{p^{2}}
\, \epsilon_{\nu \rho} p_{\nu} {\cal A}_{\rho}(p) \ ,
\ee
which can be derived for instance with dimensional regularisation.
We integrate out the continuum gauge field ${\cal A}_{\mu}$ and take the
lattice divergence of the current $J_{\mu ,x}^{5}$ in the perfect lattice
background, which leads to
\bea
\langle \hat p_{\mu} J_{\mu}^{5}(p) \rangle_{A} &=& \frac{g}{\pi}
\sum_{l \in \Z^{2}} \epsilon_{\nu \rho} \frac{p_{\nu} + 2 \pi l_{\nu}}
{(p+2 \pi l)^{2}} \Pi (p+2\pi l) (-1)^{l_{\rho}} \nn \\
&\times & \Pi_{\rho}(p+2\pi l)
G^{g}_{\rho \sigma}(p) ^{-1} A_{\sigma}(p) \ ,
\eea
where $G^{g}_{\rho \sigma}(p)$ is the perfect gauge propagator
(without previous gauge fixing in the continuum
it generalises to a tensor).
On the other hand, we build the perfect topological charge density
in agreement with the blocking recipe of Subsection 2.3, i.e.\
we block the continuum density 
$\frac{1}{\pi} \epsilon_{\mu \nu} \partial_{\mu} {\cal A}_{\nu}$
to the lattice. We find that
it coincides precisely with the lattice divergence of the perfect
current \cite{perfaxial},
\be
Q_{x} := \frac{1}{\pi} \int_{C_{x}} d^{2}y \, \epsilon_{\mu \nu} 
\partial_{\mu} {\cal A}_{\nu}(y) \  \excleq \  
\langle \delta J_{x}^{5} \rangle_{A} \ .
\ee
Therefore the perfect lattice action does indeed
reproduce the perturbative axial anomaly correctly, at any
lattice spacing.

Hence the perfect fermion constructed with a $\delta$-function RGT
is the only non-local lattice fermion that did not run into
conceptual trouble. It represents the only conceptually
successful implementation of chiral symmetry in its standard form
on the lattice. For practical purposes, however, one prefers 
the local form, which is still compatible with a modified chiral symmetry,
as we will discuss in Section~7.

\subsection{The perfect quark gluon vertex function}

The kernel function $K_{\mu}$ in the fermionic blocking scheme to
the first order in the gauge coupling is submit to the constraint
(\ref{kernelcond}), and it is difficult to find explicit solutions
for it. This issue is discussed in Ref.\ \cite{Lat96}.
In the small momentum expansion one obtains unambiguously
to the leading orders
\be
K_{\mu}(p,q) \simeq \frac{q_{\mu}}{12} \, \Big[ \, 1 + \frac{1}{120}
\Big\{ p_{\mu}^{2} - 4 \, [ \, p_{\mu}(p_{\mu}-q_{\mu}) + q_{\mu}^{2} \, ]
\, - 5 \, [ \, \vec p \, ( \, \vec p - \vec q \, ) + \vec q^{\, 2} \, ] \,
\Big\} \, \Big] \ ,
\ee
where $p = (p_{\mu}, \vec p \, ), \ q = (q_{\mu}, \vec q \, ) $.
We also discovered two full solutions, which can be evaluated numerically.
We display here one of them, denoted as the ``recursive kernel'',
in a corrected form (unfortunately this formula contains an error
in Ref.\ \cite{Lat96}),
\bea
K_{\mu}(p,q) &=& \frac{1}{4} \sum_{n \geq 0} \frac{\Pi (p)}
{\Pi (p /2^{n})} \Pi_{\mu} ((p-q) / 2^{n}) \Pi ( q / 2^{n+1}) \nn \\
& \times & \sin (q_{\mu}/2^{n+1}) {\cal K}_{\mu} ( p /2^{n+1} , 
q / 2^{n+1} ) \nn \\
{\cal K}_{\mu} (p,q) &=& \sum_{\vec l \in \{ 0,1 \}^{d-1}} 
\Big[ \prod_{\nu \neq \mu} \frac{\cos (p_{\nu} + \pi l_{\nu}/2) 
\cos(q_{\nu} - \pi l_{\nu}/2)} {( 1 + \vec l^{\ 2}) } \Big] \ , \qquad
\eea
where $\vec l$ excludes the $\mu$-component.
 
Along with eq.\ (\ref{vertex}) this provides a fully explicit
--- though somewhat complicated --- form of the perfect quark gluon
vertex function \cite{QuaGlu}. We recall that its gauge invariance is 
guaranteed by the Ward identity (\ref{Ward}).

We worked out a truncated version of the quark gluon vertex
function in coordinate space,
which we applied in simulations to be
addressed in Subsection 6.4. Its general form involves terms in the full
Clifford algebra. It simplifies drastically if we map the system 
down to $d=2$ by dimensional reduction. For that case we gave
explicit couplings, including the perturbatively perfect clover term,
at various fermion masses in Ref.\ \cite{Lat96}.

\section{The Hypercube Fermion}

\subsection{Construction of the Hypercube Fermion}

In Subsection 4.1 we have already described the truncation of the
perfect free fermion to a hypercube fermion (HF) by means of periodic
boundary conditions. We gave the couplings for the massless fermion
in Table \ref{HFtab}. We have also seen that it has excellent scaling
properties, and we will illustrate in Section 7 that its 
approximation to chirality is excellent as well.

The numerical treatment of all couplings inside a unit hypercube 
is clearly more complicated than the effort for standard 
formulations (Wilson or staggered), which only
deal with nearest neighbour couplings. However, simulations with this
generalised form are feasible and have been carried out. To this end,
the first question was how to handle the link variables to include 
the gauge interaction. If the spinors $\bar \Psi_{x}$ and
$\Psi_{y}$ are coupled in some lattice action, we can arrange
for gauge invariance of the corresponding term by connecting these
spinors over a lattice path, where the compact link variables are 
multiplied \cite{latbooks} (cf.\ Subsection 3.4).
The way to do so is ambiguous. The perfect or classically perfect
actions do actually determine the couplings to these lattice paths
(once the RGT is chosen).
But their evaluation is tedious, and a truncation of the path length
is required, which is again arbitrary.

\begin{figure}
\begin{center}
 \includegraphics[angle=0,width=.4\linewidth]{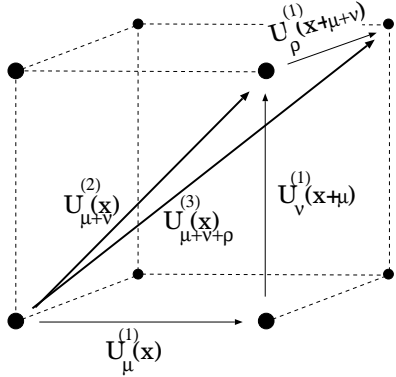}
\caption{\it{A 3d illustration for the construction of hyperlinks
in the gauging of a hypercube fermion.}}
\label{hyplink}
\end{center}
\end{figure}

The simplest approach just
includes connections over the shortest lattice paths. For most
connections in a hypercube there are several shortest paths, 
which are then all included with the same weight and averaged over. 
In this way, we introduce {\em ``hyperlinks''} which connect a lattice site 
$x$ to all the $3^{d}$ sites in a common unit hypercube with $x$.
An illustration of 2d and 3d hyperlinks, $U^{(2)}_{\mu + \nu}$ and
$U^{(3)}_{\mu + \nu + \rho}$ , is given in Figure \ref{hyplink}.
Regarding the numerical implementation, it is favourable to 
construct these hyperlinks in a given gauge configuration
recursively, i.e.\ to start with the 2d hyperlinks (over plaquette
diagonals), which then also serve as building blocks for the 3d 
hyperlinks, from which one arrives conveniently at the 4d hyperlinks 
\cite{Wupp}.

In Ref.\ \cite{Wupp} we discussed a suitable preconditioning method
for the HF. The goal is to divide the lattice into
sublattices such that a block structure in the lattice Dirac operator
$D_{x,y}(U)$ is gene\-rated. This block structure allows for the transition
to an equivalent Dirac operator with a better condition 
number. This means that the ratio between the maximal and minimal
eigenvalue of $D^{\dagger}D$ decreases, and this ratio is essential
for the computational effort in a simulation.
Moreover, the transformed operator is block diagonal, 
which also simplifies its inversion.

For the Wilson fermion, this method is usually applied with
two sublattices, which do not contain nearest neighbour sites,
as distinguished by the sign term (\ref{epsistag}). 
It is known as even-odd (or red-black) preconditioning. 
For the HF a set of $2^{d}$ sublattices is suitable, which we denoted 
as ``rainbow preconditioning''.\footnote{This approach was
also discussed for the hypercube scalar in Ref.\ \cite{WBscal}.} 
In fact, it yields gain factors between
3 and 4 in typical QCD simulations \cite{Wupp}.
(We add that also for Wilson fermions it can be profitable to deal with
a larger set of sublattices, as the locally-lexicographic symmetric 
successive overrelaxed preconditioner ({\it ll}-SSOR) shows \cite{llSSOR}.)

\subsection{Approximate rotation symmetry}

\begin{figure}
\begin{center}
 \includegraphics[angle=0,width=.6\linewidth]{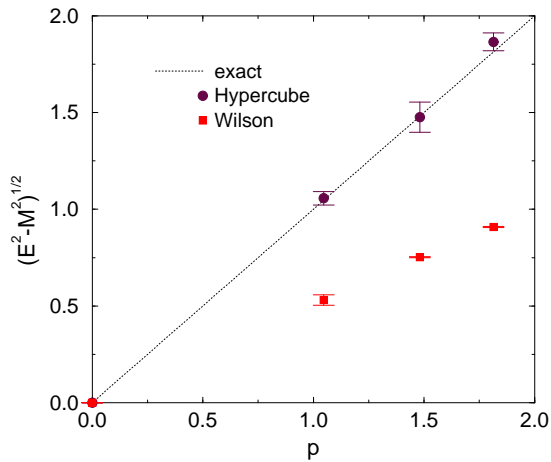}
\caption{\it{The ``speed of light'' in eq.\ (\ref{mesospeed}) determined 
from the dispersion relation of a pseudoscalar meson on the lattice.
We see that the continuum relation $c_{\rm meson} =1$ is approximated
very well for the HF, but not for the Wilson fermion \cite{Lat96}. The 
simulation was performed quenched with the Wilson gauge action at $\beta =5$.}}
\label{mesofig}
\end{center}
\end{figure}

A first simulation of this HF with the simple gauging described above was
presented in Ref.\ \cite{Lat96}. We set the bare fermion mass to zero 
and evaluated the dispersion relation for the pseudoscalar meson
with the Wilson plaquette gauge action (see e.g.\ Refs.\ \cite{latbooks}),
at $\beta =5$ in a quenched simulation on a lattice of size
$6^{3} \times 18$. The corresponding ``speed of light''
\be  \label{mesospeed}
c_{\rm meson} = \frac{\sqrt{E^{2} - M^{2}}}{ | \vec p \, |}
\ee
is shown in Figure \ref{mesofig} and compared to the result for the 
Wilson fermion. In this formula, $E$, $M$ and $\vec p$ are the energy, 
mass and 3-momentum of the pseudoscalar meson.
The continuum behaviour, $c_{\rm meson} =1$, is marked by a dotted line.
We see that it is approximated very well for the HF --- which leads to
$c_{\rm meson} = 1.04(5)$ --- in striking contrast to the Wilson fermion. 
This property corresponds to an excellent approximation
to Lorentz symmetry (resp.\ to Euclidean rotation invariance).
However, the meson mass is strongly renormalised in this case.
In lattice units it amounts to $M \simeq 3$, hence it can hardly
describe a pion. Of course, we were dealing with a very coarse
lattice. Still, this property calls for a closer look at chiral
symmetry, which we will undertake in Section 7.\\

The HF has been applied successfully 
in simulations of the Schwinger model \cite{Schwinger} with two flavours,
on a $16\times 16$ lattice at $\beta= 6$, $4$ and $2$ \cite{WBIH}.
In these simulations, the gauge configurations where generated quenched,
but the measurements did include the fermion determinant.\footnote{In 
the recent literature, it became fashionable
to denote this kind of simulation simply as ``dynamical'', although 
the fermion determinant is still treated as constant
in the generation of the configurations.}
We used the Wilson plaquette gauge action, which is perfect
for pure 2d $U(1)$ gauge theory (see Subsection 5.3).

\begin{figure}
\begin{center}
 \includegraphics[angle=0,width=.5\linewidth]{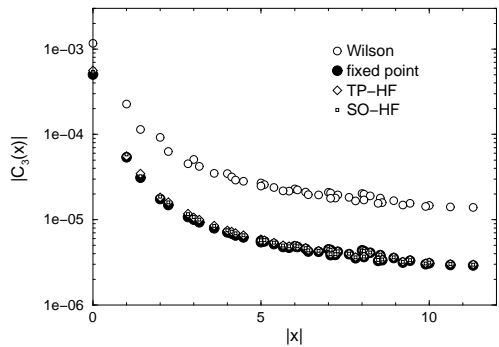}
\caption{\it{The isotropic correlation function (\ref{C3schwing})
for the Wilson fermion, a fixed point fermion and two HF versions
in the Schwinger model. We see in all cases but the Wilson fermion 
a smooth decay, which confirms the good quality of approximate rotation 
symmetry down to short distances.}}
\label{rotHFschwing}
\end{center}
\end{figure}

First we present another test of the quality of rotation symmetry.
Figure \ref{rotHFschwing} shows the decay of the correlation function
\be  \label{C3schwing}
C_{3}(x) = \langle \bar \Psi_{0} \, \sigma_{3} \Psi_{0} \cdot
\bar \Psi_{x} \, \sigma_{3} \Psi_{x} \rangle \ 
\ee
against the distance $| x |$, where $x$ are lattice sites in all 
directions.
We measured this correlator for the Wilson fermion and for several 
types of 2d HFs, which are described in Ref.\ \cite{WBIH}.
For the TP-HF, the fermionic coup\-lings correspond exactly to the 
description in Subsection 4.1, whereas the SO-HF is further
optimised with respect to the scaling behaviour.
Both variants also include a clover term.
This plot shows in addition the results obtained with the classically
perfect action constructed in Ref.\ \cite{Pany}.

The observation that the HFs and the (far more complicated)
classically perfect action display a much smoother decay of $C_{3}$
down to short distances $|x|$ approves their good approximation to
rotation symmetry.

\subsection{Spectral properties in the two-flavour Schwinger model}

\subsubsection{Applications of the hypercube fermion}

\begin{figure}
\begin{center}
 \includegraphics[angle=0,width=.48\linewidth]{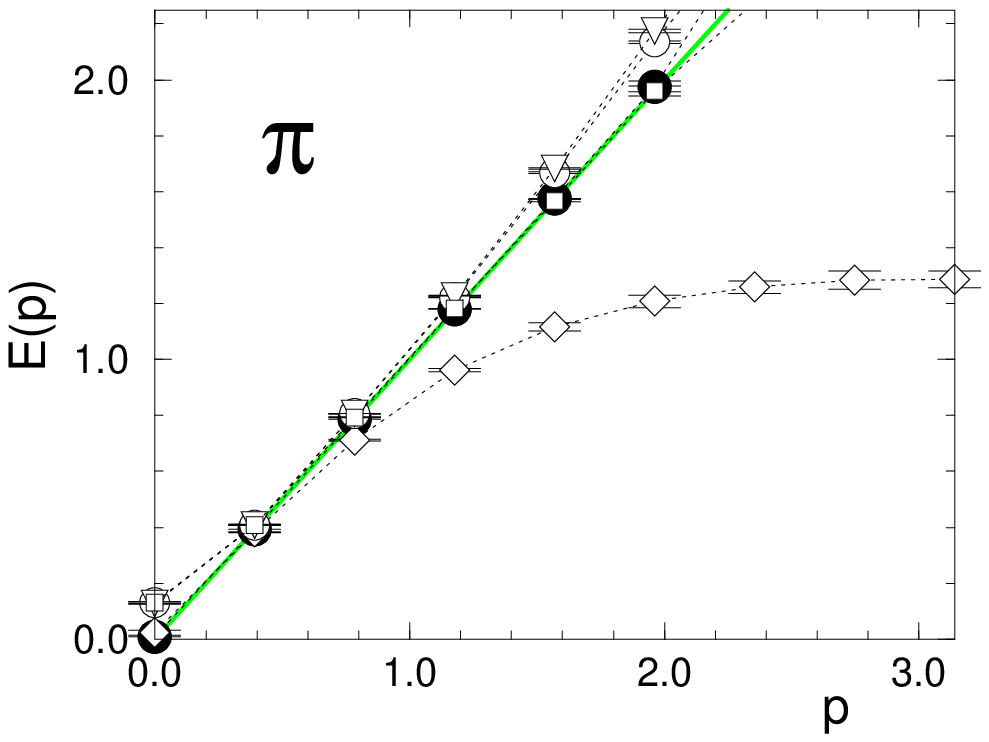}
 \includegraphics[angle=0,width=.48\linewidth]{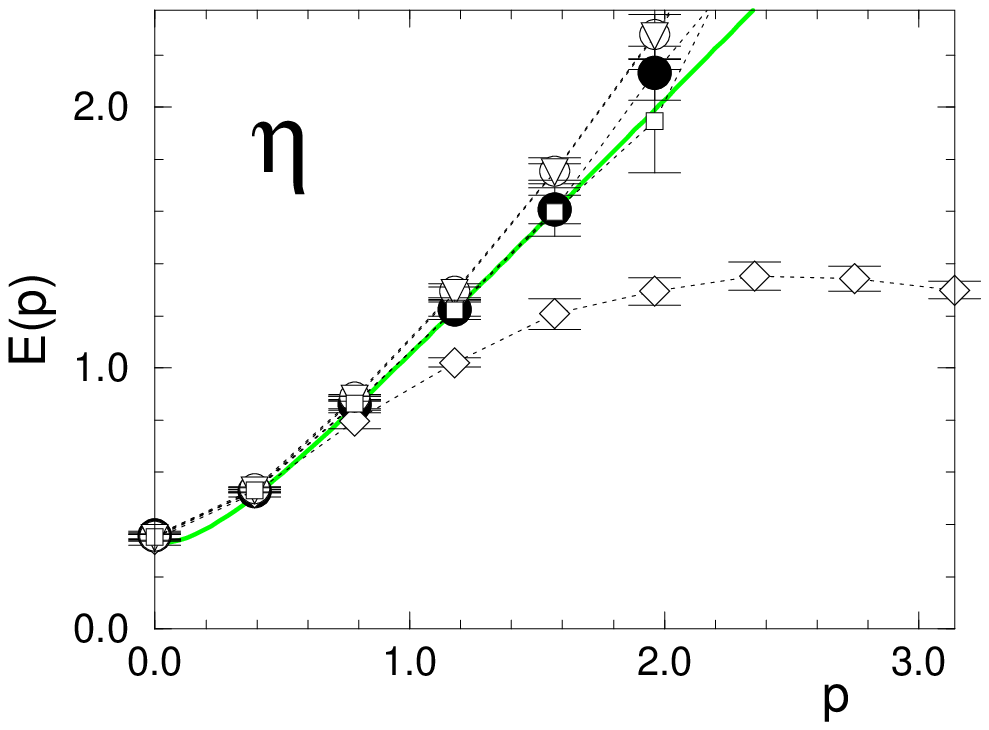}
\caption{\it{The dispersion relations for the ``pion'' and the
``$\eta$-meson'' in the Schwinger model with various lattice fermion
formulations: Wilson fermions (diamonds), the classically perfect action
(filled circles) and three types of HFs, in particular the 
scaling optimised SO-HF (little empty boxes) \cite{WBIH}. 
The solid lines mark the behaviour in the continuum.}}
\label{schwingHFdisp}
\end{center}
\end{figure}

In Ref.\ \cite{WBIH} we also
measured the dispersion relations for the states, which are
analogous to the pion and the $\eta$-meson (a general discussion of
these properties of the Schwinger model can be found in Ref.\
\cite{Sachs}). Again, for increasing momenta (in lattice units)
scaling errors due to lattice artifacts are enhanced, 
cf.\ Subsections 2.2 and 4.1.
In Figure \ref{schwingHFdisp} we show these dispersions,
which are again strongly improved for the classically perfect action
and for the HFs, in particular for the SO-HF. It is remarkable that the
latter --- which is still very simple --- scales at least as well as
the highly involved classically perfect action of Ref.\ \cite{Pany}. \\

\subsubsection{Applications of a truncated perfect staggered fermion}

\begin{figure}[h!]
\begin{center}
 \includegraphics[angle=270,width=.48\linewidth]{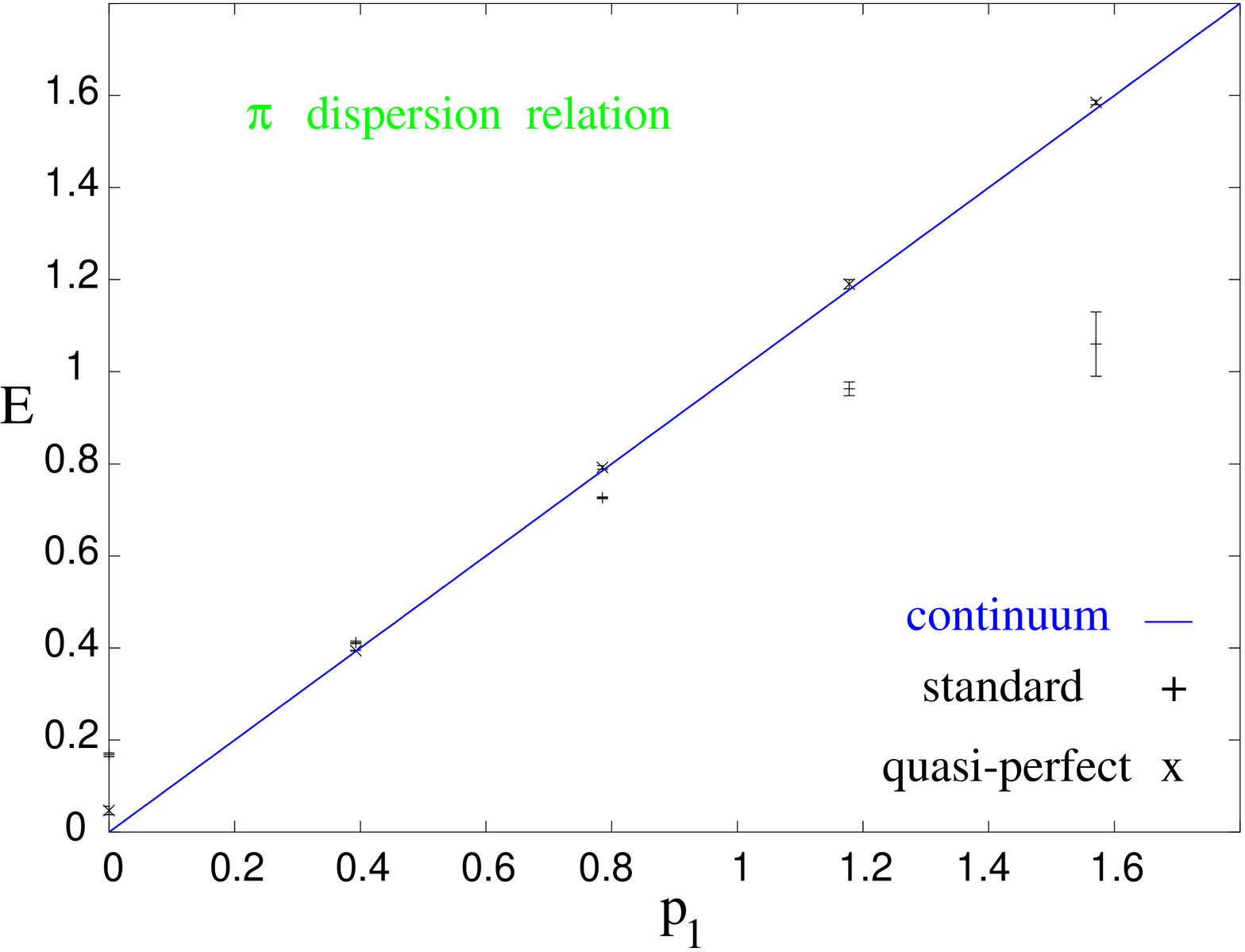}
 \includegraphics[angle=270,width=.48\linewidth]{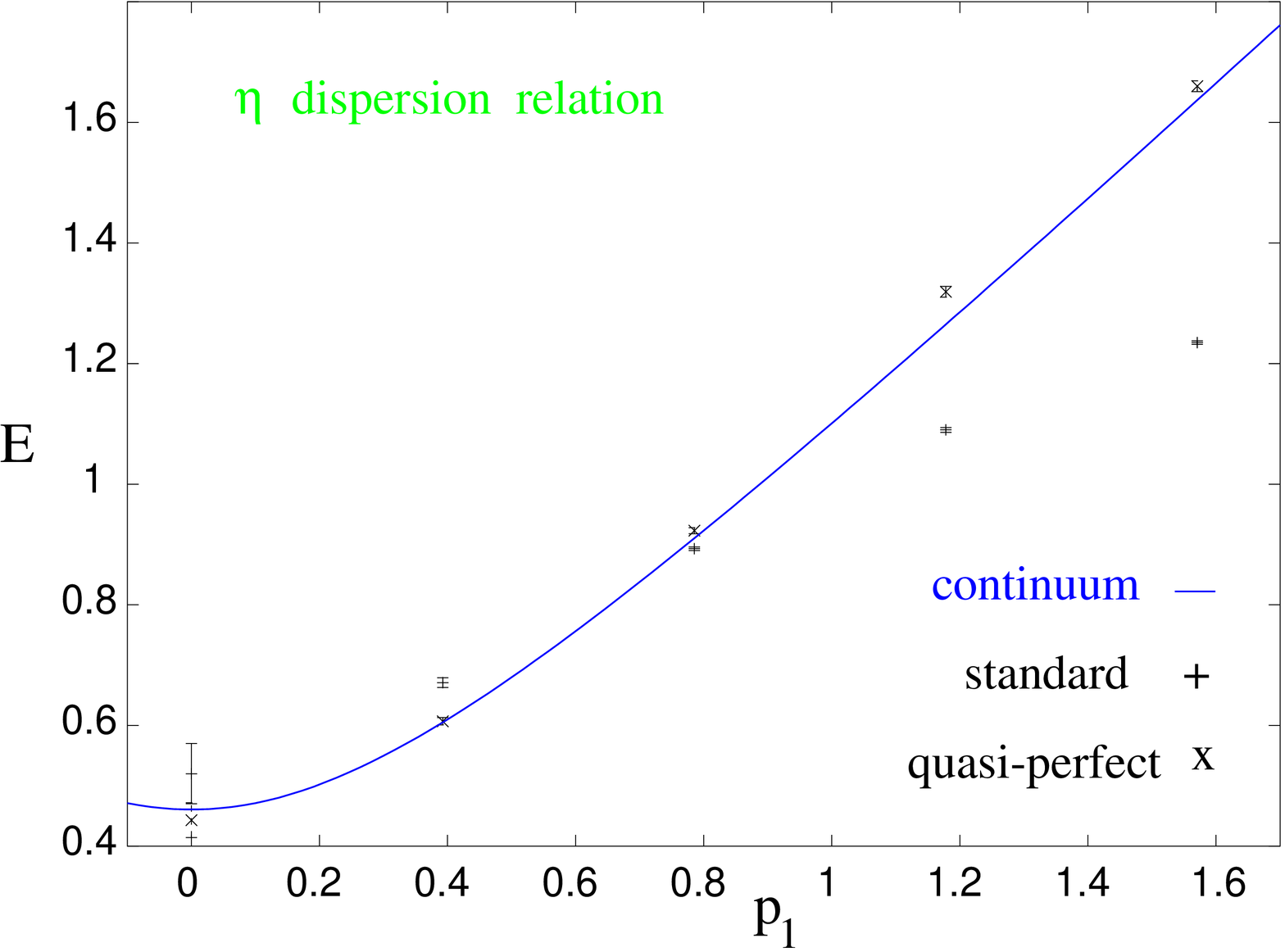}
\caption{\it{The dispersion relations for the ``pion'' and the
``$\eta$-meson'' in the 2-flavour Schwinger model with dynamical
staggered fermions.
We show results for standard staggered fermions and for
truncated perfect staggered fermions \cite{Dilg}, similar to the HF
discussed before. The solid lines mark the behaviour in the 
continuum, which is also here very close to the data for the
truncated perfect action.}}
\label{schwingstagdisp}
\end{center}
\end{figure}

\begin{figure}[h!]
\begin{center}
 \includegraphics[angle=0,width=.48\linewidth]{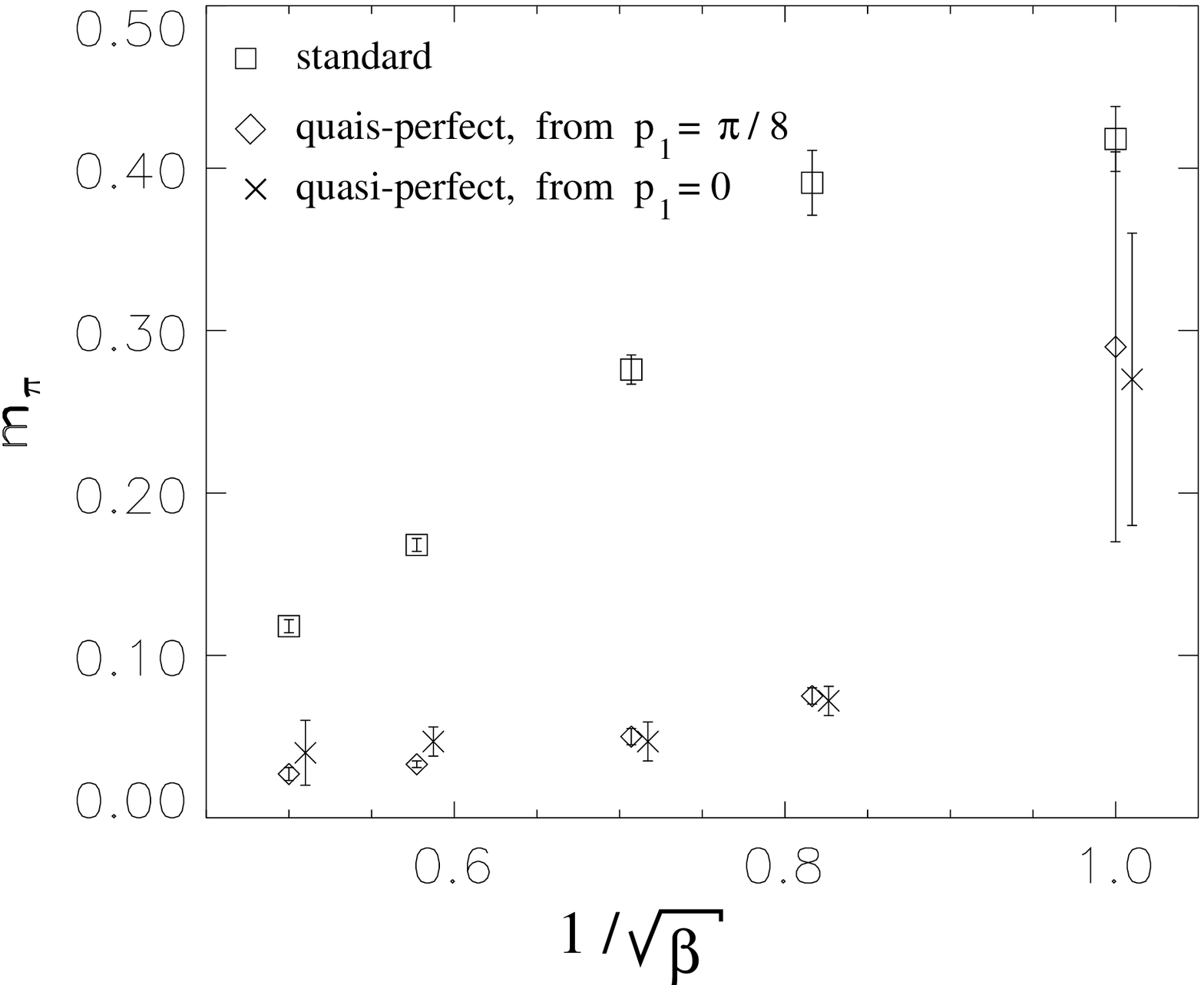}
 \includegraphics[angle=0,width=.48\linewidth]{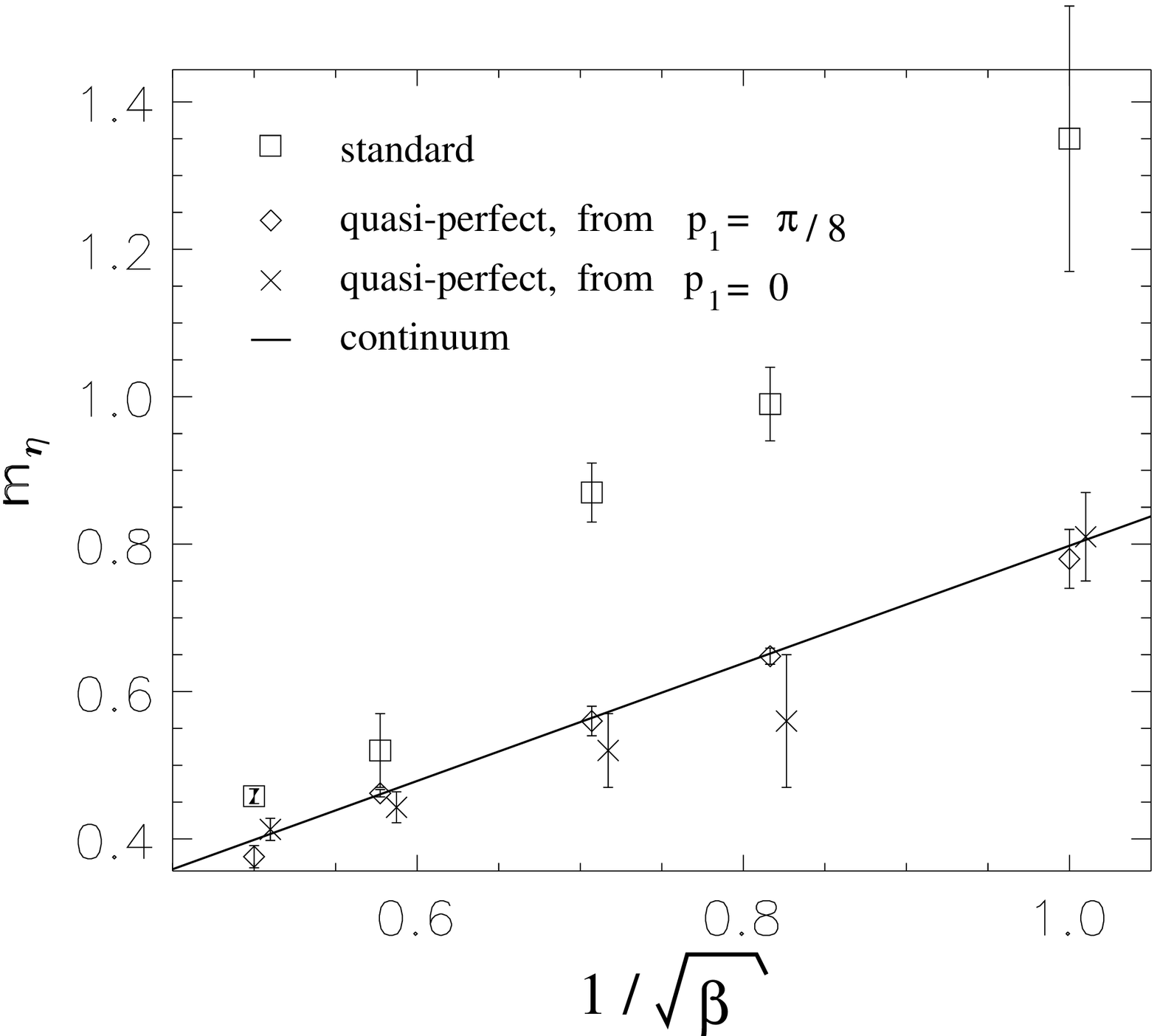}
\caption{\it{The ``meson'' masses in the Schwinger model
from dynamical simulations with different types of staggered
fermions, at various values of $\beta$,
resp.\ lattice spacings $a \propto 1 / \sqrt{\beta}$.
We confirm that the results
for the truncated perfect staggered fermion are far closer to the
continuum results, and they provide much lighter ``pions''.}}
\label{schwingstagmass}
\end{center}
\end{figure}

In Ref.\ \cite{Dilg} we constructed with a similar procedure a 
truncated  perfect staggered fermion, starting from the
formulation described in Subsection 4.2.
We applied it in Schwinger model simulations as well, and
these simulations were truly dynamical. We designed a
variant of the Hybrid Monte Carlo algorithm \cite{HMC}, which is particularly
suitable for this formulation. It uses a simplified action
(the standard staggered fermion action with fat links) for the
Molecular Dynamics, and the full quasi-perfect action in the
Metropolis accept/reject step. This provided a numerically cheap
evaluation of the force, along with a still useful acceptance rate
(as expected, the latter decreases at strong gauge coupling,
which corresponds to a large physical volume).
Again we used a $16 \times 16$ lattice
and the plaquette gauge action. At $\beta =3$ we found 
the neat $\pi$ and $\eta$ dispersion
relations shown in Figure \ref{schwingstagdisp}.
In the framework of that project, we further investigated the
``meson'' masses under variation of the gauge coupling $\beta$
(the parameters are given in detail in Ref.\ \cite{Dilg}).
The results, shown in Figure \ref{schwingstagmass}, reveal again
that the truncated perfect formulation is far more suitable to
approximate the continuum behaviour, and in particular to
realise light pions --- even on coarse lattices.

This work has pioneered on one side the construction and application
of improved staggered fermions ---
which became indeed fashion in the beginning of this
century --- as well as the use of a simplified force term in the
Hybrid Monte Carlo simulation of a complicated quasi-perfect action.

\subsection{The charmonium spectrum}

\begin{figure}[h!]
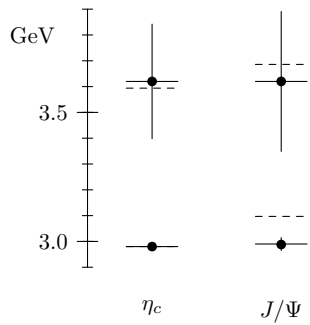

\begin{center}
\hspace*{1cm}
\def\fpsangle{0}
\epsfxsize=40mm
\fpsbox{spectrum.3}
\end{center}
\vspace{-5mm}
\caption{{\it The charmonium spectrum, measured in simulations with the HF
and a truncated perfect quark gluon vertex function \cite{cbarc}.
The experimental values are dashed,
and the ground state of $\eta_{c}$ sets the scale.}}
\label{chaspec}
\end{figure}

Regarding QCD, we performed simulations to evaluate the charmonium
spectrum employing the HF \cite{cbarc}.
In this case we used a truncated version
of the perfect quark gluon vertex function discussed in Subsection 5.4.
The result is shown in Figure \ref{chaspec}.
This was a quenched simulation
with the Wilson gauge action at $\beta = 5.6$ on a $8^{3}\times 16$ lattice.
The bare quark mass was set to $m=0.9$ 
--- that value was adapted to match the ground
state $\eta_{c}(2.98 {\rm GeV})$. Considering the fact that
only this ground state was used as an input, the agreement with the
experimental spectrum is clearly satisfactory. The inclusion of
a term $\propto \gamma_{\mu} \gamma_{\nu} \gamma_{\rho}$ in the vertex
function (see also Ref.\ \cite{Lat96})
was especially helpful for the quality of the charmonium spectrum. 

\subsection{Spectral functions at finite temperature}

\begin{figure}[h!]
\centering
\includegraphics[angle=0,width=0.49\linewidth]{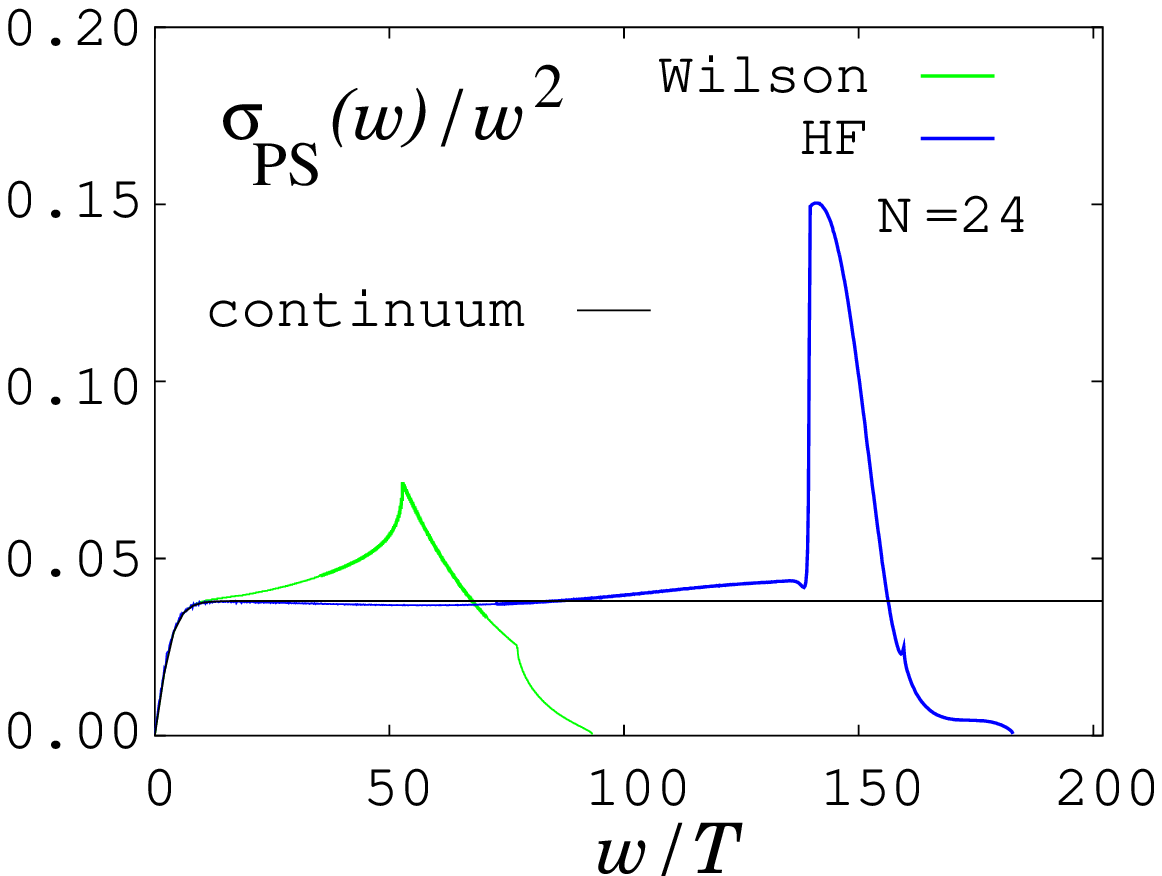}
\vspace*{2mm}
\includegraphics[angle=0,width=0.46\linewidth]{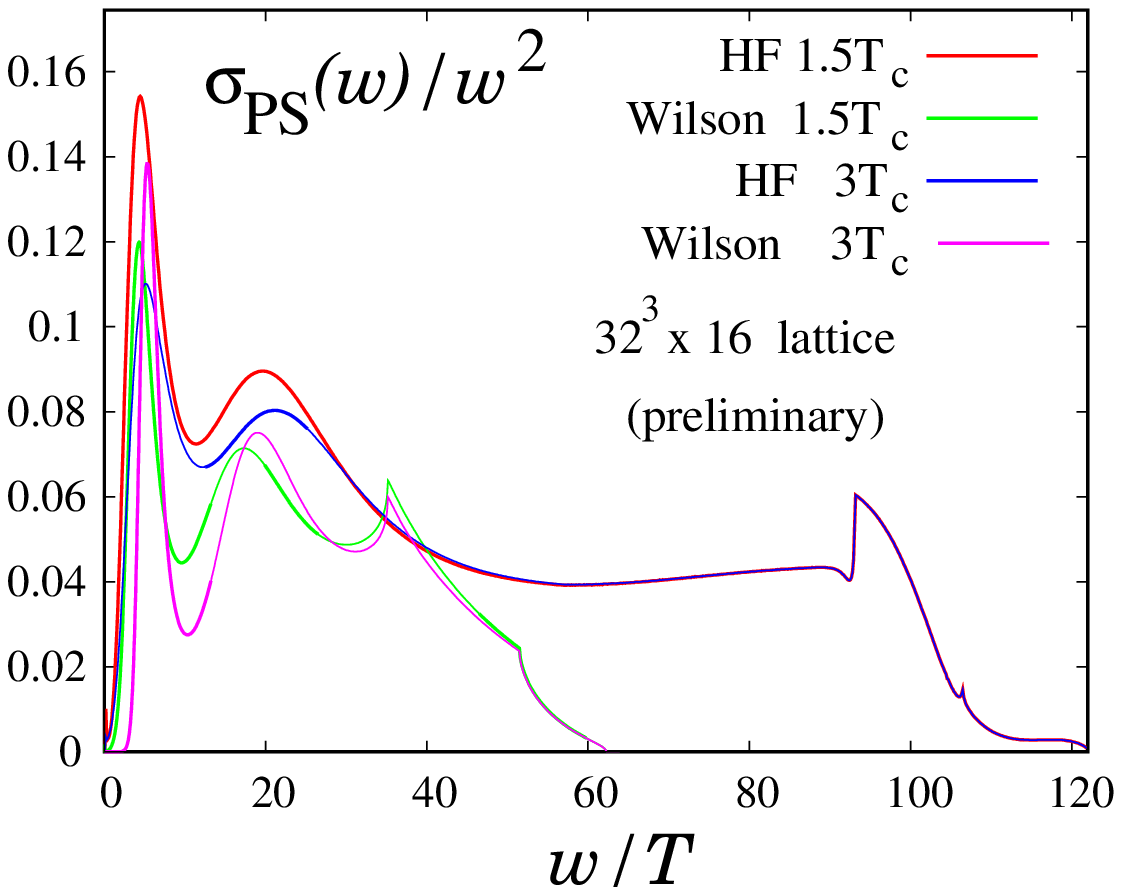}
\caption{{\it The spectral function $\sigma_{\rm PS}$,
as a function of the frequencies $\omega$,
at critical temperature
$T_{c} = \infty$ (on the left, free case) and at finite $T_{c}$
(on the right, interacting case). These results are obtained 
from lattice data using the Maximum Entropy Method for the HF 
and for the Wilson fermion.}}
\label{thermofig}
\end{figure}

At last, we add that the HF is currently being applied in
studies of the spectral functions of QCD at finite temperature \cite{Biel,Zyp}.
These spectral functions, depending on the frequencies,
are obtained from lattice data using the Maximum Entropy Method,
which was suggested for this purpose in Ref.\ \cite{MEM}.
For the HF they reveal a continuum-like behaviour up to much larger 
energies than it is the case for the Wilson fermion, see Figure
\ref{thermofig}. The basis for this virtue is that the HF moves all
the doublers to a unique cutoff scale, whereas the Wilson fermion
splits them into four (in general $d$) subsets, see Subsection 7.5.
We recall that the naive doubler species have between $1$ and $d$
momentum components $\pi /a$ in the free case. This causes
the splitting in their cutoff energy for $D_{\rm W}$ --- an effect 
that $D_{\rm HF}$ overcomes.

\section{Chiral Correction by means of the Overlap Formula}

\subsection{The Ginsparg-Wilson relation}

In Subsection 4.1 we discussed the block variable
RGT for the free fermion field. Eq.\ (\ref{fermRGT})
introduced the transformation term, which we assume (for simplicity)
to be diagonal in the spinor indices,
\be
\sum_{x,y} \Big[ \bar \Psi_{x} -
\int_{C_{x}} d^{d}u \, \bar \psi (u) \Big] (R^{-1})_{xy}
\Big[ \Psi_{y} - \int_{C_{y}} d^{d}v \, \bar \psi (v) \Big] \ .
\ee
For the bare fermion mass $\, m=0 \,$ the inverse perfect lattice
Dirac operator has the structure
\be
(D^{-1}) = \frac{- \gamma_{\mu} \rho_{\mu} }{\rho^{2}
+ \lambda^{2}} + R \ , \quad R = \frac{\lambda}{\rho^{2} + \lambda^{2}} \ , 
\ee
where we are using the notation of eq.\ (\ref{perfactcoor})
and $\rho^{2} = \rho_{\nu} \rho_{\nu}$
(in coordinate space, the products are lattice convolutions). 
We repeat 
that in the limit of a $\delta$-function RGT, $R \to 0$, the lattice action
is invariant under the standard chiral transformation (\ref{chiralrot}),
as we see from the anti-commutator $\{ D , \gamma_{5} \} = 0$.

However, we also saw that locality requires $R \neq 0$, in agreement
with the Nielsen Ninomiya No-Go Theorem \cite{NoGo}, and we found for $m=0$
the term 
\be  \label{Rdelta}
R_{xy} = \frac{1}{2} \delta_{xy}
\ee
to be optimal for locality \cite{QuaGlu}.
In this case, we have to face $\{ D , \gamma_{5} \} \neq 0$, but 
due to our requirements for $R$ in Subsection 4.1 
(eq.\ (\ref{Rnonzero})) the anti-commutator
\be  \label{GWR1}
\{ D^{-1}, \gamma_{5} \} = 2 \gamma_{5} R 
\ee
is {\em local.} The exact form of the factor $\Pi$ (which depends
on the blocking scheme, for our case it is given in eqs.\
(\ref{perfermiprop}) and (\ref{perfscalprop})) 
does not affect this relation. This already
indicates that for a given term $R$ there is a variety of solutions
to eq.\ (\ref{GWR1}).

We stressed in Section 4 --- and a long time ago, 
for instance in Ref.\ \cite{perfaxial} --- that the specific 
chiral symmetry breaking by the term $R$ cannot distort any physical
properties, in particular not those related to chirality.
Relation (\ref{GWR1}) illustrates this again, since a local term
$R$ cannot shift the poles in the propagator $D^{-1}$.
If we multiply the operator $D$ from both sides, we arrive at the
equivalent equation
\be  \label{GWR}
\{ D , \gamma_{5} \} = 2 D \gamma_{5} R D \ ,
\ee
which is now known as the {\em Ginsparg-Wilson relation} 
(GWR).\footnote{If we allow a 
general Clifford structure for the term $R$, the right-hand-side turns into
$D \{ \gamma_{5},R \} D$.}
The dimensions reveal that the right-hand side is $O(a)$ suppressed.
As we mentioned earlier, the GWR was written down in 
Ref.\ \cite{GiWi}, which was forgotten over a long
period until its re-discovery in 1997 \cite{Has97}.

We may compare this property to the Wilson operator, which breaks
chiral symmetry such that $\{ D_{{\rm W},m=0}, \gamma_{5} \}$
amounts to a local term, unlike relation (\ref{GWR1}).
If we try to insert the free operator $D_{{\rm W},m=0}$
into the GWR, we arrive at
a non-local term $R_{xy}$, which decays only as $| x-y|^{-6} $ in $d=4$,
and as $| x-y|^{-4} $ in $d=2$ \cite{Dubna}. Hence the Wilson 
operator does not solve the GWR.

We continue to use the GWR as a requirement also in the interacting 
case. For gauge interactions, $D(U)$ involves the link variables, and so
does $R$, if it couples different lattice sites. However,
we stay now with the choice (\ref{Rdelta}), for which 
the GWR takes the simple form
\be  \label{Rxy}
R_{xy} = \frac{1}{2} \, \rho \, \delta_{xy} \ , \ \rho =1 \quad 
\Rightarrow \quad
\{ D , \gamma_{5} \} =  D \gamma_{5} D \ .
\ee

M.\ L\"{u}scher observed that this relation assures the
exact invariance under a lattice modified chiral symmetry transformation
\cite{ML98}.
In the notation of eq.\ (\ref{chiralrot}), it is sufficient to consider
this modified transformation to $O(\alpha )$
(due to the Lie group structure of the chiral rotation),
\bea
\bar \Psi D \Psi & \to & \bar \Psi \Big( 1 + \alpha ( 1 -
\frac{1}{2}D ) \gamma_{5} \Big) \, D \, \Big( 1 + \alpha \gamma_{5}
( 1 - \frac{1}{2}D ) \Big) \Psi + O(\alpha^{2}) \nn \\  \label{modchirot}
&=& \bar \Psi D \Psi + \alpha \bar \Psi \Big[
\{ D , \gamma_{5} \} - D \gamma_{5} D \Big] \Psi + O(\alpha^{2}) \ .
\eea
The GWR corresponds exactly to a vanishing term in the
square bracket. In fact, it could have been also discovered
by requiring invariance under such a lattice modified
chiral symmetry.\footnote{The GWR is presented along these lines
in Ref.\ \cite{Dubna}, which also keeps a general term $R$ in the
transformation (\ref{modchirot}).
Then the transformation of $D$ reads \\
$D \to [1 + \alpha (1 - DR) \gamma_{5} ] D 
[1 + \alpha \gamma_{5} (1 - RD) ] = D + \alpha 
(\{ D , \gamma_{5} \} - D \{R, \gamma_{5} \}D) + O(\alpha^{2})$.\\
Again the invariance to $O(\alpha)$ is equivalent to the GWR,
and for local terms $R$ there is a smooth transition to the standard
chiral symmetry in the continuum limit. }
Unlike the remnant chiral symmetry of staggered
fermions, this chiral rotation involves the full number of
generators that appear in the continuum theory.
The continuum limit $a \to 0$ yields a smooth transition
to the standard form of chiral invariance.
Based on these observations, even a non-perturbative formulation 
of chiral $U(1)$ gauge theory  (cf.\ Subsections 3.3 and 3.4) 
has been worked out on the conceptual level \cite{chiU1}. 
Earlier works in this direction, based on the overlap formalism, 
are collected in Ref.\ \cite{overchiral}.\\

Perfect Dirac operators solve the GWR, but --- as we saw ---
they can in general not be constructed explicitly. A step towards
an applicable solution was the observation that also classically
perfect Dirac operators (cf.\ Subsection 2.3)
are Ginsparg-Wilson operators \cite{HLN}.
Still, a truncation of the couplings and therefore a deviation from 
exact chiral symmetry is needed in its construction, but in view
of the possibilities to build approximate solutions this is
a more accessible starting point than the (quantum) perfect action.

A fully explicit solution was presented by H.\ Neuberger \cite{Neu}.
Let us recall the properties (\ref{gamma5E}) and
start from some massless lattice Dirac operator $D_{0}$, which we
assume to be $\gamma_{5}$-Hermitian, i.e.
\be  \label{g5herm}
D^{\dagger}_{0} = \gamma_{5} D_{0} \gamma_{5} \ .
\ee
This property holds for practically all Dirac operators that have 
been considered, in particular for the Wilson operator $D_{\rm W}$.\footnote{An 
exception is the operator of the so-called ``twisted mass fermion'' 
\cite{twist}. For the staggered fermion, the analogous relation
$\epsilon (x) \rho_{\mu}(x,y) \epsilon (y) = - \rho_{\mu}(x,y)$
holds, where $\epsilon (x)$ is the sign factor (\ref{epsistag}),
and $\rho_{\mu}$ is the nearest-neighbour coupling in eq.\ (\ref{stagact}).}
Now we define the shifted operator
\be
A : = D_{0} -1 \ ,
\ee
which is unitary {\em if} \ $D_{0}$ is a GW operator,
\be
A^{\dagger} A = \gamma_{5} \Big[ D_{0} \gamma_{5} D_{0} -
\{ D_{0} , \gamma_{5} \} + \gamma_{5} \Big] \ .
\ee
Of course, we cannot assume this for a quite general $D_{0}$, and we pointed
out before that it does not hold in particular for $D_{\rm W}$.
But we can transform $A$ such that unitarity is enforced,\footnote{Note that
$A$ does in general not commute with $A^{\dagger}A$, so we have
to specify an order where to multiply the inverse square root.
This is not necessary anymore in the form given in the lower line
of eq.\ (\ref{overlap}).}
\be
A \to A_{\rm ov} := A / \sqrt{ A^{\dagger} A } \ \quad \Rightarrow
\quad A_{\rm ov}^{\dagger} A_{\rm ov} = \uno \ .
\ee
In this way, we obtain the {\em overlap Dirac operator}
\bea
D_{\rm ov} &=& 1 +  A_{\rm ov} = 1 + (D_{0} -1) 
/ \sqrt{ (D_{0}^{\dagger}-1) (D_{0} - 1) } \nn \\
&=& 1 + \gamma_{5} \frac{H}{\sqrt{H^{2}}} \ , \quad 
H := \gamma_{5} A \ ,
\label{overlap}
\eea
where $H$ is Hermitian, $H = H^{\dagger}$.
H.\ Neuberger introduced this operator \cite{Neu} with $D_{0} = D_{\rm W}$,
and we denote the corresponding overlap operator as the 
{\em Neuberger operator} $D_{\rm N}$.
 
Since the resulting lattice action $S [ \bar \Psi, \Psi, U]$ has now
a modified but exact chiral symmetry, one may be worried about the
fate of the axial anomaly. However, the anomaly is in fact captured
correctly, due to the variance of the functional measure under modified
chiral rotations \cite{ML98}, which is analogous to the continuum.
This property means an explicit realisation of the program formulated
in Ref.\ \cite{Fuji}. The axial anomaly has been computed
perturbatively by many authors, for instance in
Refs.\ \cite{GiWi,ML98,HLN,Rothe} for general Ginsparg-Wilson operators,
and specifically for $D_{\rm N}$ also in Refs.\ \cite{anomalDN}.
A proof which extends to all topological 
sectors was given for $D_{\rm N}$ in Ref.\ \cite{DA}.
In fact, this extension is non-trivial, as the considerations
in Refs.\ \cite{TWC} underline.

\subsection{Massless lattice fermions in $d=3$}

In three dimensions, there is no chiral symmetry (since there is no matrix
$\gamma_{5}$ at hand), but parity takes a r\^{o}le similar to a discrete
chiral symmetry.\footnote{We encountered a discrete chiral symmetry before 
in the Gross-Neveu model (in the continuum), see eq.\ (\ref{chi-sym})
where in view of the chiral rotation (\ref{chiralrot}) only the angles
\ $ \alpha \ {\rm mod } \ 2\pi \ \in \ \{ 0 ,\pi \}$ \ occur.
In that case, extending the 4-Fermi term to $\, (\bar \Psi \Psi)^{2}
+ (\bar \Psi i \gamma_{5} \Psi)^{2} \,$ promotes the chiral symmetry
to a continuous form (Thirring model).}
The parity operator ${\cal R}$ is then the analogue to $\gamma_{5}$,
and we can write the parity transformation on the lattice for spinors
and compact link variables as
\bea
P &:& \bar \Psi_{x} \to i \bar \Psi_{x} {\cal R} \ , \quad
\Psi_{x} \to i {\cal R} \Psi_{x} \ , \nn \\
&& U_{\mu ,x} \to U_{\mu ,x}^{P} := U_{\mu , -x}^{\dagger} \ .
\label{paritrans}
\eea
Similar to the $\gamma_{5}$-Hermiticity (\ref{g5herm}), practically
all lattice Dirac operators which are considered obey
\be
D(U)^{\dagger} = {\cal R} D(U^{P}) {\cal R}  \ .
\ee
The relation
\be  \label{parcon}
D(U) + D(U)^{\dagger} = 0 \ 
\ee
then implies parity invariance of the action 
$S = \sum_{xy} \bar \Psi_{x} D(U)_{xy} \Psi_{y}$, and fermion mass zero.
However, this condition (\ref{parcon}) is again not easy
to fulfil --- it leads to a doubling problem as the chiral
symmetry does in even dimensions. 
The doublers in the naive action can be avoided by a Wilson term,
but this term breaks the condition (\ref{parcon}) and therefore parity
symmetry, so that additive mass renormalisation sets in.

A (massless) overlap fermion in $d=3$ was introduced in Ref.\ 
\cite{Kiku}, which was actually the first place where the overlap
formula (\ref{overlap}) appeared. In Ref.\ \cite{parity} we 
considered generally a 3d Ginsparg-Wilson operator given by
the condition
\be  \label{3dGWR}
D + D^{\dagger} = D^{\dagger}D \ ,
\ee
and we studied a lattice modified parity symmetry. This modification alters
the transformations (\ref{paritrans}) such that
\be
\Psi_{x} \to i {\cal R} (1 - D) \Psi_{x} \ .
\ee
For a solution $D$ to eq.\ (\ref{3dGWR}) the lattice action is exactly
invariant under this modified parity symmetry, but the functional
measure transforms as
\be
{\cal D} \bar \Psi \, {\cal D} \Psi \ \to \
[ {\rm det}(1 -D) ]^{-1} {\cal D} \bar \Psi \, {\cal D} \Psi \ .
\ee
Once more in analogy to the chiral symmetry in even dimensions,
this transformation of the measure gives rise to the requested parity
anomaly \cite{parity}. We should clarify that this is not an anomaly
in the usual sense, which has a unique value, but it comes with 
an arbitrary integer factor (a comprehensive discussion is given
in Ref.\ \cite{CoLu}). 
Hence a successful regu\-larisation should
capture all possibilities for this integer. In the current case,
they are all captured indeed by varying either the kernel $R$ in the
GWR (which then modifies the right-hand-side of eq.\ (\ref{3dGWR})
to $ 2 D^{\dagger}RD$), or by considering
different kernels $D_{0}$ in the overlap formula (\ref{overlap}). \\

At this point we add that also a pure Abelian 3d Chern-Simons gauge theory
with the continuum action
\be
S[{\cal A}] = \int d^{3}x \, \epsilon_{\mu \nu \rho} \,  
{\cal A}_{\mu} \partial_{\nu} {\cal A}_{\rho} 
\ee
suffers from a doubling problem on the lattice due to the
occurrence of a linear momentum \cite{Peru}.
A solution can be found also here either by the perfect action formalism,
or by a formula of the overlap-type \cite{CS}.

\subsection{The overlap hypercube fermion}

We proceed to a slightly more general form of the
term $R$ in the GWR (\ref{GWR}): we now allow for a
parameter $\rho \gsim 1$ in eq.\ (\ref{Rxy}). The overlap formula
can be adapted to general forms of $R$ \cite{EPJC,Bern}. For the still
simple form that we are considering now, it reads
\bea
D_{\rm ov} &=& \rho \left( 1 + A / \sqrt{ A^{\dagger} A} \right) 
= \rho \Big( 1 + \gamma_{5} \frac{H}{\sqrt{H^{2}}} \Big) \ , \nonumber \\
A &:=& D_{0} - \rho = \gamma_{5} H \ .  \label{ovrho}
\eea
As we mentioned, the standard formulation inserts as a kernel the Wilson 
operator, $D_{0} = D_{\rm W}$, 
which is far from chiral, and which undergoes a drastic
change in the overlap formula to yield the Neuberger operator $D_{\rm N}$.
Ref.\ \cite{EPJC} suggested to consider more general possibilities
for $D_{0}$, and motivated in particular the use of a truncated perfect
operator, which is approximately chiral already. In this case, the square
root in eq.\ (\ref{ovrho}) is close to the constant $\rho$, and
the transition from $D_{0}$ to $D_{\rm ov}$ is therefore only a modest
modification. An intuitive argument for this property
is that an exact GW kernel $D_{0}$
reproduces itself identically in the overlap formula (\ref{ovrho})
(for a fixed parameter $\rho$).\footnote{In this sense, the overlap 
formula captures all solutions to the GWR. 
Attempts to construct chiral lattice Dirac operators beyond GWR solutions
were considered in Refs.\ \cite{WK}.}

In particular we are using $D_{\rm HF}$ 
as the input Dirac operator --- note that it is $\gamma_{5}$-Hermitian 
as well. Its inexact chirality is then corrected
by the overlap formula, which leads to $D_{\rm ovHF}$, the operator
of the {\em overlap-HF}, while keeping both Dirac operators similar,
\be  \label{ovapprox}
D_{\rm ovHF} \approx D_{\rm HF} \ .
\ee
The motivation for this construction is that the property 
(\ref{ovapprox}) allows us to
preserve other virtues of $D_{\rm HF}$ (beyond the approximate 
chirality) in the chirally exact formulation $D_{\rm ovHF}$.
As such virtues we are going to discuss:
\begin{itemize}

\item a high level of locality

\item approximate rotation symmetry

\item a good scaling behaviour.

\end{itemize}
Below we will summarise results for these three aspects
and comment on their meaning.\footnote{Of course, one could also
construct an approximate Ginsparg-Wilson operator directly
by starting from some short-ranged parameterisation ansatz
and tuning the couplings such that the GWR is minimally violated.
This was done in Refs. \cite{GatHip} for the Schwinger model
and for QCD.} 
Still before that we mention that
the simulation of an overlap fermion with a hypercubic kernel
requires more computational effort compared to $D_{\rm N}$.
The use of the complicated kernel by itself corresponds to an overhead 
of about a factor $15$ in QCD (without applying the preconditioning
method reviewed in Subsection 6.1). 
However we should consider that in
4d simulations, the inverse square root in the overlap formula
(\ref{ovrho}) (resp.\ the sign function $\frac{H}{\sqrt{H^{2}}}$ ) 
is approximated by polynomials. Again thanks to the relation
(\ref{ovapprox}), the convergence in the polynomial evaluation
is faster --- for a fixed precision one gains back by this property
about a factor of $3$ \cite{ovHFQCD,Stani,WBStani}. 
Hence an overhead of about a factor
$5$ remains. We are confident that this factor will be more than compensated 
by the superior properties listed above, which we will discuss in
Subsections 7.6 and 7.7.

\subsection{The axial anomaly in the continuum limit}

On the conceptual side, we were able to prove that also the {\em overlap-HF}
(obtained by inserting $D_{0}=D_{\rm HF}$ in the overlap formula, as we
just advocated) has the correct chiral anomaly in the continuum limit of all 
topological sectors \cite{DAWB}.\footnote{This may be compared to the fully
perfect action, which even displays the correct axial anomaly at finite
lattice spacing, as we discussed in Subsection 5.3.}
This proof required a number of generalisation steps 
compared to the proof that had been worked out previously for the 
Neuberger operator $D_{\rm N}$ \cite{DA}.

In Ref.\ \cite{DAWB} we gave a non-perturbative evaluation of
the continuum limit of the axial anomaly and the fermion index.
To this end, we formulated the Dirac operator $D_{\rm HF}$
on a $2n$-dimensional Euclidean lattice in a form,
which is well-suited for analytic investigations. 
We used it first to study the dependence of the 
doubler structure of $D_{\rm HF}$ on its coupling parameters. 
Then we evaluated the classical continuum 
limits of the axial anomaly and the index of the
overlap-HF operator, showing that the correct continuum expressions 
are recovered when the parameters are in the physical (doubler-free) range.
This continuum limit relies only on general pro\-perties of $D_{\rm HF}$
and not on its explicit form (in contrast to the previous evaluations 
in the Wilson case \cite{DA}).
The main new technical tool was a set of identities, 
which allowed the continuum form 
\be
\epsilon_{\mu_1\dots\mu_{2n}}\, {\rm Tr} \, F_{\mu_1 \mu_2}(x)\cdots{}
F_{\mu_{2n-1}\mu_{2n}}(x)
\ee
of the axial anomaly to be extracted, and its coefficient to be 
evaluated topologically. 

These properties are basically not specific to the HF structure. 
We expect that this proof can be extended to completely 
general overlap Dirac operators obtained by inserting an ultralocal
(and $\gamma_{5}$-Hermitian)
lattice Dirac operator $D_{0}$ (involving the full Clifford algebra 
of $\gamma$ matrices, as the operators in Refs.\ \cite{Bern,GatHip,BGR}) 
into the overlap formula.

\subsection{Approximate chirality of the hypercube fermion}

\begin{figure}[h!]
\vspace*{-10mm}
\begin{center}
  \includegraphics[angle=270,width=1.\linewidth]{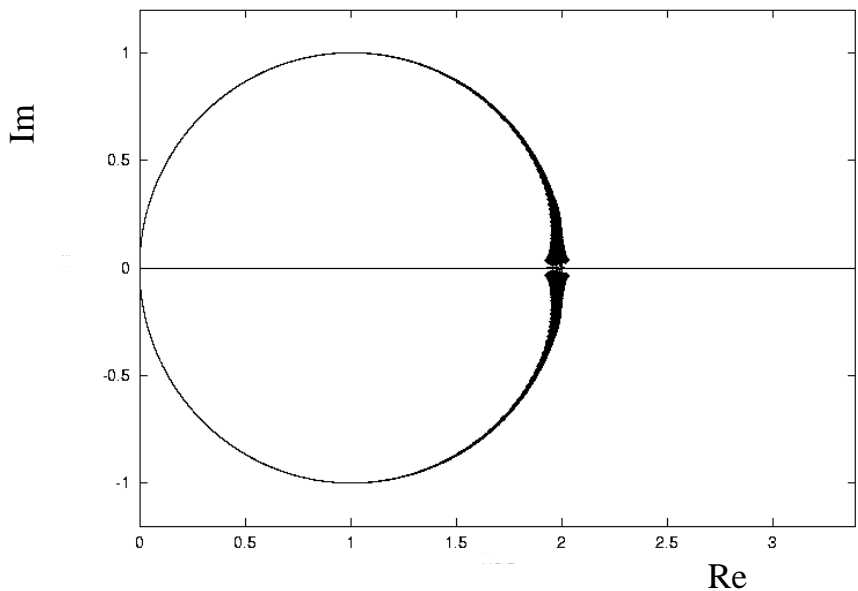}
\end{center}
\vspace*{-6mm}
\caption{{\it The spectrum of the truncated perfect, free HF in $d=4$,
on a lattice of unit spacing and infinite size.
It is very close the GW circle with centre and radius
$\rho =1$, which shows that it approximates chirality very well.}}
\label{freeHFspec}
\end{figure}

An exact solution to the GWR with a term $R$ of the form
(\ref{GWR}) and $\rho \gsim 1$ has its spectrum $\sigma (D)$
on a circle in the complex plane with centre and radius $\rho$,
as we see from the relation $A^{\dagger}A \equiv \rho^{2}$. 
We call it the {\em GW circle},
\be
\sigma (D) \subset \Big\{ z \ \Big\vert \ | z - \rho |  = \rho \Big\} \ .
\ee
This property reveals immediately the absence of additive mass
renormalisation and of ``exceptional configurations'' (with accidental
near-zero modes in quenched simulations).
In order to check how well some input operator $D_{0}$ approximates
chirality already, it is therefore a sensible criterion to evaluate
the spectrum $\sigma (D_{0})$ and to test if it is close to this
GW circle \cite{EPJC}. Let us start with the free HF in $d=4$: the spectrum 
of $D_{\rm HF}$ in infinite volume is shown in Figure \ref{freeHFspec}, 
and we see that it approximates the GW circle
(with $\rho =1$) extremely well. On the other hand, the 
spectrum of the free Wilson operator covers four circles of this
kind, so that its real part extends up to $\approx 8$.~\footnote{In 
Ref.\ \cite{EPJC} we also measured the violation
of the GWR directly, and we compared the couplings before and after
the application of the overlap formula as further criteria for the
approximate chirality of different options for $D_{0}$.} 
(We anticipated this property in Section 6 when we commented on Figure
\ref{thermofig}.)

\begin{figure}[h!]
\begin{center}
  \includegraphics[angle=270,width=0.49\linewidth]{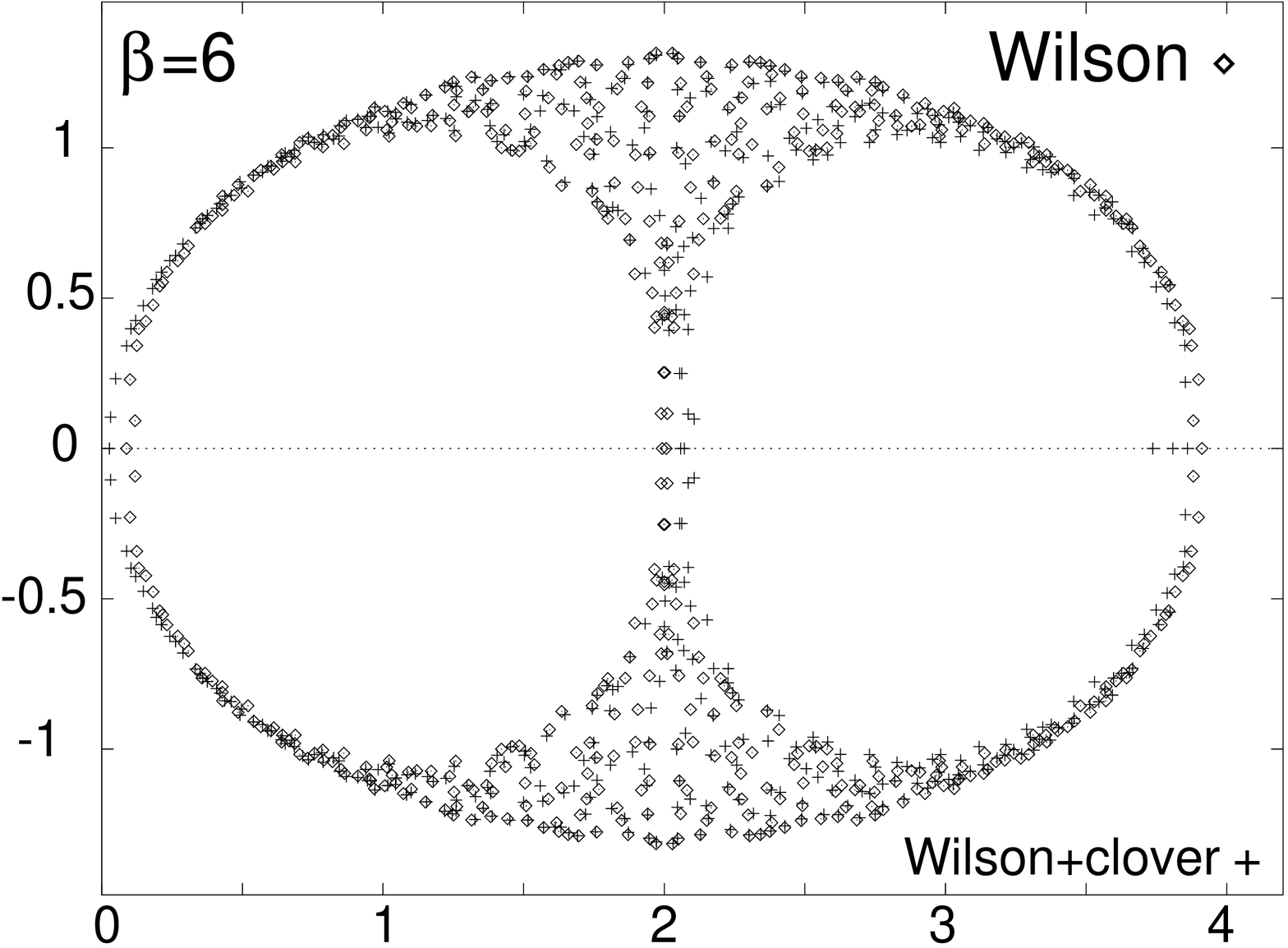}
  \includegraphics[angle=270,width=0.49\linewidth]{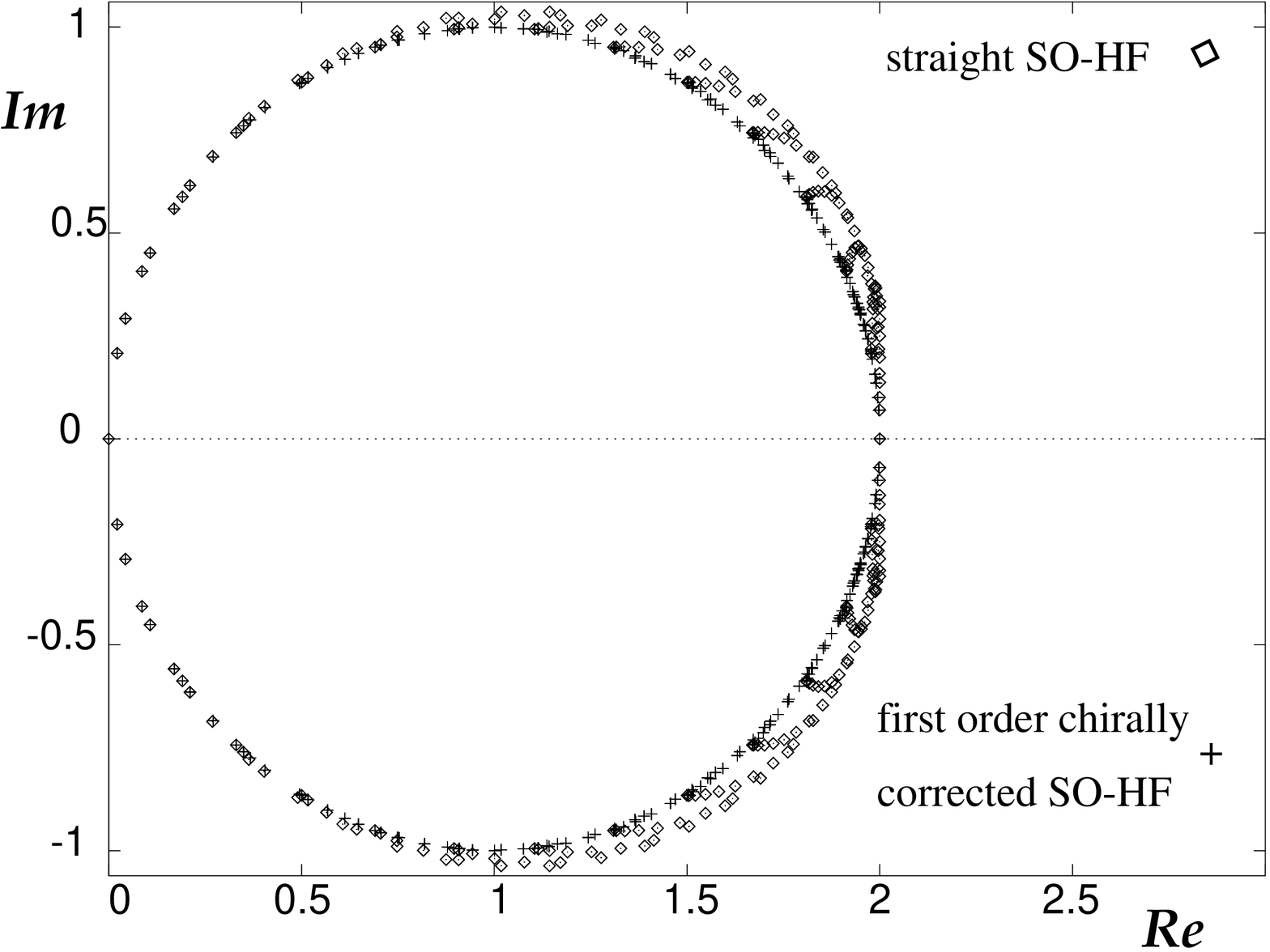}
\end{center}
\caption{{\it The spectra of the Wilson operator (with and without
a clover term with coefficient $1$) and of the HF operator for a 
typical configuration in the
Schwinger model at $\beta =6$. The Wilson spectrum deviates
strongly from the GW circle, whereas the HF spectrum approximates it
well. In the latter case we also show the result if the overlap
formula is approximated by a polynomial of first order only, 
which is sufficient to put the eigenvalues quite exactly onto the 
GW circle \cite{WBIH}.}}
\label{schwingspecfig}
\end{figure}

\begin{figure}[h!]
\begin{center}
  \includegraphics[angle=270,width=0.49\linewidth]{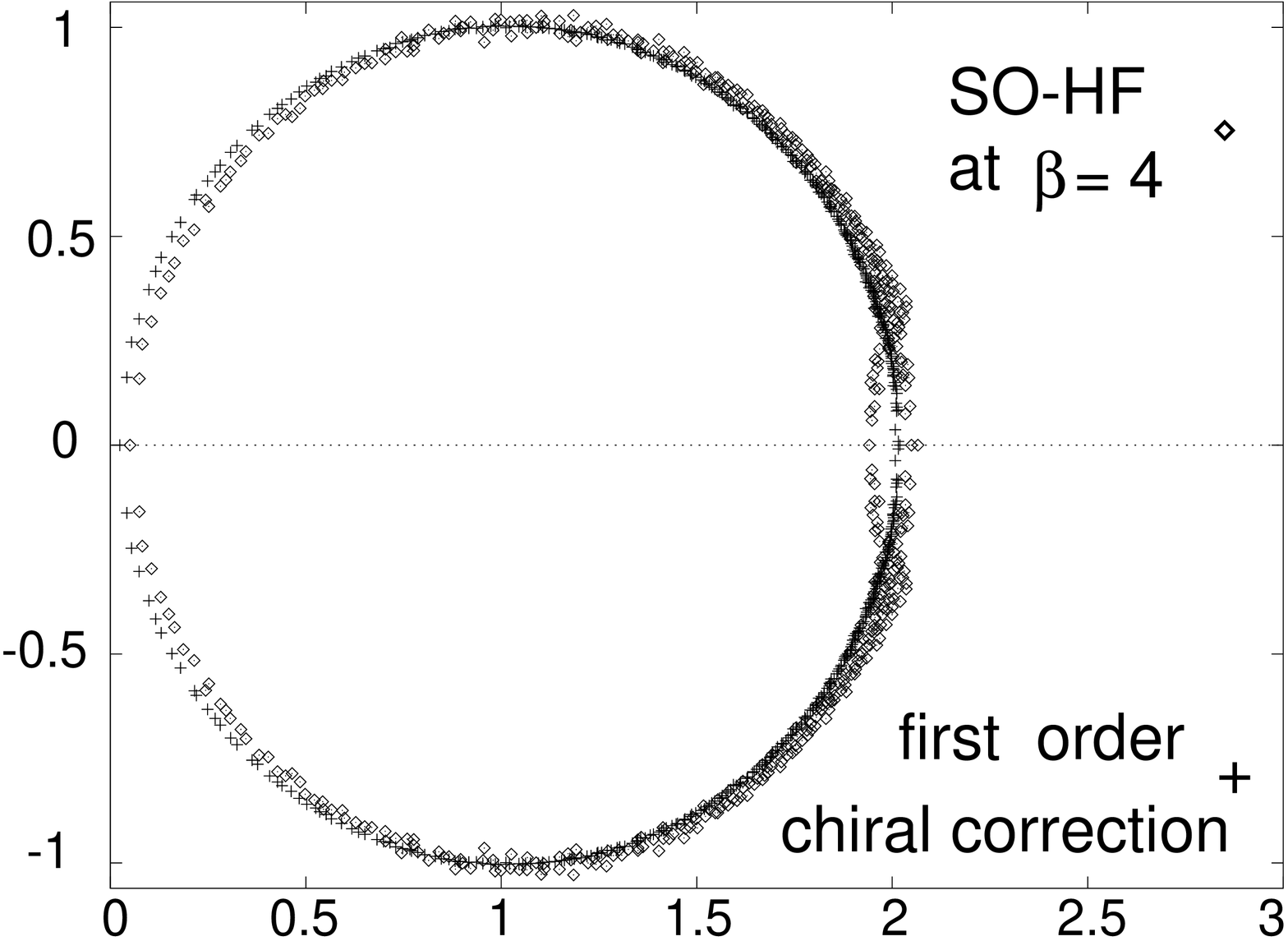}
  \includegraphics[angle=270,width=0.49\linewidth]{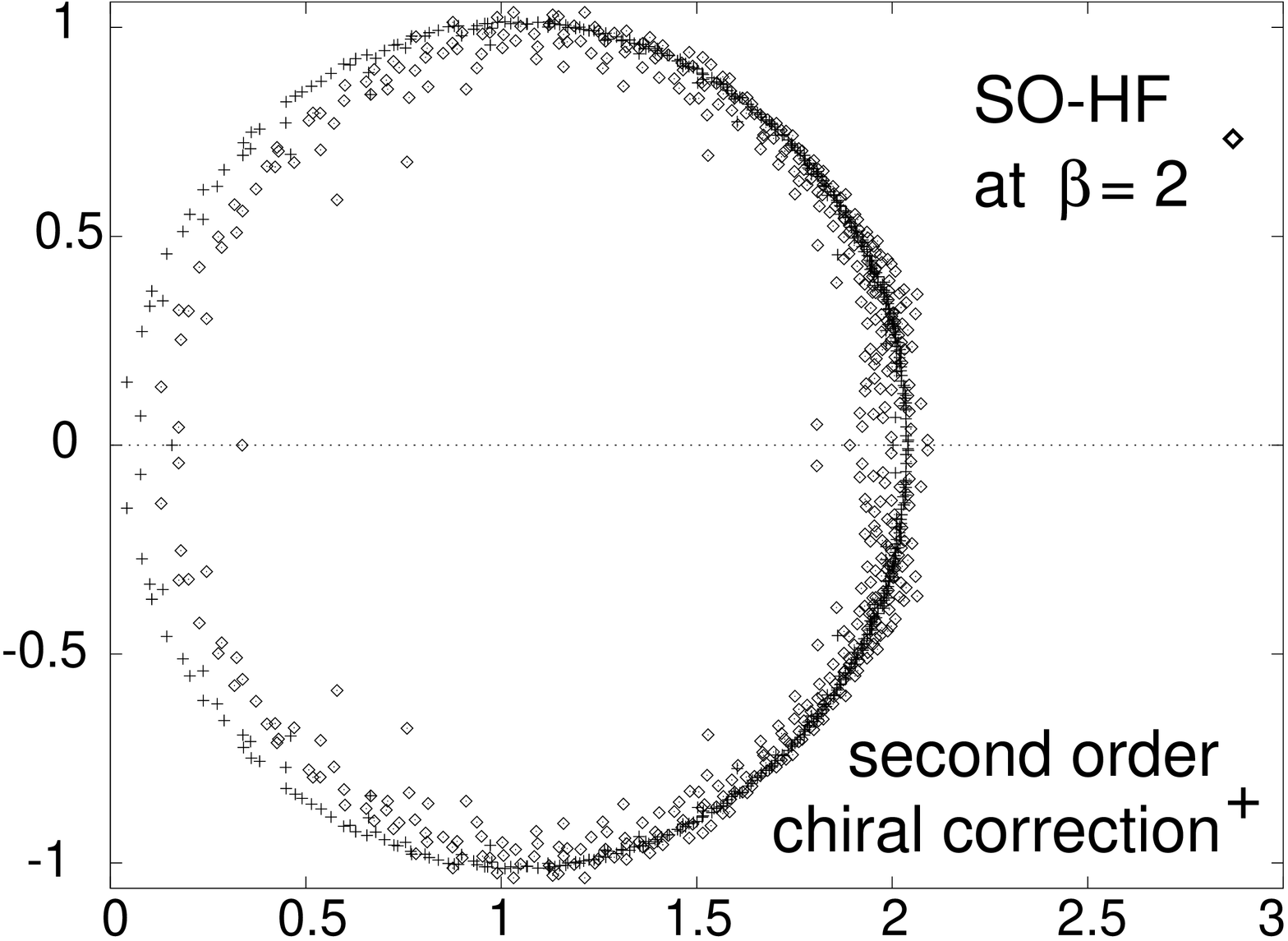}
\end{center}
\caption{{\it The spectra of the HF operator for typical configurations in the
Schwinger model at $\beta =4$ and at $\beta =2$. The GW circle is 
still approximated well. We also show a polynomial
correction with the Taylor expanded overlap formula to the first resp.\
second order.}}
\label{schwingspecfig2}
\end{figure}

To study this property in the presence of gauge interactions,
we first return to the 2-flavour Schwinger model. We considered this model 
as described in the first part of Subsection 6.3, and we show
the spectrum of a typical configuration at $\beta \equiv 1/g^{2} = 6$
for $D_{\rm W}$ and for $D_{\rm HF}$ in Figure \ref{schwingspecfig} 
\cite{WBIH}. In two dimensions, the spectrum 
of $D_{\rm W}$ covers the vicinity of two circles, 
whereas the HF is again very close to the GW circle with $\rho =1$.
As an experiment, we also show
in the latter case the spectrum after a minimal
approximation to the overlap formula: we only use its first
term in the Taylor expansion, which leads to a spectrum that can
hardly be distinguished from the exact GW circle.
As we increase the strength of the gauge coupling $1 / \sqrt{\beta}$, 
the eigenvalues spread out a bit more, but they still follow
closely the GW circle for $\beta =4$ and even $\beta =2$, as
we illustrate in Figure \ref{schwingspecfig2}.

\begin{figure}[h!]
\begin{center}
  \includegraphics[angle=270,width=0.6\linewidth]{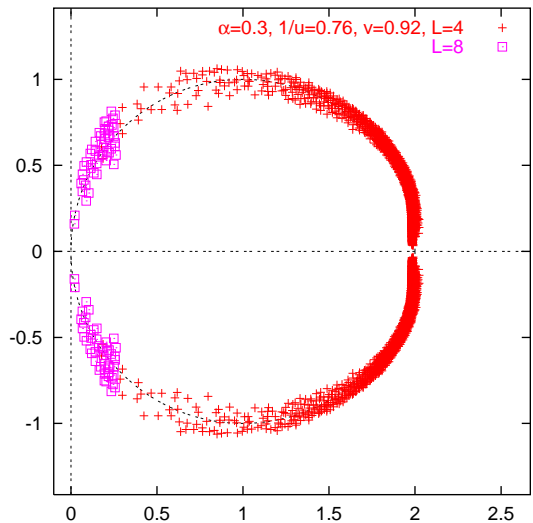}
\end{center}
\caption{{\it The spectrum of the optimised 
HF operator for a typical configuration in quenched QCD at $\beta =6$
on lattices of the sizes $4^{4}$ (crosses, full spectrum) and $8^{4}$ 
(squares, physical part of the spectrum).}}
\label{nonplusultra}
\end{figure}

In QCD this property is more tedious to achieve.
Refs.\ \cite{China,ovHFQCD} des\-cribe the construction of a
suitable HF formulation for quenched QCD at $\beta \equiv 6/g^{2} =6$. 
It starts again from the free, truncated perfect
HF and restores approximate criticality under gauge interaction by a
link amplification\footnote{This procedure is inspired by the method
of ``tadpole improvement'' \cite{LeMa}.}
$U_{x,\mu} \to u U_{x,\mu}$, $ u \gsim 1$.
Further ingredients are a separate link amplification factor $v$ for the 
vector term (which controls the imaginary part of the spectrum, with hardly
any effect on the mass renormalisation), and 
the simplest version of a ``fat link''.\footnote{A similar parameterisation 
was later adapted in the approach of Refs.\ \cite{GatHip}.}
The latter amounts to the substitutions
\be
U_{x, \mu} \to (1- \alpha ) U_{x, \mu} + \frac{\alpha}{6} \
\sum \ [ \ {\rm staples~terms} \ ]
\ee
for all compact link variables, where we chose $\alpha$
in the range $0.3 \dots 0.5$. The fat link helps to pull the
eigenvalues around real part $1$ somewhat closer to the GW 
circle \cite{China,ovHFQCD}. At some point, also a clover term
was considered, but since its optimisation led to a coefficient close to
$0$ we dropped it again. We arrived at a very satisfactory approximation
for typical configurations at $\beta =6$, as Figure \ref{nonplusultra}
shows. This plot includes the full spectrum
on a $4^{4}$ lattice, as well as the low eigenvalues on a $8^{4}$ lattice,
which fill the gap near zero (this gap is generic on small lattices).

Later on, such a construction was also
accomplished at $\beta =5.85$, i.e.\ on a coarser lattice, where
it is more difficult. Still this program could
be carried out successfully \cite{Stani,WBStani}.
In this case, we made a compromise between the criteria of a minimal
condition number of $A^{\dagger}A$ and optimal
locality of $D_{\rm ovHF}$ (to be discussed in this and the following 
Subsection).

\begin{figure}[h!]
\begin{center}
  \includegraphics[angle=270,width=0.8\linewidth]{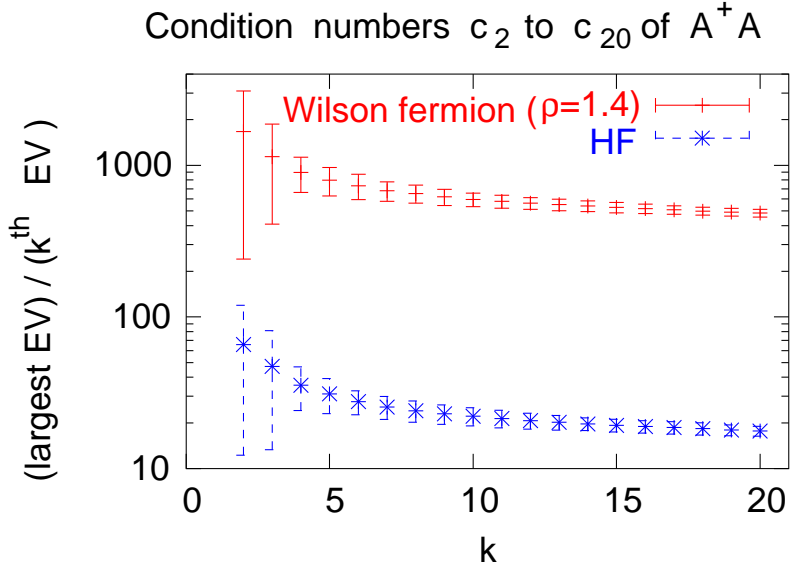}
\end{center}
\caption{{\it The condition numbers of the overlap ingredient
$A^{\dagger}A = H^{2}$, where $H$ is the Hermitian operator
$H = \gamma_{5}A$ (and $A = D_{0}-\rho$, see eq.\ (\ref{ovrho})), 
for the Neuberger operator and for the overlap-HF, in QCD on a
$12^{4}$ lattice at $\beta =6$. 
$k-1$ is the number of lowest modes which are projected
out. The condition number of the remaining operator, i.e.\ the ratio 
$c_{k} = {\rm (largest ~eigenvalue~of~}A^{\dagger}A)/
(k^{\rm th} {\rm ~eigenvalue~of~}A^{\dagger}A)$,
is about $25$ times lower for the overlap-HF \cite{ovHFQCD}, 
from which we infer a gain factor of about
$5$ in the polynomial degree. The computational effort is roughly
proportional to this degree. \newline
This gain factor is practically the same
at $\beta = 5.85$ on a $12^{3} \times 24$ lattice \cite{WBStani}.}}
\label{condfig}
\end{figure}

An immediate consequence of the approximate chirality of $D_{\rm HF}$
is that the polynomial evaluation of $D_{\rm ovHF}$ (to a fixed precision)
is faster, i.e.\ the required degree is lower, as we mentioned before.
In QCD we used Chebyshev polynomials for this purpose, which converge
exponentially as the degree rises 
(see e.g.\ Ref.\ \cite{NR}).\footnote{The ``minimax'' polynomial 
provides a slightly better approximation with the same 
degree \cite{NR}, but the Chebyshev polynomial has the
advantage that the use of huge coefficients can be circumvented
thanks to the Clenshaw recurrence formula. Regarding rational approximations,
the Zolotarev polynomial is optimal in this case \cite{Zolo}.
We add that in the Schwinger model it was possible to evaluate the overlap 
operator by diagonalising $A^{\dagger}A$,
hence no polynomial was needed in $d=2$.}
The required degree is then
proportional to the square root of the condition number of the
operator $A^{\dagger}A$. In practice one usually projects out
the lowest few modes and treats the eigenspace spanned by them 
separately --- this reduces the condition number of the remaining 
operator very significantly. Figure \ref{condfig} compares these
condition numbers for the Neuberger operator and the overlap-HF,
for QCD on a $12^{4}$ lattice at $\beta =6$, with $k-1 = 1 \dots 19$
modes projected out. We recognise for the overlap-HF 
a gain factor $\approx 25$, which we anticipated
at the end of Subsection 7.3. 
This gain factor is essentially due to the reduction of the
maximal $A^{\dagger}A$ eigenvalue, and it
persists practically unchanged at
$\beta =5.85$; details can be found in Ref.\ \cite{WBStani} (Table 1). 

\subsection{Locality and rotation symmetry}

\begin{figure}[h!]
\begin{center} \hspace*{1mm}
  \includegraphics[angle=270,width=0.47\linewidth]{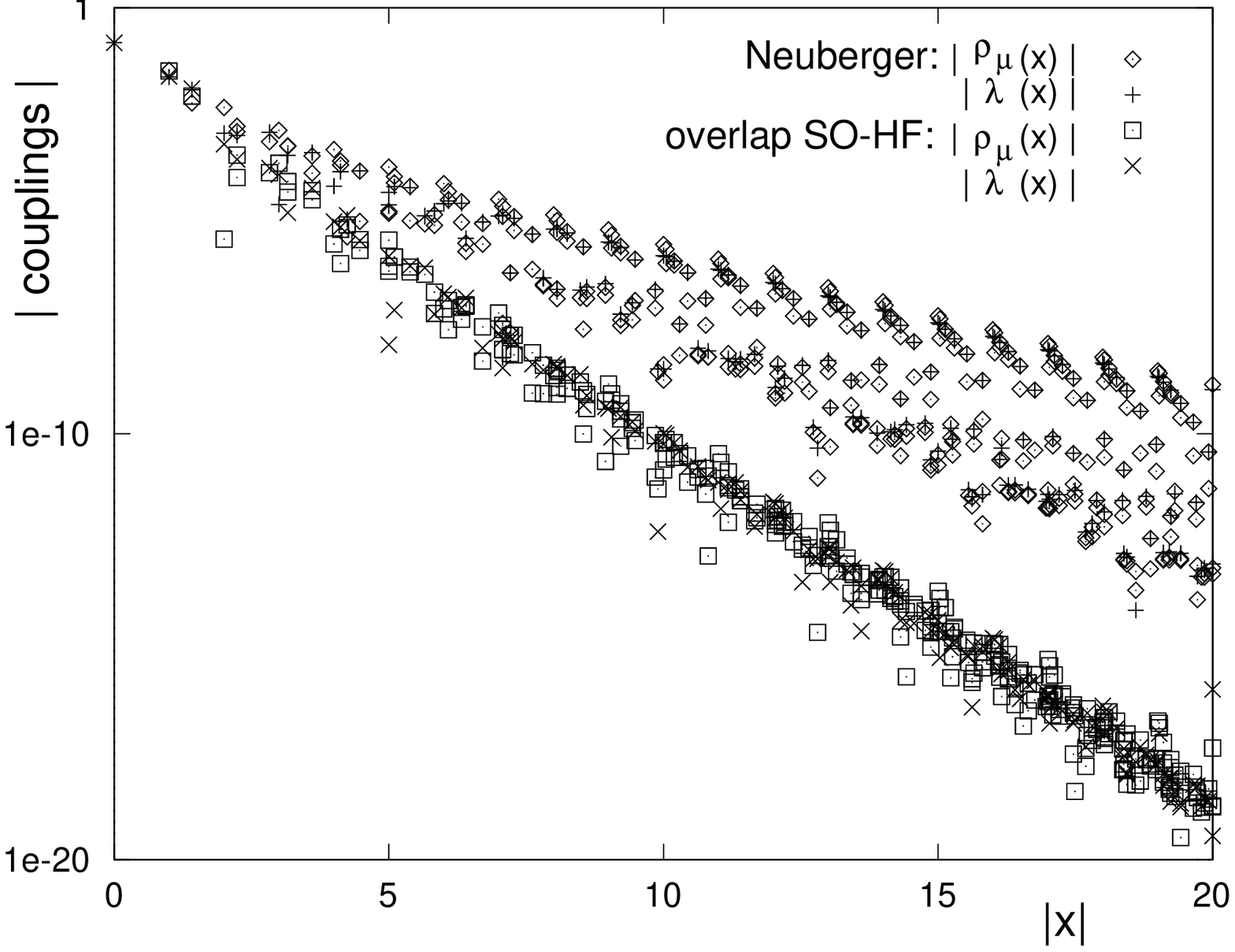} \vspace*{2mm} \\
  \includegraphics[angle=0,width=0.5\linewidth]{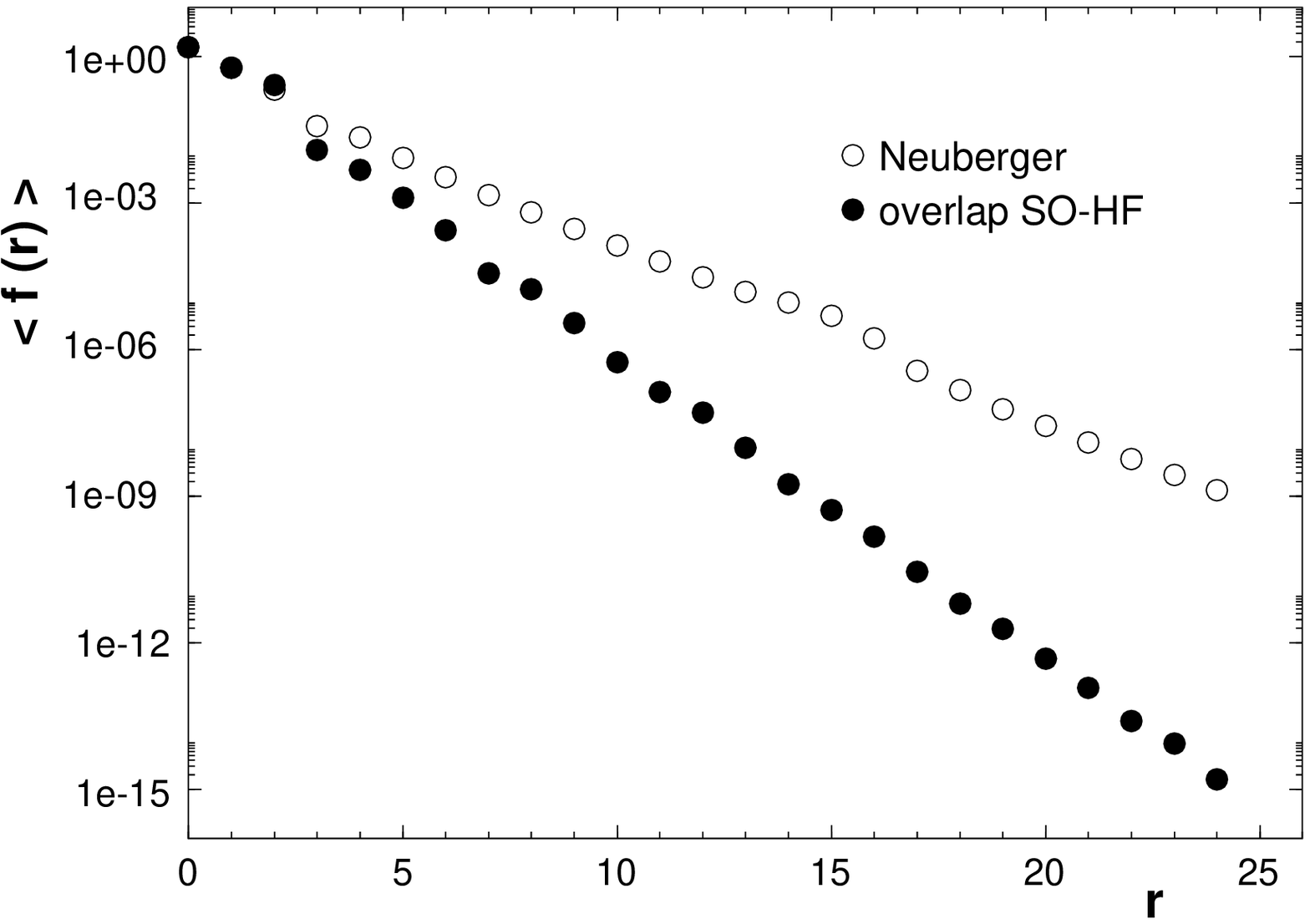}  \vspace*{-5mm}
\end{center}
\caption{{\it The locality of the overlap-HF vs.\ the standard Neuberger
operator in $d=2$. On top we show the decay of the free couplings
in the vector term $\rho_{\mu}$ and in the scalar term $\lambda$
(in the notation $D_{\rm ov} = \rho_{\mu} \gamma_{\mu} + \lambda$),
against the Euclidean distance $|x|$.
We see that the overlap-HF couplings follow a much faster exponential
decay, indicating a higher level of locality. Moreover, the couplings
in the Neuberger operator are much more spread out, which reveals
a better approximate rotation symmetry for the overlap-HF. \newline
Below we compare the locality in the Schwinger model at
$\beta =6$, measured according to eq.\ (\ref{localfun}) 
(in the taxi driver metrics). 
We see that the overlap-HF is still by far more local.}}
\label{locality2d}
\end{figure}

Now we move on to the point where the overlap operator is already
evaluated to a high accuracy of at least $10^{-12}$ (in many case,
the precision was also set to  $10^{-15}$ or $10^{-16}$). 
First we compare
the level of locality of the overlap-HF to the standard formulation
$D_{\rm N}$. A strong gain in this level was first observed for the
free fermion in Ref.\ \cite{EPJC}, which was one of the motivations
to generalise the overlap operator, as described in 
Subsection 7.3.\footnote{Ultralocality, i.e.\ the limitation
of the couplings to a finite range on the lattice (cf.\
footnote \ref{ultralocfn}), is impossible for any Ginsparg-Wilson operator
in $d > 1$, as considerations in the free case show \cite{ultraloc}.}
In Figure \ref{locality2d} we show this property in $d=2$: on top
we see that the couplings in $D_{\rm ovHF}$ decay much faster than
in $D_{\rm N}$. Since the  $D_{\rm ovHF}$ couplings follow closely a
single exponential curve, this plot also illustrates an improved
rotation symmetry. Both properties can be understood based on the
ultralocality and the good rotation symmetry of $D_{\rm HF}$,
together with relation (\ref{ovapprox}). (Of course, ultralocality
also holds for $D_{\rm W}$, but the relation (\ref{ovapprox}) does not,
hence this property is not approximately inherited in $D_{\rm N}$.)
In fact, the level of locality is related to the minimal separation
of a $D_{0}$ eigenvalue from $\rho$ (the centre of the GW circle) \cite{HJL}; 
this implies a link between the quests for locality of an overlap
operator and approximate chirality of its kernel $D_{0}$.
The optimal parameters have roughly the same trend for these criteria,
but they are not identical (we mentioned in the previous Subsection
that we made a compromise between maximal locality of $D_{\rm ovHF}$
and minimal condition number of $A^{\dagger}A$).

In the lower plot in Figure \ref{locality2d} we see that the
superior locality of $D_{\rm ovHF}$ persists in the Schwinger model at
$\beta = 6$ \cite{WBIH}. In this case we measure the locality in the way 
suggested in Ref.\ \cite{HJL}: we put a unit source $\eta_{y}$ 
at one site $y$, and we consider all sites $x$ separated 
from $y$ by a taxi driver distance
$r = \sum_{\mu} | x_{\mu} - y_{\mu}| := \Vert x-y \Vert_{1}$.
Then we identify the maximum
of the norm $\Vert D_{xy} \eta_{y} \Vert$, which we denote as
$f(r)$,
\be  \label{localfun}
f(r) = \ ^{\rm max}_{~ \, x} \Big\{ \Vert D_{xy}(U) \eta_{y} \ \Big\vert
\ \Vert x-y \Vert_{1} = r \Big\} \ . 
\ee
The exponential decay of $\langle f(r)\rangle $ in $r$
is a compelling criterion to demonstrate locality. For $\beta =6$ we see in 
Figure \ref{locality2d} (below) that this decay
is much faster for $D_{\rm ovHF}$ than for $D_{\rm N}$.

\begin{figure}[h!]
\begin{center}
\hspace*{3mm} \includegraphics[angle=270,width=0.56\linewidth]{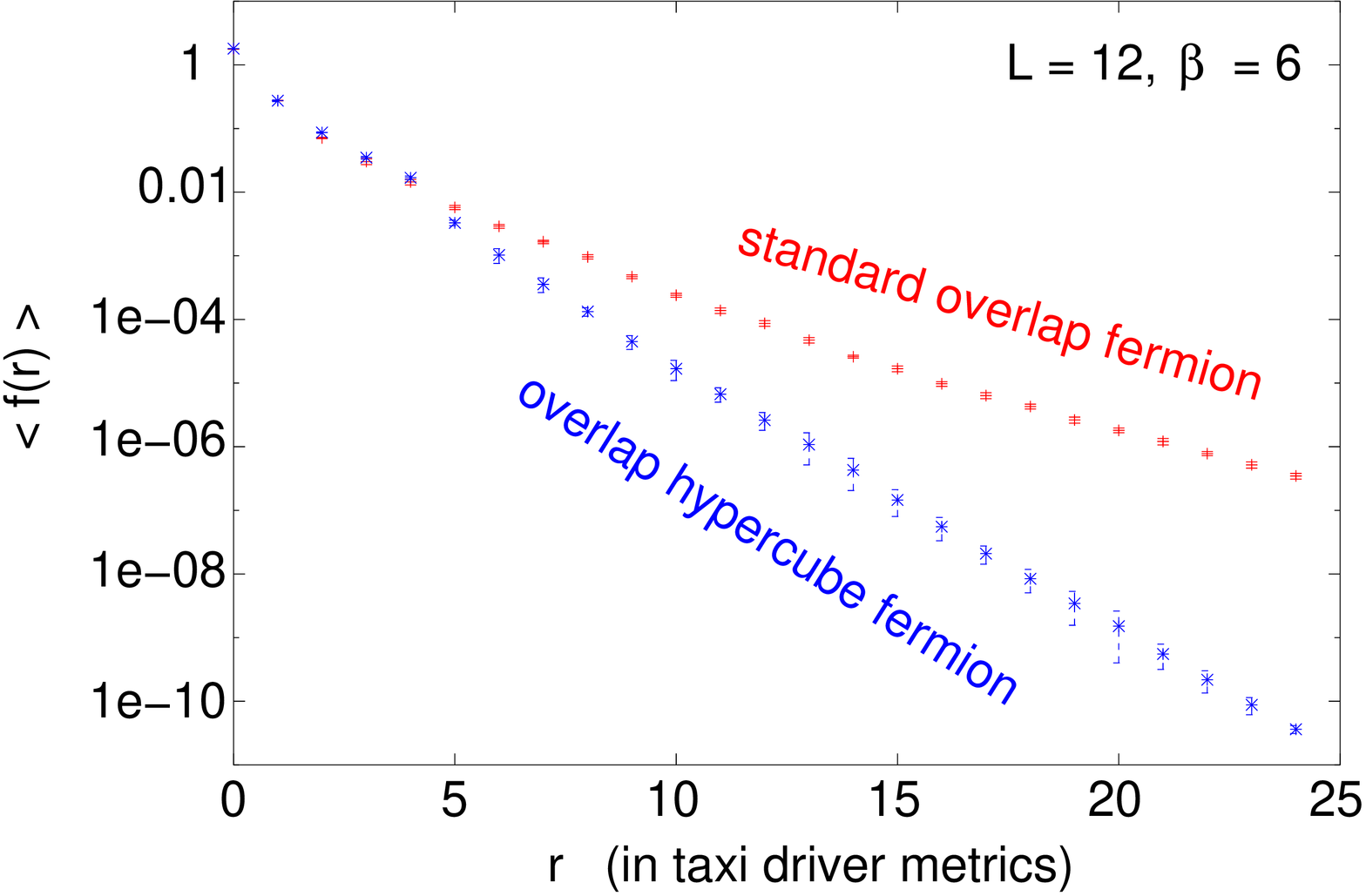}
  \includegraphics[angle=270,width=0.62\linewidth]{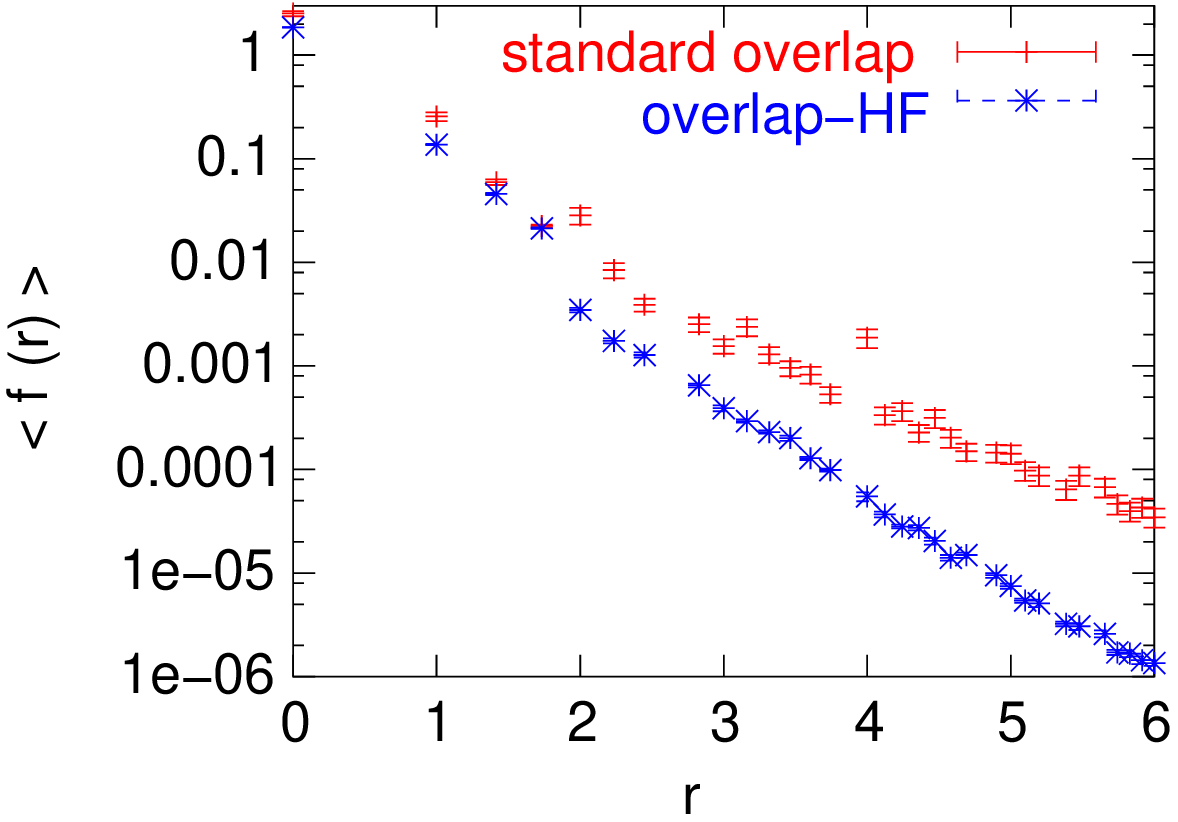}
\end{center}
\vspace*{-4mm}
\caption{{\it The locality of the overlap-HF vs.\ the standard
operator $D_{\rm N}$ in QCD. On top the decay is compared in the taxi driver
metrics at $\beta =6$, where we find a gain factor of almost $2$
in the exponent of the decay \cite{ovHFQCD}. \newline
The plot below refers to $\beta = 5.85$ in the Euclidean
metrics, which also provides a comparison of the quality of rotation 
symmetry \cite{WBStani}.}}
\vspace*{-2mm}
\label{localQCD}
\end{figure}

We proceed to QCD and first illustrate that we obtain again a higher degree
of locality for the overlap-HF at $\beta =6$ \cite{ovHFQCD} and at
$\beta = 5.85$ \cite{Stani,WBStani}, see Figure \ref{localQCD}. 
On top, at $\beta =6$ (which corresponds to a physical
lattice spacing of about $a \simeq 0.093 ~{\rm fm}$)\footnote{For the 
physical units in quenched QCD, we always refer to the
{\em Sommer scale} \cite{Sommer} in this work.\label{Sommerfn}}
we still use the taxi driver metrics,
but below ($\beta = 5.85$, corresponding to 
$a \simeq 0.123~{\rm fm}$) we switch to the Euclidean metrics.
The observation that the decay for the overlap-HF is not only faster,
but in the Euclidean metrics also smoother, confirms again that
our overlap-HF formulation is both, more local and to a better approximation
rotation invariant than the standard formulation. We add that the
quality of rotation symmetry was also tested directly in the Schwinger
model with the procedure shown before for the HF in Figure \ref{rotHFschwing}.
For the overlap-HF a smooth decay of the isotropic correlator $C_{3}$
was found, similar to the HF,
affirming an improved rotation symmetry \cite{WBIH}.

\begin{figure}[h!]
\begin{center}
  \includegraphics[angle=270,width=0.65\linewidth]{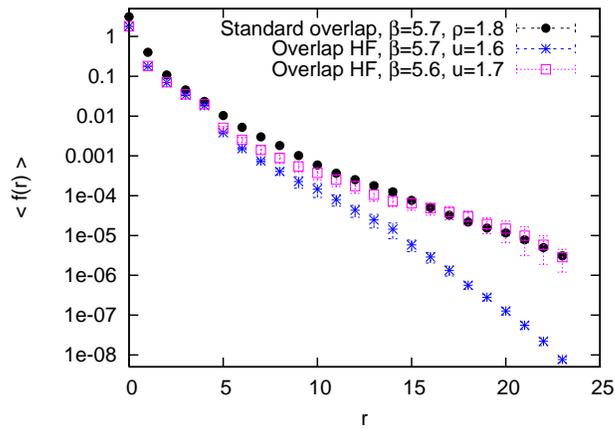}
\end{center}
\vspace*{-4mm}
\caption{{\it The locality of the overlap-HF 
(with a link amplification factor $u$ and $\rho =1$) vs.\ the Neuberger
operator $D_{\rm N}$, in QCD at strong coupling (in taxi driver metrics). 
At $\beta = 5.7$ $D_{\rm N}$ (with an optimised
parameter $\rho =1.8$) is still local, but at $\beta = 5.6$ its locality ---
and therefore its validity as a lattice Dirac operator --- collapses.
The overlap-HF is local in both cases. At $\beta = 5.7$ its locality
is still stronger than the one of $D_{\rm N}$ at $\beta = 6$
and $\rho =1.4$ (which is optimal for locality in that case \cite{HJL}).
\newline
These measurements were done on a $12^{3} \times 24$ lattice, and
the anisotropy causes the bending down at large $r$.}}
\label{localstrong}
\end{figure}

At last we turn to strong gauge couplings, which correspond to rough 
configurations and therefore to coarse lattices. Generally, the overlap
formula is only applicable to generate a valid lattice Dirac operator
up to a certain coupling strength, where locality collapses.\footnote{For
a theoretical discussion of this issue we refer to Ref.\ \cite{ShaGo}.}
We see in Figure \ref{localstrong} that the Neuberger operator
is still local at $\beta =5.7$ (corresponding to
$a \simeq 0.17 ~{\rm fm}$), but at $\beta =5.6$
no exponential decay can be observed anymore (for any parameter $\rho$). 
In contrast, the overlap-HF
(where only the link amplification factor is adapted compared to
the formulation at $\beta = 5.85$) is local in both cases, and at
$\beta= 5.7$ the function $\langle f(r)\rangle $ still exhibits a 
remarkably fast decay.

This superior locality is essentially due to the HF
structure of the overlap kernel. By means of fat links alone
the locality of the overlap operator can also be improved, but
only marginally \cite{DuHoWe} 
(assuming the optimal value of $\rho$ in each case). 

\subsection{The scaling behaviour}

\begin{figure}[h!]
\begin{center}
\includegraphics[angle=270,width=0.6\linewidth]{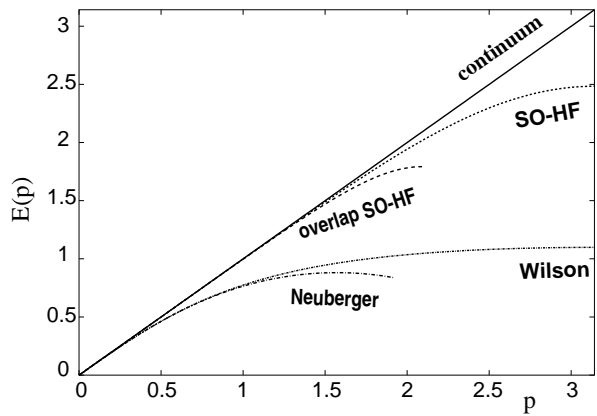}
\end{center}
\caption{{\it The dispersion relation of the free, massless
overlap-HF (scaling optimised version), compared to the continuum 
and to the standard overlap formulation $D_{\rm N}$. These
dispersions end when the argument of the square root becomes negative.
To provide an overview
we also include the dispersion for the kernel 
operators $D_{\rm HF}$ and $D_{\rm W}$.}}
\label{freedispov}
\end{figure}

\begin{figure}[h!]
\begin{center}
  \includegraphics[angle=270,width=0.49\linewidth]{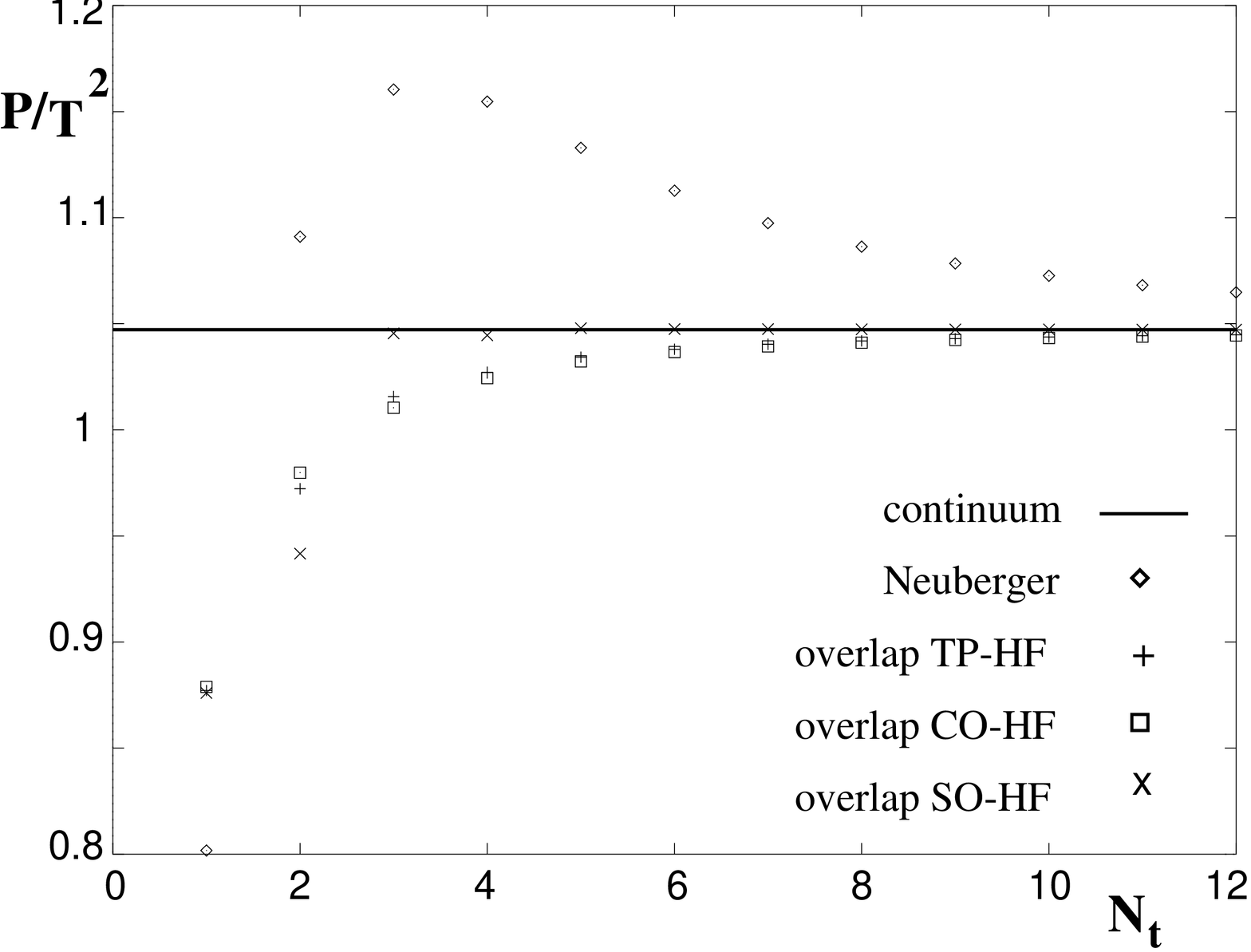}
  \includegraphics[angle=270,width=0.49\linewidth]{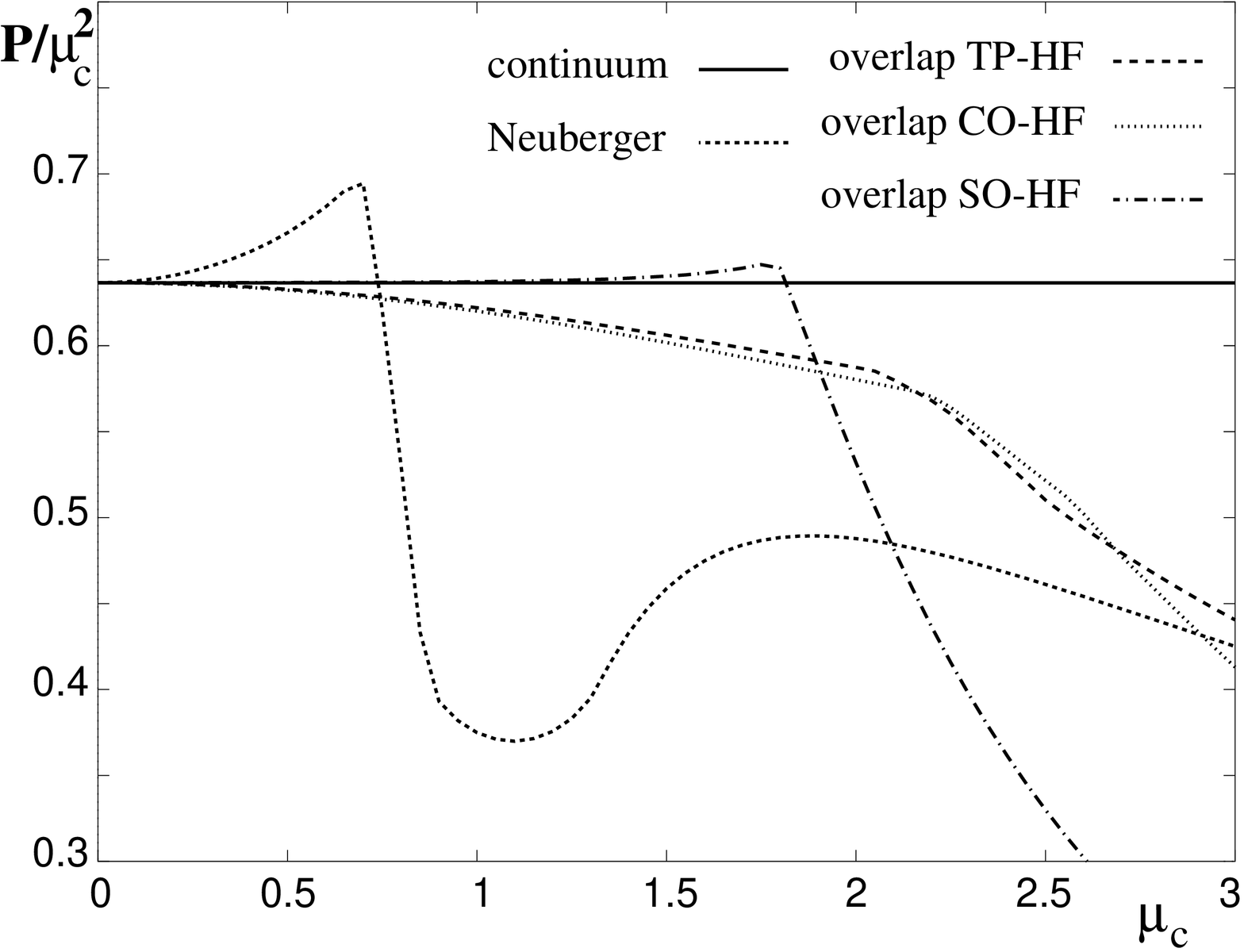}
\end{center}
\caption{{\it The thermodynamic scaling ratios 
${\rm pressure}/{\rm (temperture)}^{2}$ and 
${\rm pressure}/{\rm (chemical~potential)}^{2}$ for
free overlap fermions in $d=2$. 
The hierarchy of the scaling behaviour
is confirmed in all respects.}}
\label{freethermov}
\end{figure}

Again referring to the perfect action background of the HF and to relation
(\ref{ovapprox}), we also expect a good scaling behaviour for the
overlap-HF. For the free overlap-HF, this is clearly confirmed by the
dispersion relation, which we show for momenta $\vec p = (p_{1},0,0)$
in Figure \ref{freedispov}. In contrast,
$D_{\rm N}$ scales worse than the Wilson operator $D_{\rm W}$
in this case.\footnote{On the other hand, all GW fermions are still
free of $O(a)$ artifacts in the interacting case, in contrast to
$D_{\rm W}$.} Qualitatively the same behaviour is observed for
massive overlap fermions \cite{WBIH}.
We also repeated the thermodynamic scaling
tests described earlier (in Subsection 4.1, before applying the overlap 
formula). The results in $d=2$, for three version of the overlap-HF, are
by far improved compared to the standard overlap operator, see
Figure \ref{freethermov} \cite{WBIH}. Here we incorporated the
chemical potential also for the overlap fermions according to the
prescription (\ref{chempotsub}); for an alternative method and first
simulations, see Ref.\ \cite{BloWet}.

\begin{figure}[h!]
\begin{center}
  \includegraphics[angle=0,width=0.49\linewidth]{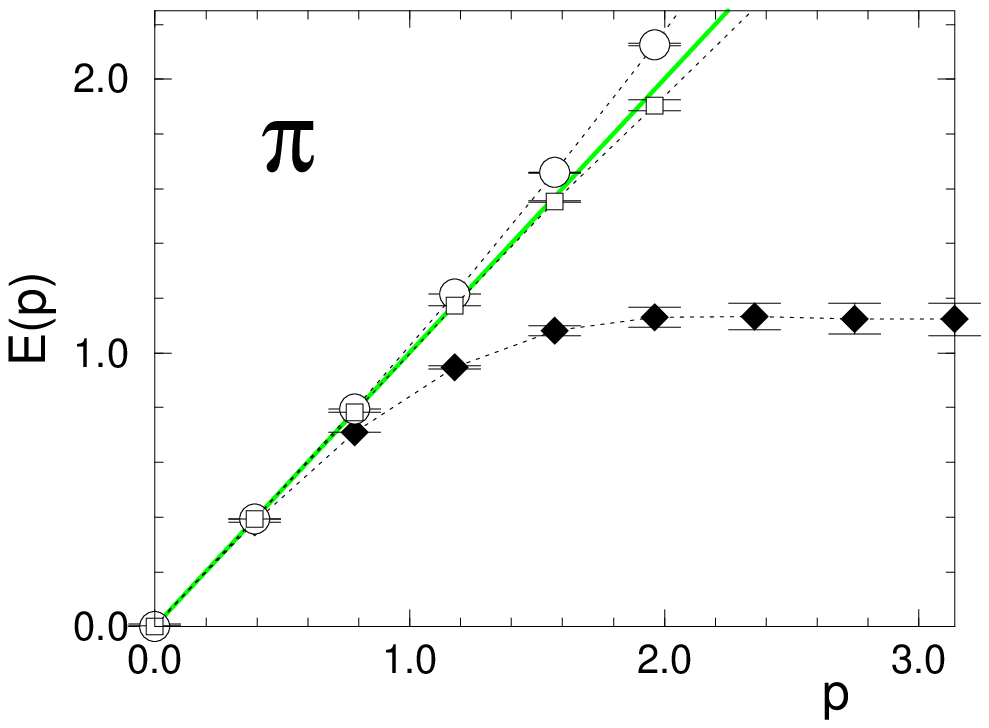}
  \includegraphics[angle=0,width=0.49\linewidth]{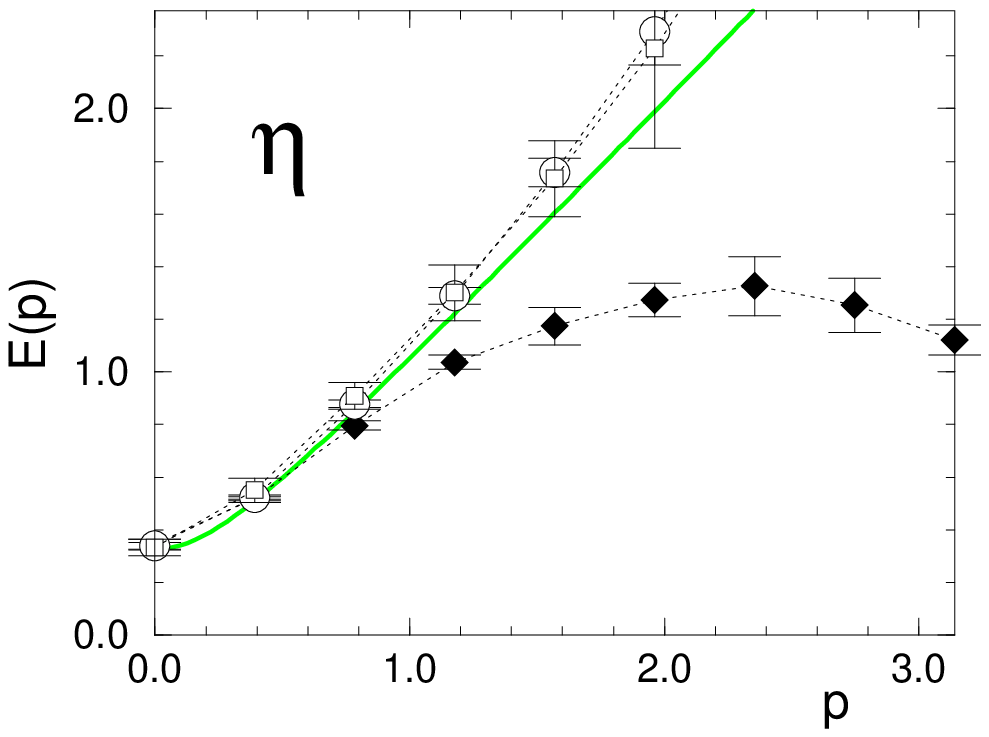}
\end{center}
\caption{{\it The mesonic dispersion relation in the Schwinger model
with two types of overlap-HFs (open circles and squares). 
Both, the ``pion'' (on the left)
and the ``$\eta$-particle'' (on the right) display a scaling which is by far
improved for the overlap-HFs compared to the standard overlap fermion 
(diamonds). Once more we obtain the best scaling by inserting the (scaling
optimised) SO-HF into the overlap formula (squares) \cite{WBIH}.}}
\label{mesodispov}
\end{figure}

In the interacting case, we reconsidered the ``meson'' dispersion relations
in the Schwinger model (cf.\ Subsection 6.3), this time for exact
Ginsparg-Wilson operators. Also here we observe a scaling behaviour which is by 
far better for the overlap-HF than for the Neuberger operator, as Figure
\ref{mesodispov} shows. Further scaling tests in the Schwinger model with the
dynamical HF and the quenched re-weighted overlap-HF can be found in 
Ref.\ \cite{CJNP2} (they were compared to the scaling with
dynamical Wilson fermions, which has also been investigated in
Ref.\ \cite{Magda}). 
In QCD, a systematic scaling test is tedious and
still outstanding, but the toy model results summarised here raise optimism
also in that respect.

\subsection{The link to domain wall fermions}

Finally we remark that the overlap fermion is equivalent to the
{\em domain wall fermion} \cite{DWF} in the limit of an infinite number
$L_{s}$ of layers in a fifth direction.\footnote{The meaning of this
fifth direction is technical rather than geometrical, because it does
not involve link variables. So that direction has no scale; what one needs
is its thermodynamic limit $L_{s} \to \infty$.}
It is then a practical issue if one
tries to approximate this limit by a satisfactory number $L_{s}$, or by
a sufficiently precise polynomial approximation to the overlap formula.
Also in the case of the domain wall fermions, the standard formulation
inserts a Wilson kernel. Replacing it by a HF kernel, as suggested
in Ref.\ \cite{EPJC}, could lead to similar improvements as we 
demonstrated in this Section for the overlap fermion. 
The improved condition number of the kernel operator manifests
itself here in a lower demand for $L_{s}$ (for some required precision
of chirality). Furthermore the gains in locality, approximate rotation symmetry
and scaling are expected here in practically the same form as in the
overlap case.

\section{Relating QCD Simulations to Chiral Perturbation Theory}

\subsection{Chiral Perturbation Theory}

We start this Section with a few general remarks on Chiral
Perturbation Theory ($\chi$PT), an effective low energy theory,
which we are going to relate to our QCD simulation results.

When a continuous, global symmetry breaks spontaneously,
we obtain a continuous set of degenerate vacuum states.
Expanding around one selected vacuum, one distinguishes between excitations
to higher energy (which are identified with massive particles)
and fluctuations, which preserve the ground state energy.
The subgroups of the energy conserving symmetry group can either transfer 
the selected vacuum to a different vacuum state, or leave it simply invariant.
The number of generators relating different vacuum states corresponds 
--- according to the Goldstone Theorem \cite{Gold} --- 
to the number of massless Nambu-Goldstone
bosons (NGBs) involved. At low energy, the NGBs can be described 
by an effective theory as fields in the coset space of the
spontaneous symmetry breaking (SSB). Such effective theories still apply
if we add a small explicit symmetry breaking; we then deal with light
quasi-NGBs, which dominate the low energy physics. The effective Lagrangian
${\cal L}_{\rm eff}$ contains terms of the quasi-NGB fields,
which obey the original symmetry, as well as the (explicit) symmetry 
breaking terms.
All these terms are hierarchically ordered according to some low energy
counting rules for the momenta and the quasi-NGB masses.
A simple example for such an expansion is outlined in Subsection 8.4.1.

This concept is very general, but it was introduced in the framework of
chiral symmetry breaking in QCD \cite{XPT}. At zero quark masses the left-
and right-handed spinors decouple (see eq.\ (\ref{Lagm})), so 
${\cal L}_{\rm QCD}$ is invariant under their independent rotation.
QCD is then assumed to exhibit a chiral SSB of the pattern
\begin{equation}
SU(N_{f})_{L} \otimes SU(N_{f})_{R} \ \to \ SU(N_{f})_{L+R} \ ,
\end{equation}
where $N_{f}$ is the number of quark flavours involved. $\chi$PT is the 
corresponding low energy effective theory \cite{XPT,preg,epsreg1}. 
Following the general prescription, it deals with fields in the
SSB coset space, $U(x) \in SU(N_{f})$.\footnote{Details about the
symmetry groups involved and further aspects are reviewed 
extensively for instance in Ref.\ \cite{Ulf}.}

A small quark mass supplements a slight explicit symmetry breaking.
The quasi-NGBs are then identified with the light mesons, i.e.\
the pions $\pi^{+},\, \pi^{0},\, \pi^{-}$ for $N_{f}=2$ 
--- and for $N_{f}=3$ also the kaons and $\eta$-particles.

In view of our lattice study, we have to put the system into a finite volume;
we choose its shape as $V = L^{3} \times T$ $\ (T \geq L)$. 
We will refer to the formulations of $\chi$PT in two
regimes, with different counting rules for the terms in 
${\cal L}_{\rm eff}[U]$. The usual case --- to be addressed in
Subsection 8.2 --- is characterised by a large volume,
$L m_{\pi} \gg 1$, where $m_{\pi}$ is the
pion mass, i.e.\ the lightest mass involved, which corresponds to the inverse
correlation length. This is the {\em $p$-regime}, where finite size
effects are suppressed, and one expands in the meson momenta and masses
($p$-expansion) \cite{preg}.

The opposite situation, $L m_{\pi} < 1$, is denoted as the 
{\em $\epsilon$-regime}. In that setting, an expansion in the meson momenta is
not straightforward, due to the dominant r\^{o}le of the zero modes.
However, the functional integral over these modes can be performed
by means of collective variables \cite{epsreg1}. There is a large gap to the
higher modes, which can then be expanded again, along with the 
meson masses ($\epsilon$-expansion) \cite{epsreg1,epsreg2,HasLeu}.
We will address that regime extensively in Subsection 8.4.

In both regimes, the leading order of the effective Lagrangian 
(in Euclidean space) reads
\begin{eqnarray}  \label{Leff}
{\cal L}_{\rm eff} [U] &=& \frac{F_{\pi}^2}{4} \, {\rm Tr} 
[ \partial_{\mu} U^{\dagger}
\partial_{\mu} U ] - \frac{1}{2} \Sigma \, {\rm Tr} 
[ {\cal M} (U + U^{\dagger}) ] + \dots \ , \nonumber \\
&& U \in SU(N_{f}) \ , \qquad {\cal M} = 
\left( \begin{array}{ccc} m_{u} && \\ & m_{d} & \\ && (m_{s}) \end{array} 
\right) \ .
\end{eqnarray}
Throughout Section 8 we consider only the $u$- and $d$-quark, and we
assume their masses to be degenerated; we denote them by $m_{q}$.
The coefficients to the terms in ${\cal L}_{\rm eff}$ are the 
Low Energy Constants (LECs), and we recognise $F_{\pi}$ and $\Sigma$ as 
the leading LECs. 
Experimentally the pion decay constant was
measured as $F_{\pi} \simeq 92.4 ~{\rm MeV}$. 
$\Sigma$ is not directly accessible in experiments, 
but its value is assumed to be in the range
$(250 ~{\rm MeV})^{3} \dots (300 ~{\rm MeV})^{3}$.
For instance, in the one flavour case a value around
$(270 ~{\rm MeV})^{3}$ was recently obtained based on a
large $N_{c}$ expansion \cite{ASV}.

The LECs are of physical importance, but they enter
$\chi$PT as free parameters. For a theoretical prediction of their 
values one has to return to the fundamental theory, which is QCD in 
this case. Due to the notorious lack of analytic tools
for QCD at low energy, the evaluation of the LECs is a challenge
for lattice simulations.

The LECs in Nature correspond to their values in (practically) infinite 
volume, and the $p$-regime is close to this situation. 
However, 
these phenomenological values of the LECs can also be determined {\em in
the $\epsilon$-regime,} in spite of the strong finite size effects.
Actually one makes use exactly of the finite size effects
to extract the physical LECs.
Generally, we need a long Compton wave length for the pions, $1/m_{\pi}$,
and in view of lattice simulations in the $p$-regime 
we have to use an even larger box length $L$.
In this respect, it looks very attractive to work in the $\epsilon$-regime
instead, where we can get away with a small volume \cite{GHLW}. 

However, such simulations face conceptual problems, which delayed their
realisation until this century: 
first, to realise light pions the lattice
fermion formulation should keep track of the chiral symmetry.
Second, the $\epsilon$-regime has the peculiarity that 
the topology is important (in accordance with the
importance of the zero modes) \cite{LeuSmi}.
$\chi$PT predictions for expectation values
often differ when restricted to distinct topological sectors, 
so it would be a drastic loss of information to sum them up.
This distinction requires a sound definition of the topological
charge on the lattice. 
In both respects,
the use of Ginsparg-Wilson fermions is ideal, due to the specific
properties explained in Subsections 7.1 and 8.3.

\subsection{Simulations in the $p$-regime}

Usually, for $\chi$PT in a finite spatial box $L^{3}$,
one expects the meson momenta $p$ to be small, so that
\be  \label{4piFpi}
p \sim \frac{2\pi}{L} \ll 4 \pi F_{\pi} \ .
\ee
The term $4 \pi F_{\pi}$ takes a r\^{o}le analogous to $\Lambda_{\rm QCD}$.
Regarding the counting rules for the momenta and the pion mass, the condition
$ L \gg 1 / m_{\pi}$ allows for an application of the {\em $p$-expansion}
\cite{preg}.
It expands in the following dimensionless ratios,
which are expected to be small and counted in the same order,
\be
\frac{1}{L F_{\pi}} \sim \frac{p}{\Lambda_{\rm QCD}} \sim
\frac{m_{\pi}}{\Lambda_{\rm QCD}} \ .
\ee
In this Subsection we present quenched
simulation results in the $p$-regime.
In Ref.\ \cite{WBStani}
we applied the overlap-HF (described in Section 7) at $\beta = 5.85$ on a
lattice of size $12^{3} \times 24$, which corresponds to a physical 
volume of $V \simeq (1.48 ~{\rm fm})^{3} \times 2.96~{\rm fm}$ (where
we use again the Sommer scale \cite{Sommer}, cf.\ footnote \ref{Sommerfn}).
We evaluated 100 propagators for each of the bare quark masses
$$
a m_q = 0.01, \ 0.02, \ 0.04, \ 0.06, \ 0.08 \quad {\rm and} \quad 0.1 \ 
$$
(in physical units: $16.1~{\rm MeV} \dots 161~{\rm MeV}$).
At this point we re-introduce a general lattice spacing $a$.
We will see that the smallest mass in this set is at the
edge of the $p$-regime --- even smaller quark masses will be considered
in the $\epsilon$-regime (Subsection 8.4). 
Part of the observables presented in the current
Subsection were also measured on the same lattice with the Neuberger
operator $D_{\rm N}$ at $\rho =1.6$ (a preferred value at
$\beta =5.85$) \cite{XLF}.

We include $m_{q}$ in the overlap operator (\ref{overlap}) in the 
usual way,
\begin{equation}
D_{\rm ov}(m_{q}) = \Big( 1 - \frac{a m_{q}}{2 \rho} 
\Big) D_{\rm ov} + m_{q} \ ,
\end{equation}
which leaves the largest real overlap Dirac eigenvalue ($2\rho /a$) 
invariant.
$m_{q}$ re\-presents the bare mass for the quark flavours $u$ and $d$.

We first evaluate the pion mass in three different ways:

\begin{itemize}

\item $m_{\pi,PP}$ is obtained from the decay of the pseudoscalar correlation
function $\langle P(x) P(0) \rangle$, with 
$P(x) = \bar \psi_{x} \gamma_{5} \psi_{x}$.

\item $m_{\pi,AA}$ is extracted from the decay of the axial-vector correlation
function $\langle A_{4}(x) A_{4}(0) \rangle$, with 
$A_{4}(x) = \bar \psi_{x} \gamma_{5} \gamma_{4} \psi_{x}$.

\item $m_{\pi,PP-SS}$ is obtained from the decay of the difference
$$
\langle P(x) P(0) - S(x) S(0) \rangle \ , \quad {\rm where} \quad
S(x) = \bar \psi_{x} \psi_{x}
$$ 
is the scalar density.
This subtraction is useful at small $m_{q}$, where configurations
with zero modes ought to be strongly suppressed by the fermion determinant.
In our quenched study, this
suppression does not happen as it should, but the above subtraction in the
observable eliminates the zero mode contributions, which are mostly unphysical.

\end{itemize}

\begin{figure}[h!]
  \centering
  \includegraphics[angle=270,width=0.64\linewidth]{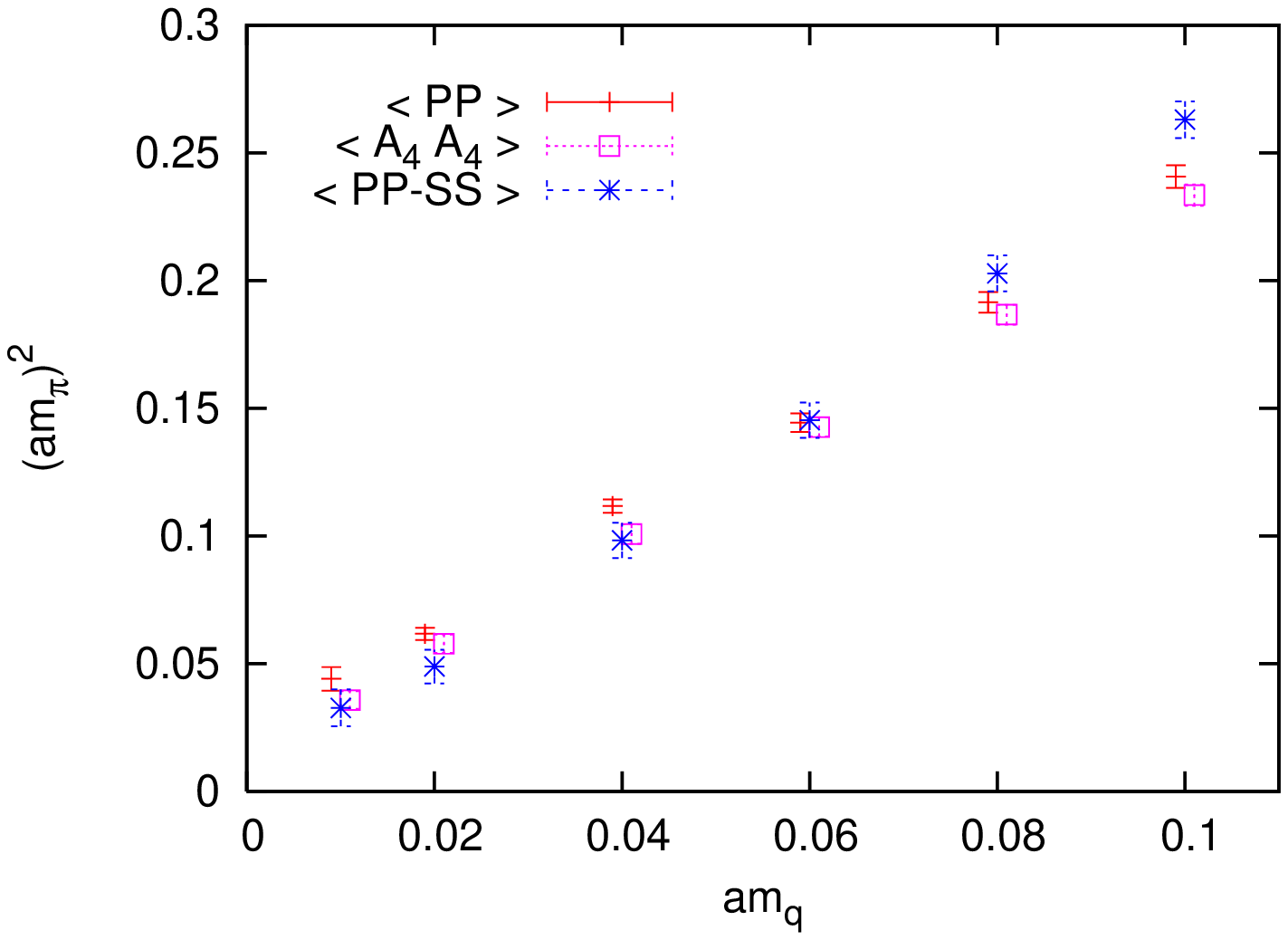}
  \includegraphics[angle=270,width=.64\linewidth]{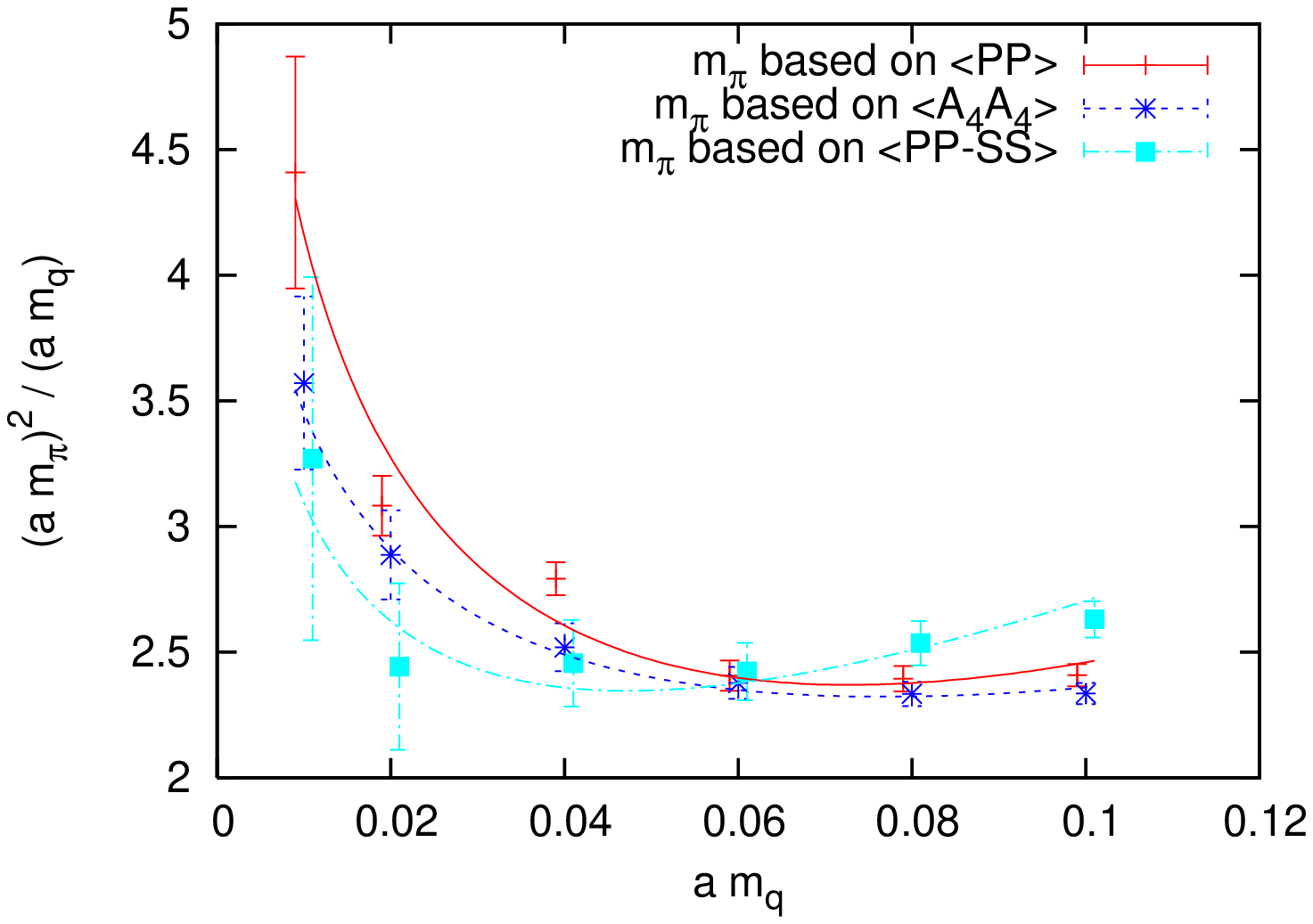}
\caption{{\it On top: The pion mass evaluated from overlap-HFs in the
$p$-regime in three different ways, as described in Subsection 8.2. \newline
Below: The pion masses fitted to the formula (\ref{chilog}), which is 
expected due to the logarithmic quenching artifacts
in the absence of an additive mass renormalisation.}}
\label{pimass}
\end{figure}

The results in Figure \ref{pimass} (on top) show that the
pion masses follow to a good 
approximation the expected behaviour $m_{\pi}^{2} \propto m_{q}$.
Deviations occur at the smallest masses, where we observe the hierarchy
\be
m_{\pi,PP} > m_{\pi,AA} > m_{\pi,PP-SS} \ ,
\ee
in agreement with 
Ref.\ \cite{Bern}.
This shows that the scalar density subtraction is in fact profitable,
since it suppresses the distortion of the linear behaviour down
to the lightest pion mass in Figure \ref{pimass}, 
\be
m_{\pi ,PP-SS} (a m_{q}=0.01) \simeq (289 \pm 32) ~ {\rm MeV} \ .
\ee
That mass corresponds to a ratio $ L / \xi \approx 2$, which confirms
that we are leaving the $p$-regime around this point. 
Based on the moderate quark masses in Figure \ref{pimass}, we find a
very small intercept in the
chiral extrapolation,
\be
m_{\pi , PP-SS} (m_{q} \to 0) = (-2 \pm 24) ~ {\rm MeV} \ .
\ee
\hspace*{-3mm}
Due to quenching, one expects at small quark masses a logarithmic
behaviour of the form
\begin{equation}  \label{chilog}
\frac{a m_{\pi}^{2}}{m_{q}} = C_{1} + C_{2} \ln a m_{q} + C_{3} a m_{q} \ ,
\quad (C_{1},C_{2},C_{3} :~ {\rm constants}) \ .
\end{equation}
Corresponding results are given for instance in Refs.\ 
\cite{Bern2,Schier}. Figure \ref{pimass} (below)
shows the fits of our data
to eq.\ (\ref{chilog}), which works best for $m_{\pi,AA}$.
\begin{figure}[h!]
  \centering
  \includegraphics[angle=270,width=.5\linewidth]{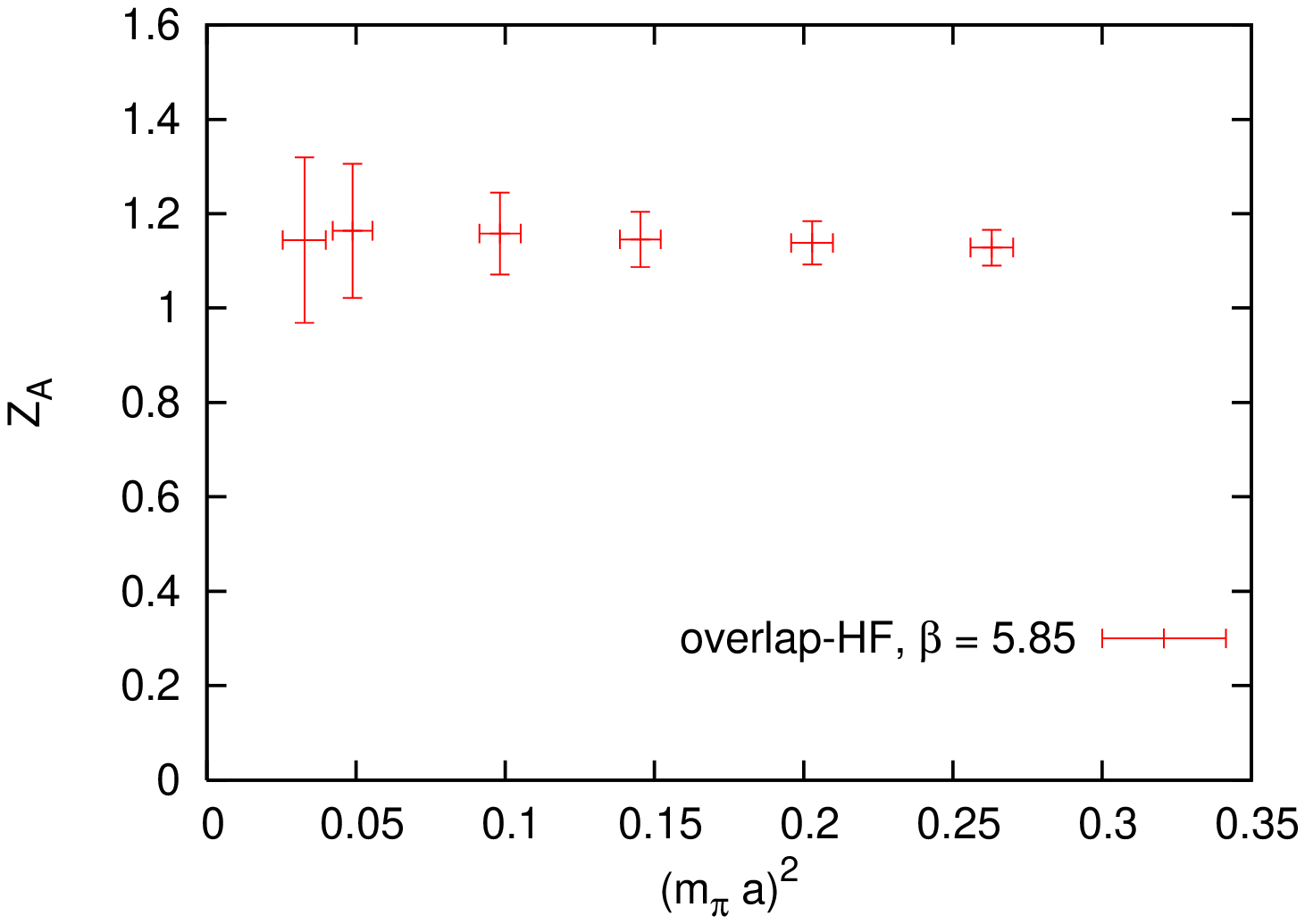} \hspace*{-3mm}
  \includegraphics[angle=270,width=.5\linewidth]{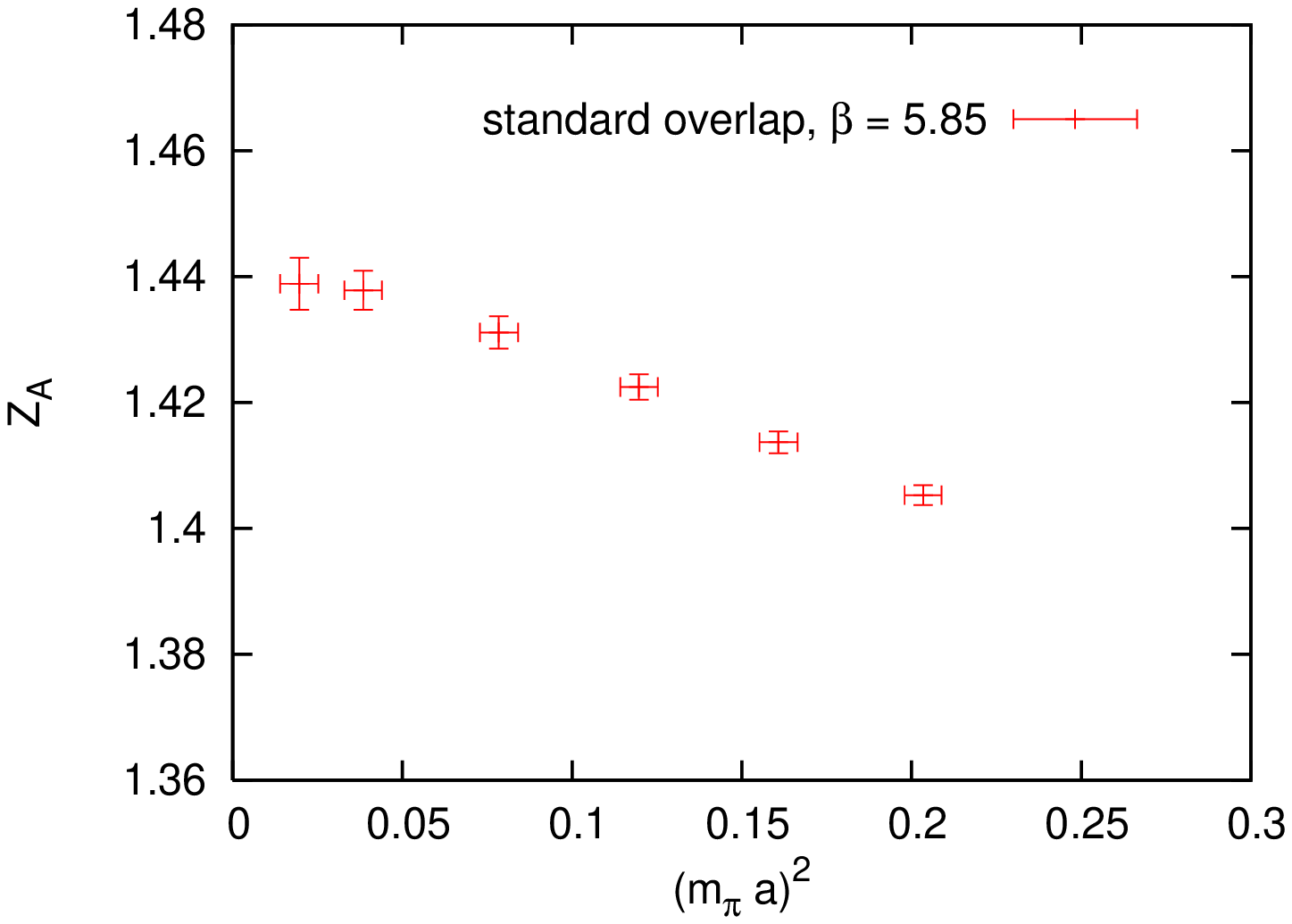}
 \caption{{\it The axial current renormalisation constant $Z_{A}$
evaluated from $m_{\rm PCAC}$ according to eq.\ (\ref{PCAC}). 
For $D_{\rm ovHF}$ (on the left) we find $Z_{A} \approx 1$ \cite{WBStani}, 
in contrast to the result with the standard overlap operator 
$D_{\rm N}$ \cite{XLF} (on the right). The chiral extrapolations are
given in eq.\ (\ref{ZAeq}).}}
\label{PCACfig}
\end{figure}

On the same lattice we also measured the $\rho$-meson mass in the $p$-regime
using the overlap operators $D_{\rm N}$ \cite{XLF} and $D_{\rm ovHF}$
\cite{WBStani}, as well as
the quark mass according to the PCAC relation,
\begin{equation}  \label{PCAC}
m_{\rm PCAC} = \frac{ \sum_{\vec x} \langle
(\partial_{4} A_{4}^{\dagger} (x)) P(0) \rangle }
{ \sum_{\vec x} \langle P^{\dagger} (x) P(0) \rangle } \ ,
\end{equation}
where we used a symmetric nearest-neighbour difference for $\partial_{4}$.
It determines the axial-current renormalisation constant
$Z_{A} = m_{q} / m_{\rm PCAC}$. For the overlap-HF this constant
is close to $1$ \cite{WBStani}, see Figure \ref{PCACfig} (on the left),
which is favourable in view of the link to perturbation theory.
This is in contrast to the large $Z_{A}$ value found for the
standard overlap operator \cite{XLF}, 
see Figure \ref{PCACfig} (on the right).
A chiral extrapolation leads to
\begin{equation}  \label{ZAeq}
Z_{A} = 1.17(2) \quad {\rm for~}D_{\rm ovHF} \ , \quad
Z_{A} = 1.448(4) \quad {\rm for~}D_{\rm N} \ .
\end{equation}
Regarding $D_{\rm N}$, consistent results were reported later
in Refs.\ \cite{JapZA,Babich}, and (somewhat surprisingly)
at $\beta =6, \ \rho =1.4$ it even rises to $Z_{A} \simeq 1.55$
\cite{Babich}. When one uses the improved L\"{u}scher-Weisz gauge 
action \cite{LWg},
the value of $Z_{A}$ for $D_{\rm N}$ is still in that range \cite{Schier}.
The application of fat links, however, helps to reduce $Z_{A}$ \cite{DuHoWe}.

\begin{figure}[h!]
  \centering
\includegraphics[angle=270,width=0.7\linewidth]{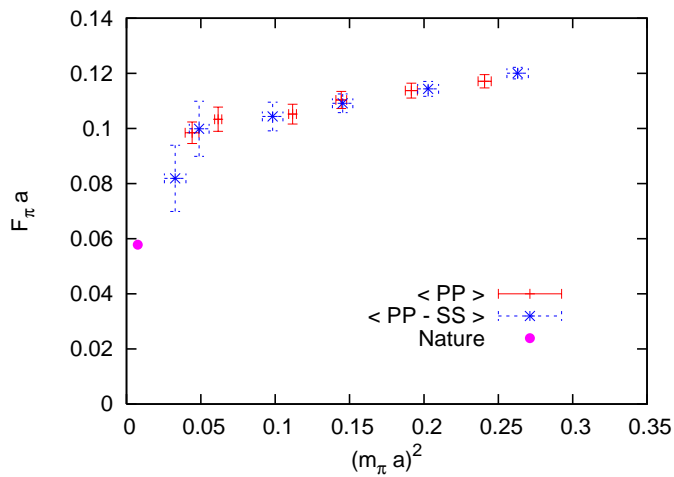}
\caption{{\it The pion decay constant based on a matrix element evaluation 
in the $p$-regime --- given by eq.\ (\ref{Fpip}) --- using the overlap-HF
\cite{WBStani}.}}
\label{Fpifig}
\end{figure}

As a further observable in the $p$-regime, we measured the pion decay constant
by means of the relation
\begin{equation}  \label{Fpip}
F_{\pi} = \frac{2 m_{q}}{m_{\pi}^{2}} \ \left| \langle 0 |
P | \pi \rangle \right| \ ,
\end{equation}
based on $P(x)P(0)$, and based on $P(x)P(0) - S(x)S(0)$. 
The results for the operator $D_{\rm ovHF}$ \cite{WBStani} 
are given in Figure \ref{Fpifig}. In particular the value at 
$am_{q}=0.01$ (the lightest quark mass in this plot)
is significantly lower for the case of the scalar 
subtraction. Hence this subtraction pushes the result towards the
phenomenological value. 

However, a chiral extrapolation based on such data for $F_{\pi}$
cannot be reliable.
An extrapolated value of $F_{\pi,PP}$  would come out clearly too large,
as it is also the case for $D_{\rm N}$ \cite{XLF}.
But in particular the instability of $F_{\pi ,PP-SS}$ 
at our lightest pion masses in the $p$-regime (obtained at
$a m_{q} = \ 0.01$ and $0.02$) calls for a clarification by
yet smaller quark masses. We did consider still much smaller values of $m_{q}$
in the same volume. As the results for the pion masses suggest, we are thus
leaving the $p$-regime. For the tiny masses $a m_{q} \leq 0.005$
we enter in fact the $\epsilon$-regime, where observables like $F_{\pi}$ 
have to be evaluated in completely different manners, see Subsection 8.4.

\subsection{The distribution of topological charges}

In  the next Subsection we are going to present result in the $\epsilon$-regime
of QCD. In this Subsection we focus on topological charges,
which play an essential r\^{o}le the $\epsilon$-regime \cite{LeuSmi}
(as we mentioned before).

As we already outlined when commenting on Figure 3, it is
a priori not obvious how to introduce topological sectors
on the lattice. However, if one deals with
Ginsparg-Wilson fermions, a sound definition is given by adapting
the Atiyah-Singer Theorem from the continuum and defining
the topological charge of a lattice gauge configuration
by the fermionic index $\nu$ \cite{HLN},
\be
{\rm topological~charge} \, \excleq \, \nu := n_{+} - n_{-} \ ,
\ee
where $n_{\pm}$ is the number of zero modes with positive/negative
chirality. These numbers are unambiguously determined once
a Ginsparg-Wilson Dirac operator is fixed (and in practice
only chirality positive or chirality negative zero modes occur
in one general non-trivial configuration\footnote{A cancellation
between zero modes of both chiralities is manifest for the
free fermion, but in a random gauge background this coincidence
has a probability of measure zero \cite{MLpriv};
it disappears under small variations in the gauge configuration
without topological protection.

The case of an undefined index also exists, but it has again
probability of measure zero. In the overlap formula (\ref{overlap})
this corresponds to a zero denominator. Such configurations are
``exceptional'' in the topological sense \cite{geocharge}, though not in 
the sense of accidental (near) zeros of $D$ due to quenching, 
cf.\ Subsection 7.5.}). 
However, for a given gauge configuration, 
the index for different Ginsparg-Wilson operators
does not need to agree. Albeit the level of agreement should be high
for smooth configurations, i.e.\ it should 
--- and it does\footnote{For instance, we observed at $\beta = 6.15$ on 
a $16^{4}$ lattice that the index $\nu$ of $D_{\rm N}$ is very stable as $\rho$ 
rises from $1.3$ to $1.7$; this changes less than $2 \%$ of the indices.}
--- increase for rising values of $\beta$. 

\begin{figure}[h!]
  \hspace*{-2mm}
  \includegraphics[angle=270,width=.35\linewidth]{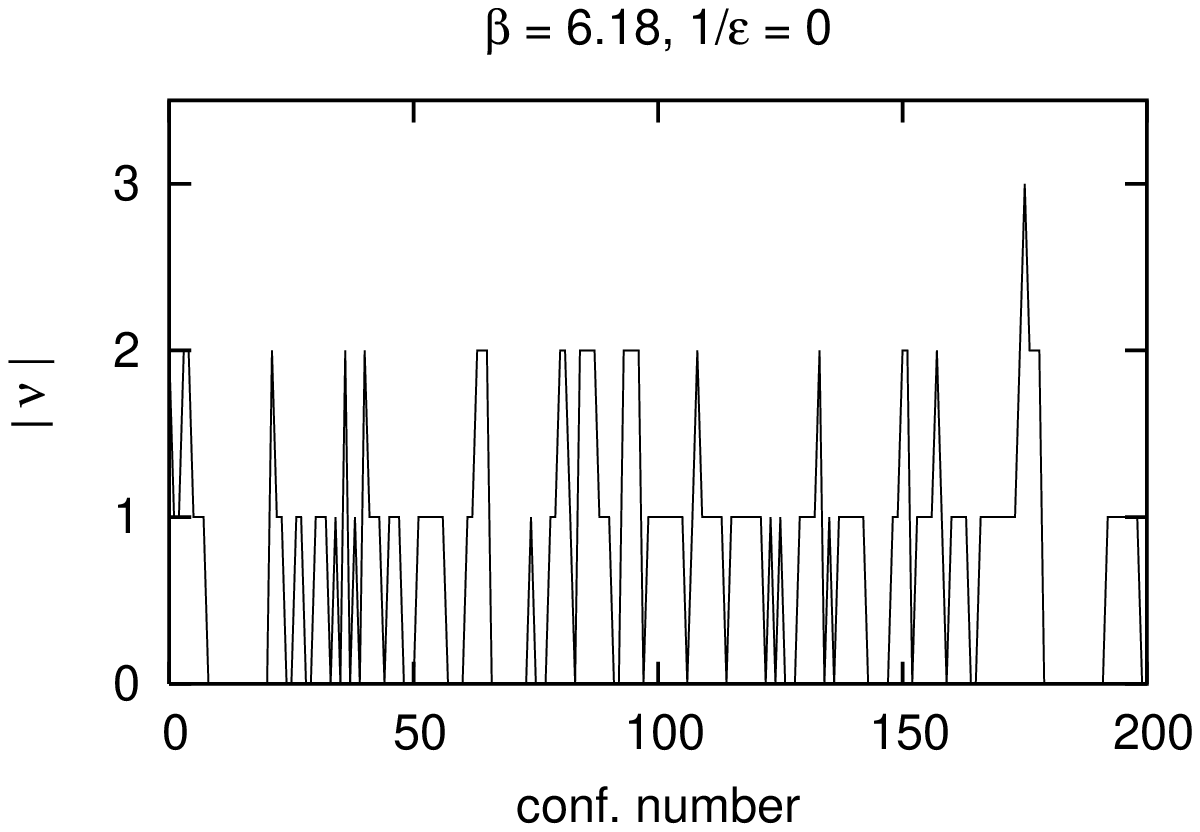}
  \hspace*{-5mm}
  \includegraphics[angle=270,width=.35\linewidth]{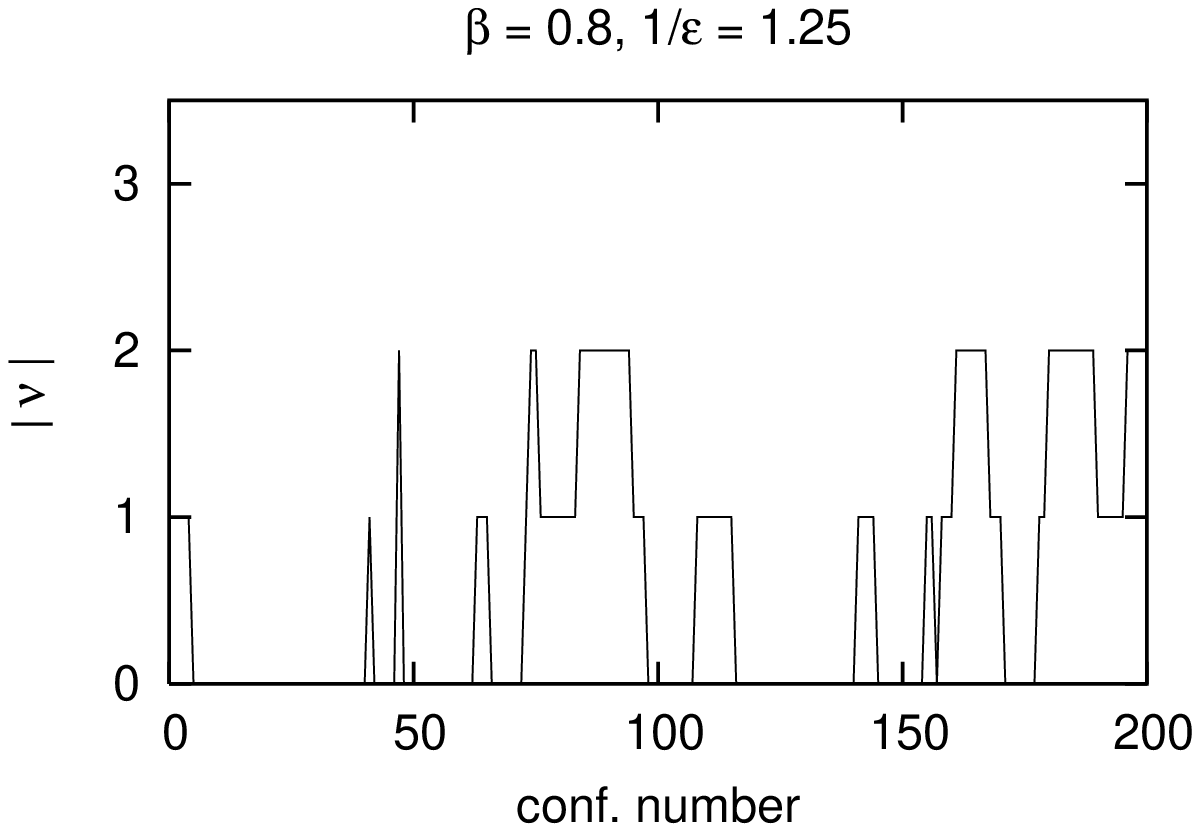}
  \hspace*{-5mm}
  \includegraphics[angle=270,width=.35\linewidth]{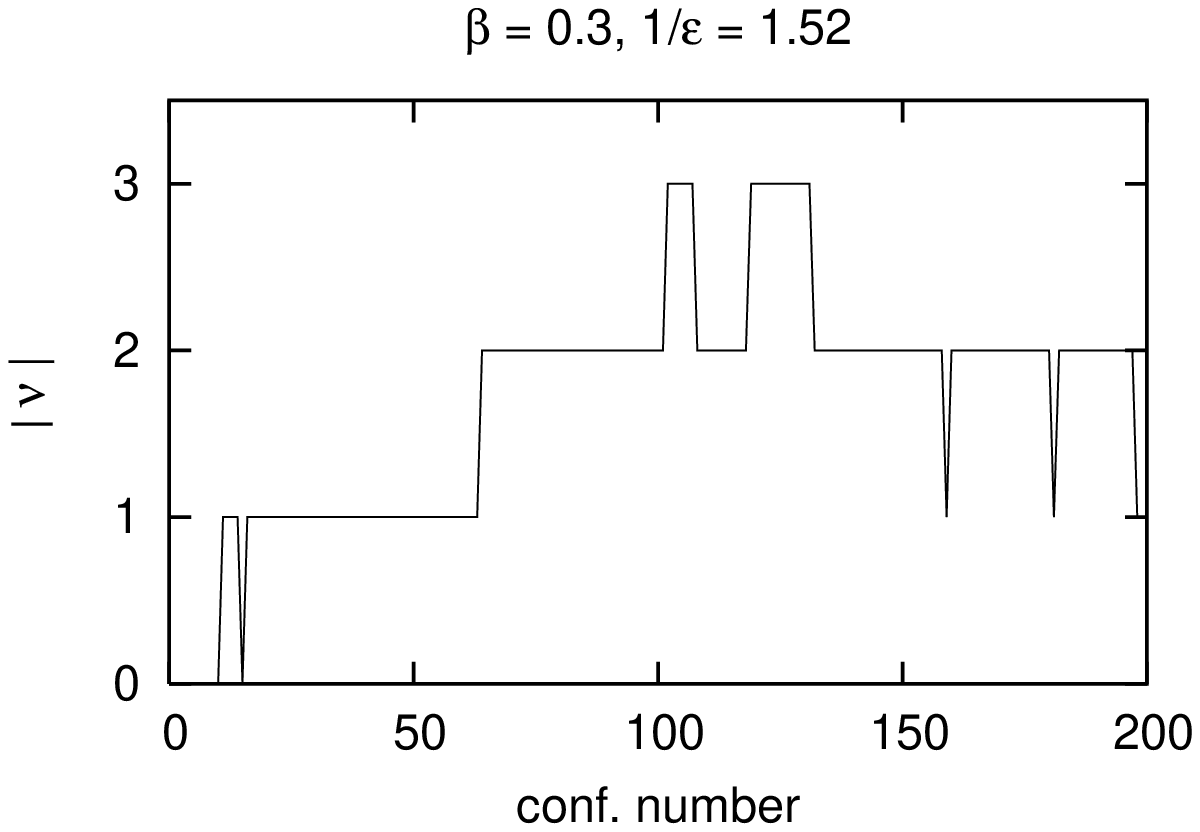}
  \caption{{\it Histories of the topological charge $| \nu |$ on a 
$16^{4}$ lattice for three sets of parameters, which correspond
approximately to the same physical scale, $a \approx 0.08~{\rm fm}$
\cite{topogaugeproc}.}}
\label{Qhisto} 
\end{figure}

In view of the LEC evaluation in the $\epsilon$-regime, numerical
measurements inside a specific topological sector --- and a confrontation
with the analytic predictions in this sector --- are in principle sufficient.
This requires the collection of a large number of
configurations in a specific sector. The (hyperbolic) ``topology
conserving gauge actions'' $S_{\varepsilon}^{\rm hyp}$
\cite{chiU1,topogauge1,topogaugeproc,topogauge2,topogaugedyn} are designed 
to facilitate this task,
\begin{equation}  \label{hypact}
S_{\varepsilon}^{\rm hyp} (U_{P}) = 
\left\{ \begin{array}{cc}
\frac{S_{P}(U_{P})}{ 1 -  S_{P}(U_{P}) / \varepsilon } &
{\rm ~~for~~}  S_{P}(U_{P}) < \varepsilon \\
+ \infty & {\rm otherwise} \end{array} \right. 
\end{equation}
where $S_{P}(U_{P}) = S_{\infty}^{\rm hyp} (U_{P})$ 
is the Wilson plaquette gauge action, and $U_{P}$ are the plaquette
variables \cite{latbooks}. For 
$\varepsilon \leq 1/[6 (2 + \sqrt{2})] \simeq 0.049$ 
topological
transitions cannot occur under continuous deformations of the gauge 
configuration \cite{Neubound}. But in practice we have to relax $\varepsilon$ 
to larger values to allow for reasonably strong
fluctuations. For strong
gauge couplings we can then arrange for a useful physical lattice
spacing. Examples illustrating the increased topological 
stability in the course of a Monte Carlo history are shown in 
Figure \ref{Qhisto}.

\begin{figure}[h!]
  \centering
\includegraphics[angle=270,width=.7\linewidth]{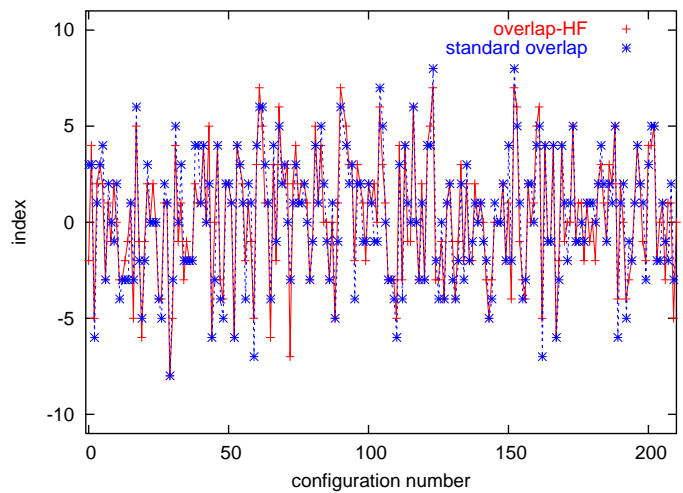}
\caption{{\it Index histories for $D_{\rm ovHF}$ (see Section 7)
and for the standard overlap operator
$D_{\rm N}$ (at $\rho =1.6$) for the same set of QCD configurations.}}
\label{tophistory}
\end{figure}

\begin{figure}[h!]
  \centering
\hspace*{-5mm}
\includegraphics[angle=270,width=.52\linewidth]{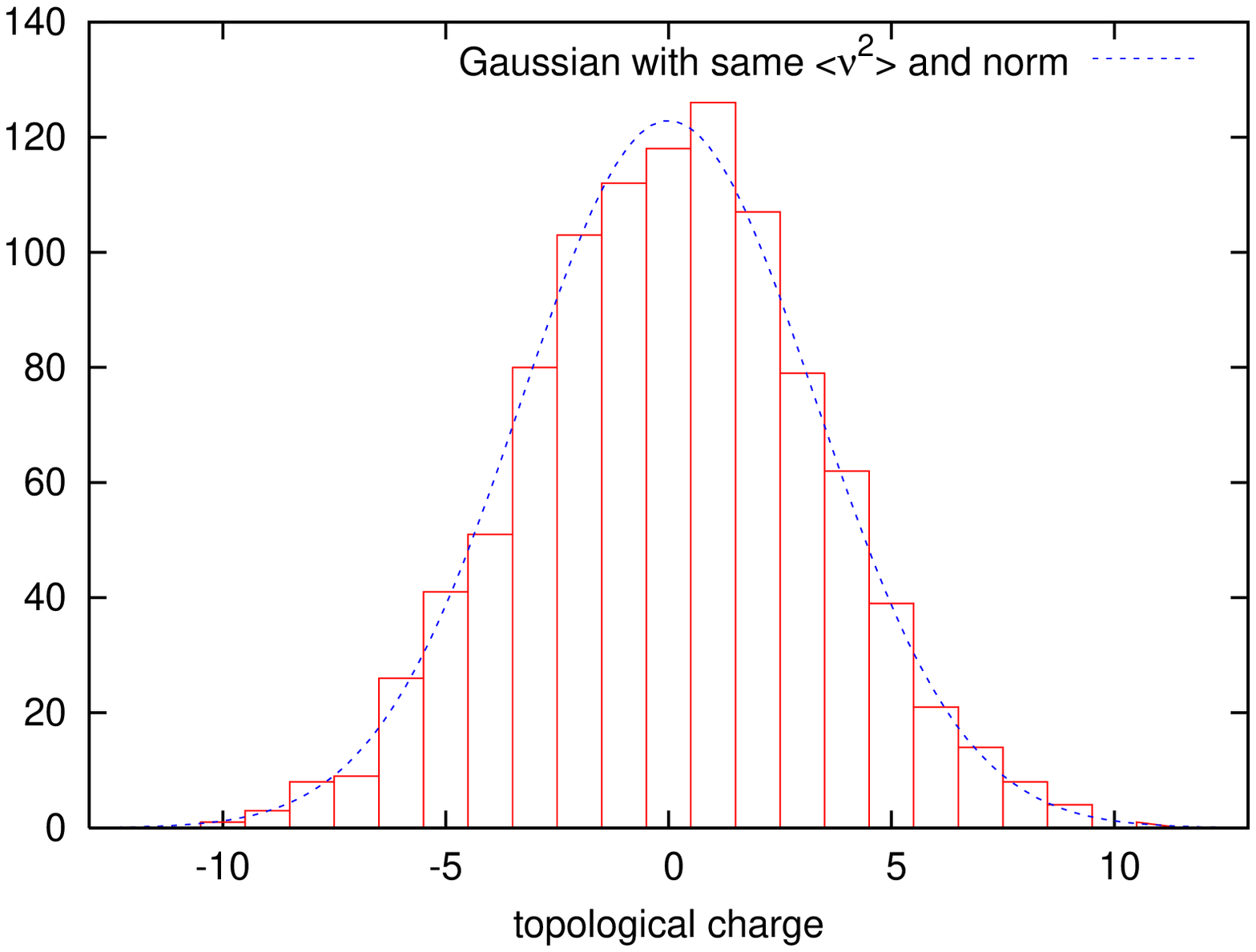}
\hspace*{-5mm}
\includegraphics[angle=270,width=.52\linewidth]{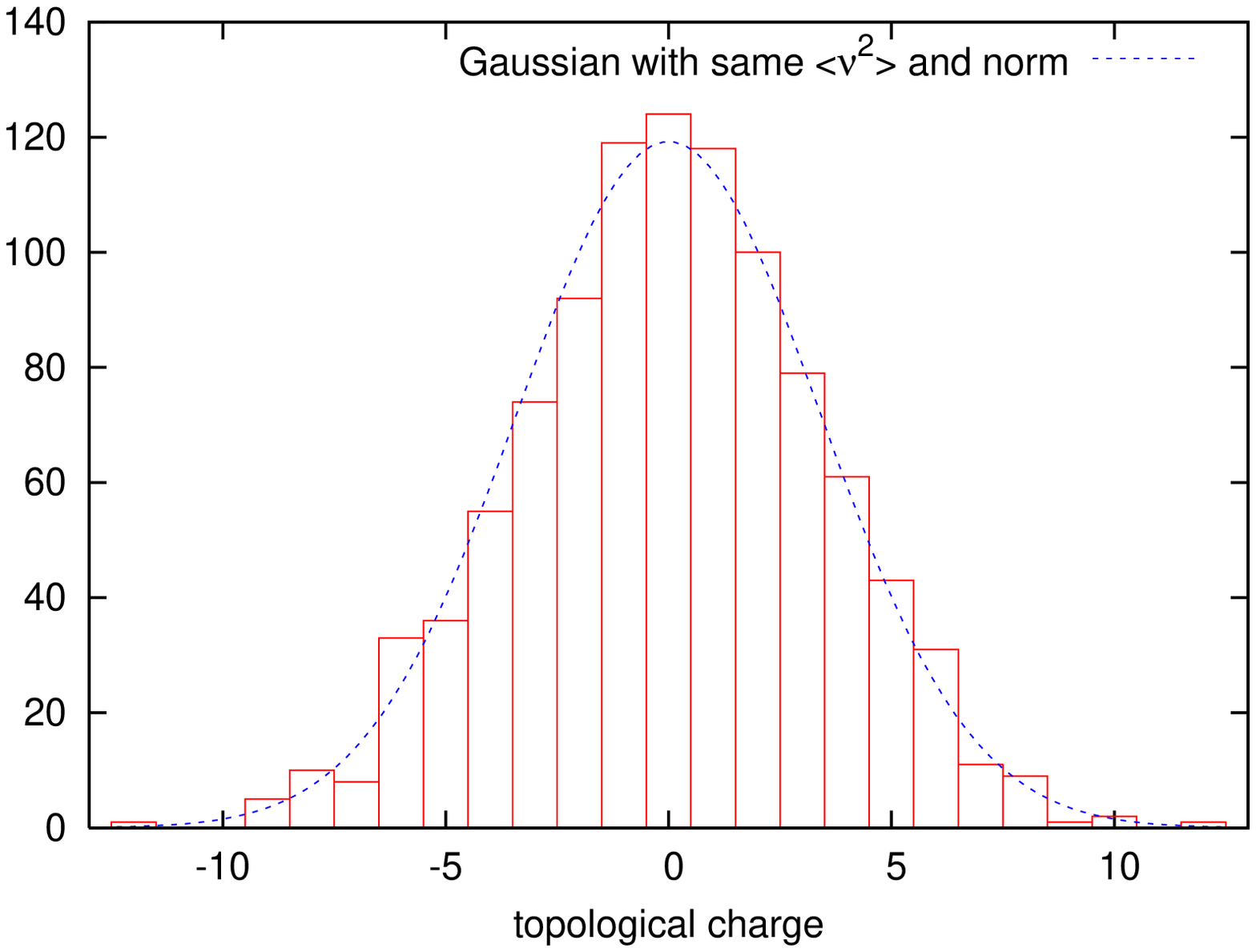}
\caption{{\it The histogram of $D_{\rm ovHF}$ indices 
(on the left) and of $D_{\rm N}$ indices (on the right),
on a $12^{3} \times 24$ lattice at $\beta = 5.85$.
In both cases, 1013 configurations are included \cite{WBStani}.}}
\label{indhistogram}
\end{figure}

Let us now return to the Wilson gauge action, which allows
us to investigate also the statistical distribution of the
topological charges.
Again at $\beta = 5.85$ on a $12^{3} \times 24$ lattice 
we compared the charges for the overlap-HF operator, 
and for the standard overlap operator 
$D_{\rm N}$ at $\rho = 1.6$ \cite{WBStani}. 
As an example, the histories of about 200 indices for the same configurations
are compared in Figure \ref{tophistory}. Of course, these two types of 
indices are considerably correlated, but only about $40 \%$ really coincide.
Nevertheless both follow well the expected Gaussian
distributions $\propto \exp(- {\rm const.} \cdot \nu^{2})$, 
with a width $\approx 3.3$, 
see Figure \ref{indhistogram}.
This width fixes the topological susceptibility (cf.\ eq.\ (\ref{topsus}))
\be  \label{chitop}
\chi_{\rm top} = \frac{1}{V} \langle \nu^{2} \rangle \ ,
\ee
which is of importance to explain the heavy mass of the $\eta'$-meson
\cite{WiVe}. 

\begin{figure}[h!]
  \centering
\includegraphics[angle=270,width=.7\linewidth]{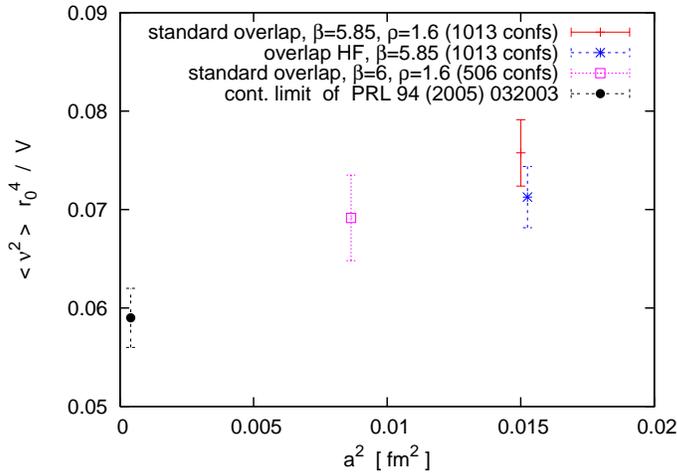}
\caption{{\it The topological susceptibility
measured by indices of $D_{\rm ovHF}$ and of $D_{\rm N}$,
in a volume $V = (1.48~{\rm fm})^{3} \times 
2.96~{\rm fm}$, with two different lattice spacings $a$.
Our data --- given in detail in  Ref.\ \cite{WBStani} ---
agree well with the continuum extrapolation 
in Ref.\ \cite{DDGP}.}}
\label{toposus}
\end{figure}

In Figure \ref{toposus} we present our results with $D_{\rm ovHF}$
and $D_{\rm N}$ on the lattice referred to so far, plus a result
for $D_{\rm N}$ at $\beta =6$ in the same physical volume
(lattice size $16^{3} \times 32$). We also mark the
continuum extrapolation according to Ref.\ \cite{DDGP}, which is
fully consistent with our results. That measurement
of $\chi_{\rm top}$ was based on $D_{\rm N}$ indices on $L^{4}$ 
lattices.\footnote{A compilation of earlier lattice results 
for $\chi_{\rm top}$ (with various methods) is given in Ref.\ \cite{Bern}.}
The resulting value for $\chi_{\rm top}$ is compatible with the Witten-Veneziano
scenario that much of the $\eta'$ mass is generated by a $U(1)$ anomaly
\cite{WiVe}.

\subsection{The $\epsilon$-regime}

We first repeat that the $\epsilon$-regime of QCD is characterised by 
a relatively small volume, i.e.\ the correlation length $\xi$ exceeds the linear
box size $L$. Together with the requirement (\ref{4piFpi}) this
amounts to the condition\footnote{In such small volumes the notion of
$m_{\pi}$ is problematic, but this inequality can be interpreted by
referring to the would-be pion mass in an extended volume.}
\be
\frac{1}{m_{\pi}} > L \gg \frac{1}{4\pi F_{\pi}} \ .
\ee
In such a box, the $p$-expansion of
$\chi$PT fails, in particular because of the importance of the
zero modes. However, the latter can be treated separately by means of
collective variables, and the higher modes --- along with the pion mass ---
are then captured by the {\em $\epsilon$-expansion} \cite{epsreg1}.
One now counts the ratios
\be  \label{epsexp}
\frac{m_{\pi}}{\Lambda_{\rm QCD}} \sim \frac{p^{2}}{\Lambda_{\rm QCD}^{2}}
\sim \frac{1}{( L F_{\pi})^{2}}
\ee
as small quantities in the same order.

This setting, where pions are squeezed into a tiny box,
cannot be considered physical.
Nevertheless there is a strong motivation for its numerical study:
the point is that the finite size effects are parametrised by the
LECs of the effective chiral Lagrangian as they occur in infinite volume. Hence
the physical values of the LECs can (in principle) be evaluated even in such
an unphysically small box, as we mentioned in the introduction to Section 8.

Unfortunately, most results for the LECs in the $\epsilon$-regime
are obtained in the quenched approximation so far, hence they are
affected by (mostly logarithmic) finite size effects \cite{loga}.
So the final results
by this method still have to wait for the feasibility of QCD simulations
with dynamical, chiral quarks. These prospects will be commented on
in Subsection 9.2.

\subsubsection{A 3-loop calculation in the $\epsilon$-expansion}

The chiral symmetry breaking for two flavours
is locally isomorphic to orthogonal groups,
\be
SU(2) \otimes SU(2) \to SU(2) \quad \sim \quad O(4) \to O(3) \ .
\ee
Generally, the spontaneous symmetry breaking
$O(N) \to O(N-1)$ generates $N-1$ NGBs.\footnote{There is, however,
no other symmetry breaking $O(N) \to O(n)$, $N>n$, which is locally isomorphic 
to $SU(N_{f}) \otimes SU(N_{f}) \to SU(N_{f})$ for any $N_{f}$.} 
They can be described
by the non-linear $\sigma$-model and studied with the formalism of
$\chi$PT. In $d=3$ and $d=4$ the corresponding Lagrangian involves
an infinite string of terms, which can be ordered by the power of the 
momenta involved. In a volume $V = L^{d}$ (with periodic boundary
conditions) we count the momenta as
$\partial_{\mu} = O (L^{-1})$, and we obtain the leading terms
\bea
{\cal L}^{\rm (sym)}[ \vec S ] &=& \frac{F^{2}}{2} \partial_{\mu} \vec S
\partial_{\mu} \vec S + \frac{1}{2} g_{4}^{(1)} \partial^{2} \vec S
\partial^{2} \vec S + \frac{1}{4} g_{4}^{(2)} 
(\partial_{\mu} \vec S \partial_{\mu} \vec S )^{2} \nn \\
&+& \frac{1}{4} g_{4}^{(3)} (\partial_{\mu} \vec S \partial_{\nu} 
\vec S )^{2} + O(L^{-6}) \ , \qquad \vec S^{\, 2}(x) \equiv 1 \ , \quad
\eea
where we attach an independent LEC to each term.
We may also add terms that break the $O(N)$ symmetry explicitly
through a small, constant magnetic field $\vec H$, which adopts
the r\^{o}le of the light quark masses,
\bea
- {\cal L}^{\rm (sb)}[ \vec S ] &=& \Sigma (\vec H \vec S ) +
h_{2,0}^{(1)}(\vec H \vec S)^{2} + h_{2,0}^{(2)}(\vec H \vec H)^{2} \nn \\
& + & h_{1,2}^{(1)}(\vec H \vec S) \, 
(\partial_{\mu} \vec S \partial_{\mu} \vec S )
+ h_{1,2}^{(2)}(\vec H \partial^{2} \vec S) + \dots \quad .
\eea
Then we assemble the total Lagrangian ${\cal L} = {\cal L}^{\rm (sym)} +
{\cal L}^{\rm (sb)}$.

We count $H := | \vec H | = O(V^{-1})$, 
so that the NGBs with $m^{2} = \Sigma H / F^{2}$ feel the finite size strongly,
$ m L \ll 1$. Then the partition function can be $\epsilon$-expanded in 
the dimensionless ratio $\epsilon = L^{2-d}/F^{2}$ in $d=3$ and $4$.

To compute the free energy ${\cal F} = - \ln Z$ one first has to find a 
way to handle the zero modes. We mentioned before that this can be achieved 
by using collective variable \cite{epsreg1}. 
In the present case, one considers the magnetisation
\be
\vec M := \int d^{d}x \, \vec S (x) = | \vec M | \, \Omega \, \vec e \ ,
\ee
where $\vec e = \vec H / | \vec H| $ is a fixed unit vector
and $\Omega \in O(N)$ is integrated over in the functional 
integral. (The corresponding
unitary integration for quark flavours is discussed in Refs.\
\cite{Hansen}.)
Here the vector field $\vec S$ can be decomposed into one dominant
component plus fluctuations ${\underline{\pi}}(x) = (\pi_{1}(x), \dots
\pi_{N-1}(x))$, $\ \pi_{i}(x) = O(L^{1 - d/2})$,
which are treated perturbatively. In this way, the partition
function can be evaluated order by order.

This was first carried out to 2 loops in Ref.\ \cite{HasLeu}, where
the functional measure was treated with
the Faddeev-Popov procedure (the constraint
$\delta ( \vec S^{\, 2} -1)$ is implemented in an exponential form).
A different method was applied in
Ref.\ \cite{WBDiss}, which was based on the Polyakov measure
\cite{Poly}. That measure was originally introduced in the framework 
of string theory and it captures the above constraint by 
integrating ${\cal D} {\underline{\pi}}$ 
over the $\epsilon$-expanded elements of a metrics in flavour space. 
In this way we computed the partition function to
3 loops. In both cases dimensional regularisation \cite{BolGiam}
was used. It turned out that the free energy is in fact
perturbatively {\em renormalisable order by order} (although the
number of required counter terms increases rapidly in each order).
This property is highly
non-trivial due to the requirement that the renormalised LECs
must not pick up any volume dependence. This is realised
both, in $d=3$ and in $d=4$, for the large number of terms occurring
to the 3-loop order, 
thanks to numerous cancellations between
forbidden contributions. These calculations represent therefore
a sensitive test for the validity of the $\epsilon$-expansion 
scheme, as well as the methods used for the treatment of the
functional measure.

Before chiral lattice fermions became available, the program
of evaluating LECs through simulations in the $\epsilon$-regime was 
tested in the framework of this spin model \cite{epsO4}.

\subsubsection{The chiral condensate}

Chiral Random Matrix Theory (RMT)
conjectures predictions for the low
lying eigenvalues, ordered as $\lambda_{n}$, $n = 1,2,3 \dots$
(excluding possible zero eigenvalues)
of the Dirac operator in the $\epsilon$-regime; 
for a review, see Ref.\ \cite{VerWet}.
More precisely, it conjectures densities of
the dimensionless variables $\Sigma V \lambda_{n}$,
where $\Sigma$ is the chiral condensate in the effective 
Lagrangian (\ref{Leff}). Here we focus on
the variable $z := \Sigma V \lambda_{1,P}$,
where $\lambda_{1,P}$ emerges from the leading non-zero eigenvalue
$\lambda_{1}$ if the spectral circle of the overlap operator
is mapped stereographically onto the imaginary axis,
$\lambda_{1,P} = | \lambda_{1} / (1 - \lambda_{1} / 2 \rho )|$.

These RMT predictions depend on $| \nu |$, the absolute
value of the topological charge. Here we make use
of the explicit formulae \cite{lowEV} for the density of the first
non-zero (re-scaled) eigenvalues $z$ in the sectors
$|\nu |$, which we denote by $\rho_{1}^{(| \nu |)}(z)$.
For the lowest eigenvalues, the particular density 
$\rho_{1}^{(0)}$ was first confirmed by staggered fermion simulations
(results are summarised in Ref.\ \cite{VerWet}).
But in those studies the charged 
sectors\footnote{In that case, a topological charge has 
to be introduced by some traditional method like cooling \cite{cool}.}
yielded the very same density, in {\em contradiction} to RMT. 

\begin{figure}[h!]
  \centering
\includegraphics[angle=270,width=.49\linewidth]{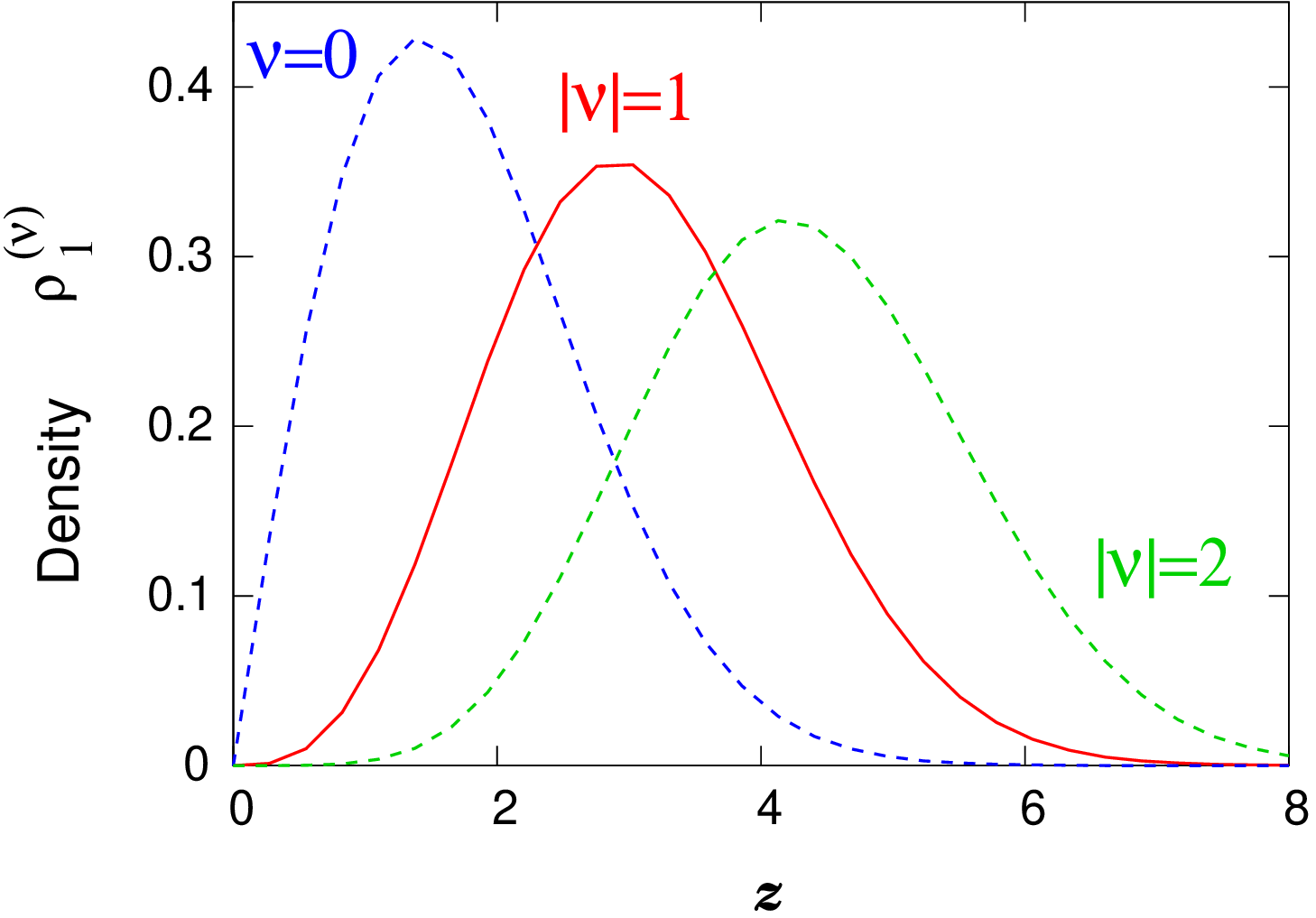}
\includegraphics[angle=270,width=.49\linewidth]{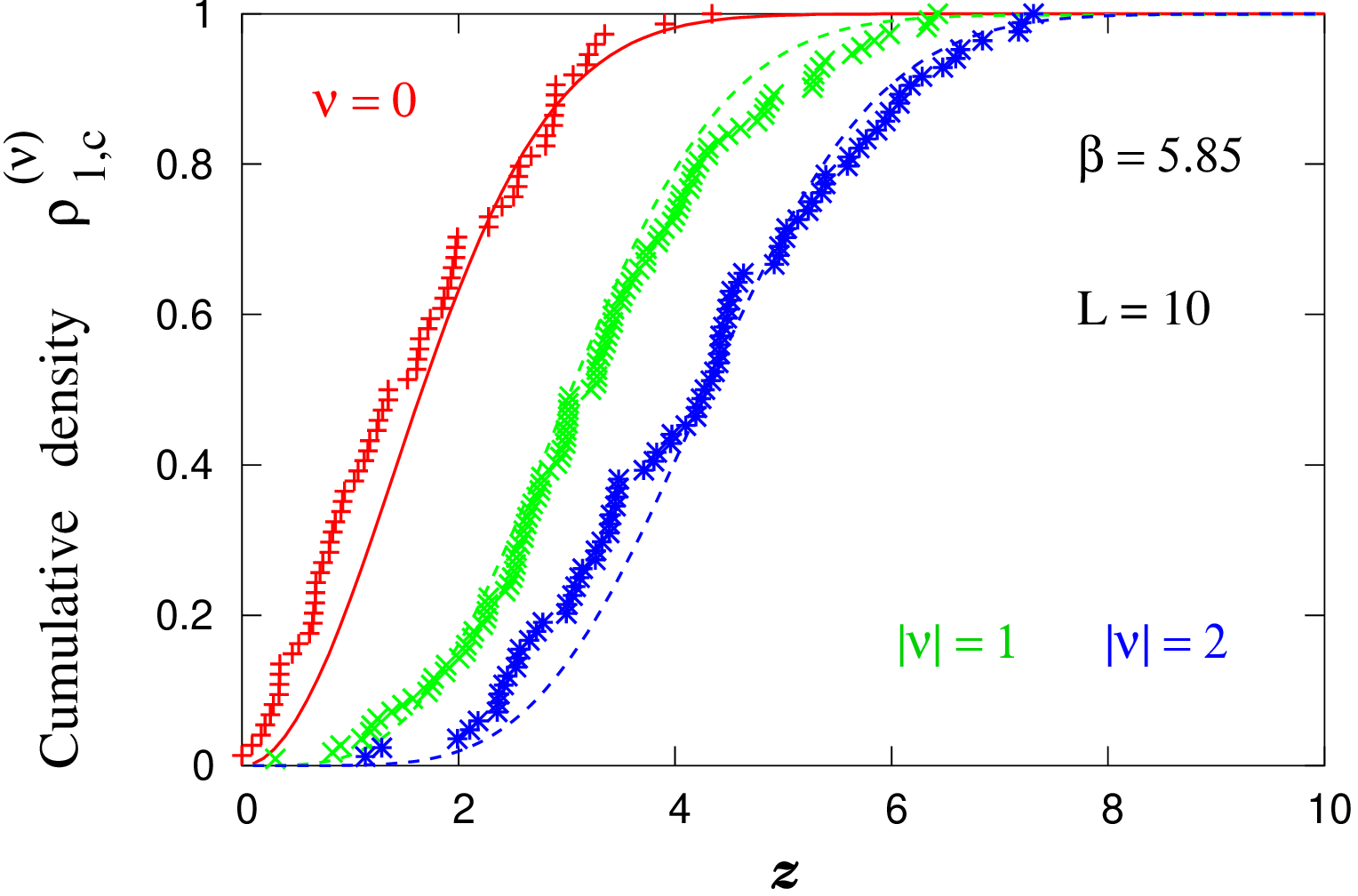}
\caption{{\it On the left: RMT predictions for the leading non-zero
Dirac eigenvalue in the topological sectors with charge 
$| \nu | = 0, \ 1$ and $2$. \newline
On the right: RMT predictions (lines) and simulation
results for the corresponding cumulative densities. 
The numerical data were obtained with $D_{\rm N}$
on a $10^{4}$ lattice at $\beta = 5.85$ (see first work in Ref.\
\cite{RMT}), and they do essentially 
follow the RMT predictions. The confidence level for the agreement
of cumulative densities has been verified with the Kolmogorov-Smirnov
test \cite{NR}.}}
\label{RMTfig1}
\end{figure}

\begin{figure}[h!]
  \centering
\includegraphics[angle=270,width=.7\linewidth]{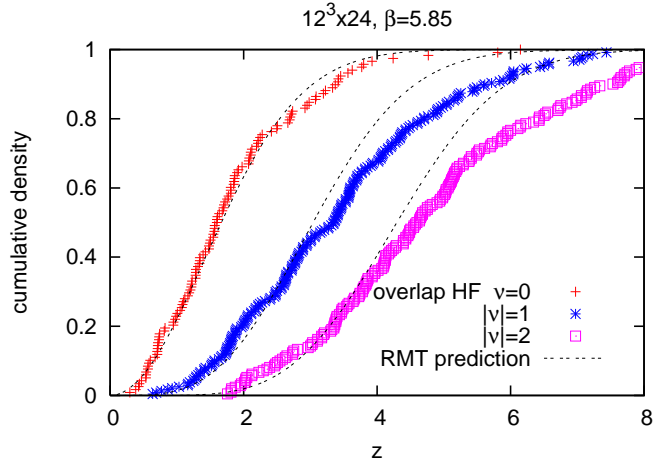}
\caption{{\it The cumulative density of the (M\"{o}bius projected) 
lowest Dirac eigenvalue $\lambda_{1,P}$ of the overlap-HF operator, 
in the topological sectors $|\nu | =0$, $1$ and $2$. 
We compare the chiral RMT predictions to our data
for $z= \Sigma V \lambda_{1,P}$
with $\Sigma^{1/3} = 298 ~ {\rm MeV}$ --- the optimal value in the
neutral sector ($\nu = 0$). This value works well up to $z \approx 3$
for all topological sectors, i.e.\ well beyond the Thouless value
$z_{\rm Thouless} \lsim 1$, which is often considered a theoretical bound
for the applicability of these predictions.}}
\label{SigmafigovHF}
\end{figure}

The distinction between the topological
sectors was first observed to hold for $D_{\rm N}$, following the RMT
predictions to a good precision \cite{RMT}, if the linear box 
size exceeds a lower limit of about $L \gsim 1.1~{\rm fm}$ 
(the exact limit depends  on the criterion, of course).\footnote{Meanwhile 
a topological splitting has also been observed
to set in for staggered fermions if the link variables are strongly smeared
\cite{staggtop}.} The predictions for the densities of the leading
non-zero eigenvalues in the sectors $| \nu | = 0, \ 1$ and $2$ are shown
in Figure \ref{RMTfig1} on the left; we see that zero modes repel
the finite eigenvalues. On the right-hand-side of Figure \ref{RMTfig1}
we present results
for the corresponding cumulative density with $D_{\rm N}$ on a $10^{4}$
lattice at $\beta = 5.85$ (box length $\simeq 1.23~{\rm fm}$).
Once the predicted densities $\rho_{1}^{(| \nu |)}$ (marked by lines)
are well reproduced,
we can read off the value of $\Sigma$, which is the only free
fitting parameter for all topological sectors. We proceed to larger
lattices and show our result for $D_{\rm ovHF}$ on a $12^{3} \times 24$ 
lattice at $\beta = 5.85$ in Figure \ref{SigmafigovHF} \cite{WBStani}.
The optimal fit shown in this plot, and its
counterpart for the Neuberger operator $D_{\rm N}$, yield
\be  \label{SigmaRMT}
\Sigma^{1/3} = 298(4) ~ {\rm MeV} \quad ({\rm from} ~ D_{\rm ovHF}) \ , \
\Sigma^{1/3} = 301(4) ~ {\rm MeV} \quad ({\rm from} ~ D_{\rm N}) \ .
\ee 
These fits focus on the lowest eigenvalues resp.\ energies, where
chiral RMT is most reliable.
Clearly, in this range the neutral sector ($\nu =0$) dominates.
In the case of $D_{\rm N} \, $, the charged sectors $| \nu | =1$ and $2$ 
alone would favour a different $\Sigma$ value \cite{WBStani}. 
Such ambiguities are quite strong in the results with smeared staggered
fermions \cite{staggtop}. In the case of $D_{\rm ovHF}$,
however, a unique $\Sigma$ works well for all the three sectors
$| \nu | = 0,1,2$, up to about $z \approx 3$,
as Figure \ref{SigmafigovHF} shows.
This range extends well beyond the Thouless value
$z_{\rm Thouless} = F_{\pi}^{2} \sqrt{V} \lsim 1$, 
which is often understood as a threshold for the RMT applicability.

\begin{figure}[h!]
  \centering
\includegraphics[angle=270,width=.7\linewidth]{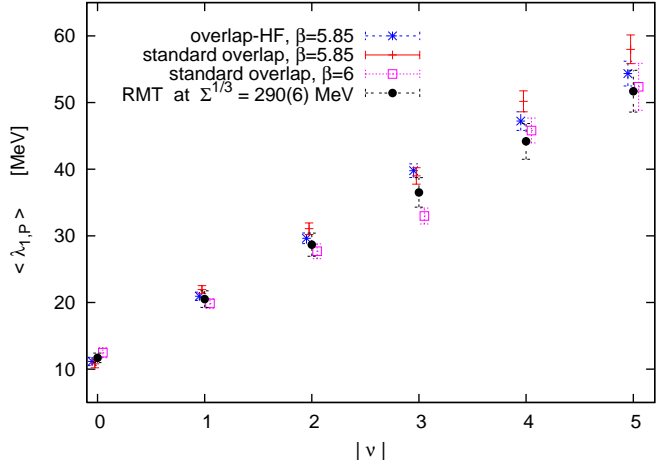}
\caption{{\it The mean values of the first non-zero Dirac eigenvalue
(in physical units) in the charge sectors $|\nu | =0 \dots 5$.
All these numerical results \cite{WBStani} agree with chiral RMT if 
we choose $\Sigma^{1/3} = 290(6) ~ {\rm MeV}$.}}
\label{l1mean}
\end{figure}

As an alternative approach to test the agreement of our data
with the chiral RMT, and to extract a value for $\Sigma$, we now consider
the mean values of the leading non-zero Dirac eigenvalues $\lambda_{1}$
in all the charge sectors up to $| \nu |=5$. In physical units,
the results $\langle \lambda_{1,P} \rangle$ agree very well
for different overlap operators and lattice spacings
--- see Figure \ref{l1mean} ---
although this consideration extends beyond very low energy. Each single
result for $\langle \lambda_{1,P} \rangle_{| \nu |}$ 
can then be matched to the RMT value for a specific choice of
$\Sigma$. It is very remarkable that all these 18 results are
compatible with RMT if we choose
\begin{equation}
\Sigma = (290(6) ~ {\rm MeV})^{3} \ ,
\end{equation}
as Figure \ref{l1mean} also shows.

A renormalisation procedure for $\Sigma$ obtained in this
way is discussed in Ref.\ \cite{WenWit}. However, we will only use 
this quenched lattice result as a fitting
input in Subsection 8.4.3, so we stay with the bare condensate
$\Sigma$ for our fixed lattice parameters.\\
We add that there are also first applications of this technique
with dynamical overlap fermions \cite{dynSigma}.

\subsubsection{The pion decay constant determined from the axial current correlator}

In order to relate our quenched simulation results to the 
effective low energy theory,
we now refer to {\em quenched $\chi$PT}. In that framework,
mesonic correlation functions were calculated to the first order in
Refs.\ \cite{corre1,corre2}. The vector current correlation function 
vanishes, while the scalar and pseudoscalar correlators involve
already in the first order additional, quenching specific LECs,
which obstruct the access to the physical LECs in the
Lagrangian (\ref{Leff}). Therefore we first focus on the axial-vector
correlator, which only depends on $\Sigma $ and $F_{\pi}$ in the
first order. In particular we are going
to compare our data to the quenched $\chi$PT prediction
in a (periodic) volume $V= L^{3} \times T$ \cite{corre2},
for the topological sectors with charge $\pm \nu$,
\bea   \label{AA}
Z_{A}^{2} \cdot \la A_{4}(t) \, A_{4}(0) \ra_{| \nu |} &=& 
2  \left(\frac{F_{\pi}^{2}}{T} 
 + 2 m_{q} \, \Sigma_{\vert \nu \vert}(z_{q}) \, 
T  \, h_{1}(\tau )\right) \ , \\
h_{1}(\tau ) &=& \frac{1}{2} \Big( \tau^{2} - \tau + \frac{1}{6} \Big) 
\ , \qquad \tau = \frac{t}{T} \ , \nn \\
\Sigma_{\nu}(z_{q}) &=& \Sigma \left( z_{q} \Big[ I_{\nu}(z_{q}) K_{\nu}(z_{q}) + 
I_{\nu +1}(z_{q}) K_{\nu -1}(z_{q}) \Big] + \frac{\nu}{z_{q}} \right) \nn \ ,
\eea
where
\be
A_{4}(t) = a^{3} \sum_{\vec x} \bar \psi (t ,\vec x) \gamma_{5}
\gamma_{4} \psi (t , \vec x ) \qquad \quad (t > 0)
\ee
is the bare axial-vector current at 3-momentum $\vec p = \vec 0$.
$I_{\nu}$ and $K_{\nu}$ are modified Bessel functions, and
$z_{q} := \Sigma V m_{q}$ (in analogy to the dimensionless variable $z$
in Subsection 8.4.2).

It is remarkable that this prediction in the 
$\epsilon$-regime has the shape of a {\em parabola} with a minimum
at $t = T/2$. This is in qualitative contrast to the ${\tt cosh}$
behaviour, which is standard in large volumes. $\Sigma$
affects both, the curvature and the minimum of this parabola,
whereas $F_{\pi}$ only appears in the additive constant --- that
feature is helpful for its evaluation.

A first comparison of this curve to lattice data was presented
in Ref.\ \cite{AApap}, using $D_{\rm N}$ at $\beta =6$, $\rho = 1.4$,
$am_{q} = 0.01$ on lattice sizes $10^{3} \times 24$
and $12^{4}$. For the anisotropic volume the linear
size of $L \simeq 0.93~{\rm fm}$ turned out to be too small:
the data for $\, \la A_{4}(t) \, A_{4}(0) \ra_{1,2} \,$ were 
incompatible with the parabola of eq.\ (\ref{AA})
for any positive $\Sigma$.
This observation is consistent with the lower bound for $L$ that we
also found for the agreement of the microscopic spectrum with chiral 
RMT ($L \gsim 1.1~{\rm fm}$, see Subsection 8.4.3). 

Another observation in that study was that the corresponding Monte Carlo
history in $\nu =0$ is plagued by strong spikes, giving rise to large
statistical errors. A huge statistics of
$O(10^{4})$ topologically neutral configurations
would be required for conclusive results. 
These spikes occur for the configurations
with a tiny (non-zero) Dirac eigenvalue $\lambda_{1,P}$.
It agrees again with chiral
RMT that such configurations are most frequent in the topologically
neutral sector, see Figure \ref{RMTfig1}.
(A method called ``low mode averaging''
was designed and applied to alleviate this problem \cite{LMA}.)

However, without applying that method we obtained a decent agreement
with the prediction (\ref{AA}) in the $12^{4}$ lattice mentioned above
($V \simeq (1.12 ~{\rm fm})^{4}$) in the sector
$| \nu | = 1$ \cite{AApap}. In view of the leading LECs, it seems 
unfortunately impossible to extract a value of $\Sigma$ from such data
(although it is encoded in $z_{q}$), since 
the theoretical curvature depends on it only in an extremely weak 
way.\footnote{Only in the sector $\nu =0$ the sensitivity to $\Sigma$ is
significant, but there we run into the statistical problem
mentioned before.}
On the other hand, $F_{\pi}$ can be extracted quite well from the 
vicinity of the minimum at $t = T/2$, but the value found in Ref.\
\cite{AApap} was too large.

Next a study of that kind appeared in Ref.\ \cite{JapZA},
which also used $D_{\rm N}$, at $\beta = 5.85$ (and $\rho =1.6$), now
on a $10^{3} \times 20$ lattice. This work analysed
the sectors $| \nu |=0$ and 1.
As a reason for this limitation the authors referred to the
condition $| \nu | \ll \langle \nu^{2} \rangle$. As we mentioned in
Subsection 8.4.2, one expects $\langle \nu^{2} \rangle \propto V$
(up to artifacts), see eq.\ (\ref{chitop}), 
hence this limitation was imposed by the volume.

\begin{figure}[h!]
  \centering
\includegraphics[angle=270,width=.67\linewidth]{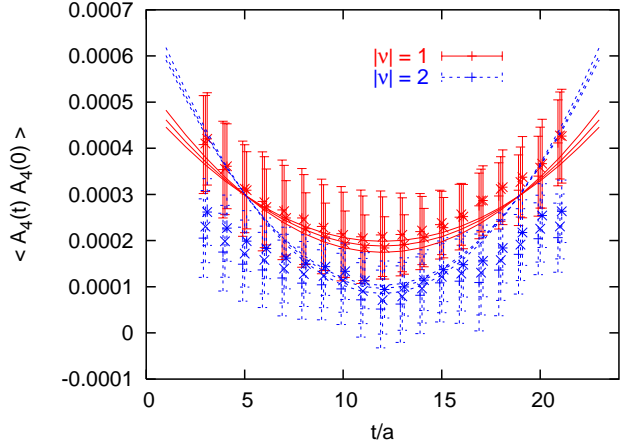}
\caption{{\it Lattice data obtained with $D_{\rm ovHF}$
for the axial-current correlation functions,
measured sepa\-rately in the topological sectors $| \nu | = 1$ and $2$.
The curves are global fits (over both sectors)
to the quenched $\chi$PT formula
(\ref{AA}), for each of our three masses in the $\epsilon$-regime.
They single out the values of $F_{\pi}$ given in Table \ref{epstabFpiAA}.}}
\label{AAfig}
\end{figure}

In Ref.\ \cite{WBStani} we presented again results 
at $\beta = 5.85$ on a $12^{3} \times 24$
lattice, where the volume admits $| \nu | = 2$.
We measured for both, $D_{\rm ovHF}$ and $D_{\rm N}$,
the axial-vector correlators at the masses $am_{q}=0.001$, $0.003$ and
$0.005$, which are safely in the $\epsilon$-regime.
We fitted the data to eq.\ (\ref{AA})
by inserting the chirally extrapolated factors $Z_{A}$ ($1.17$ for
$D_{\rm ovHF}$ \cite{WBStani} and $1.45$ for $D_{\rm N}$
\cite{XLF}, see Subsection 8.2), along with 
the $\Sigma$ values in eq.\ (\ref{SigmaRMT}).
For each of the overlap operators we performed --- at each of the quark
masses --- a global fit over the topological sectors that we considered.
The result for $D_{\rm ovHF}$ is shown in Figure \ref{AAfig}.
It revealed for the first time
a convincing distinction between the sectors $| \nu | = 1$ and
$| \nu | = 2$ --- this predicted topological splitting
could not be observed for $D_{\rm N}$
up to now. For $D_{\rm N}$ at $am_{q} =0.005$ we also include
the neutral sector; as expected it has clearly larger errors than the 
charged sectors, but it is helpful nevertheless to reduce the 
error on $F_{\pi}$ in the global fit.
The values for $F_{\pi}$ obtained in this way for $D_{\rm ovHF}$
and for $D_{\rm N}$ are in accurate agreement, as Table 
\ref{epstabFpiAA} shows.
\begin{table}
\begin{center}
\begin{tabular}{|c||c|c|}
\hline
$a m_{q}$ & $D_{\rm ovHF}$ & $D_{\rm N}$ \\
\hline
\hline
$0.001$ & $F_{\pi} = (110 \pm 8) ~ {\rm MeV}$ 
& $F_{\pi} =(109 \pm 11) ~ {\rm MeV}$ \\
\hline
\hline
$0.003$ & $F_{\pi} =(113 \pm 7) ~ {\rm MeV}$ & 
$F_{\pi} =(110 \pm 11) ~ {\rm MeV}$ \\
\hline
\hline
$0.005$ & $F_{\pi} = (115 \pm 6) ~ {\rm MeV}$ & 
$F_{\pi} = (111 \pm 4) ~ {\rm MeV}$ \\
\hline
\end{tabular}
\end{center}
\caption{{\it Our results for the pion decay constant $F_{\pi}$, 
evaluated in the $\eps$-regime
based on the axial-current correlation function (\ref{AA}).
These results are obtained at $\beta = 5.85$ on a $12^{3} \times 24$ lattice.
The statistics involved $100$ propagators for $D_{\rm N}$ at $a m_{q}=0.005$,
and $50$ propagators in all other cases.
$F_{\pi}$ was determined from fits to the quenched
$\chi$PT formula (\ref{AA}) in the range \ $t / a \in [11,13]$.}}
\label{epstabFpiAA}
\end{table}

\subsubsection{The pion decay constant based on the zero modes}

At last we review our results based on an alternative method to evaluate
$F_{\pi}$ in the $\epsilon$-regime \cite{WBStani}. 
This method was introduced in Ref.\ \cite{zeromodes}, 
and it involves solely the zero mode contributions to the
pseudoscalar correlation function. Here one works directly
{\em in} the chiral limit. Let us briefly summarise
the main idea of this approach.

Ref.\ \cite{zeromodes} computed
the chiral Lagrangian to the next-to-next-to-lead\-ing order
in quenched $\chi$PT, ${\cal L}_{\rm q\chi PT}^{(2)}$.
It can be written in a form that involves an auxiliary scalar field
$\Phi_{0}$, which is coupled to the quasi Nambu-Goldstone field $U$
by a new LEC denoted as $K$.
The auxiliary field also contributes
\be
{\cal L}^{(2)} [ \Phi_{0} ] =
\frac{\alpha_{0}}{2 N_{c}} \partial_{\mu} \Phi_{0} \partial_{\mu} \Phi_{0} 
+ \frac{m_{0}^{2}}{2 N_{c}} \Phi_{0}^{2}
\ee
to ${\cal L}_{\rm q\chi PT}^{(2)}$, which brings in $\alpha_{0}$ and $m_{0}$
as another two quenching specific LECs, in addition to $K$.
The field $\Phi_{0}$ supplements the quenching effects;
in the dynamical case it decouples form the field $U$.

It is ambiguous how to count these additional terms in the
quenched $\epsilon$-expansion. Ref.\ \cite{zeromodes} assumes
the action terms with the coefficients $\alpha_{0}$ and $\, K \sqrt{N_{c}} \,$ 
to be of $O(1)$, whereas the one with $m_{0}$ is in $O(\epsilon )$.
The last assumption is somewhat unusual
(for instance, it differs from the framework of Subsection
8.5.1). Nevertheless it is an acceptable possibility, which simplifies
this approach since it removes
the auxiliary mass term from the dominant order. If one further
defines the dimensionless parameter
\be  \label{defalpha}
\alpha := \alpha_{0} - \frac{4 N_{c}^{2} K F_{\pi} }{\Sigma} \ ,
\ee
then only the LECs $F_{\pi}$ and $\alpha$ occur in this order.

For $N_{f}$ valence quark flavours, this approach considers the correlation
function of the pseudoscalar density $P(x)$ (defined in Subsection 8.2), 
which is decomposed into a connected plus a disconnected part.
In a spectral decomposition of the propagators one obtains
the residuum in terms of the zero modes,
\bea
^{~ \lim}_{m_{q} \to 0} \ (m_{q}V)^{2} \langle P(x) P(0) \rangle_{\nu} &=&
N_{f} C^{(1)}_{| \nu |}(x) + N_{f}^{2} C^{(2)}_{| \nu |}(x) \nn \\
{\rm connected:~~~}
C^{(1)}_{| \nu |}(x) &=& - \langle v_{j}^{\dagger} (x) v_{k}(x) \cdot
v_{k}^{\dagger} (0) v_{j}(0) \rangle_{| \nu |} \nn \\
{\rm disconnected:~~~}
C^{(2)}_{| \nu |}(x) &=& \langle v_{j}^{\dagger} (x) v_{j}(x) \cdot
v_{k}^{\dagger} (0) v_{k}(0) \rangle_{| \nu |} \ .
\eea
The vectors $v_{j}$ denote the (exact) zero modes of the
Ginsparg-Wilson operator at $m_{q}=0$.
In the terms for $C^{(i)}_{| \nu |}$ these zero modes are summed over.

Next we consider the spatial integral
$\int d^{3}x \, P(x) P(0)$. Now the above procedure for the correlation function
leads to functions $C^{(i)}_{| \nu |} (t)$, $i=1,2$, which are given explicitly
in Ref.\ \cite{zeromodes}. In principle, these functions could be measured
and fitted to the predictions in order to determine $F_{\pi}$ and $\alpha$.
In practice, however, it is much more promising to consider just 
the leading term in the expansion at $t = T/2$,
\be  \label{Taylor}
\frac{V}{L^{2}} \frac{d}{dt} C^{(i)}_{| \nu |} (t) |_{t = T/2} =
D^{(i)}_{| \nu |} s + O(s^{3}) \ , \quad s = t - \frac{T}{2} \ , 
\quad i = 1,2 \ .
\ee
The explicit slope functions $D^{(i)}_{| \nu |}$ in a volume $V= L^{3}\times T$ 
are given in Refs.\ \cite{zeromodes,Stani,WBStani}.
(They also involve a shape coefficient, which we computed for our anisotropic
volume according to the prescription in Ref.\ \cite{HasLeu}.)

\begin{figure}[ht!]
  \centering
\includegraphics[angle=270,width=.7\linewidth]{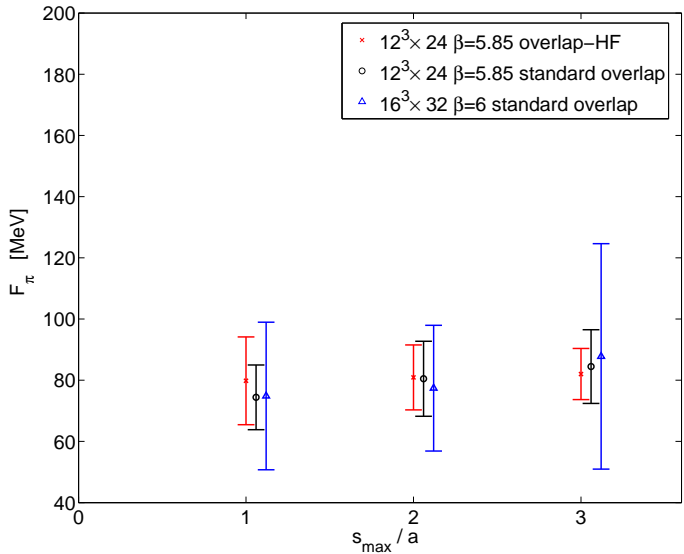}
\caption{{\it The results for $F_{\pi}$ based on a global fit
of our data to a quenched $\chi$PT prediction for the
zero mode contributions to the pseudoscalar correlations
function. 
We show the $F_{\pi}$ results of a two parameter fit (for $F_{\pi}$
and $\alpha$) over the ranges
$s \in [T/2-s_{\rm max},T/2 + s_{\rm max}]$.}}
\label{Fpizerofig}
\end{figure}

We evaluated the LECs $F_{\pi}$ and $\alpha$ from fits to the
linear term in eq.\ (\ref{Taylor}).
For each of our lattice sizes and for each type of overlap operator
we performed a global fit over the topological sectors
$\vert \nu \vert =1$ and $2$, in a fitting range $s_{\rm max} \, $.
The slopes tend to be stable over a variety of fitting ranges 
$s \in [ - s_{\rm max}, s_{\rm max} ]$,
$s_{\rm max} = a,\, 2a ,\, 3a \dots $. 
The deduced optimal values for $F_{\pi}$ 
are shown in Figures \ref{Fpizerofig}, 
and the values for $F_{\pi}$ and $\alpha$ at 
$s_{\rm max}/a = 1$ are given in Table \ref{epstabzero}. 
We see that the results for different lattice spacings
and overlap Dirac operators are in good agreement.
Considering also $\alpha (s_{\rm max})$, we found
the most stable plateau for $D_{\rm ovHF}$ \cite{WBStani}. 
\begin{table}[h!]
\begin{center}
\begin{tabular}{|c||c|c|c|}
\hline
Dirac operator & $D_{\rm ovHF}$ & $D_{\rm N}$ & $D_{\rm N}$ \\
\hline
$\beta$ & $5.85$ & $5.85$ & $6$ \\
\hline
lattice size & $12^{3}\times 24$ & $12^{3}\times 24$ & $16^{3}\times 32$ \\
\hline
\hline
$F_{\pi}$ & $(80 \pm 14)$ MeV & $(74 \pm 11)$ MeV & $(75 \pm 24)$ MeV \\
\hline
$\alpha$  & $-17 \pm 10$ & $-19 \pm 8$ & $-21 \pm 15$ \\
\hline
\end{tabular}
\end{center}
\caption{{\it Our results in the $\eps$-regime for the pion
decay constant $F_{\pi}$ --- along with the quenching specific 
LEC $\alpha$ given in eq.\ (\ref{defalpha}) ---
based on the zero mode contributions to the
pseudoscalar correlation function. We give results
for the fitting range $s_{\rm max}=a$, which
is most adequate in the light of eq.\ (\ref{Taylor}).}}
\label{epstabzero}
\end{table}

The values that we now obtain for $F_{\pi}$ is below
those of Section 8.4.3, which used a different observable and
a different $\epsilon$-counting rule for the quenched terms.
In fact, the $F_{\pi}$ results in Table \ref{epstabzero}
are close to the phenomenological value
(the latter is shifted down to $\approx 86 ~ {\rm MeV}$ if one extrapolates to
the chiral limit \cite{CoDu}).
Our values for $F_{\pi}$ and $\alpha$ obtained from the zero modes are
somewhat below the values reported in Ref.\ \cite{zeromodes} based on the 
same method. We suspect that the anisotropic
shape of our volumes, $T = 2L$, could be the main source of this 
deviation \cite{Mikkopriv}.

\section{Epilogue}

The central concept in this work are block variable RGTs (renormalisation
group transformations) applied to lattice regularised quantum field 
theories. We described this method for various types of field theoretic
models, and focused in particular on the limit obtained under iterated
RGTs. This leads to lattice formulations with fascinating properties:
in particular the symmetries and the scaling quantities of the continuum
theories can --- in principle --- be reproduced exactly on the lattice.
This amazing feature also includes exact supersymmetry and precise topological
sectors on the lattice. The corresponding perfect lattice actions were 
constructed
explicitly for the case of free particle. For interacting fields
this can in general only be achieved approximately (as an exception we
considered the Gross-Neveu model in the large $N$ limit). 
We discussed perturbatively perfect actions in various models, 
as well as issues of truncation, parameterisation and gauging.
We also demonstrated that different kinds of anomalies are represented
correctly in this way.

We summarised a variety of simulation results based on such approximately
perfect actions. They reveal a number of properties which are superior
to the standard lattice formulations, in particular an improved symmetry 
and scaling behaviour. The chiral symmetry of massless 
fermions can be rendered exact by means of the overlap formula.
This provides a lattice formulation, which performs very well in 
toy model simulations, and it is currently being applied to QCD.
%

\subsection{The status of overlap-HF applications in QCD with light quarks}

In Section 7 we reviewed our construction of overlap hypercube 
Dirac operators $D_{\rm ovHF}$, which are especially
suitable at lattice spacings of $a \simeq 0.093 ~{\rm fm}$
and $a \simeq 0.123 ~{\rm fm}$. In both cases, they display a strongly 
improved locality  compared to the standard overlap operator $D_{\rm N}$.
Hence $D_{\rm ovHF}$ defines chiral fermions on coarser lattices.

Section 8 summarised quenched simulations with $D_{\rm ovHF}$ 
and with $D_{\rm N}$ in a volume 
$V \simeq (1.48 ~ {\rm fm})^{3} \times (2.96 ~ {\rm fm})$
at $\beta = 5.85$ and at $\beta =6$.

Subsection 8.2 dealt with the $p$-regime, where we measured
the meson masses $m_{\pi}$ and $m_{\rho}$ (though only the former
was presented here, for $m_{\rho}$ we refer to Ref.\ \cite{WBStani}), 
the quark mass $m_{\rm PCAC}$ 
(based on the axial Ward identity)
and the pion decay constant $F_{\pi}$ at bare quark masses
ranging from $16.1 ~ {\rm MeV}$ to $161 ~ {\rm MeV}$. The results
for $m_{\pi}$ and $m_{\rho}$ are similar for $D_{\rm ovHF}$ 
and $D_{\rm N}$.
On the other hand, for $D_{\rm ovHF}$ the quark mass
$m_{\rm PCAC}$ is much closer to $m_{q}$ than in the
standard overlap formulation. This implies 
$Z_{A} \approx 1$, which is favourable for the
connection to perturbation theory.
Regarding $F_{\pi}$, it turned out that the data obtained in the
$p$-regime can hardly be extrapolated to the chiral limit.\\

In Subsection 8.3 we discussed a way to stabilise the topological sector
in a Monte Carlo history. We also considered
the distribution of a large number of topological charges 
defined by the fermion indices of $D_{\rm ovHF}$ or of $D_{\rm N}$.
We found histograms which approximate well a Gaussian distribution,
consistent with the conservation of parity invariance.
The resulting topological susceptibility is in good agreement
with the literature, and with the Witten-Veneziano scenario.

In Subsections 8.4 we proceeded to the $\epsilon$-regime, where we
first summarised a 3-loop calculation in the framework of the
$\epsilon$-expansion.
Numerically we determined a value for the chiral condensate
from the distribution of the lowest eigenvalues. For both, 
$D_{\rm ovHF}$ and $D_{\rm N}$ we identified $\Sigma$ close to
$(300 ~ {\rm MeV})^{3}$. 

We evaluated $F_{\pi}$ in the $\epsilon$-regime in two ways, from the 
axial-current correlation and from the zero mode contributions to the 
correlation of the pseudoscalar density. These two methods handle the
$\epsilon$-counting of the quenched terms differently, and they yield
different values for $F_{\pi}$. The axial-current method leads
to $F_{\pi} \approx 110 ~{\rm MeV}$, which is consistent
with part of the quenched results in the literature.
The zero mode method 
leads to a lower $F_{\pi}$,  
in the vicinity of the phenomenological value. The result of Ref.\ 
\cite{LMA} --- empolying yet another method, based on
the $\Delta I = 1/2$ rule, still in the $\epsilon$-regime --- 
is in between. \\ 

From the current results we conclude that the methods applied here 
do have the potential to evaluate at 
least the leading LECs from lattice simulations in the $\epsilon$-regime.
The quenched data match
the analytical predictions qualitatively (if the volume is not too small)
and --- in the setting we considered ---
they lead to results in the magnitude of the LECs in Nature.
However, the quenched results are volume dependent and in addition
ambiguous: different methods yield different values.

For values that can be confronted with 
phenomenology in detail, simulations with
dynamical quarks will be needed. In particular the $\epsilon$-regime
requires then dynamical Ginsparg-Wilson fermions. 
This regime is promising in view of the lattice size.
Also the option to extract physical information from single
topological sectors is attractive, since it is very difficult
to change the sector frequently in the course of Hybrid Monte Carlo
histories. The question is
if one is able to handle 
sufficiently small quark masses in dynamical simulations.

\subsection{Prospects for dynamical simulations with chiral fermions}

We just pointed out that
the results for light quarks can be linked successfully to $\chi$PT
(Chiral Perturbation Theory), but to some extent they are obstructed up to
now by the quenched approximation. Quenching has been necessary so far 
in QCD simulations with chiral quarks due to limitations in the 
computational resources.

We hope to overcome this limitation ---
and the systematic errors that it causes ---
in the foreseeable future, and to be able to proceed to simulations
with dynamical chiral quarks. However, in addition to powerful machines
this step also requires new algorithmic tools, which are currently under
consideration. Also in this respect overlap-HFs
open up new perspectives.

Due to the similarity with the hypercubic kernel,  
a low polynomial of this kernel can be used as a numerically 
cheap way to evaluate the fermionic force in the
Hybrid Monte Carlo (HMC, \cite{HMC}) algorithm.
This approach is conceptually guided by the algorithm that we
used for the simulation of quasi-perfect staggered fermions (see Ref.\ 
\cite{Dilg} and Subsection 6.3.2). Also for the overlap-HF formulations
of Ref.\ \cite{WBIH} it is under investigation in the two-flavour
Schwinger model. First
the extreme case of using directly $D_{\rm HF}$ in the force term was
studied in Ref.\ \cite{CJNP1}, which reported a
decreasing acceptance rate for an increasing volume.
However, it turned out that the acceptance rate can be improved
by an order of magnitude by correcting the fermionic
force term at least to a low
accuracy like $0.005$, which allows for efficient HMC simulations
with very light (degenerate) overlap-HFs \cite{dynovHF}.
Of course, the overlap operator has be to very precise in the 
Metropolis accept/reject step (we set it to $10^{-16}$).  
We verified the acceptance rate as well as algorithmic requirements, 
namely area conservation and reversibility. 

In our simulations at $\beta =5$ on a $16 \times 16$ lattice \cite{dynovHF}
(with the plaquette gauge action)
we confirmed the similarity between the kernel and the overlap 
operator by considering the spectra, and we observed 
a high level of locality for this formulation.
The characteristic decay hardly changes in the fermion mass range
$0.03 \dots 0.24$, and the locality is by far superior to the
standard overlap operator.
As an observable we evaluate the chiral condensate at light fermion masses,
based on the ratio between low lying Dirac 
eigenvalues in different topological sectors (the relevant formulae 
were taken from Ref.\ \cite{WGW}).
This represents one of the first measurements with dynamical
overlap fermions. Our results for the chiral condensate at various
masses are in excellent agreement with analytic 
predictions, which were obtained with
bosonisation and low energy approximations \cite{ccschwing}.

In QCD, simulations with dynamical overlap fermions
are still in an early stage; for recent status reports we refer
to Refs.\ \cite{Cundy}.\footnote{After the first version of this
work was written, a remarkable study appeared in Ref.\ \cite{Hide}.
It evaluates the chiral condensate from simulations with
dynamical overlap fermions in the topologically trivial sector,
applying the technique described in Subsection 8.4.2.}
Regarding the approximately chiral
fermion formulations 
explained in this work, dynamical HFs are currently investigated,
and interesting results for their phase diagram are available
already \cite{Stani06}. It provides access to a mass ratio
$m_{\pi} / m_{\rho} \leq 0.8$ at the thermal
crossover, which is not the case for Wilson fermions.
In this context we add that the truncated classically
perfect action of Ref.\ \cite{Bern}, and the ``chirally improved''
formulation of Refs.\ \cite{GatHip},
are currently applied in dynamical simulations as well \cite{Berndyn}.

Also the topology conserving gauge actions --- that we
discussed in Subsection 8.3 --- appears to be helpful in simulations
with dynamical quarks due to the suppression of small plaquette
value, as suggested in Ref.\ \cite{topogaugeproc} and tested
in Ref.\ \cite{topogaugedyn}.

The quenched studies reported here show that we have
methods at hand, which are applicable for instance for the
evaluation of Low Energy Constants in the chiral Lagrangian from
first principles. These constants play a prominent r\^{o}le in QCD
at low energy, hence their determination is a major challenge and
a sensitive test for QCD, to be addressed with dynamical Ginsparg-Wilson 
fermions.

\section{Further Fields of Research}

In my diploma thesis I investigated mathematical and mechanical
models for certain brain activities and the resulting neural
signals. More precisely, I analysed numerically and analytically
the time evolution of jerky eye motions and the required coordination 
of nerve signals, particularly in view of the non-Abelian
dynamics.

For short periods I was involved in two experiments at CERN
that observed the electroweak mixing angle (CHARM II) and neutral kaon
decays (CP-LEAR).\\

In a work with J.J.\ Giambiagi
we discussed solutions of the spherically symmetric wave equation and 
Klein-Gordon equation in an arbitrary real number of spatial and temporal 
dimensions \cite{wave}. 
Starting from a given solution, we presented various procedures 
to generate further solutions in the same or in different dimensions. The 
transition from odd to even or non-integer dimensions could be performed by 
fractional derivation or integration. 
We also discussed the 
analytic continuation to arbitrary real powers of the D'Alembert 
operator. 
Finally we worked out operators which transform a time 
into a space coordinate and vice versa, and we
commented on a possible application to black holes.\\

In a completely different project we addressed statistical mechanics 
and investigated the phase diagram of the 2d O(3) lattice model 
with a $\theta $-term \cite{merons}. The simulation made use
of the Wolff cluster algorithm \cite{wolff} and required
the construction of an improved estimator, which could only be
achieved by a new method. The clusters carry half integer
topological charge and most of the charged clusters
can be identified with merons. 
We confirmed to a high accuracy the prediction by
I.\ Affleck et al. \cite{aff}, who expected a second order phase transition
at $\theta = \pi$ due to analogies to Haldane's conjecture 
about quantum spin chains \cite{haldane}. The traditional picture
explains this phase transition by the neutralisation of vortex-antivortex 
pairs. In the framework of our ``meron cluster algorithm'',
this picture could be formulated precisely, without breaking
the O(3) symmetry.

Further algorithmic elements were added in Ref.\ \cite{Brecht}.
Recently another meron cluster simulation was performed 
successfully for the $XY$ spin chain \cite{Boyer}.\\

For a few years, I have been working in a collaboration
which studied large $N$ reduced matrix models as
candidates for a non-perturbative formulation of string theory.
In particular, we have simu\-lated variants of the so-called 
IKKT model (or IIB matrix model) \cite{IKKT}, 
which corresponds to a dimensional 
reduction of super Yang-Mills theory to zero dimensions.
This model is formulated in 10 (target) dimensions and it has a complex 
action, which can hardly be simulated. 
We studied its 4d counterpart \cite{IKKT4d}, which
has a real positive action, 
as well as the 10d version with a quenched phase \cite{IKKT10d}
(a summary is given in Ref.\ \cite{Bang}).
The space-time coordinates arise dynamically, and we found
numerically that their distribution does not
support the scenario of spontaneous Lorentz symmetry breaking, 
which could single out a subset of extended dimensions. 
(This question was also addressed in Refs.\ \cite{Bielefeld}.)
We did see, however,
a well-defined large $N$ scaling, and we observed the area law
for the Wilson loop to hold in some regime. \\

In the context of large $N$ reduced matrix models,
we are now studying field theory on spaces with a non-commutative (NC) 
geometry. This was the subject of the Ph.D.\ thesis of my 
collaborators A.\ Bigarini, F.\ Hofheinz, Y.\ Susaki and 
J.\ Volkholz. A (fuzzy) lattice version of 2d $U(1)$ gauge theory on a
NC plane can be mapped onto a twisted Eguchi-Kawai model \cite{TEK,AMNS}.
This mathematically rigorous map enables
Monte Carlo simulations.
We observed numerically the large $N$ ``double 
scaling'' of 1- and 2-point functions of the Wilson loops.
This corresponds to a simultaneous continuum and infinite volume
limit at fixed non-commutativity. Thus we obtained the 
first fully non-perturbative evidence in favour of the
renormalisability of a NC field theory \cite{NCQED2,Frank}. 
The area law for the Wilson loop holds at small area as in the commutative 
$U(N \to \infty )$ model \cite{GrWi}, but at large area we 
observed instead a linearly rising complex phase.
This is consistent with an Aharonov-Bohm type picture
known from the Seiberg-Witten map to string theory, and from
2d models for the quantum Hall effect. The difference is, however, 
that we discovered this feature as a dynamical effect.

For pure $U(N)$ gauge theories in a commutative plane, the expectation
values of Wilson loops only depend on their (oriented) area, i.e.\ they are
invariant under area preserving diffeomorphisms. This property
enables analytic solutions. Hence the question arises if the same
holds for NC 2d gauge theories. First perturbative calculations
seemed to confirm this \cite{NCshape1}, but proceeding to higher
accuracy a reduction of the symmetry group to $SL(2,R)$ was observed
\cite{NCshape2}. However, detailed numerical measurements show that 
non-perturbatively this symmetry is broken completely, including the 
$SL(2,R)$ subgroup \cite{NCshapenp}. Hence analytic
solutions are unlikely in that case, but the theories may display 
a rich structure.

Next our numerical pilot studies of NC field theory addressed
the $\lambda \phi^{4}$ model in two and three dimensions 
\cite{NCphi4,Frank,Antonio}.
The results reveal an interesting phase diagram, and in particular
the occurrence of a new ``striped phase'', as it was first
conjectured in by Gubser and Sondhi \cite{GuSo}.
Its existence shows that the famous UV/IR mixing of divergences
in NC field theory is not an artifact of perturbation theory, but
it belongs to the very nature of the system. In three dimensions
(with a commutative Euclidean time) we could again safely 
extrapolate our results to the double scaling limit. 
In particular the dispersion relation stabilises, and it
reveals an IR divergence in part of the phase diagram.
Although this striped phase involves the spontaneous breaking 
of Poincar\'{e} symmetry, it could persist in the double scaling limit 
of this model in $d=2$ \cite{aristocats,Antonio}, where we also studied 
a NC linear $\sigma$-model with 3 flavours \cite{Antonio}.

Our next focus in this framework was
NC QED in $d=4$ \cite{NCQED4} (with two NC directions).
We measured the photon dispersion relation, which is
distorted compared to the commutative world. 
A non-perturbative result for this 
distortion is of phenomenological interest: it was suggested
to be used to extract bounds
on the non-commutativity parameter from the confrontation with
experimental data, obtained in particular from cosmic gamma rays 
\cite{cosmo}. For instance, the Gamma-ray Large Area Space Telescope 
(GLAST) project
(see e.g.\ Ref.\ \cite{GLAST}) was scheduled to be launched in September 
2007  and to monitor gamma rays in the range from 20 MeV to 1 TeV. 
A relative delay for photons
emitted simultaneously by Gamma Ray Bursts or blazar flares could hint
at a NC geometry --- or the absence of a detectable delay could
impose a narrow bound on the possible extent of non-commutativity.

On the other hand, perturbation theory showed that
the NC photon is asymptotically free \cite{asyfreeU1}, but it
suffers from IR instability \cite{LLT}, unless one proceeds
to its supersymmetric version. If one is not willing to
put SUSY as another hypothetical concept on top, this property seems to
be a true disaster for the scenario of a NC world.
However, we observed non-perturbatively at moderate coupling 
a phase with broken translation symmetry, which
appears to be IR stable \cite{NCQED4}. 
Several observables follow a scaling law, which leads to this broken phase 
in the simultaneous UV and IR limit at constant non-commutativity.
Thus the double scaling limit leads to a phase which was not visible
in perturbation theory, and which could accommodate the NC photon.\\

We applied similar methods to study the ``fuzzy sphere'' formulation 
\cite{Madore} of the 3d $\lambda \phi^{4}$ theory \cite{JHEPfuzzy}. 
Further details about that study can be found in the Ph.D.\ thesis
by J.\ Medina, quoted in Ref.\ \cite{fuzzysimu} (that reference
also collects further simulations results on the fuzzy sphere). 
The fuzzy sphere regularises quantum field theories with a finite
set of degrees of freedom. Hence the hope is that it might be useful
for Monte Carlo simulations, as an alternative to the lattice.
A potential virtue is that it preserves a number of continuum
symmetries, which are broken by the standard lattice
formulations. In the $\lambda \phi^{4}$ model
this formulation turned out to be tedious to simulate, 
but it may be in business for issues where the lattice 
faces severe difficulties, like SUSY 
\cite{SUSY,Alessandra,SUSYlatexact}. Therefore
we now proceeded to the simulation of a fuzzy sphere 
formulation of a 2d Wess-Zumino type model \cite{SUSYform}.
(Related approaches were already presented in 
Refs.\ \cite{FuzzySUSY,FuzzySUSYnum}.)
This discretisation is indeed numerically tractable, which
enabled us to explore the phase diagram \cite{prep}
with a moderate computational effort.
Again the ordered regime is divided into a phases of
a uniform and non-uniform order, as the coefficients of the
polarisation tensors show.

\vspace*{1cm}

\newpage


\noindent
{\underline{\bf Acknowledgements :}}\\

{\small First I would like to thank Prof.\ K.\ Hepp at ETH Z\"{u}rich
and Prof.\ H.\ Leutwyler and the Universit\"{a}t Bern, who
supervised my diploma work and my Ph.D.\ thesis, respectively,
as well as Prof.\ M.\ M\"{u}ller-Preu\ss ker, who is the head of
the group at the Humboldt-Universit\"{a}t zu Berlin, where I wrote
this ``Habilitationsschrift''.  In this group I enjoyed the contact
with him, with Prof.\ D.\ Ebert, Dr.\ E.M.\ Ilgenfritz and
other colleagues. I am grateful for the system administration by
Drs.\ B.\ Bunk and A.\ Sternbeck, and for the assistance by 
our excellent secretary Mrs.\ S.\ Richter. I have also benefited
from inspiring questions and remarks by the participants of my lectures at
Humboldt-Universit\"{a}t and at the Universit\"{a}t Potsdam.

At this point, I would also like to acknowledge the leaders of
the further groups where I have been working, namely Profs.\  
J.A.\ Helay\"{e}l-Neto, P.\ Hoyer, J.\ Negele, G. Schierholz
and K.\ Schilling.\\

In the course of my research I collaborated with about 50 persons
from 16 countries and 5 continents, which illustrates the cosmopolitan
nature of research in Theoretical Physics. It is a pleasure
to thank them all. In particular, the following
collaborators had a long-lasting and significant influence on the 
work that we reviewed here:

\noindent
{\em David Adams, Richard Brower,
Shailesh Chandrasekharan, Hermann Dilger, Erich Focht,
Ivan Hip, Jun Nishimura, Kostas Orginos, Mauro Papinutto,
Luigi Scorzato, Stanislav Shcheredin, Jan Volkholz and 
Uwe-Jens Wiese.} \\

I would also like to thank explicitly the following collaborators in the
context of this and further fields of research (addressed in Section 10):\\
J.\ Ambj\o rn, K.N.\ Anagnostopuolos, A.\ Bigarini, T.\ Chiarappa, 
N.\ Eicker, J.J.\ Giambiagi, F.\ Hofheinz, K.\ Jansen, Th.\ Lippert, 
J.\ Medina, K.-I.\ Nagai, D.\ O'Connor, M.\ Panero, A.\ Shindler,
Y. Susaki, C.\ Urbach and U.\ Wenger.

In addition to my direct collaborators I am indebted to a many persons 
for helpful communications, for instance A.P.\ Balachandran,
R.\ Burkhalter, S.\ Catterall,
P.\ Damgaard, S.\ D\"{u}rr, M.\ Hasenbusch, P.\ Hasenfratz, M.\ Laine, 
E.\ Laermann, C.B.\ Lang, E.\ Lopez,
M.\ L\"{u}scher, X.\ Martin, F.\ Niedermayer, M.\ Peardon,
R.\ Sommer, R.J.\ Szabo, J.\ Verbaarschot, P.\ Weisz and T.\ Wettig.\\
The QCD simulations that we discussed in Sections 7 and 8 were mostly
performed on the IBM p690 clusters of the ``Norddeutscher Verbund 
f\"ur Hoch- und H\"ochstleistungsrechnen'' (HLRN). 
In this context I would like to thank V.\ Linke and H.\ St\"{u}ben
for their support.\\
I further acknowledge comments received on this work
by E.\ Laermann, M.\ M\"{u}ller-Preu\ss ker and S.\ Shcheredin.\\

Finally I thank my wife Marlene and my sons Diego and 
Philippe for their moral support and patience.
}

\end{document}